\documentstyle[aps,eqsecnum,twocolumn,floats,psfig]{revtex}
\pssilent

\begin{document}
\draft
\title{Random-matrix theory of quantum transport}
\author{C. W. J. Beenakker}
\address{Instituut-Lorentz, University of Leiden\\
2300 RA Leiden, The Netherlands\bigskip\\
\parbox{14cm}{\rm
This is a review of the statistical properties of the scattering matrix of a
mesoscopic system. Two geometries are contrasted: A quantum dot and a
disordered wire. The quantum dot is a confined region with a chaotic classical
dynamics, which is coupled to two electron reservoirs via point contacts. The
disordered wire also connects to two reservoirs, either directly, or via a
point contact or tunnel barrier. One of the two reservoirs may be in the
superconducting state, in which case conduction involves Andreev reflection at
the interface with the superconductor. In the case of the quantum dot, the
distribution of the scattering matrix is Dyson's circular ensemble for
ballistic point contacts, or the Poisson kernel for point contacts containing a
tunnel barrier. In the case of the disordered wire, the distribution of the
scattering matrix is obtained from the Dorokhov-Mello-Pereyra-Kumar equation,
which is a one-dimensional scaling equation. The equivalence is discussed with
the non-linear $\sigma$ model, which is a supersymmetric field theory of
localization. The distribution of scattering matrices is applied to a variety
of physical phenomena, including universal conductance fluctuations, weak
localization, Coulomb blockade, sub-Poissonian shot noise, reflectionless
tunneling into a superconductor, and giant conductance oscillations in a
Josephson junction.
{\em To be published in Rev.\ Mod.\ Phys.\ (1997).}
}}
\maketitle
\tableofcontents

\section{Introduction}
\label{intro}

\subsection{Preface}
\label{preface}

Random-matrix theory deals with the statistical properties of large matrices
with randomly distributed elements. The probability distribution of the
matrices is taken as input, from which the correlation functions of eigenvalues
and eigenvectors are derived as output. From the correlation functions one then
computes the physical properties of the system. Random-matrix theory was
developed into a powerful tool of mathematical physics in the 1960's, notably
by Wigner, Dyson, Mehta, and Gaudin. (Their work is described in detail in a
monograph by Mehta, 1991.) The original motivation for this research was to
understand the statistics (in particular the distribution of spacings) of
energy levels of heavy nuclei, measured in nuclear reactions (Wigner, 1957).
(Many of the early papers have been collected in a book by Porter, 1965.) Later
the same techniques were applied to the level statistics of small metal
particles, in order to describe the microwave absorption by granular metals
(Gor'kov and Eliashberg, 1965). Much of the work on level statistics in nuclear
and solid-state physics has been reviewed by Brody {\em et al.} (1981).

In recent years there has been a revival of interest in random-matrix theory,
mainly because of two developments. The first was the discovery that the
Wigner-Dyson ensemble applies generically to chaotic systems (Bohigas,
Giannoni, and Schmit, 1984; Berry, 1985). (For reviews of the random-matrix
theory of quantum chaos, see Bohigas, 1990; Gutzwiller, 1990; Haake, 1992.) The
second was the discovery of a relation between universal properties of large
random matrices and universal conductance fluctuations in disordered conductors
(Imry, 1986a; Altshuler and Shklovski\u{\i}, 1986). This led to the development
of a random-matrix theory of quantum transport. An influential review of the
early work was provided by Stone, Mello, Muttalib, and Pichard (1991). The
field has matured rapidly since then, and the need was felt for an up-to-date
review, in particular for physicists from outside the field. The present
article was written with this need in mind.

The random-matrix theory of quantum transport is concerned with mesoscopic
systems, at the borderline between the microscopic and the macroscopic world.
On the one hand, they are sufficiently small that electrons maintain their
quantum mechanical phase coherence, so that a classical description of the
transport properties is inadequate. On the other hand, they are sufficiently
large that a statistical description is meaningful. Quantum interference leads
to a variety of new phenomena. (For reviews, see Altshuler, Lee, and Webb,
1991; Beenakker and Van Houten, 1991; Datta, 1995; Imry, 1996.) Some of the
phenomena are ``universal'', in the sense that they do not depend on the sample
size or the degree of disorder --- at least within certain limits.
Random-matrix theory relates the universality of transport properties to the
universality of correlation functions of transmission eigenvalues. A
particularly attractive feature of this approach is its generality: Since it
addresses the entire probability distribution of the transmission matrix, it
applies to a whole class of transport properties --- not just to the
conductance. By including Andreev reflection one can treat hybrid structures
containing normal metals and superconductors. Furthermore, since the approach
is non-perturbative, it provides a unified description of both the metallic and
the localized regimes.

There exists at this moment a complete description of the statistics of the
transmission matrix for two types of geometries: The first is a confined
geometry, the second a wire geometry. The confined geometry consists of a metal
grain through which a current is passed via two point contacts. Such a system
is sometimes called a ``quantum dot'', to emphasize the quantum mechanical
phase coherence of the electrons. The wire geometry should have an aspect ratio
length/width $\gg 1$. These two geometries are considered separately in Secs.\
\ref{dots} and \ref{wires}, as far as normal metals are concerned. The new
effects which appear due to superconductivity are the subject of Sec.\
\ref{junctions}. [There is some overlap between Sec.\ \ref{junctions} and an
earlier review by the author (Beenakker, 1995).] In Sec.\ \ref{conclusion} we
identify directions for future research and discuss some outstanding problems,
in particular the extension of the random-matrix approach to thin-film and bulk
geometries (having length $\lesssim$ width). Section \ref{intro} is devoted to
an introduction, containing background material and an overview of things to
come.

\subsection{Statistical theory of energy levels}
\label{statisticalE}

The random-matrix theory of quantum transport is a statistical theory of the
transmission eigenvalues of an open system. In contrast, the random-matrix
theory established by Wigner and Dyson addresses the statistics of energy
levels of a closed system. In this subsection we briefly consider the
Wigner-Dyson ensemble of random Hamiltonians, and discuss its fundamental
ingredient: The hypothesis of geometrical correlations. We also introduce two
topics which we will need later on: Transitions between ensembles of different
symmetry and Brownian motion of energy levels.

\subsubsection{Wigner-Dyson ensemble}
\label{WDensemble}

Wigner and Dyson studied an ensemble of $N\times N$ Hermitian matrices ${\cal
H}$, with probability distribution of the form
\begin{equation}
P({\cal H})=c\exp[-\beta\,{\rm Tr}\,V({\cal H})]\label{WDPH}
\end{equation}
($c$ is a normalization constant). If $V({\cal H})\propto {\cal H}^{2}$, the
ensemble is called Gaussian. Wigner (1957, 1967) concentrated on the Gaussian
ensemble because it has independently distributed matrix elements (since ${\rm
Tr}\,{\cal H}^{2}={\rm Tr}\,{\cal H}{\cal H}^{\dagger}=\sum_{ij}|{\cal
H}_{ij}|^{2}$), and this simplifies some of the calculations. To make contact
with the Hamiltonian of a physical system, the limit $N\rightarrow\infty$ is
taken. It turns out that spectral correlations become largely independent of
$V$ in this limit, provided one stays away from the edge of the spectrum. This
is the celebrated {\em universality\/} of spectral correlations, about which we
will say more in Sec.\ \ref{correlationfunctions}.

The symmetry index $\beta$ counts the number of degrees of freedom in the
matrix elements. These can be real, complex, or real quaternion\footnote{A
quaternion $q$ is a $2\times 2$ matrix which is a linear combination of the
unit matrix and the three Pauli spin matrices: $q=a\openone+{\rm
i}b\sigma_{x}+{\rm i}c\sigma_{y}+{\rm i}d\sigma_{z}$. The quaternion is called
real if the coefficients $a$, $b$, $c$, and $d$ are real numbers.}
numbers, corresponding to $\beta=1,2$, or 4, respectively. Since the
transformation ${\cal H}\rightarrow U{\cal H}U^{-1}$, with $U$ an orthogonal
($\beta=1$), unitary ($\beta=2$), or symplectic\footnote{
A symplectic matrix is a unitary matrix with real quaternion elements.}
($\beta=4$) matrix leaves $P({\cal H})$ invariant, the ensemble is called
orthogonal, unitary, or symplectic. Physically, $\beta=2$ applies to the case
that time-reversal symmetry is broken, by a magnetic field or by magnetic
impurities. In the presence of time-reversal symmetry, one has $\beta=1$ if the
electron spin is conserved, and $\beta=4$ if spin-rotation symmetry is broken
(by strong spin-orbit scattering). This classification, due to Dyson (1962d),
is summarized in Table \ref{table_threefoldH}.

\begin{table}[tb]
\caption[]{
Summary of Dyson's threefold way. The Hermitian matrix ${\cal H}$ (and
its matrix of eigenvectors $U$) are classified by an index
$\beta\in\{1,2,4\}$, depending on the presence or absence of
time-reversal (TRS) and spin-rotation (SRS) symmetry.
}\label{table_threefoldH}
\begin{tabular}{ccccc}
$\beta$ & TRS & SRS & ${\cal H}_{nm}$ & $U$\\
\hline\\
1 & Yes & Yes & real & orthogonal \\
2 & No & irrelevant & complex & unitary \\
4 & Yes & No & real quaternion & symplectic
\end{tabular}
\end{table}

We would like to deduce from $P({\cal H})$ what the distribution is of the
eigenvalues and eigenvectors of ${\cal H}$. Let $\{E_{n}\}$ denote the set of
eigenvalues and $U$ the matrix of eigenvectors, so that\footnote{If $\beta=4$,
the eigenvalue--eigenvector decomposition is ${\cal H}=U{\rm
diag}\,(E_{1}\openone,E_{2}\openone,\ldots E_{N}\openone)U^{\dagger}$, so that
each of the $N$ distinct eigenvalues is twofold degenerate (Kramers'
degeneracy).}
${\cal H}=U{\rm diag}\,(E_{1},E_{2},\ldots E_{N})U^{\dagger}$. Since ${\rm
Tr}\,V({\cal H})=\sum_{n}V(E_{n})$ depends only on the eigenvalues, the
distribution (\ref{WDPH}) is independent of the eigenvectors. This means that
$U$ is uniformly distributed in the unitary group (for $\beta=2$), and in the
orthogonal or symplectic subgroups (for $\beta=1$ or 4). To find the
distribution $P(\{E_{n}\})$ of the eigenvalues we need to multiply $P({\cal
H})$ with the Jacobian $J$ which relates an infinitesimal volume element
$d\mu({\cal H})$ in the space of Hermitian matrices to the corresponding volume
elements $d\mu(U)$, $dE_{n}$ of eigenvectors and eigenvalues,
\begin{equation}
d\mu({\cal H})=Jd\mu(U)\prod_{i}dE_{i}.\label{dmuHdE}
\end{equation}
The Jacobian depends only on the eigenvalues (Porter, 1965),
\begin{equation}
J(\{E_{n}\})=\prod_{i<j}|E_{i}-E_{j}|^{\beta}.\label{jacobianE}
\end{equation}
The resulting eigenvalue distribution takes the form
\begin{equation}
P(\{E_{n}\})=c\prod_{i<j}|E_{i}-E_{j}|^{\beta}\prod_{k}\exp[-\beta
V(E_{k})].\label{WDPE}
\end{equation}

\begin{figure}[tb]
\centerline{
\psfig{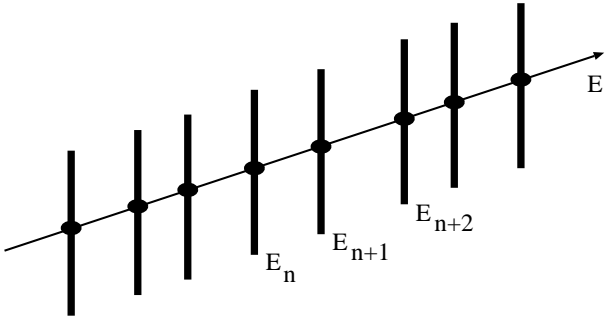}
}%
\medskip
\caption[]{
Schematic illustration of the Coulomb gas. The eigenvalues are represented by
classical particles at positions $E_{1},E_{2},\ldots E_{N}$ along a line. The
logarithmic eigenvalue repulsion is represented by the Coulomb interaction
between identical parallel line charges attached to the particles.
}\label{fig_coulomb}
\end{figure}

The distribution (\ref{WDPE}) has the form of a Gibbs distribution in
statistical mechanics,
\begin{mathletters}
\label{PGibbsab}
\begin{eqnarray}
P(\{E_{n}\})&=&c\exp\Bigl[-\beta\Bigl(\sum_{i<j}u(E_{i},E_{j})+
\sum_{i}V(E_{i})\Bigr)\Bigr],\nonumber\\
\label{PGibbsa}\\
u(E,E')&=&-\ln|E-E'|.\label{PGibbsb}
\end{eqnarray}
\end{mathletters}%
The symmetry index $\beta$ plays the role of inverse temperature. One can
imagine that the eigenvalues are classical particles on a line, at the points
$E_{1},E_{2},\ldots E_{N}$. They repel each other with a logarithmic pair
potential $u$ and are prevented from escaping to infinity by a potential $V$.
(For the Gaussian ensemble, $V$ is a parabolic potential well.) This system is
called a ``Coulomb gas'', because the logarithmic repulsion is the Coulomb
interaction between two identical parallel line charges (see Fig.\
\ref{fig_coulomb}). The idea of representing the eigenvalue repulsion by a
fictitious force is due to Wigner (1957) and Dyson (1962b). It greatly helps
our intuition.

\subsubsection{Geometrical correlations}
\label{geometrical}

The fundamental hypothesis\footnote{
This viewpoint of what is fundamental in the Wigner-Dyson theory differs from
the conventional viewpoint (Porter, 1965) that the two basic assumptions are:
(1) statistical independence of the matrix elements; (2) invariance of the
ensemble with respect to orthogonal, unitary, or symplectic transformations of
${\cal H}$. The assumptions of independence and invariance imply an unnecessary
restriction to the Gaussian ensemble.}
of the Wigner-Dyson ensemble is that {\em spectral correlations are
geometrical}. Geometrical means that they are due to the Jacobian
(\ref{jacobianE}), which relates volume elements in matrix and eigenvalue
space. Microscopic details of the system enter only via the potential $V$,
which does not by itself create any correlations between the eigenvalues. If
there were some other source of correlations, then the interaction $u$ between
the eigenvalues would deviate from the logarithmic repulsion (\ref{PGibbsb}).
The hypothesis of geometrical correlations is appealing because of its
simplicity. Is it correct? In this review we will address that question for the
transmission eigenvalues of an open system, where the answer was not known
until recently. It is instructive to contrast this with what is known about the
energy levels of a closed system.

Gor'kov and Eliashberg (1965) used the Wigner-Dyson ensemble to study the
electronic properties of small metal grains. Theoretical justification came
with the supersymmetric field theory of Efetov (1982, 1983). Assuming diffusive
motion of the electrons inside the grain, he obtained the same correlation
function of the energy level density as in the Wigner-Dyson ensemble.
Subsequently, Altshuler and Shklovski\u{\i} (1986) showed that, for energy
separations $|E-E'|$ greater than the Thouless energy $E_{\rm c}$, the
correlation function deviates from random-matrix theory. The characteristic
energy scale $E_{\rm c}=\hbar D/L^{2}$ is inversely proportional to the time it
takes for an electron to diffuse, with diffusion coefficient $D$, across a
particle of size $L$. It represents the finite width of the energy levels of an
open system. The results of the diagrammatic perturbation theory of Altshuler
and Shklovski\u{\i} were rederived by Argaman, Imry, and Smilansky (1993),
using a more intuitive semiclassical method. It follows from these microscopic
theories that the repulsion between the energy levels has the logarithmic form
(\ref{PGibbsb}) of the Wigner-Dyson ensemble for $|E-E'|\ll E_{\rm c}$. For
$|E-E'|\gg E_{\rm c}$ the interaction potential decays as a power law, and
actually becomes weakly attractive in three dimensions (Jalabert, Pichard, and
Beenakker, 1993).

There is surprisingly little direct experimental evidence for Wigner-Dyson
statistics in a metal grain. Sivan {\em et al.} (1994) measured the level
spacing in a small confined region in a semiconductor (a ``quantum dot'').
Their results where consistent with Wigner-Dyson statistics for the low-lying
excitations. Because of electron-electron interactions, the single-particle
excitation spectrum is broadened and merges into a continuum for energies
further than $E_{\rm c}$ from the Fermi level (Sivan, Imry, and Aronov, 1994;
Altshuler, Gefen, Kamenev, and Levitov, 1996).

The Wigner-Dyson ensemble of random Hamiltonians applies not just to an
ensemble of disordered metal grains, but to any quantum mechanical system which
is sufficiently complex. A necessary requirement is that there are no other
constants of the motion than the energy, to avoid crossings of energy levels.
In classical mechanics, such a system is called non-integrable or
chaotic.\footnote{
When speaking of ``chaotic'' systems, we intend that the classical motion is
non-integrable in the entire phase space (no stable periodic orbits). This is
known as ``hard'' chaos or ``global'' chaos. Each trajectory then uniformly
explores the entire phase space, on a time scale set by the ``ergodic'' time
$\tau_{\rm ergodic}$. Neighboring trajectories diverge exponentially in time
$\propto\exp(-t/\tau_{\rm ergodic})$. It is easy to realize hard chaos in
disordered systems, but not in ballistic systems. Generically, the phase space
will contain both regions of chaotic and integrable motion (``soft'' chaos).
Hard chaos has been demonstrated for special geometries in two-dimensional
ballistic systems known as ``billiards''.}
Impurity scattering is one way of making the system chaotic, but not the only
one. Scattering by the boundaries is often sufficient to destroy all constants
of the motion (unless the boundaries have some spatial symmetry). The notion of
statistics and averaging is different if the chaos is due to impurity
scattering or to boundary scattering. An ensemble of disordered metal grains
can be formed by changing the microscopic configuration of the impurities.
Alternatively, one could consider a single grain and replace the ensemble
average by a spectral average, {\em i.e.} by an average over the energy levels.
Theory is easier for ensemble averages, wheras experimentally a spectral
average is more accessible. The assumption of ergodicity is the assumption that
ensemble and spectral averages are equivalent.

Wigner-Dyson statistics of the energy levels has been demonstrated numerically
for a variety of non-integrable systems without disorder, such as a particle
moving on a billiard table (Bohigas, Giannoni, and Schmit, 1984), hydrogen in a
magnetic field (Freidrich and Wintgen, 1989), and models of strongly
interacting electrons (Poilblanc {\em et al.}, 1993). An early analytical
calculation, using periodic-orbit theory, was provided by Berry (1985). A
complete theoretical justification, such as Efetov's theory for a disordered
grain, was hampered for a long time by the lack of a natural ensemble in the
absence of disorder. This obstacle was finally overcome by Andreev, Agam,
Simons, and Altshuler (1996). Using a supersymmetric field theory for ballistic
motion (Muzykantski\u{\i} and Khmel'nitski\u{\i}, 1995) they could show that
spectral averages in a chaotic billiard agree with Wigner-Dyson statistics.

\subsubsection{Transition between ensembles}
\label{transitions}

We have talked about time-reversal symmetry as being broken or not. In reality,
a weak magnetic field does not break time-reversal symmetry completely. There
is a smooth transition from the orthogonal or symplectic ensembles to the
unitary ensemble. We discuss the transition from the Gaussian orthogonal
ensemble (GOE) to the Gaussian unitary ensemble (GUE), following Pandey and
Mehta (1983; Mehta and Pandey, 1983; Mehta, 1991).

The complex Hermitian $M\times M$ matrix
\begin{equation}
{\cal H}={\cal H}_{0}+{\rm i}\alpha{\cal A}\label{PMH}
\end{equation}
is decomposed into a real symmetric matrix ${\cal H}_{0}$ and a real
antisymmetric matrix ${\cal A}$ with imaginary weight ${\rm i}\alpha$. (Here we
denote the matrix dimension by $M$ instead of $N$ to avoid a confusion of
notation later on in this review.) The two matrices ${\cal H}_{0}$ and ${\cal
A}$ are independently distributed with the same Gaussian distribution, so that
the distribution of ${\cal H}$ is
\begin{equation}
P({\cal H})\propto\exp\left(-\sum_{i,j}\left[\frac{({\rm Re}\,{\cal
H}_{ij})^{2}}{4v^{2}}+\frac{({\rm Im}\,{\cal
H}_{ij})^{2}}{4v^{2}\alpha^{2}}\right]\right).\label{PMHdist}
\end{equation}
The variance $v^{2}$ determines the mean level spacing $\delta=\pi v/\sqrt{M}$
at the center of the spectrum for $M\gg 1$ and $\alpha\ll 1$. [To have the same
mean level spacing for all $\alpha$, one should replace $v^{2}$ by
$v^{2}(1+\alpha^{2})^{-1}$.] The distribution of ${\cal H}$ interpolates
between the GOE for $\alpha=0$ and the GUE for $\alpha=1$. The transition is
effectively complete for $\alpha\ll 1$. Indeed, the spectral correlations on
the energy scale $\delta$ are those of the GUE when the effective strength
$v\alpha$ of the term in Eq.\ (\ref{PMH}) which breaks time-reversal symmetry
exceeds $\delta$, hence when $\alpha\gtrsim 1/\sqrt M$.

To relate the parameter $\alpha$ to the magnetic field $B$, we consider the
shift $\delta E_{i}$ of the energy levels for $\alpha\ll 1$. On the one hand,
from the Hamiltonian (\ref{PMH}) one obtains, to leading order in $\alpha$,
\begin{equation}
\delta E_{i}=\alpha^{2}\sum_{j\neq i}\frac{{\cal
A}_{ij}^{2}}{E_{i}-E_{j}}.\label{deltaEi}
\end{equation}
In order of magnitude, $|\delta E_{i}|\simeq\alpha^{2}v^{2}/\delta\simeq
M\alpha^{2}\delta$. On the other hand, the typical curvature of the energy
levels around $B=0$ is given by the Thouless energy: $|\delta E_{i}|\simeq
E_{\rm c}(e\Phi/h)^{2}$, where $\Phi$ is the magnetic flux through the system.
Taken together, these two estimates imply
\begin{equation}
M\alpha^{2}\simeq\left(\frac{e\Phi}{h}\right)^{2}\frac{E_{\rm
c}}{\delta}.\label{alphaestimate}
\end{equation}
The GOE--GUE transition is completed on the energy scale $E$ if $|\delta
E_{i}|\gtrsim E$, hence if $\Phi\gtrsim(h/e)\sqrt{E/E_{\rm c}}$. Since
$\delta\ll E_{\rm c}$ in a metal, it requires much less than a flux quantum to
break time-reversal symmetry on the scale of the level spacing.

Microscopic justification for the probability distribution (\ref{PMHdist}) has
been provided by Dupuis and Montambaux (1991) (for a disordered ring) and by
Bohigas {\em et al.} (1995) (for a chaotic billiard). The precise relation
between $\alpha$ and $B$ depends on the geometry of the system, and on whether
it is disordered or ballistic. For a {\em disordered\/} two-dimensional disk or
three-dimensional sphere (radius $R$ much greater than mean free path $l$) the
relation between $\alpha$ and $\Phi=\pi R^{2}B$ is (Frahm and Pichard, 1995a):
\begin{equation}
M\alpha^{2}=\left(\frac{e\Phi}{h}\right)^{2}\frac{\hbar v_{\rm
F}l}{R^{2}\delta}\times\left\{
\begin{array}{cc}
\pi/4&{\rm disk},\\
2\pi/15&{\rm sphere}.
\end{array}
\right.\label{alphadisordered}
\end{equation}
Here $v_{\rm F}$ is the Fermi velocity. For a {\em ballistic\/} disk or sphere
($R\ll l$), which is chaotic because of diffuse boundary scattering, the
relation is instead
\begin{equation}
M\alpha^{2}=\left(\frac{e\Phi}{h}\right)^{2}\frac{\hbar v_{\rm
F}}{R\delta}\times\left\{
\begin{array}{cc}
4/3&{\rm disk},\\
8\pi/45&{\rm sphere}.
\end{array}
\right.\label{alphaballistic}
\end{equation}
For a ballistic two-dimensional billiard (area $A$) with a chaotic shape,
Bohigas {\em et al.} (1995) find $M\alpha^{2}=c(e\Phi/h)^{2}\hbar v_{\rm
F}/\delta\sqrt{A}$, with $c$ a numerical coefficient depending only on the
shape of the billiard. In each case, $M\alpha^{2}\propto E_{\rm c}$ in
accordance with Eq.\ (\ref{alphaestimate}), the Thouless energy being given by
$E_{\rm c}\simeq\hbar v_{\rm F}R^{-2}\,{\rm min}(l,R)$.

\subsubsection{Brownian motion}
\label{BrownianE}

\begin{figure}[tb]
\centerline{
\psfig{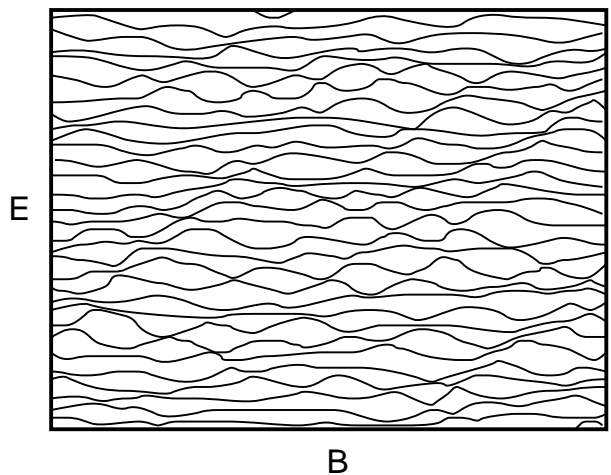}
}%
\medskip
\caption[]{
Illustration of the magnetic-field dependence of energy levels in a chaotic
system (magnetic field $B$ and energy $E$ in arbitrary units). This plot is
based on a calculation of the spectrum of the hydrogen atom in a strong
magnetic field by Goldberg {\em et al.} (1991).
}\label{fig_brownian}
\end{figure}

In the previous subsection we considered the magnetic-field dependence of the
energy levels around $B=0$, to investigate the transition from the orthogonal
to the unitary ensemble. Once the transition is completed, the level
distribution becomes $B$-independent. Individual energy levels still fluctuate
as a function of $B$ in some random way (see Fig.\ \ref{fig_brownian}). These
spectral fluctuations are a realization of the Brownian-motion process
introduced by Dyson (1962c, 1972), as a dynamical model for the Coulomb gas. A
review of this topic has been written by Altshuler and Simons (1995). Since it
is not directly related to transport, we restrict ourselves here to the basics.

Following Lenz and Haake (1990; Haake, 1992), we consider the Hamiltonian
\begin{equation}
{\cal H}={\rm e}^{-\tau}{\cal H}_{0}+\left(1-{\rm e}^{-2\tau}\right)^{1/2}{\cal
H}_{\rm GUE},\label{Htau}
\end{equation}
which interpolates between the $M\times M$ complex Hermitian matrices ${\cal
H}_{0}$ and ${\cal H}_{\rm GUE}$ as the parameter $\tau$ increases from 0 to
$\infty$. The matrix ${\cal H}_{0}$ is a fixed matrix, while ${\cal H}_{\rm
GUE}$ varies randomly over the GUE. The resulting distribution of ${\cal H}$ is
\begin{mathletters}
\label{PH0HGUE}
\begin{eqnarray}
&&P({\cal H},\tau)=\frac{1}{(1-{\rm e}^{-2\tau})^{M^{2}/2}}\,P_{\rm
GUE}\left(\frac{{\cal H}-{\rm e}^{-\tau}{\cal H}_{0}}{(1-{\rm
e}^{-2\tau})^{1/2}}\right),\nonumber\\
\label{PH0HGUEa}\\
&&P_{\rm GUE}({\cal H})\propto\exp(-c\,{\rm Tr}\,{\cal H}^{2}).\label{PH0HGUEb}
\end{eqnarray}
\end{mathletters}%
The coefficients of ${\cal H}_{0}$ and ${\cal H}_{\rm GUE}$ in Eq.\
(\ref{Htau}) are chosen such that the mean level spacing
$\delta=\pi(2Mc)^{-1/2}$ of ${\cal H}$ is $\tau$-independent.

The distribution (\ref{PH0HGUE}) satisfies the Fokker-Planck equation
\begin{equation}
c\frac{\partial}{\partial\tau}P=\sum_{\mu}\frac{\partial}{\partial{\cal
H}_{\mu}}\left(c{\cal H}_{\mu}+D_{\mu}\frac{\partial}{\partial{\cal
H}_{\mu}}\right)P\label{PHtauFP}
\end{equation}
in the $M^{2}$ independent variables $\{{\cal H}_{\mu}\}=\{{\cal H}_{ii}$,
${\rm Re}\,{\cal H}_{ij}$, ${\rm Im}\,{\cal H}_{ij}$, $1\leq i<j\leq M\}$. The
diffusion coefficient $D_{\mu}$ equals $1/2$ for the diagonal elements ${\cal
H}_{ii}$ and $1/4$ for the off-diagonal elements ${\rm Re}\,{\cal H}_{ij}$,
${\rm Im}\,{\cal H}_{ij}$. Integrating out the eigenvectors of ${\cal H}$, one
obtains from Eq.\ (\ref{PHtauFP}) a Fokker-Planck equation for the distribution
$P(\{E_{n}\},\tau)$ of the eigenvalues $E_{n}$,
\begin{equation}
c\frac{\partial}{\partial\tau}P=\sum_{i}\frac{\partial}{\partial
E_{i}}\left(cE_{i}+\sum_{j\neq
i}\frac{1}{E_{j}-E_{i}}+\frac{1}{2}\frac{\partial}{\partial
E_{i}}\right)P.\label{PEtauFP}
\end{equation}
The implication of Eq.\ (\ref{PEtauFP}) is that the energy levels $E_{i}(\tau)$
execute a Brownian motion in fictitious time $\tau$.

To relate $\tau$ to $B$, we first relate $\tau$ to the parameter $\alpha$ of
the previous subsection, since we already know how to relate $\alpha$ to $B$.
For infinitesimal $\tau$ the Hamiltonian (\ref{Htau}) can be written as
\begin{equation}
{\cal H}={\cal H}_{0}+\sqrt{2\tau}\,\bigl({\cal H}_{\rm GOE}+{\rm i}{\cal
A}\bigr).\label{Htau1}
\end{equation}
Here ${\cal H}_{\rm GOE}$ and ${\cal A}$ are, respectively, real symmetric and
real antisymmetric matrices, having independent Gaussian distributions with the
same variance. Equivalently, one can use a purely antisymmetric perturbation of
${\cal H}_{0}$ and double its variance:
\begin{equation}
{\cal H}={\cal H}_{0}+2{\rm i}\sqrt{\tau}\,{\cal A}.\label{Htau2}
\end{equation}
Comparison with Eq.\ (\ref{PMH}) leads to the relation (Frahm, 1995b)
\begin{equation}
\Delta\alpha=2\sqrt{\tau}\label{alphatau}
\end{equation}
between the fictitious time $\tau$ of the Brownian motion and an increment
$\Delta\alpha$ of the Pandey-Mehta Hamiltonian in the absence of time-reversal
symmetry ({\em i.e.} for $M\alpha^{2}\gg 1$). Since $\alpha\propto\Phi$
according to Eq.\ (\ref{alphaestimate}), one finds that $\tau$ is related to
the flux increment $\Delta\Phi$ by $M\tau\simeq(e\Delta\Phi/h)^{2}E_{\rm
c}/\delta$.

Microscopic justification for the Brownian-motion model has been provided by
Beenakker (1993b, Beenakker and Rejaei, 1994b), through a comparison of the
correlation functions obtained from Eq.\ (\ref{PEtauFP}) with those obtained
for a disordered metal grain by Szafer and Altshuler (1993) and Simons and
Altshuler (1993, Altshuler and Simons, 1995). The model has one fundamental
limitation: Brownian motion correctly describes level correlations at any two
values of $B$, but does not describe how levels at three or more values of $B$
are correlated. The reason is that Brownian motion is a Markov process, meaning
that it has no memory: The distribution $P$ at time $\tau+\Delta\tau$ is fully
determined by the distribution at time $\tau$. Knowledge of $P$ at earlier
times is irrelevant for the evolution at later times. The true level dynamics,
in contrast, is no Markov process; It does have a memory. To see this, it
suffices to take a look at Fig. \ref{fig_brownian}: The energy levels evolve
smoothly as a function of magnetic field, hence their location at $B+\Delta B$
is not independent from that at $B-\Delta B$ if $\Delta B$ is small enough. As
a consequence, the correlator of two densities $\langle
n(B_{1})n(B_{2})\rangle$ [with $n(B)=\sum_{n}\delta\biglb(E-E_{n}(B)\bigrb)$]
can be obtained from the Fokker-Planck equation (\ref{PEtauFP}), but the
correlator of three densities $\langle n(B_{1})n(B_{2})n(B_{3})\rangle$ can
not.

\subsection{Statistical theory of transmission eigenvalues}
\label{statisticalT}

\subsubsection{Scattering and transfer matrices}
\label{SandM}

\begin{figure}[tb]
\centerline{
\psfig{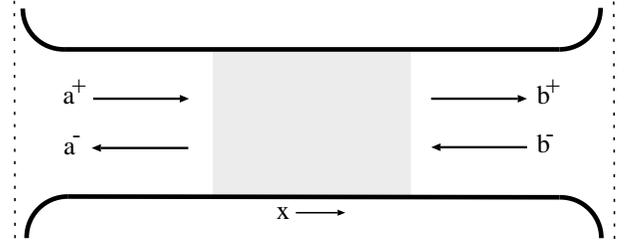}
}%
\medskip
\caption[]{
Disordered region (dotted) connected by ideal leads to two electron reservoirs
(to the left and right of the dashed lines). The scattering matrix $S$ relates
the amplitudes $a^{+},b^{-}$ of incoming waves to the amplitudes $a^{-},b^{+}$
of outgoing waves, while the transfer matrix $M$ relates the amplitudes
$a^{+},a^{-}$ at the left to the amplitudes $b^{+},b^{-}$ at the right.
}\label{fig_scattering}
\end{figure}

The scattering theory of electronic conduction is due to Landauer (1957, 1987),
Imry (1986b), and B\"{u}ttiker (1986b, 1988b). It provides a complete
description of transport at low frequencies, temperatures, and voltages, under
circumstances that electron-electron interactions can be neglected. (For an
overview of the great variety of experiments in which the theory has been
tested, see Beenakker and Van Houten, 1991.) A mesoscopic conductor is modeled
by a phase-coherent disordered region connected by ideal leads (without
disorder) to two electron reservoirs (see Fig.\ \ref{fig_scattering}).
Scattering in the phase-coherent region is elastic. All inelastic scattering is
assumed to take place in the reservoirs, which are in equilibrium at
temperature zero and electrochemical potential (or Fermi energy) $E_{\rm F}$.
The ideal leads are ``electron waveguides'', introduced to define a basis for
the scattering matrix of the disordered region.

The wavefunction $\psi$ of an electron in a lead at energy $E_{\rm F}$
separates into a longitudinal and a transverse part,
\begin{equation}
\psi_{n}^{\pm}(\vec{r}\,)=\Phi_{n}(y,z)\exp(\pm{\rm i}k_{n}x).\label{psilead}
\end{equation}
The integer $n=1,2,\ldots N$ labels the propagating modes, also referred to as
scattering channels. Mode $n$ has a real wavenumber $k_{n}>0$ and transverse
wavefunction $\Phi_{n}$. (We assume, for simplicity of notation, that the two
leads are identical.) The normalization of the wavefunction (\ref{psilead}) is
chosen such that it carries unit current. A wave incident on the disordered
region is described in this basis by a vector of coefficients
\begin{equation}
c^{\rm in}\equiv(a_{1}^{+},a_{2}^{+},\ldots
a_{N}^{+},b_{1}^{-},b_{2}^{-},\ldots b_{N}^{-}).\label{cindef}
\end{equation}
The first set of $N$ coefficients refers to the left lead, and the second set
of $N$ coefficients to the right lead in Fig.\ \ref{fig_scattering}. Similarly,
the reflected and transmitted wave has vector of coefficients
\begin{equation}
c^{\rm out}\equiv(a_{1}^{-},a_{2}^{-},\ldots
a_{N}^{-},b_{1}^{+},b_{2}^{+},\ldots b_{N}^{+}).\label{coutdef}
\end{equation}
The scattering matrix $S$ is a $2N\times 2N$ matrix which relates these two
vectors,
\begin{equation}
c^{\rm out}=Sc^{\rm in}.\label{coutScin}
\end{equation}
It has the block structure
\begin{equation}
S=\left(\begin{array}{cc}
r&t'\\t&r'
\end{array}\right),\label{Srt}
\end{equation}
with $N\times N$ reflection matrices $r$ and $r'$ (reflection from left to left
and from right to right) and transmission matrices $t$ and $t'$ (transmission
from left to right and from right to left).

Current conservation implies that $S$ is a unitary matrix:
$S^{-1}=S^{\dagger}$. It is a consequence of unitarity that the four Hermitian
matrices $tt^{\dagger}$, $t't'^{\dagger}$, $1-rr^{\dagger}$, and
$1-r'r'^{\dagger}$ have the same set of eigenvalues $T_{1},T_{2},\ldots T_{N}$.
Each of these $N$ transmission eigenvalues is a real number between 0 and 1.
The scattering matrix can be written in terms of the $T_{n}$'s by means of the
polar decomposition (Mello, Pereyra, and Kumar, 1988; Martin and Landauer,
1992)
\begin{equation}
S=\left(\begin{array}{cc}
U&0\\0&V
\end{array}\right)
\left(\begin{array}{cc}
-\sqrt{1-{\cal T}}&\!\!\sqrt{\cal T}\\
\sqrt{\cal T}&\!\!\sqrt{1-{\cal T}}
\end{array}\right)
\left(\begin{array}{cc}
U'&0\\0&V'
\end{array}\right).
\label{polarS}
\end{equation}
Here $U,V,U',V'$ are four $N\times N$ unitary matrices and ${\cal T}={\rm
diag}\,(T_{1},T_{2},\ldots T_{N})$ is an $N\times N$ diagonal matrix with the
transmission eigenvalues on the diagonal.\footnote{
The transmission eigenvalues for $\beta=4$ are twofold degenerate: ${\cal
T}={\rm diag}\,(T_{1}\openone,T_{2}\openone,\ldots T_{N}\openone)$. Compare the
footnote on Kramers' degeneracy of the energy levels in Sec.\
\protect\ref{WDensemble}.}

If time-reversal symmetry is broken ($\beta=2$), unitarity is the only
constraint on $S$. The presence of time-reversal symmetry imposes additional
constraints. If both time-reversal and spin-rotation symmetry are present
($\beta=1$), then $S$ is unitary and symmetric: $S=S^{\rm T}$, hence $U'=U^{\rm
T}$, $V'=V^{\rm T}$. (The superscript T indicates the transpose of the matrix.)
If time-reversal symmetry is present but spin-rotation symmetry is broken
($\beta=4$), then $S$ is unitary and self-dual: $S=S^{\rm R}$, hence $U'=U^{\rm
R}$, $V'=V^{\rm R}$. (The superscript R indicates the dual\footnote{
The dual $Q^{\rm R}$ of a matrix $Q$ with quaternion elements
$Q_{nm}=a_{nm}\openone+{\rm i}b_{nm}\sigma_{x}+{\rm i}c_{nm}\sigma_{y}+{\rm
i}d_{nm}\sigma_{z}$ has elements $Q^{\rm R}_{nm}=a_{mn}\openone-{\rm
i}b_{mn}\sigma_{x}-{\rm i}c_{mn}\sigma_{y}-{\rm i}d_{mn}\sigma_{z}$.}
of a quaternion matrix.)

The {\em scattering matrix\/} relates incoming to outgoing states. The {\em
transfer matrix\/} relates states in the left lead to states in the right lead.
A wave in the left lead is given by the vector of coefficients
\begin{equation}
c^{\rm left}\equiv(a_{1}^{+},a_{2}^{+},\ldots
a_{N}^{+},a_{1}^{-},a_{2}^{-},\ldots a_{N}^{-}).\label{cleftdef}
\end{equation}
The first set of $N$ coefficients refers to incoming waves, the second set of
$N$ coefficients to outgoing waves. Similarly, a wave in the right lead has
vector of coefficients
\begin{equation}
c^{\rm right}\equiv(b_{1}^{+},b_{2}^{+},\ldots
b_{N}^{+},b_{1}^{-},b_{2}^{-},\ldots b_{N}^{-}).\label{crightdef}
\end{equation}
The transfer matrix $M$ is a $2N\times 2N$ matrix which relates these two
vectors,
\begin{equation}
c^{\rm right}=Mc^{\rm left}.\label{crightMcleft}
\end{equation}
The scattering and transfer matrices are equivalent descriptions of the
disordered region. A convenient property of the transfer matrix is the {\em
multiplicative\/} composition rule: The transfer matrix of a number of
disordered regions in series (separated by ideal leads) is the product of the
individual transfer matrices. The scattering matrix, in contrast, has a more
complicated composition rule (containing a matrix inversion). By expressing the
elements of $M$ in terms of the elements of $S$ one obtains  the polar
decomposition of the transfer matrix (Mello, Pereyra, and Kumar, 1988; Mello
and Pichard, 1991),
\begin{equation}
M=\left(\begin{array}{cc}
V&0\\0&V'^{\dagger}
\end{array}\right)
\left(\begin{array}{cc}
\sqrt{{\cal T}^{-1}}&\sqrt{{\cal T}^{-1}-1}\\
\sqrt{{\cal T}^{-1}-1}&\sqrt{{\cal T}^{-1}}
\end{array}\right)
\left(\begin{array}{cc}
U'&0\\0&U^{\dagger}
\end{array}\right),
\label{polarM}
\end{equation}
in terms of the same $N\times N$ matrices used in Eq.\ (\ref{polarS}).

Current conservation imposes a ``pseudo-unitarity'' constraint on the transfer
matrix:
\begin{equation}
S^{-1}=S^{\dagger}\Longleftrightarrow\Sigma
M^{-1}\Sigma=M^{\dagger},\label{MSigma}
\end{equation}
where $\Sigma$ is a diagonal matrix with $\Sigma_{nn}=1$ for $1\leq n\leq N$
and $\Sigma_{nn}=-1$ for $N+1\leq n\leq 2N$. As a consequence, the matrix
product $MM^{\dagger}$ and its inverse $(MM^{\dagger})^{-1}=\Sigma
MM^{\dagger}\Sigma$ have the same set of eigenvalues, or in other words, the
eigenvalues of $MM^{\dagger}$ come in inverse pairs. We denote the $2N$
eigenvalues of $MM^{\dagger}$ by $\exp(\pm 2x_{n})$, with $x_{n}\geq 0$
($n=1,2,\ldots N$). By comparing Eqs.\ (\ref{polarS}) and (\ref{polarM}) one
obtains an algebraic relation between the transfer and transmission matrices
(Pichard, 1984),
\begin{equation}
\left[2+MM^{\dagger}+\bigl(MM^{\dagger}\bigr)^{-1}\right]^{-1}=\frac{1}{4}
\left(\begin{array}{cc}
tt^{\dagger}&0\\0&t'^{\dagger}t'
\end{array}\right),\label{MMttrelation}
\end{equation}
which implies that the exponent $x_{n}$ is related to the transmission
eigenvalue $T_{n}$ by
\begin{equation}
T_{n}=\frac{1}{\cosh^{2}x_{n}}.\label{Txrelation}
\end{equation}

An altogether different representation of the scattering matrix is the
eigenvalue--eigenvector decomposition
\begin{equation}
S=\Omega\,{\rm diag}\bigl({\rm e}^{{\rm i}\phi_{1}},{\rm e}^{{\rm
i}\phi_{2}},\ldots{\rm e}^{{\rm
i}\phi_{2N}}\bigr)\Omega^{\dagger}.\label{SOmegaphi}
\end{equation}
The real numbers $\phi_{n}$ are the scattering phase shifts. (There is a
twofold Kramers' degeneracy of the $\phi_{n}$'s for $\beta=4$.) The $2N\times
2N$ unitary matrix $\Omega$ has real elements for $\beta=1$, complex elements
for $\beta=2$, and real quaternion elements for $\beta=4$. (Hence $\Omega$ is
orthogonal for $\beta=1$ and symplectic for $\beta=4$.) The symmetry properties
of the scattering matrix are summarized in Table \ref{table_threefoldS}. The
decomposition (\ref{SOmegaphi}) mixes states at the left of the disordered
region with those at the right, and therefore does not distinguish between
transmission and reflection. This is why the polar decomposition (\ref{polarS})
is more suitable for a transport problem. A statistical theory of scattering
phase shifts was developed by Dyson (1962a), in the early days of random-matrix
theory. Dyson's ensemble of random scattering matrices, known as the {\em
circular ensemble}, turns out to be the appropriate ensemble for conduction
through a quantum dot, as we will discuss in Sec.\ \ref{dots}.

\begin{table}[tb]
\caption[]{
Symmetry of the scattering matrix $S$ and its matrix of eigenvectors
$\Omega$, for the three values of $\beta$.
}\label{table_threefoldS}
\begin{tabular}{ccccc}
$\beta$ & $S$ & $\Omega$\\
\hline\\
1 & unitary symmetric & orthogonal \\
2 & unitary & unitary \\
4 & unitary self-dual & symplectic
\end{tabular}
\end{table}

\subsubsection{Linear statistics}
\label{linearstatistics}

The transmission eigenvalues determine a variety of transport properties. First
of all the conductance $G=\lim_{V\rightarrow 0}\bar{I}/V$, defined as the ratio
of the time-averaged electrical current $\bar{I}$ through the conductor and the
voltage difference $V$ between the two electron reservoirs --- in the limit of
vanishingly small voltage. This is the limit of linear response, to which we
restrict ourselves in this review. At zero temperature, the conductance is
given by
\begin{equation}
G=G_{0}\sum_{n=1}^{N}T_{n},\;\;G_{0}\equiv\frac{2e^{2}}{h}.\label{Landauer}
\end{equation}
Equation (\ref{Landauer}) is known as the Landauer formula, because of
Landauer's pioneering 1957 paper. It was first written down in this form by
Fisher and Lee (1981). For an account of the controversy surrounding this
formula, which has now been settled, we refer to Stone and Szafer (1988). The
factor of two in the definition of the conductance quantum $G_{0}$ is due the
two-fold spin degeneracy in the absence of spin-orbit scattering. In the
presence of spin-orbit scattering, there is a two-fold Kramers' degeneracy in
zero magnetic field. In the presence of both spin-orbit scattering and a
magnetic field, one has a reduced conductance quantum $G_{0}=e^{2}/h$ with
twice the number of transmission eigenvalues.

The discreteness of the electron charge causes time-dependent fluctuations of
the current $I(t)=\bar{I}+\delta I(t)$, which persist down to zero temperature.
These fluctuations are known as shot noise. The power spectrum of the noise has
the zero-frequency limit
\begin{equation}
P=4\int_{0}^{\infty}\!\!dt\,\overline{\delta I(t+t_{0})\delta
I(t_{0})},\label{Pdef}
\end{equation}
where the overline indicates an average over the initial time $t_{0}$ in the
correlator. The shot-noise power is related to the transmission eigenvalues by
(B\"{u}ttiker, 1990)
\begin{equation}
P=P_{0}\sum_{n=1}^{N}T_{n}(1-T_{n}),\;\;P_{0}\equiv 2eVG_{0}.\label{PTrelation}
\end{equation}
Equation (\ref{PTrelation}) is the multi-channel generalization of formulas by
Khlus (1987) and Lesovik (1989).

More generally, we will study transport properties of the form
\begin{equation}
A=\sum_{n=1}^{N}a(T_{n}).\label{linearstatistic}
\end{equation}
The quantity $A$ is called a linear statistic on the transmission eigenvalues.
The word ``linear'' indicates that $A$ does not contain products of different
eigenvalues, but the function $a(T)$ may well depend non-linearly on $T$ --- as
it does in the case of the shot-noise power (\ref{PTrelation}), where $a(T)$
depends quadratically on $T$. The conductance (\ref{Landauer}) is special
because it is a linear statistic with a linear dependence on $T$. Other linear
statistics (with $a(T)$ a rational or algebraic function) appear if one of the
two electron reservoirs is in the superconducting state (see Sec.\
\ref{junctions}).

\subsubsection{Geometrical correlations}
\label{geometrical2}

The analogue for random scattering matrices of the Wigner-Dyson ensemble of
random Hamiltonians is an ensemble of unitary matrices where all correlations
between the transmission eigenvalues are geometrical. Here ``geometrical''
means due to the Jacobian $J$ which relates the volume elements in the polar
decomposition (\ref{polarS}),
\begin{equation}
d\mu(S)=J\prod_{\alpha}d\mu(U_{\alpha})\prod_{i}dT_{i}.\label{dmuJ}
\end{equation}
The set $\{U_{\alpha}\}$ is the set of independent unitary matrices in Eq.\
(\ref{polarS}): $\{U_{\alpha}\}=\{U,V\}$ if $\beta=1$ or 4;
$\{U_{\alpha}\}=\{U,U',V,V'\}$ if $\beta=2$. The Jacobian depends only on the
transmission eigenvalues,\footnote{
For a calculation of the Jacobian (\protect\ref{jacobianT}) from scattering
matrix to transmission eigenvalues, see Baranger and Mello (1994), Jalabert,
Pichard, and Beenakker (1994), and Jalabert and Pichard (1995). For an earlier,
closely related, calculation of the Jacobian from transfer matrix to
transmission eigenvalues, see Muttalib, Pichard, and Stone (1987), Mello,
Pereyra, and Kumar (1988), and Zanon and Pichard (1988).}
\begin{equation}
J(\{T_{n}\})=\prod_{i<j}|T_{i}-T_{j}|^{\beta}\prod_{k}T_{k}^{-1+\beta/2}.
\label{jacobianT}
\end{equation}
The analogue of the Wigner-Dyson distribution (\ref{WDPH}),
\begin{equation}
P(S)=c\exp[-\beta\,{\rm Tr}\,f(tt^{\dagger})],\label{WDPS}
\end{equation}
yields upon multiplication by $J$ a distribution of the $T_{n}$'s analogous to
Eq.\ (\ref{WDPE}),
\begin{equation}
P(\{T_{n}\})=c\prod_{i<j}|T_{i}-T_{j}|^{\beta}\prod_{k}T_{k}^{-1+\beta/2}
\exp[-\beta f(T_{k})].\label{WDPT}
\end{equation}

Muttalib, Pichard, and Stone (1987), and Pichard, Zanon, Imry, and Stone (1990)
have based a statistical theory of transmission eigenvalues on the distribution
(\ref{WDPT}). (Their theory is reviewed by Stone, Mello, Muttalib, and Pichard,
1991.) To make contact with their work, we perform the change of variables
\begin{equation}
T_{n}=\frac{1}{1+\lambda_{n}}.\label{lambdadef}
\end{equation}
Since $T_{n}$ lies between 0 and 1, the variable $\lambda_{n}$ ranges from 0 to
$\infty$. The distribution (\ref{WDPT}) transforms to
\begin{mathletters}
\label{PGibbscde}
\begin{eqnarray}
P(\{\lambda_{n}\})&=&c\exp\Bigl[-\beta\Bigl(\sum_{i<j}
u(\lambda_{i},\lambda_{j})
+ \sum_{i}V(\lambda_{i})\Bigr)\Bigr],\nonumber\\
\label{PGibbsc}\\
u(\lambda,\lambda')&=&-\ln|\lambda-\lambda'|,\label{PGibbsd}\\
V(\lambda)&=&[N-\case{1}{2}(1-2/\beta)]\ln(1+\lambda)\nonumber\\
&&\mbox{}+f\biglb((1+\lambda)^{-1}\bigrb).\label{PGibbse}
\end{eqnarray}
\end{mathletters}%
Equation (\ref{PGibbscde}) has the same form as the Gibbs distribution
(\ref{PGibbsab}) in the Wigner-Dyson ensemble, with the difference that the
$\lambda_{n}$'s can only take on positive values --- while the $E_{n}$'s are
free to range over the whole real axis. All microscopic information about the
conductor (its size and degree of disorder) is contained in the confining
potential $V(\lambda)$. The hypothesis of geometrical correlations does not
specify this function. Muttalib, Pichard, and Stone (1987) have shown that the
probability distribution (\ref{PGibbscde}) maximizes the entropy of the
ensemble subject to the constraint of a given mean density $\rho(\lambda)$ of
the $\lambda_{n}$'s. The function $V(\lambda)$ is the Lagrange multiplier for
this constraint. The Wigner-Dyson ensemble can similarly be interpreted as the
ensemble of maximum entropy for a given mean density of states (Balian, 1968).

The correlation functions implied by the probability distribution
(\ref{PGibbscde}) have been studied for a variety of potentials $V(\lambda)$ by
Stone, Mello, Muttalib, and Pichard (1991), Slevin, Pichard, and Mello (1991),
Chen, Ismail, and Muttalib (1992), Muttalib, Chen, Ismail, and Nicopoulos
(1993), and Slevin, Pichard, and Muttalib (1993). It was originally believed
that precise agreement with the microscopic theory of a disordered wire could
be obtained if only $V(\lambda)$ were properly chosen (Mello and Pichard,
1989). We now know that this is not correct (Beenakker, 1993a): {\em The true
eigenvalue repulsion is not logarithmic}. In other words, there exist
correlations between the transmission eigenvalues over and above those induced
by the Jacobian. As we will discuss in Sec.\ \ref{Trepulsion}, the hypothesis
of geometrical correlations is valid for $T_{n}$'s close to unity. However, it
overestimates the repulsion of smaller $T_{n}$'s (Beenakker and Rejaei, 1993,
1994a). The appearance of random-matrix ensembles with a non-logarithmic
eigenvalue repulsion is a distinctive feature of the random-matrix theory of
quantum transport.

An implication of the non-logarithmic repulsion is that the true ensemble is
not of maximum entropy, at least not in the sense of Muttalib, Pichard, and
Stone (1987). We make this qualification because, unlike in statistical
mechanics, there is not a single definition of the entropy of a random-matrix
ensemble. Slevin and Nagao (1993, 1994) have constructed an alternative maximum
entropy ensemble, in which the repulsion is logarithmic in the variables
$x_{n}$ (recall that $T_{n}=1/\cosh^{2}x_{n}$). The true repulsion, however, is
not logarithmic in any variable. It is not known whether there exists some
maximum entropy principle which would produce the correct ensemble for a
disordered wire.

\subsection{Correlation functions}
\label{correlationfunctions}

The established method to compute correlation functions of eigenvalues in the
Wigner-Dyson ensemble is the method of orthogonal polynomials (Mehta, 1991).
This method works for any dimensionality $N$ of the random matrix, but requires
a logarithmic repulsion $u(\lambda,\lambda')=-\ln|\lambda-\lambda'|$ of the
eigenvalues. Moreover, although in principle one can assume an arbitrary
confining potential, in practice one is restricted in the choice of
$V(\lambda)$. (One needs to be able to construct a basis of polynomials which
are orthogonal with weight function ${\rm e}^{-\beta V}$.) For applications to
quantum transport one requires a method that is not restricted to particular
$u$ and $V$, but the large-$N$ limit is often sufficient. The method of
functional derivatives was developed for such applications (Beenakker, 1993a,
1993c, 1994a). A similar method (for the case of logarithmic repulsion) has
been developed in connection with matrix models of quantum gravity (Makeenko,
1991).

\subsubsection{Method of functional derivatives}
\label{functionalderivatives}

We consider the two-point correlation function
\begin{equation}
K(\lambda,\lambda')=\left\langle\sum_{i,j}
\delta(\lambda-\lambda_{i})\delta(\lambda'-\lambda_{j})
\right\rangle-\rho(\lambda)\rho(\lambda').\label{Kdef}
\end{equation}
Here $\rho(\lambda)=\langle\sum_{i}\delta(\lambda-\lambda_{i})\rangle$ is the
mean eigenvalue density and $\langle\cdots\rangle$ denotes the average with
probability distribution (\ref{PGibbscde}). Explicitly,
\begin{eqnarray}
&&\rho(\lambda)=\frac{\int\! d{\lambda}_{1}\cdots\int
\! d{\lambda}_{N}\,{\rm e}^{-\beta W}\sum_{i}\delta(\lambda-\lambda_{i})}{\int
\! d{\lambda}_{1}\cdots\int\! d{\lambda}_{N}\,{\rm e}^{-\beta W}},
\label{rhomean}\\
&&W(\{\lambda_{n}\})=\sum_{i<j}u(\lambda_{i},\lambda_{j})+
\sum_{i}V(\lambda_{i}).\label{Wdef}
\end{eqnarray}
The interaction potential $u(\lambda,\lambda')$ may or may not be logarithmic.
By differentiating Eq.\ (\ref{rhomean}) we obtain an exact relationship between
the two-point correlation function and the functional derivative of the mean
density with respect to the confining potential,
\begin{equation}
K(\lambda,\lambda')=-\frac{1}{\beta}\,
\frac{\delta\rho(\lambda)}{\delta V(\lambda')}.
\label{KdeltarhodeltaV}
\end{equation}

To evaluate this functional derivative we must know how the density depends on
the potential. This is a classic problem in random-matrix theory. In the
large-$N$ limit the solution is given by the integral equation (Wigner, 1957)
\begin{equation}
V(\lambda)+\int_{\lambda_{-}}^{\lambda_{+}}\!d\lambda'\,
u(\lambda,\lambda')\rho(\lambda')=
{\rm constant},\label{wignerint}
\end{equation}
where ``constant'' means independent of $\lambda$ inside the interval
$(\lambda_{-},\lambda_{+})$ where $\rho>0$. The boundaries $\lambda_{\pm}$ of
the spectrum can be either fixed or free. A fixed boundary is independent of
$V$. (An example is the constraint $\lambda>0$.) A free boundary is to be
determined selfconsistently from Eq.\ (\ref{wignerint}), by requiring that
$\rho$ vanishes at the boundary. A free boundary thus depends on $V$. Equation
(\ref{wignerint}) has the ``mechanical equilibrium'' interpretation that the
density $\rho$ adjusts itself to the potential $V$ in such a way that the total
force at any point vanishes. The support of $\rho$ is therefore an
equipotential. Finite-$N$ corrections to Eq.\ (\ref{wignerint}) are smaller by
an order $N^{-1}$ for $\beta=1$ or 4, and by an order $N^{-2}$ for $\beta=2$
(Dyson, 1972; see Appendix \ref{Dysonexpansion}). A rigorous proof, containing
precise conditions on $u$ and $V$, has been given by Boutet de Monvel, Pastur,
and Shcherbina (1995).

Variation of Eq.\ (\ref{wignerint}) gives\footnote{
Variation of the boundary $\lambda_{\pm}$ of the spectrum gives an additional
contribution
$\pm\delta\lambda_{\pm}\rho(\lambda_{\pm})u(\lambda,\lambda_{\pm})$. This
contribution vanishes, either because $\delta\lambda_{\pm}=0$ (fixed boundary)
or because $\rho(\lambda_{\pm})=0$ (free boundary). Variation of the
$\lambda$-independent right-hand-side of Eq.\ (\protect\ref{wignerint}) gives
some other $\lambda$-independent constant, not necessarily equal to zero.}
\begin{mathletters}
\label{variation}
\begin{equation}
\delta V(\lambda)+\int_{\lambda_{-}}^{\lambda_{+}}
\!d\lambda'\,u(\lambda,\lambda')\delta\rho(\lambda')=
{\rm constant},\label{variationa}
\end{equation}
with the constraint
\begin{equation}
\int_{\lambda_{-}}^{\lambda_{+}}\!d\lambda\,\delta\rho(\lambda)=0
\label{variationb}
\end{equation}
\end{mathletters}%
(since the variation of $\rho$ is to be carried out at constant $N$). The
inverse of Eq.\ (\ref{variation}) is
\begin{mathletters}
\label{variationinverse}
\begin{eqnarray}
&&\delta\rho(\lambda)=-\int_{\lambda_{-}}^{\lambda_{+}}\!d\lambda'\,u^{\rm
inv}(\lambda,\lambda')\delta V(\lambda'),\label{variationinversea}\\
&&\int_{\lambda_{-}}^{\lambda_{+}}\!d\lambda''\,u(\lambda,\lambda'')u^{\rm
inv}(\lambda'',\lambda')=\delta(\lambda-\lambda')-
\frac{1}{\lambda_{+}-\lambda_{-}}.\nonumber\\
\label{variationinverseb}
\end{eqnarray}
\end{mathletters}%
Equation (\ref{variationinverseb}) means that the integral kernel $u^{\rm inv}$
is the inverse of $u$ for functions $f(\lambda)$ restricted by $\int
d\lambda\,f=0$.

Combination of Eqs.\ (\ref{KdeltarhodeltaV}) and (\ref{variationinverse})
yields a relation between the two-point correlation function and the inverse of
the interaction potential (Beenakker, 1993a),
\begin{equation}
K(\lambda,\lambda')=\frac{1}{\beta}\,u^{\rm
inv}(\lambda,\lambda').\label{Kuinvrelation}
\end{equation}
This relation is universal in that it does not contain the confining potential
explicitly. There is an implicit dependence on $V$ through $\lambda_{\pm}$ in
Eq.\ (\ref{variationinverse}), but this can be neglected far from a free
boundary. There exists a variety of other demonstrations of such insensitivity
of correlation functions to the choice of confining potential (Kamien,
Politzer, and Wise, 1988; Ambj\o{}rn and Makeenko, 1990; Ambj\o{}rn,
Jurkiewicz, and Makeenko, 1990; Pastur, 1992; Br\'{e}zin and Zee, 1993, 1994;
Eynard, 1994; Forrester, 1995; Hackenbroich and Weidenm\"{u}ller, 1995; Morita,
Hatsugai, and Kohmoto, 1995; Kobayakawa, Hatsugai, Kohmoto, and Zee, 1995;
Freilikher, Kanzieper, and Yurkevich, 1996; Ambj\o{}rn and Akemann, 1996).

A universal two-point correlation function implies universal fluctuations of
linear statistics, as we discuss in the next subsection.

\subsubsection{Universal conductance fluctuations}
\label{UCF}

\begin{figure}[tb]
\centerline{
\psfig{figure=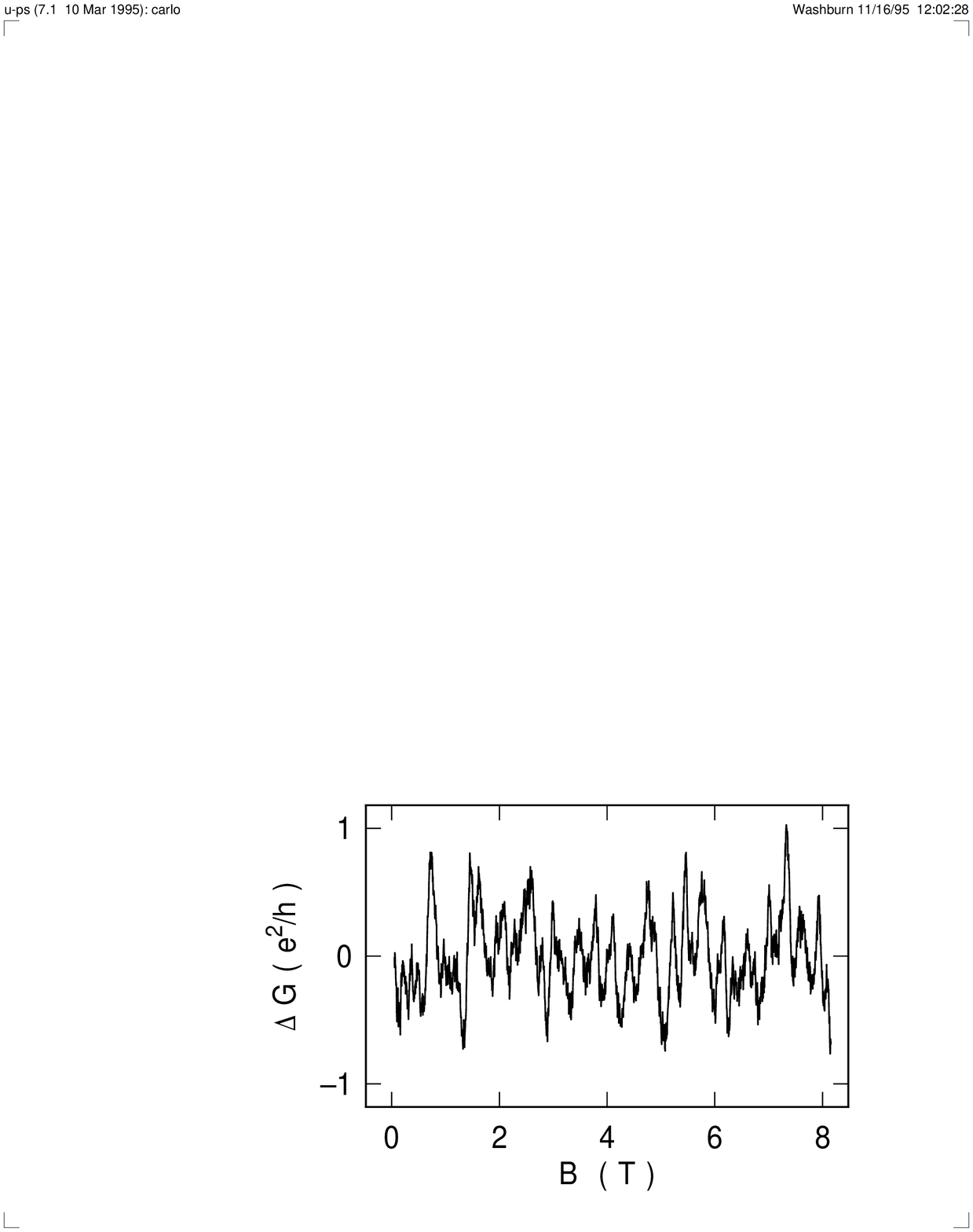,width=
8cm,clip=t,bbllx=170pt,bblly=55pt,bburx=525pt,bbury=293pt}
}%
\medskip
\caption[]{
Fluctuations as a function of perpendicular magnetic field of the conductance
of a 310~nm long and 25~nm wide Au wire at 10~mK. The trace appears random, but
is completely reproducible from one measurement to the next. The
root-mean-square of the fluctuations is $0.3\,e^{2}/h$, which is not far from
the theoretical result $\sqrt{1/15}\,e^{2}/h$ [Eq.\ (\protect\ref{VarG215})
with $\beta=2$ due to the magnetic field and a reduced conductance quantum of
$e^{2}/h$ due to the strong spin-orbit scattering in Au]. After Washburn and
Webb (1986).
}\label{fig_UCFAu}
\end{figure}

Quantum interference leads to significant sample-to-sample fluctuations in the
conductance at low temperatures. These fluctuations can also be observed in a
single sample as a function of magnetic field, since a small change in field
has a similar effect on the interference pattern as a change in impurity
configuration. Experimental data by Washburn and Webb (1986) for a Au wire at
10~mK is shown in Fig.\ \ref{fig_UCFAu}. The fluctuations are {\em not\/}
time-dependent noise, but completely reproducible. Such a magnetoconductance
trace is called a ``magnetofingerprint'', because the pattern is specific for
the particular sample being studied. Notice that the magnitude of the
fluctuations is of order $e^{2}/h$. This is not accidental.

The universality of the conductance fluctuations was discovered theoretically
by Altshuler (1985) and Lee and Stone (1985). There are two aspects to the
universality: (1) The variance ${\rm Var}\,G$ of the conductance is of order
$(e^{2}/h)^{2}$, independent of sample size or disorder strength; (2) ${\rm
Var}\,G$ decreases by precisely a factor of two if time-reversal symmetry is
broken by a magnetic field. The Altshuler--Lee--Stone theory is a diagrammatic
perturbation theory for a disordered metal. Two classes of diagrams, cooperons
and diffusons, contribute equally to the variance in the presence of
time-reversal symmetry. A magnetic field suppresses the cooperons but leaves
the diffusons unaffected, hence the factor-of-two reduction. (We are assuming
here, for simplicity, that there is no spin-orbit scattering.) The variance
${\rm Var}\,G/G_{0}$ of the conductance (in units of the conductance quantum
$G_{0}=2e^{2}/h$) is a number of order unity which is weakly dependent on the
shape of the conductor. For a wire geometry (length much greater than width) at
zero temperature, the variance is
\begin{equation}
{\rm Var}\,G/G_{0}=\frac{2}{15}\beta^{-1}.\label{VarG215}
\end{equation}
There is no dependence on the mean free path $l$, the wire length $L$, or the
number of transverse modes $N$, provided $l\ll L\ll Nl$. That is to say, the
wire should be much longer than the mean free path but much shorter than the
localization length. The Altshuler--Lee--Stone theory has been tested in many
experiments (for reviews, see Altshuler, Lee, and Webb, 1991; Beenakker and Van
Houten, 1991).

Shortly after the discovery of the universality of conductance fluctuations, an
explanation was given in terms of the repulsion of energy levels (Altshuler and
Shklovski\u{\i}, 1986) or of transmission eigenvalues (Imry, 1986a). Imry's
argument contrasts ``closed'' and ``open'' scattering channels. Most
transmission eigenvalues in a disordered conductor are exponentially small.
These are the closed channels. A fraction $l/L$ of the total number $N$ of
transmission eigenvalues is of order unity. These are the open channels. Only
the open channels contribute effectively to the conductance: $G/G_{0}\equiv
N_{\rm eff}\approx Nl/L$. Fluctuations in the conductance can be interpreted as
fluctuations in the number $N_{\rm eff}$ of open channels. The alternative
argument of Altshuler and Shklovski\u{\i} is based on Thouless' (1977)
relationship $N_{\rm eff}\approx E_{\rm c}/\delta$. (The Thouless energy
$E_{\rm c}$ was defined in Sec.\ \ref{geometrical}; $\delta$ is the mean level
spacing.) Conductance fluctuations can be interpreted as fluctuations in the
number of energy levels in an energy range $E_{\rm c}$. If the transmission
eigenvalues or energy levels were uncorrelated, one would estimate that
fluctuations in $N_{\rm eff}$ would be of order $\sqrt{N_{\rm eff}}$. This
would imply that ${\rm Var}\,G/G_{0}$ would be of order $N_{\rm eff}$ --- which
is $\gg 1$. The fact that the variance is of order unity is a consequence of
the strong suppression of the fluctuations in $N_{\rm eff}$ by eigenvalue
repulsion.

This argument can be made quantitative. Take a linear statistic
\begin{equation}
A=\sum_{n=1}^{N}a(\lambda_{n}).\label{Alambdadef}
\end{equation}
For the conductance, we would have $a(\lambda)=(1+\lambda)^{-1}$ [{\em cf.}
Eqs.\ (\ref{Landauer}) and (\ref{lambdadef})]. The average of $A$,
\begin{equation}
\langle A\rangle=\int_{\lambda_{-}}^{\lambda_{+}}\!\!d\lambda\,
a(\lambda)\rho(\lambda),\label{Aaverage}
\end{equation}
diverges for $N\rightarrow\infty$. We can identify $\langle A\rangle\equiv
N_{\rm eff}$. The variance ${\rm Var}\,A=\langle A^{2}\rangle-\langle
A\rangle^{2}$ is obtained from a double integration of the two-point
correlation function (\ref{Kdef}),
\begin{equation}
{\rm Var}\,A=\int_{\lambda_{-}}^{\lambda_{+}}\!\!d\lambda
\int_{\lambda_{-}}^{\lambda_{+}}\!\!d\lambda'\,
a(\lambda)a(\lambda')K(\lambda,\lambda').\label{VarAKdef}
\end{equation}
For independent $\lambda_{n}$'s, we would expect ${\rm Var}\,A$ to be of order
$N_{\rm eff}$, so that it too would diverge with $N$. Instead, Eq.\
(\ref{Kuinvrelation}) implies that
\begin{equation}
{\rm Var}\,A=\frac{1}{\beta}\int_{\lambda_{-}}^{\lambda_{+}}\!\!d\lambda
\int_{\lambda_{-}}^{\lambda_{+}}\!\!d\lambda'\, a(\lambda)a(\lambda')u^{\rm
inv}(\lambda,\lambda'),\label{VarAuinv}
\end{equation}
with corrections of order $1/N_{\rm eff}$. This tells us that ${\rm Var}\,A$
for large $N$ is independent of $N$, provided the interaction potential
$u(\lambda,\lambda')$ is $N$-independent. Moreover, ${\rm Var}\,A\propto
1/\beta$ if $u$ is $\beta$-independent. These are the two aspects of
universality mentioned above. Let us illustrate this general result by two
examples (Beenakker, 1993a, 1993c).

The first example is the Wigner-Dyson ensemble (\ref{PGibbsab}), with a
logarithmic repulsion. The eigenvalues are free to vary over the whole real
axis, hence the end points $\lambda_{\pm}$ of the spectrum are free boundaries.
Let us assume that the function $a(\lambda)$ is non-zero only for $\lambda$ in
the bulk of the spectrum, so that the integrals from $\lambda_{-}$ to
$\lambda_{+}$ may be replaced by integrals from $-\infty$ to $+\infty$. To
determine the functional inverse of
$u(\lambda,\lambda')=-\ln|\lambda-\lambda'|$ in the bulk of the spectrum, we
need to solve the integral equation
\begin{equation}
-\int_{-\infty}^{\infty}\!d\lambda''\,\ln|\lambda-\lambda''|\,u^{\rm
inv}(\lambda'',\lambda')= \delta(\lambda-\lambda').\label{lninverse}
\end{equation}
This is readily solved by Fourier transformation, with the result
\begin{equation}
u^{\rm
inv}(\lambda,\lambda')=-\frac{1}{\pi^{2}}\,\frac{\partial}{\partial\lambda}
\frac{\partial}{\partial\lambda'} \ln|\lambda-\lambda'|.\label{lnuinv}
\end{equation}
Substitution into Eq.\ (\ref{VarAuinv}) yields a formula for the variance of a
linear statistic,
\begin{mathletters}
\label{VarAWDresult}
\begin{eqnarray}
{\rm Var}\,A&=&-\frac{1}{\beta\pi^{2}}\int_{-\infty}^{\infty}
\!\!d\lambda\int_{-\infty}^{\infty}
\!\!d\lambda'\left(\frac{da(\lambda)}{d\lambda}\right)
\left(\frac{da(\lambda')}{d\lambda'}\right)\nonumber\\
&&\mbox{}\times\ln|\lambda-\lambda'|,
\label{VarAWDresulta}
\end{eqnarray}
or in an equivalent Fourier representation,
\begin{eqnarray}
&&{\rm Var\,}A=\frac{1}{\beta\pi^{2}}\int_{0}^{\infty}\!\!dk\,
|a(k)|^{2}k,\label{VarAFDresultb}\\
&&a(k)=\int_{-\infty}^{\infty}\!\!d\lambda\,{\rm e}^{{\rm
i}k\lambda}a(\lambda).\label{VarAFDresultc}
\end{eqnarray}
\end{mathletters}%
Equation (\ref{VarAWDresult}) was first derived for the Gaussian ensemble,
$V(\lambda)\propto\lambda^{2}$, by Dyson and Mehta (1963; Mehta, 1991). Note
that ${\rm Var}\,A$ diverges logarithmically for a step function
$a(\lambda)=\theta(\lambda-\lambda_{\rm c})$. More generally, if $a(\lambda)$
changes abruptly on the scale of the eigenvalue spacing, its variance does not
have a universal $N\rightarrow\infty$ limit. All physical quantities which we
will consider, however, are smooth functions of $\lambda$.

The second example is the ensemble (\ref{PGibbscde}) of Muttalib, Pichard, and
Stone (1987), relevant for transport properties. The repulsion is still
logarithmic, but the eigenvalues are constrained by $\lambda_{n}>0$. Thus
$\lambda_{-}=0$ is a fixed lower bound of the spectrum. There is also a free
upper bound at some $\lambda_{+}\gg 1$, which does not affect transport
properties and can be ignored. (Recall that large $\lambda$ corresponds to
small $T$.) Instead of Eq.\ (\ref{lninverse}) we now have the integral equation
\begin{equation}
-\int_{0}^{\infty}\!d\lambda''\,\ln|\lambda-\lambda''|\,u^{\rm
inv}(\lambda'',\lambda')= \delta(\lambda-\lambda'),\label{lninverse2}
\end{equation}
which can be solved by Mellin transformation. (The Mellin transform is a
Fourier transform with respect to the variable $\ln\lambda$.) The result is
\begin{equation}
u^{\rm
inv}(\lambda,\lambda')=-\frac{1}{\pi^{2}}\,\frac{\partial}{\partial\lambda}
\frac{\partial}{\partial\lambda'} \ln\left|\frac{\surd\lambda-\surd\lambda'}
{\surd\lambda+\surd\lambda'}\right|.\label{lnuinv2}
\end{equation}
Instead of Eq.\ (\ref{VarAWDresult}) we obtain the formula (Beenakker, 1993a,
1993c; see also Basor and Tracy, 1993; Jancovici and Forrester, 1994)
\begin{mathletters}
\label{VarAWDresult2}
\begin{eqnarray}
{\rm Var}\,A&=&-\frac{1}{\beta\pi^{2}}\int_{0}^{\infty}
\!\!d\lambda\int_{0}^{\infty}
\!\!d\lambda'\left(\frac{da(\lambda)}{d\lambda}\right)
\left(\frac{da(\lambda')}{d\lambda'}\right)\nonumber\\
&&\mbox{}\times\ln\left|\frac{\surd\lambda-\surd\lambda'}
{\surd\lambda+\surd\lambda'}\right|,
\label{VarAWDresult2a}
\end{eqnarray}
or equivalently,
\begin{eqnarray}
&&{\rm Var\,}A=\frac{1}{\beta\pi^{2}}\int_{0}^{\infty}\!\!dk\,
|\tilde{a}(k)|^{2}k\tanh(\pi k),\label{VarAFDresult2b}\\
&&\tilde{a}(k)=\int_{0}^{\infty}\!\!d\lambda\,\lambda^{{\rm
i}k-1}a(\lambda).\label{VarAFDresult2c}
\end{eqnarray}
\end{mathletters}%
The difference between Eqs.\ (\ref{VarAWDresult}) and (\ref{VarAWDresult2})
originates entirely from the positivity constraint on $\lambda$ in the
transport problem.

Substitution of $a(\lambda)=(1+\lambda)^{-1}$ into Eq.\ (\ref{VarAWDresult2})
yields the variance of the conductance
\begin{equation}
{\rm Var}\,G/G_{0}=\frac{1}{8}\beta^{-1},\label{VarG18}
\end{equation}
which differs slightly, but significantly, from Eq.\ (\ref{VarG215}). This was
the first demonstration that the eigenvalue repulsion in a disordered wire
could not be precisely logarithmic (Beenakker, 1993a).

The variance ${\rm Var}\,A$ is the second cumulant of the distribution function
$P(A)$. What about higher-order cumulants? Politzer (1989) has shown that the
cumulants of order three and higher of a linear statistic $A$ vanish in the
large-$N$ limit. This means that $P(A)$ tends to a Gaussian distribution in
that limit. Politzer's argument is that the linearity of the relation
(\ref{wignerint}) between $\rho$ and $V$ implies that for each $p\geq 3$ the
functional derivative $\delta^{p-1}\rho/\delta V^{p-1}$ vanishes, and hence
that the $p$-point correlation function as well as the $p$-th cumulant vanish.
Only the two-point correlation function (proportional to $\delta\rho/\delta V$)
and the second cumulant survive the large-$N$ limit.

\subsection{Overview}
\label{overview}

The two questions which the random-matrix theory of quantum transport addresses
are: {\em First:\/} What is the ensemble of scattering matrices? {\em
Second:\/} How to obtain from it the statistics of transport properties? In
this article we review the answer to both questions for the two geometries
where the answer is known: a quantum dot and a disordered wire.

The quantum dot is the easiest of the two geometries. For the first question we
rely on Efetov's demonstration that the Hamiltonian of a disordered metal grain
is distributed according to the Wigner-Dyson ensemble (\ref{WDPH}). The
corresponding distribution of scattering matrices follows upon coupling the
bound states inside the grain to propagating modes outside. If the coupling is
via quantum point contacts, the scattering matrix is distributed according to
the circular ensemble. (A quantum point contact is a narrow opening, much
smaller than the mean free path, with a quantized conductance of $NG_{0}$.) The
circular ensemble is defined by
\begin{equation}
P(S)={\rm constant},\label{PSconstant}
\end{equation}
that is to say, the scattering matrix $S$ is uniformly distributed in the
unitary group --- subject only to the constraints imposed by time-reversal
and/or spin-rotation symmetry. The corresponding distribution of the
transmission eigenvalues is of the form (\ref{WDPT}), with $f(T_{n})\equiv 0$.
Hence the eigenvalue repulsion in a quantum dot is logarithmic. Correlation
functions of the transmission eigenvalues can be computed either by exact
integration over the unitary group (which is practical for small $N$), or using
the large-$N$ method of Sec.\ \ref{functionalderivatives}. In particular, the
limit $N\rightarrow\infty$ of the variance of a linear statistic is given by
Eq.\ (\ref{VarAWDresult2}). The circular ensemble does not say how the
scattering matrices at different energies or magnetic field values are
correlated. For that information one needs to return to the underlying
Hamiltonian ensemble.

Historically, the latter approach came first: Verbaarschot, Weidenm\"{u}ller,
and Zirnbauer (1985), and Iida, Weidenm\"{u}ller, and Zuk (1990a, 1990b)
computed correlators of scattering matrix elements, and moments of the
conductance, directly from the Hamiltonian ensemble. More recently, the entire
distribution function of the transmission eigenvalues was determined starting
from the ensemble of scattering matrices (Baranger and Mello, 1994; Jalabert,
Pichard, and Beenakker, 1994). The equivalence of the two approaches has been
established by Brouwer (1995). Both random-matrix approaches agree with
Efetov's (1982, 1983) supersymmetric field theory of a disordered metal grain.
There is considerable numerical and analytical evidence that they apply
generically to any chaotic cavity, regardless of whether the chaos is due to
impurity or to boundary scattering (Bohigas, Giannoni, and Schmit, 1984;
Andreev, Agam, Simons, and Altshuler, 1996).

\begin{figure}[tb]
\centerline{
\psfig{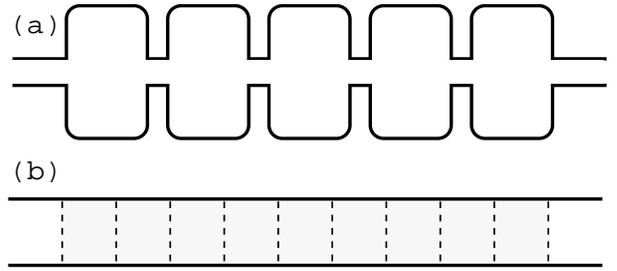}
}%
\medskip
\caption[]{
Two ways to construct a conductor with the geometry of a wire: (a) Strongly
scattering cavities, coupled in series via ideal leads; (b) Weakly disordered
segments in series. On long length scales, the two geometries have equivalent
statistical properties. The number of scattering channels $N$ is determined by
the width of the ideal leads in case (a) and by the width of the disordered
segments in case (b).
}\label{fig_qdotwire}
\end{figure}

Once transport through a single quantum dot is understood, a logical next step
is to connect many quantum dots in series, so that they form a wire (Fig.\
\ref{fig_qdotwire}a). Iida, Weidenm\"{u}ller, and Zuk (1990a, 1990b;
Weidenm\"{u}ller, 1990) and Altland (1991) computed the mean and variance of
the conductance for such a model. An altogether different approach was taken
earlier by Dorokhov (1982) and by Mello, Pereyra, and Kumar (1988). The wire is
divided into weakly scattering segments (short compared to the mean free path
$l$), so that the effect of adding a new segment can be determined by
perturbation theory (Fig.\ \ref{fig_qdotwire}b). The result is a differential
equation for the evolution with increasing wire length $L$ of the distribution
function of the variables $\lambda_{n}=(1-T_{n})/T_{n}$:
\begin{eqnarray}
&&l\frac{\partial}{\partial
L}P(\lambda_{1},\lambda_{2},\ldots\lambda_{N},L)\nonumber\\
&&\hspace{5mm}=\frac{2}{\beta N+2-\beta}\sum_{n=1}^{N}
\frac{\partial}{\partial\lambda_{n}}\lambda_{n}(1+\lambda_{n})
J\frac{\partial}{\partial\lambda_{n}}\frac{P}{J}.\label{DMPK0}
\end{eqnarray}
The Jacobian
\begin{equation}
J(\{\lambda_{n}\})=\prod_{i<j}|\lambda_{i}-\lambda_{j}|^{\beta}
\label{jacobianlambda}
\end{equation}
relates volume elements in the polar decomposition (\ref{polarM}) of the
transfer matrix,
\begin{equation}
d\mu(M)=J\prod_{\alpha}d\mu(U_{\alpha})\prod_{i}d\lambda_{i}.\label{dmuM}
\end{equation}
The evolution equation (\ref{DMPK0}) is known as the
Dorokhov-Mello-Pereyra-Kumar (DMPK) equation. For some time it was believed
that the solution to Eq.\ (\ref{DMPK0}) was of the form (\ref{PGibbscde}). The
exact solution (Beenakker and Rejaei, 1993, 1994a) of the DMPK equation for
$\beta=2$ showed that this is not the case, and that the eigenvalue repulsion
implied by Eq.\ (\ref{DMPK0}) is not logarithmic, as it is in Eq.\
(\ref{PGibbscde}).

The model of quantum dots in series of Iida, Weidenm\"{u}ller, and Zuk (1990a,
1990b) reduces on large length scales to a supersymmetric field theory known as
the one-dimensional non-linear $\sigma$ model (Mirlin, M\"{u}ller-Groeling, and
Zirnbauer, 1994). This model was originally derived by Efetov and Larkin
(1983), starting from a Hamiltonian with randomly distributed impurities. A
later derivation, due to Fyodorov and Mirlin (1991, 1994), uses a banded random
matrix to model the Hamiltonian of the disordered wire. The DMPK equation and
the $\sigma$ model of one-dimensional localization originated almost
simultaneously in the early eighties, and at the same institute (Dorokhov,
1982, 1983; Efetov and Larkin, 1983). Nevertheless, work on both approaches
proceeded independently over the next decade. The equivalence of the DMPK
equation and the $\sigma$ model was finally demonstrated in 1996, by Brouwer
and Frahm. This review is based on the DMPK equation. The $\sigma$ model is
reviewed extensively in a monograph by Efetov (1996).

\begin{figure}[tb]
\centerline{
\psfig{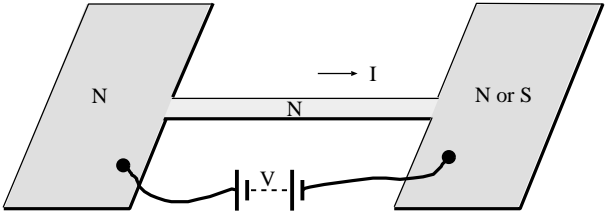}
}%
\medskip
\caption[]{
A current $I$ is passed through a conductor by connecting it to two electron
reservoirs (shaded) at a voltage difference $V$. The conductor and one of the
two reservoirs are normal metals (N), while the other reservoir may be in the
superconducting state (S). }\label{fig_reservoirs}
\end{figure}

In order to study electronic transport through a quantum dot or a disordered
wire, it has to be connected to two electron reservoirs (see Fig.\
\ref{fig_reservoirs}). A current is passed through the system by bringing the
reservoirs out of equilibrium. In Secs.\ \ref{dots} and \ref{wires} we assume
that both reservoirs are in the normal state. In Sec.\ \ref{junctions} we
consider the case that one of the two reservoirs is a superconductor. At the
interface between the normal metal and the superconductor a peculiar scattering
process occurs, discovered in 1964 by Andreev. This scattering process, known
as Andreev reflection, converts dissipative current in the normal metal into
dissipationless supercurrent in the superconductor. Andreev reflection modifies
the quantum interference effects existing in the normal state, and introduces
new effects as well. Random-matrix theory is particularly suited to contrast
the two cases of normal and superconducting reservoirs, because the same
scattering-matrix ensembles can be used. [For reviews devoted solely to
normal-metal--superconductor junctions, see Klapwijk (1994), Beenakker (1995),
and Van Wees and Takayanagi (1997).] If both reservoirs are superconductors,
then the system is a Josephson junction which supports a current in
equilibrium. This current is a thermodynamic, rather than a (non-equilibrium)
transport property, and will not be considered here. [For a review of the
scattering-matrix approach to the theory of the Josephson effect, see Beenakker
(1992b).]

This review was written in an attempt to provide a complete coverage of the
present status of the random-matrix theory of quantum transport. Mindful of my
own limitations, I apologize to those whose works I have overlooked or not
sufficiently appreciated. No attempt was made to include other theories of
transport, nor random-matrix theories of other than transport properties.
Moreover, the adjective ``quantum'' is meant to exclude classical waves. Many
of the effects described here have optical analogues which can be studied by
the same random-matrix techniques. This provides an interesting opportunity for
future research, which we will touch on in Sec.~\ref{conclusion}.

\section{Quantum dots}
\label{dots}

A cavity of sub-micron dimensions, etched in a semiconductor is called a
quantum dot. Quantum mechanical phase-coherence strongly affects its electronic
properties, hence the adjective ``quantum''. We consider the generic case that
the classical motion in the cavity can be regarded as chaotic, on time scales
long compared to the ergodic time $\tau_{\rm ergodic}$. As discussed in Sec.\
\ref{statisticalE}, the Hamiltonian of this closed system is then distributed
according to the Wigner-Dyson ensemble, on energy scales small compared to the
Thouless energy $E_{\rm c,closed}\simeq\hbar/\tau_{\rm ergodic}$. In order of
magnitude, $E_{\rm c,closed}\simeq(\hbar v_{\rm F}/L^{2})\,{\rm min}\,(l,L)$ in
a cavity of linear dimension $L$, mean free path $l$, and Fermi velocity
$v_{\rm F}$. It does not matter for Wigner-Dyson statistics whether motion
inside the cavity is ballistic ($L\ll l$) or diffusive ($L\gg l$). The material
inside the quantum dot is assumed to be a good metal, which means that $E_{\rm
c,closed}$ should be much greater than the mean level spacing $\delta$. The
Fermi wave length $\lambda_{\rm F}$ in a good metal is much smaller than $l$,
so that the wave functions are extended --- rather than localized.

The transport properties of the quantum dot can be measured by coupling it to
two electron reservoirs, and bringing them out of equilibrium. This open system
can still be regarded as chaotic, if the coupling is sufficiently weak that the
mean dwell time\footnote{
The mean dwell time in a chaotic cavity is given by $2\pi\hbar/\tau_{\rm
dwell}= \delta\sum_{n}\Gamma_{n}$, where $\Gamma_{n}$ is the tunnel probability
of mode $n$ through a point contact. For example, in the case of two ballistic
point contacts containing $N_{1},N_{2}$ modes, one has $2\pi\hbar/\tau_{\rm
dwell}=(N_{1}+N_{2})\delta$.}
$\tau_{\rm dwell}$ of an electron exceeds $\tau_{\rm ergodic}$. In terms of
energies, this condition can be written as $E_{\rm c,open}\ll E_{\rm
c,closed}$, where $E_{\rm c,open}\simeq\hbar/\tau_{\rm dwell}$ is the Thouless
energy of the open system. The ratio $E_{\rm c,open}/\delta$ is of the order of
the conductance $G$ of the quantum dot in units of $e^{2}/h$. While we do
require $\delta\ll E_{\rm c,closed}$, we do not restrict the relative magnitude
of $\delta$ and $E_{\rm c,open}$. Under the condition $E_{\rm c,open}\ll E_{\rm
c,closed}$, transport quantities are insensitive to microscopic properties of
the quantum dot, such as the shape of the cavity and the degree of disorder. In
particular, just as in the closed system, it does not matter whether the motion
is ballistic or diffusive inside the cavity. This universality does not extend
to the contacts to the reservoirs: It matters whether the coupling is via
ballistic point contacts or via tunnel barriers. We will see that the
distribution of the scattering matrix is given by the circular ensemble for
ballistic contacts, and by the Poisson kernel for tunneling contacts.

Throughout most of this section we will assume non-interacting electrons. This
is justified if capacitive charging of the quantum dot relative to the
reservoirs is insignificant, which it is if the coupling is via ballistic point
contacts, but usually not if the coupling is via tunnel barriers.

\subsection{Transport theory of a chaotic cavity}
\label{chaotic}

A random-matrix theory of transport through a chaotic cavity can be based
either on an ensemble of scattering matrices or on an ensemble of Hamiltonians.
We introduce these two approaches separately, and then discuss their
relationship and microscopic justification.

\subsubsection{Circular ensemble of scattering matrices}
\label{circular}

Bl\"{u}mel and Smilansky (1990) found that the correlations of the phase shifts
$\phi_{n}$ for chaotic scattering are well described by the distribution
function
\begin{equation}
P(\{\phi_{n}\})\propto\prod_{n<m}\left|\exp({\rm i}\phi_{n})-\exp({\rm
i}\phi_{m})\right|^{\beta}
\label{Pphicircular}
\end{equation}
of the circular ensemble (for a review, see Smilansky, 1990). The circular
ensemble was introduced by Dyson (1962a) as a mathematically more tractable
alternative to the Gaussian ensemble. Baranger and Mello (1994) and Jalabert,
Pichard, and Beenakker (1994) based a transport theory on the circular
ensemble. For this purpose one needs to know the statistics of the transmission
eigenvalues $T_{n}$, which are not directly related to the scattering phase
shifts $\phi_{n}$. (The relationship involves both the eigenvalues and the
eigenfunctions of the scattering matrix.)

The calculation of $P(\{T_{n}\})$ starts from the defining property of the
circular ensemble, which is that the scattering matrix $S$ is {\em uniformly\/}
distributed in the unitary group, subject only to the symmetry and self-duality
constraints imposed by time-reversal and spin-rotation symmetry (see Sec.\
\ref{SandM}). Uniformity is defined with respect to a measure $d\mu(S)$ which
is invariant under multiplication: $d\mu(S)=d\mu(USV)$ for arbitrary unitary
matrices $U,V$ such that the product $USV$ still satisfies the constraints
imposed on $S$. (This requires $V=U^{\rm T}$ for $\beta=1$ and $V=U^{\rm R}$
for $\beta=4$.) This measure is known as the ``invariant measure'' or ``Haar
measure'' (Hamermesh, 1962). The probability distribution in the circular
ensemble is thus given by
\begin{equation}
P(\{x_{n}\})\prod_{i}dx_{i}=\frac{1}{\cal V}\,d\mu(S),\label{PScircular}
\end{equation}
where ${\cal V}=\int d\mu(S)$ is the volume of the matrix space and $\{x_{n}\}$
is a set of independent variables which parameterizes $S$. The general method
to compute the invariant measure in a given parameterization is to consider the
change $dS$ associated with an infinitesimal change $dx_{n}$ in the $x_{n}$'s.
The invariant arclength ${\rm Tr}\,dSdS^{\dagger}$ defines the metric tensor
$g_{ij}$ according to
\begin{equation}
{\rm Tr}\,dSdS^{\dagger}=\sum_{i,j}g_{ij}dx_{i}dx_{j}.\label{metrictensordef}
\end{equation}
The determinant ${\rm Det}\,g$ then yields the invariant measure
\begin{equation}
d\mu(S)=|{\rm Det}\,g|^{1/2}\,\prod_{i}dx_{i},\label{invariantmeasuredef}
\end{equation}
and hence the distribution $P(\{x_{n}\})\propto|{\rm Det}\,g|^{1/2}$.

In the scattering phase-shift representation the measure takes the form (Dyson,
1962a)
\begin{equation}
d\mu(S)=\prod_{n<m}\left|\exp({\rm i}\phi_{n})-\exp({\rm
i}\phi_{m})\right|^{\beta}\,d\mu(U)\prod_{i}d\phi_{i},\label{dmuSphi}
\end{equation}
where $U$ is the matrix of eigenvectors which diagonalizes the scattering
matrix: $(U^{-1}SU)_{nm}=\delta_{nm}\exp({\rm i}\phi_{n})$. The matrix $U$ is
orthogonal, unitary, or symplectic, for $\beta=1,2$, or 4 respectively. (If
$\beta=4$ each eigenvalue is two-fold degenerate, and the products in Eq.\
(\ref{dmuSphi}) include only the distinct eigenvalues.) The invariant measure
(\ref{dmuSphi}) implies that the eigenvectors and eigenvalues of $S$ are
distributed independently. The matrix of eigenvectors $U$ is uniformly
distributed in the orthogonal, unitary, or symplectic group. The eigenvalues
$\exp({\rm i}\phi_{n})$ are distributed according to Eq.\ (\ref{Pphicircular}).
This ensemble is called ``circular'' because the eigenvalue density is constant
on the unit circle in the complex plane. The adjective orthogonal, unitary, or
symplectic is added to distinguish the cases $\beta=1,2$, or 4. Note that this
name derives from the matrix of eigenvectors $U$ --- not from the scattering
matrix $S$ ({\em cf.} Table \ref{table_threefoldS}). For example, the circular
orthogonal ensemble for $\beta=1$ (abbreviated COE) is the ensemble of
uniformly distributed, unitary symmetric matrices. The circular symplectic
ensemble (CSE, $\beta=4$) contains the unitary self-dual matrices, and the
circular unitary ensemble (CUE, $\beta=2$) contains all unitary matrices.

It is sometimes useful to be able to write averages over the COE and CSE as
averages over the CUE.\footnote{
There exists also a relationship between averages over the COE and CSE, for
which we refer to Brouwer and Beenakker (1996a).}
For the COE, consisting of unitary symmetric matrices, this is achieved by the
representation $S=UU^{\rm T}$. Averaging $S$ over the COE is then equivalent to
averaging $U$ over the CUE:
\begin{equation}
\langle f(S)\rangle_{S\in {\rm COE}}=\langle f(UU^{\rm T})\rangle_{U\in {\rm
CUE}}.\label{COECUE}
\end{equation}
For the CSE one first needs to represent the $N\times N$ quaternion matrix $S$
by a $2N\times 2N$ complex matrix $U$. We denote this representation by $S\cong
U$. The dual of $S$ is $S^{\rm R}\cong {\cal C}^{\rm T}U^{\rm T}{\cal C}$,
where ${\cal C}$ is a $2N\times 2N$ matrix with zero elements, except for
${\cal C}_{2i-1,2i}=1$, ${\cal C}_{2i,2i-1}=-1$ ($i=1,2,\ldots N$):
\begin{equation}
{\cal C}=\left(
\begin{array}{cccccccc}
0&1&&&&&&\\
-1&0&0&&&&&\\
&0&0&1&&&&\\
&&-1&0&0&&&\\
&&&\cdot&\cdot&\cdot&&\\
&&&&\cdot&\cdot&\cdot&\\
&&&&&0&0&1\\
&&&&&&-1&0
\end{array}
\right).\label{calCdef}
\end{equation}
Note that ${\cal C}^{\rm T}=-{\cal C}$ and ${\cal C}^{2}=-1$. A self-dual
matrix is represented by $S\cong U{\cal C}^{\rm T}U^{\rm T}{\cal C}$. Averaging
$S$ over the CSE is equivalent to averaging $U$ over the CUE:
\begin{equation}
\langle f(S)\rangle_{S\in {\rm CSE}}=\langle f(U{\cal C}^{\rm T}U^{\rm T}{\cal
C})\rangle_{U\in {\rm CUE}}.\label{CSECUE}
\end{equation}
Averages over the CUE amount to an integration over the unitary group. A few
integration formulas which we will need are collected in Appendix
\ref{UNintegrate}.

The representation of $S$ in terms of the set of transmission eigenvalues
$\{T_{n}\}$ is the polar decomposition (\ref{polarS}). The corresponding
measure is (Baranger and Mello, 1994; Jalabert, Pichard, and Beenakker, 1994)
\begin{equation}
d\mu(S)=\prod_{n<m}|T_{n}-T_{m}|^{\beta}\prod_{k}T_{k}^{-1+\beta/2}
\prod_{\alpha}d\mu(U_{\alpha})\prod_{i}dT_{i},\label{dmuST}
\end{equation}
where $\{U_{\alpha}\}$ is the set of independent unitary matrices in Eq.\
(\ref{polarS}). The polar decomposition (\ref{polarS}) assumes that the two
leads attached to the cavity support the same number of transverse modes, so
that the transmission matrices $t$ and $t'$ are square matrices. More
generally, one can consider the case that the number of modes $N_{1}$ and
$N_{2}$ in the two leads is different, so that $t$ and $t'$ are rectangular
matrices. The two matrix products $tt^{\dagger}$ and $t't'^{\dagger}$ contain a
common set of ${\rm min}(N_{1},N_{2})$ non-zero transmission eigenvalues. Only
these appear in the invariant measure, which in comparison with Eq.\
(\ref{dmuST}) contains an extra factor
$\prod_{k}T_{k}^{\frac{1}{2}\beta|N_{2}-N_{1}|}$ in the exponent (Brouwer,
1994). The resulting probability distribution of the transmission eigenvalues
is
\begin{equation}
P(\{T_{n}\})\propto\prod_{n<m}|T_{n}-T_{m}|^{\beta}\prod_{k}T_{k}^{\frac{1}{2}
\beta(|N_{2}-N_{1}|+1-2/\beta)}. \label{PTcircular}
\end{equation}

\subsubsection{Poisson kernel}
\label{Poissonkernel}

The circular ensemble is the ``most random'' ensemble of scattering matrices.
It would seem a natural choice for a chaotic cavity, which one could call the
``most random'' conductor. As we will see in Sec.\ \ref{microscopicOK}, this
choice has a justification from microscopic theory. The notion of a ``most
random'' ensemble can be made quantitative by associating an information
entropy ${\cal S}=-\int d\mu(S)\, P(S)\ln P(S)$ with the probability
distribution $P(S)$. The most random ensemble is then the ensemble which
maximizes ${\cal S}$, subject to certain constraints (Balian, 1968). For the
circular ensemble the only constraints are the symmetry ($\beta=1$) or
self-duality ($\beta=4$) of $S$. Mello, Pereyra, and Seligman (1985), and more
recently Baranger and Mello (1996a), have considered the additional set of
constraints
\begin{equation}
\int d\mu(S)\,S^{p}P(S)=\bar{S}^{p},\;p=1,2,\ldots,\label{constraintSbar}
\end{equation}
where $\bar{S}$ is a given sub-unitary matrix. (Sub-unitary means that the
eigenvalues of $\bar{S}\bar{S}^{\dagger}$ are $\leq 1$.) The distribution which
maximizes the entropy subject to these constraints is
\begin{equation}
P(S)\propto|{\rm Det}(1-\bar{S}^{\dagger}S)|^{-\beta(N_{1}+N_{2}-1+2/\beta)}.
\label{Poissonkerneldef}
\end{equation}
The circular ensemble is the special case $\bar{S}=0$, so that $P(S)={\rm
constant}$. Equation (\ref{Poissonkerneldef}) generalizes the circular ensemble
to non-zero average scattering matrix $\bar{S}$.

The distribution (\ref{Poissonkerneldef}) is known in the mathematical
literature as the {\em Poisson kernel\/} (Hua, 1963). It was introduced into
random-matrix theory by Krieger (1965), and first applied to a chaotic cavity
by Doron and Smilansky (1992). The name originates from the problem of
determining the analytic function $V(\bar{S})$ of sub-unitary matrices
$\bar{S}$, from the knowledge of $V(S)$ for unitary $S$. This problem is the
multi-dimensional generalization of the two-dimensional electrostatic problem
of computing the potential inside a cylinder from the values it takes on the
surface. The solution $f(\bar{S})=\int d\mu(S)\,f(S)P(S)$ is called Poisson's
formula in the electrostatic context. Equation (\ref{constraintSbar}) is known
as the {\em analyticity-ergodicity constraint\/} (Mello, 1995). The name refers
to the analyticity requirement that $S$ has poles only in the lower half of the
complex-energy plane, and to the ergodicity assumption that ensemble averages
equal spectral averages. Together, these two conditions imply $\langle
S^{p}\rangle=\langle S\rangle^{p}$ (with $p$ a positive integer), so that a
single matrix $\bar{S}$ determines all positive moments of $S$.

\begin{figure}[tb]
\vspace*{-1cm}
\psfig{figure=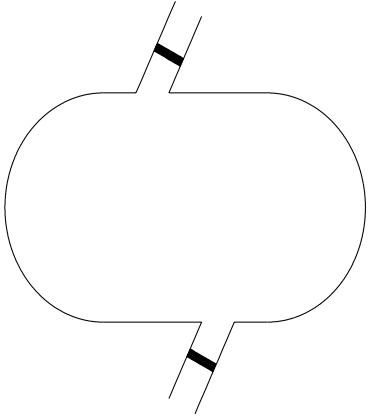,width= 8cm}
\vspace*{-1cm}
\caption[]{
Chaotic cavity (the stadium billiard) coupled to two reservoirs via narrow
leads containing tunnel barriers. The distribution of the scattering matrix is
given by the Poisson kernel (\protect\ref{Poissonkerneldef}), which reduces to
the circular ensemble in the absence of tunnel barriers in the leads.
}\label{fig_dotleads}
\end{figure}

The circular ensemble is appropriate for a chaotic cavity that is coupled to
the leads by means of ballistic point contacts (``ideal'' leads), since the
only property of the coupling which enters is the number of modes $N_{1}$,
$N_{2}$. More generally, one can consider non-ideal leads, containing tunnel
barriers (see Fig.\ \ref{fig_dotleads}). Assume that the segment of the lead
between the tunnel barrier and the cavity is long enough, so that the
scattering matrices $S_{1}$, $S_{2}$ of barriers 1 and 2, as well as the
scattering matrix $S_{0}$ of the cavity are well defined. The scattering matrix
$S$ of the whole structure is obtained by multiplying the three transfer
matrices corresponding to $S_{1}$, $S_{0}$, and $S_{2}$. Brouwer (1995) has
shown that {\em if\/} $S_{0}$ is distributed according to the circular
ensemble, {\em then\/} $S$ is distributed according to the Poisson kernel. The
average scattering matrix in this case is
\begin{equation}
\bar{S}=\left( \begin{array}{cc}
r_{1}&0\\0&r_{2}
\end{array}\right),\label{barSr1r2}
\end{equation}
with $r_{1}$ and $r_{2}$ the reflection matrices of barriers 1 and 2 for
electrons incident from the reservoirs. The eigenvalues of
$1-\bar{S}\bar{S}^{\dagger}$ are the tunnel probabilities $\Gamma_{n}$ through
the leads.

\subsubsection{Gaussian ensemble of Hamiltonians}
\label{GaussianH}

The Hamiltonian approach to transport through a chaotic cavity goes back to
work in the sixties on nuclear reactions (Mahaux and Weidenm\"{u}ller, 1969).
The Hamiltonian of the cavity connected to leads by tunnel barriers is
represented by
\begin{eqnarray}
H&=&\sum_{a}|a\rangle E_{\rm F}\langle a|+\sum_{\mu,\nu}|\mu\rangle{\cal
H}_{\mu\nu}\langle\nu|\nonumber\\
&&\mbox{}+\sum_{\mu,a}\bigl(|\mu\rangle W_{\mu a}^{\vphantom{\ast}}\langle
a|+|a\rangle W_{\mu a}^{\ast}\langle\mu|\bigr).\label{HAMHAM}
\end{eqnarray}
The set $\{|a\rangle\}$ ($a = 1,2,\ldots N$, with $N=N_{1}+N_{2}$ the total
number of propagating modes in the leads) forms a basis of scattering states in
the leads at the Fermi energy $E_{\rm F}$. The set of bound states in the
isolated cavity is denoted by $\{|\mu\rangle\}$ ($\mu = 1,2,\ldots M$). The
finite number $M$ is artificial, and will eventually be taken to infinity. The
matrix elements ${\cal H}_{\mu\nu}$ form a Hermitian $M\times M$ matrix ${\cal
H}$, with real ($\beta=1$), complex ($\beta=2$), or real quaternion ($\beta=4$)
elements. The coupling constants $W_{\mu a}$ form a real (complex, real
quaternion) $M\times N$ matrix $W$, which is assumed to be independent of
energy. The $N\times N$ scattering matrix $S$ associated with the Hamiltonian
(\ref{HAMHAM}) equals\footnote{
The Hamiltonian (\ref{HAMHAM}) gives $S=1$ in the case $W=0$ of an isolated
cavity. A more general Hamiltonian would give $S=S_{0}$, with $S_{0}$ a
``background'' scattering matrix that does not couple to the cavity (Nishioka
and Weidenm\"{u}ller, 1985). This more general case amounts to the
transformation $S\rightarrow USV$, where the unitary matrices $U$ and $V$ are
independent of ${\cal H}$.}
\begin{eqnarray}
S&=&1-2\pi{\rm i}W^{\dagger}(E_{\rm F}-{\cal H}+{\rm i}\pi
WW^{\dagger})^{-1}W\nonumber\\
&=&\frac{1+{\rm i}\pi W^{\dagger}({\cal H}-E_{\rm F})^{-1}W}{1-{\rm i}\pi
W^{\dagger}({\cal H}-E_{\rm F})^{-1}W}.\label{SHeq}
\end{eqnarray}
One verifies that, for $\beta=1,2,4$, the matrix $S$ is unitary symmetric,
unitary, and unitary self-dual, respectively.

The Hamiltonian ${\cal H}$ of a chaotic cavity is distributed according to the
Gaussian ensemble,
\begin{equation}
P({\cal H})\propto\exp\left(-\beta(\pi/2\delta)^{2}M^{-1}\,{\rm Tr}\,{\cal
H}^2\right).\label{GaussEns}
\end{equation}
The coefficient $\delta$ equals the mean level spacing at the Fermi level in
the limit $M\rightarrow\infty$ (Mehta, 1991). The average scattering matrix
$\bar{S}$ in this limit is given by (Verbaarschot, Weidenm\"{u}ller, and
Zirnbauer, 1985)
\begin{equation}
\bar{S}=\frac{M\delta-\pi^{2}W^{\dagger}W}{M\delta+\pi^{2}W^{\dagger}W}.
\label{SbarSupSym}
\end{equation}
Comparison with Eq.\ (\ref{barSr1r2}) shows that the eigenvalue $w_{n}$ of the
coupling-matrix product $W^{\dagger}W$ is related to the tunnel probability
$\Gamma_{n}$ of mode $n$ in the lead by
\begin{mathletters}
\label{w2Gamma}
\begin{eqnarray}
&&\Gamma_{n}=\frac{4M\delta\pi^{2}w_{n}}{(M\delta+\pi^{2}w_{n})^{2}},
\label{w2Gammaa}\\
&&w_{n}=\frac{M\delta}{\pi^{2}\Gamma_{n}} \left(2-\Gamma_{n}\pm
2\sqrt{1-\Gamma_{n}}\right).\label{w2Gammab}
\end{eqnarray}
\end{mathletters}%
Notice that $\Gamma_{n}$ does not determine $w_{n}$ uniquely.

The approach of coupling $M$ eigenstates of the cavity to $N$ scattering
channels in the leads introduces a large number of coupling constants $W_{\mu
a}$, while a much smaller number of parameters $\Gamma_{n}$ determine the
transport properties at the Fermi level. This is why the scattering-matrix
approach is more convenient than the Hamiltonian approach, in cases that the
energy dependence of the transport properties is not required. The equivalence
of the two approaches is discussed in the next subsection, together with the
microscopic justification.

\subsubsection{Justification from microscopic theory}
\label{microscopicOK}

A microscopic justification for the Gaussian ensemble has been provided by
Efetov (1982, 1983) for a disordered metal grain, and by Andreev, Agam, Simons,
and Altshuler (1996) for a chaotic billiard ({\em cf.} Sec.\
\ref{geometrical}). A microscopic justification for the circular ensemble and
the Poisson kernel has been provided indirectly by a demonstration of the
equivalence with the Gaussian ensemble (Brouwer, 1995; see also Lewenkopf and
Weidenm\"{u}ller, 1991). We present an outline of Brouwer's equivalence proof.
It proceeds in two steps. The first step is to show that, in the limit
$M\rightarrow\infty$, the Gaussian distribution (\ref{GaussEns}) can be
replaced by the Lorentzian distribution
\begin{equation}
P({\cal H})\propto{\rm Det}\,\bigl[(M\delta/\pi)^{2}+{\cal
H}^{2}\bigr]^{-(\beta M+2-\beta)/2}.\label{LorEns}
\end{equation}
The second step is to show that, for any $M\geq N$, the distribution $P(S)$ of
the scattering matrix obtained from Eq.\ (\ref{LorEns}) is the Poisson kernel
(\ref{Poissonkerneldef}).

The replacement of (\ref{GaussEns}) by (\ref{LorEns}) is allowed because the
eigenvector and eigenvalue distributions of the Gaussian and the Lorentzian
ensemble are equal on a fixed energy scale, in the limit $M\rightarrow\infty$.
The equivalence of the eigenvector distributions is obvious: The distribution
of ${\cal H}$ depends solely on the eigenvalues for both the Lorentzian and the
Gaussian ensemble, so that the eigenvector distribution is uniform for both
ensembles. The equivalence of the distribution of the eigenvalues is proven by
an explicit comparison of the $p$-point correlation functions. (These can be
computed exactly in both the Gaussian and the Lorentzian ensembles, using the
method of orthogonal polynomials.)

The technical reason for working with the Lorentzian ensemble instead of with
the Gaussian ensemble is the invariance property that, if ${\cal H}$ has a
Lorentzian distribution, then its inverse ${\cal H}^{-1}$ as well as any
submatrix of ${\cal H}$ have a Lorentzian distribution. This property makes it
particularly easy to compute the distribution of the scattering matrix, for any
$M\geq N$. The resulting distribution has the form of a Poisson kernel,
\begin{eqnarray}
&&P(S)\propto|{\rm Det}(1-\bar{S}^{\dagger}S)|^{-\beta N-2+\beta},
\label{Poissonkernel2}\\
&&\bar{S}=\frac{M\delta-{\rm i}\pi E_{\rm F}-\pi^{2}W^{\dagger}W}{M\delta-{\rm
i}\pi E_{\rm F}+\pi^{2}W^{\dagger}W}. \label{SbarSupSym2}
\end{eqnarray}
Equation (\ref{SbarSupSym}) for the average scattering matrix is recovered from
Eq.\ (\ref{SbarSupSym2}) in the limit $M\rightarrow\infty$.

The conclusion is that the Poisson kernel for the distribution of scattering
matrices (and in particular the circular ensemble, to which the Poisson kernel
reduces for $\bar{S}=0$) is equivalent to the Lorentzian ensemble of
Hamiltonians for any $M\geq N$. The Lorentzian ensemble, in turn, is equivalent
in the limit $M\rightarrow\infty$ to the Gaussian ensemble, which for a chaotic
cavity has been derived from microscopic theory. This provides the microscopic
justification for the random-matrix theory of transport through a quantum dot.

\subsection{Weak localization}
\label{WLdot}

Consider a chaotic cavity with two small holes of the same size. An electron
which is injected through one of the holes will exit either through the same
hole (reflection) or through the other hole (transmission). Classically,
chaotic motion in the cavity implies that the transmission and reflection
probabilities are equal. Quantum mechanically, the transmission probability is
slightly smaller than the reflection probability. This effect is known as
``weak localization'', after the analogous effect in disordered metals. In a
semiclassical formulation, the enhancement of the reflection probability is due
to the constructive interference of pairs of time-reversed trajectories
(Baranger, Jalabert, and Stone, 1993a, 1993b; Argaman, 1995, 1996; Aleiner and
Larkin, 1996). A magnetic field breaks the time-reversal symmetry, thereby
destroying the constructive interference and equalizing the transmission and
reflection probabilities. The magnitude of the weak-localization effect in a
quantum dot was first computed by Iida, Weidenm\"{u}ller, and Zuk (1990a,
1990b), using the Hamiltonian approach described in Sec.\ \ref{GaussianH}. The
calculation is easier using the scattering-matrix approach of Sec.\
\ref{circular} (Baranger and Mello, 1994; Jalabert, Pichard, and Beenakker,
1994). Using the latter approach, we discuss the weak-localization correction
to the conductance and the generalization to other transport properties.

\subsubsection{Conductance}
\label{WLdotG}

In the absence of time-reversal symmetry, the scattering matrix $S$ of a
chaotic cavity is uniformly distributed over the unitary group.  This is the
circular unitary ensemble (CUE, $\beta=2$). The average of the scattering
probability $|S_{nm}|^{2}$ follows from Eq.\ (\ref{UU}),
\begin{equation}
\langle |S_{nm}|^{2}\rangle_{\rm CUE}=\int
d\mu(S)\,S^{\vphantom{\ast}}_{nm}S^{\ast}_{nm}=\frac{1}{N_{1}+N_{2}}.
\label{SCUEaverage}
\end{equation}
(The integral is over ${\cal U}(N_{1}+N_{2})$ with invariant measure $d\mu(S)$,
normalized such that $\int d\mu(S)=1$, where $N_{1}$ and $N_{2}$ are the number
of modes in the leads connected to contacts 1 and 2.) In the CUE, scattering
between two different modes ($n\neq m$) is equally probable as between two
identical modes ($n=m$). In the presence of time-reversal symmetry, $S$ is
unitary and symmetric (assuming no spin-orbit scattering). This is the circular
orthogonal ensemble (COE, $\beta=1$). The average of $|S_{nm}|^{2}$ follows
from Eqs.\ (\ref{COECUE}) and (\ref{UUUU}),
\begin{eqnarray}
\langle |S_{nm}|^{2}\rangle_{\rm COE}&=&\int
d\mu(U)\,\sum_{k,k'=1}^{N_{1}+N_{2}}U^{\vphantom{\ast}}_{nk}
U^{\vphantom{\ast}}_{mk} U^{\ast}_{nk'}U^{\ast}_{mk'}\nonumber\\
&=&\frac{1+\delta_{nm}}{N_{1}+N_{2}+1}.\label{SCOEaverage}
\end{eqnarray}
Scattering from mode $n$ back to mode $n$ is now twice as probable as from mode
$n$ into another mode $m$. The absence of spin-orbit scattering is essential.
In the circular symplectic ensemble (CSE, $\beta=4$) one obtains from Eqs.\
(\ref{CSECUE}) and (\ref{UUUU}) the average\footnote{
The absolute value $|Q|$ of a quaternion number $Q$ (represented by a $2\times
2$ matrix) is defined by $|Q|^{2}=\frac{1}{2}\,{\rm Tr}\,QQ^{\dagger}$.}
\begin{eqnarray}
\langle |S_{nm}|^{2}\rangle_{\rm CSE}&=&\frac{1}{2}\sum_{p,q=0}^{1}\int
d\mu(U)\,\bigl(U{\cal C}U^{\rm T}{\cal C}\bigr)_{2n-p,2m-q}\nonumber\\
&&\hspace{2cm}\times\bigl(U^{\ast}{\cal C}U^{\dagger}{\cal
C}\bigr)_{2n-p,2m-q}\nonumber\\
&=&\frac{2-\delta_{nm}}{2N_{1}+2N_{2}-1}.\label{SCSEaverage}
\end{eqnarray}
[The integration is over ${\cal U}(2N_{1}+2N_{2})$.] Scattering between the
same mode is now less probable than between different modes. Equations
(\ref{SCUEaverage})--(\ref{SCSEaverage}) can be summarized in one
$\beta$-dependent expression,
\begin{equation}
\langle
|S_{nm}|^{2}\rangle=\frac{1-(1-2/\beta)\delta_{nm}}{N_{1}+N_{2}-1+2/\beta}.
\label{SCEaverage}
\end{equation}

According to the Landauer formula (\ref{Landauer}), the conductance $G$ of the
quantum dot is obtained from the scattering probabilities $|S_{nm}|^{2}$ by
summing $n$ over all $N_{1}$ modes in lead 1, and summing $m$ over all $N_{2}$
modes in lead 2,
\begin{equation}
G=G_{0}\sum_{n=1}^{N_{1}}\sum_{m=N_{1}+1}^{N_{1}+N_{2}}|S_{nm}|^{2},
\label{GSnm}
\end{equation}
where $G_{0}=2e^{2}/h$. Substitution of Eq.\ (\ref{SCEaverage}) into Eq.\
(\ref{GSnm}) yields the average conductance (Baranger and Mello, 1994)
\begin{equation}
\langle
G/G_{0}\rangle=\frac{N_{1}N_{2}}{N_{1}+N_{2}-1+2/\beta}.\label{GSnmaveragea}
\end{equation}
For $N_{1}\gg 1$, $N_{2}\gg 1$ we may expand
\begin{equation}
\langle
G/G_{0}\rangle=\frac{N_{1}N_{2}}{N_{1}+N_{2}}+\left(1-\frac{2}{\beta}\right)
\frac{N_{1}N_{2}}{(N_{1}+N_{2})^{2}}.\label{GSnmaverageb}
\end{equation}
The first term in Eq.\ (\ref{GSnmaverageb}) is the classical series conductance
$G_{\rm series}=G_{0}(N_{1}^{-1}+N_{2}^{-1})^{-1}$ of the two contact
conductances $N_{1}G_{0}$ and $N_{2}G_{0}$. The second term is the
weak-localization correction $\delta G$. For the case of two identical contacts
one has simply
\begin{equation}
\delta G/G_{0}=\frac{1}{4}\left(1-\frac{2}{\beta}\right),\;\;N_{1}=N_{2}\gg
1.\label{deltaGSnm}
\end{equation}

\subsubsection{Other transport properties}
\label{WLdotother}

To compute the weak-localization correction for other transport properties, one
needs the density of the transmission eigenvalues. We use the parameterization
$T_{n}=1/(1+\lambda_{n})$ and write the probability distribution
(\ref{PTcircular}) of the $N_{\rm min}\equiv{\rm min}\,(N_{1},N_{2})$ non-zero
transmission eigenvalues in the form of a Gibbs distribution:
\begin{mathletters}
\label{PGibbsCU}
\begin{eqnarray}
&&P(\{\lambda_{n}\})\propto\exp\left[\beta\sum_{i<j}
\ln|\lambda_{i}-\lambda_{j}|
-\beta\sum_{i}V(\lambda_{i})\right],\nonumber\\
\label{PGibbsCUa}\\
&&V(\lambda)=\case{1}{2}\bigl(N_{1}+N_{2}-1+2/\beta\bigr)\ln(1+\lambda).
\label{PGibbsCUb}
\end{eqnarray}
\end{mathletters}%
The density $\rho(\lambda)=\langle\sum_{n}\delta(\lambda-\lambda_{n})\rangle$
of the $\lambda$'s is determined for $N_{\rm min}\gg 1$ by the integral
equation (\ref{Dysonint}). We decompose $\rho=\rho_{0}+\delta\rho$ into a
contribution $\rho_{0}$ of order $N_{\rm min}$ and a correction $\delta\rho$ of
order unity. Similarly, we decompose the potential $V=V_{0}+\delta V$ into two
terms: $V_{0}=\case{1}{2}(N_{1}+N_{2})\ln(1+\lambda)$, $\delta
V=-\case{1}{2}(1-2/\beta)\ln(1+\lambda)$.

The leading-order contribution $\rho_{0}$ satisfies
\begin{equation}
\int_{0}^{\lambda_{\rm
c}}\!d\lambda'\,\frac{\rho_{0}(\lambda')}{\lambda-\lambda'}
=\frac{d}{d\lambda}V_{0}(\lambda),\label{DysonintCU}
\end{equation}
where the singular integral is the principal value. [Equation
(\ref{DysonintCU}) is the derivative with respect to $\lambda$ of Eq.\
(\ref{wignerint}).] The density $\rho_{0}$ vanishes for
$\lambda\geq\lambda_{\rm c}$. The general solution of this integral equation,
with normalization $\int\rho_{0}\,d\lambda=N_{\rm min}$, is (Mikhlin, 1964)
\begin{eqnarray}
\rho_{0}(\lambda)&=&\pi^{-2}[\lambda(\lambda_{\rm c}-\lambda)]^{-1/2}\left(\pi
N_{\rm min}\vphantom{\int_{0}^{\lambda_{\rm c}}}\right.\nonumber\\
&&\left.\mbox{}-\int_{0}^{\lambda_{\rm c}}\!\!d\lambda'\,
\frac{[\lambda'(\lambda_{\rm c}-\lambda')]^{1/2}}{\lambda-\lambda'}
\frac{d}{d\lambda'}V_{0}(\lambda')\right).\label{Mikhlinsol}
\end{eqnarray}
The free boundary $\lambda_{\rm c}$ is to be determined from
$\rho_{0}(\lambda_{\rm c})=0$. The resulting density is
\begin{mathletters}
\label{rholambdaCU}
\begin{eqnarray}
&&\rho_{0}(\lambda)=\frac{1}{\pi}(N_{1}N_{2})^{1/2}\frac{1}{1+\lambda}
\left(\frac{1}{\lambda}-\frac{1}{\lambda_{\rm
c}}\right)^{1/2},\label{rholambdaCUa}\\
&&\lambda_{\rm c}=\frac{4N_{1}N_{2}}{(N_{1}-N_{2})^{2}},\label{rholambdaCUb}
\end{eqnarray}
\end{mathletters}%
in agreement with a calculation using a different method by Nazarov (1995a).
For $N_{1}=N_{2}$ one may put $\lambda_{\rm c}\rightarrow\infty$, and the
density simplifies to (Jalabert, Pichard, and Beenakker, 1994; Baranger and
Mello, 1994)
\begin{equation}
\rho_{0}(\lambda)=(N_{1}/\pi)(1+\lambda)^{-1}\lambda^{-1/2}.
\label{rholambdaCUc}
\end{equation}

Linearization of Eq.\ (\ref{Dysonint}) around $\rho_{0}$ yields an equation for
$\delta\rho$,
\begin{eqnarray}
\int_{0}^{\lambda_{\rm
c}}\!d\lambda'\,\frac{\delta\rho(\lambda')}{\lambda-\lambda'}
&=&\frac{d}{d\lambda}\delta
V(\lambda)-\case{1}{2}(1-2/\beta)\frac{d}{d\lambda}\ln\rho_{0}(\lambda)
\nonumber\\
&=&\case{1}{4}(1-2/\beta)\left[\lambda^{-1}-(\lambda-\lambda_{\rm
c})^{-1}\right].\label{DysonintCU2}
\end{eqnarray}
The solution is a delta-function peak at the two ends of the spectrum, with the
same weight but opposite sign,
\begin{equation}
\delta\rho(\lambda)=\case{1}{4}(1-2/\beta)\bigl[\delta(\lambda-0^{+})-
\delta(\lambda-\lambda_{\rm c}+0^{+})\bigr].\label{deltarhoCU}
\end{equation}

Transforming back from $\lambda$ to $T$, one obtains the density of
transmission eigenvalues $\rho(T)=\rho(\lambda)d\lambda/dT$. The average of a
linear statistic $A=\sum_{n}a(T_{n})$ then follows upon integration,
\begin{mathletters}
\label{ACUTc}
\begin{eqnarray}
&&\langle A\rangle=\frac{N_{1}+N_{2}}{2\pi}\int_{T_{\rm
c}}^{1}\!dT\,\left(\frac{T-T_{\rm c}}{1-T}\right)^{1/2}
\frac{a(T)}{T}\nonumber\\
&&\hspace{2.5cm}\mbox{}+\case{1}{4}(1-2/\beta)\bigl[a(1)-a(T_{\rm
c})\bigr],\label{ACUaverage}\\
&&T_{\rm c}=\left(\frac{N_{1}-N_{2}}{N_{1}+N_{2}}\right)^{2}.\label{Tcdef}
\end{eqnarray}
\end{mathletters}%
One verifies that the result (\ref{GSnmaverageb}) for the conductance is
recovered for $a(T)=T$. As an example of another transport property, we take
the shot-noise power (\ref{PTrelation}). Substitution of $a(T)=T(1-T)$ into
Eq.\ (\ref{ACUTc}) yields
\begin{equation}
\langle P/P_{0}\rangle=\frac{(N_{1}N_{2})^{2}}{(N_{1}+N_{2})^{3}}
-\left(1-\frac{2}{\beta}\right)\frac{N_{1}N_{2}(N_{1}-N_{2})^{2}}
{(N_{1}+N_{2})^{4}}.\label{PCUresult}
\end{equation}
The weak-localization correction vanishes if $N_{1}=N_{2}$ (Jalabert, Pichard,
and Beenakker, 1994).

\subsubsection{Tunnel barriers}
\label{WLdottunnel}

In the presence of tunnel barriers in the leads, the distribution of the
scattering matrix $S$ is given by the Poisson kernel (\ref{Poissonkerneldef}).
Eq.\ (\ref{barSr1r2}) relates the ensemble-averaged scattering matrix $\bar{S}$
which appears in the Poisson kernel to the reflection matrices of the tunnel
barriers. (The circular ensemble corresponds to $\bar{S}=0$.) The eigenvalue
$\Gamma_{n}$ of $1-\bar{S}\bar{S}^{\dagger}$ is the tunnel probability of mode
$n$ in the lead. The fluctuating part $\delta S\equiv S-\bar{S}$ of $S$ can be
parameterized as
\begin{equation}
\delta S=A(1-UB)^{-1}UC,\label{eq:Sparam}
\end{equation}
where $U$ is a unitary matrix and the matrices $A$,$B$,$C$ are such that the
matrix
\begin{equation}
S_{\rm barrier}=\left(
\begin{array}{cc}
\bar{S}&A\\C&B
\end{array}
\right)\label{eq:SigmaTdef}
\end{equation}
is unitary. In zero magnetic field one should require furthermore that $U$ and
$S_{\rm barrier}$ are symmetric (or self-dual in the presence of spin-orbit
scattering). The usefulness of the parameterization (\ref{eq:Sparam}) is that
$U$ is distributed according to the circular ensemble, for any choice of
$A,B,C$ (Hua, 1963; Friedman and Mello, 1985a; Brouwer, 1995). Physically, $U$
corresponds to the scattering matrix of the cavity without the tunnel barriers
in the leads and $S_{\rm barrier}$ corresponds to the scattering matrix of the
tunnel barriers in the absence of the cavity.

The parameterization (\ref{eq:Sparam}) reduces the problem of computing the
average conductance to an integration of $U$ over the unitary group. The result
of the integration will depend on $\bar{S}$, but not on $A,B,C$. Because the
conductance (\ref{GSnm}) is a rational function of $U$, the integration can not
be carried out in closed form. For $N\Gamma\gg 1$ a perturbative calculation is
possible (Brouwer and Beenakker, 1996a). The result is
\begin{eqnarray}
&&\langle G/G_{0}\rangle=\frac{g_{1}^{\vphantom{2}}g_{1}'} {g_{1}\vphantom{'}+
g_{1}'}+\left(1-\frac{2}{\beta}\right)
\frac{g_{2}^{\vphantom{2}}g_{1}'^{2}+g_{2}'g_{1}^{2}}
{(g_{1}^{\vphantom{2}}+g_{1}')^{3}}, \label{eq:Gtunnel}\\
&&g_{p}^{\vphantom{2}}=\sum_{n=1}^{N_{1}}\Gamma_{n}^{p},\;\;
g_{p}'=\sum_{n=N_{1}+1}^{N_{1}+N_{2}}\Gamma_{n}^{p}. \label{eq:gkdef}
\end{eqnarray}
The first term in Eq.\ (\ref{eq:Gtunnel}) is the classical series conductance
of the two tunnel conductances $G_{0}g_{1}$ and $G_{0}g_{1}'$. The term
proportional to $1-2/\beta$ is the weak-localization correction. In the absence
of tunnel barriers one has $g_{p}=N_{1}$, $g_{p}'=N_{2}$, and Eq.\
(\ref{GSnmaverageb}) is recovered. In the case of two identical tunnel barriers
($N_{1}=N_{2}$, $\Gamma_{n}=\Gamma_{n+N_{1}}$ for $n=1,2,\ldots N_{1}$), Eq.\
(\ref{eq:Gtunnel}) simplifies to (Iida, Weidenm\"{u}ller, and Zuk, 1990a,
1990b)
\begin{equation}
\langle G/G_{0}\rangle=\frac{g_{1}}{2}+
\left(1-\frac{2}{\beta}\right)\frac{g_{2}}{4g_{1}}. \label{eq:GtunnelId}
\end{equation}
If all $\Gamma_{n}$'s are equal to $\Gamma$, Eq.\ (\ref{eq:GtunnelId})
simplifies further to $\langle G/G_{0}\rangle=\case{1}{2}N_{1}\Gamma+
\case{1}{4}(1-2/\beta)\Gamma$.

\subsubsection{Magnetoconductance}
\label{magnetoG}

A weak magnetic field suppresses the weak-localization correction to the
average conductance. In the absence (presence) of spin-orbit scattering, the
magnetoconductance consists of a dip (peak) around $B=0$ of magnitude $\delta
G\simeq e^{2}/h$ and width $B_{\rm c}$. The flux $\Phi_{\rm c}$ through the
particle at magnetic field $B_{\rm c}$ is of order $\Phi_{\rm
c}\simeq(h/e)(E_{\rm c,open}/E_{\rm c,closed})^{1/2}$. This is the flux at
which time-reversal symmetry is broken on the energy scale $E_{\rm c,open}$
({\em cf.} Sec.\ \ref{transitions}). Up to a numerical coefficient of order
unity, one has
\begin{equation}
\Phi_{\rm c}\simeq\frac{h}{e}\left(\frac{\tau_{\rm ergodic}}{\tau_{\rm
dwell}}\right)^{1/2}\simeq\frac{h}{e}\left(\frac{N\Gamma L^{2}\delta}{\hbar
v_{\rm F}\,{\rm min}(l,L)}\right)^{1/2},\label{Phicapproximate}
\end{equation}
where $N$ is the total number of modes in the point contacts, $\Gamma$ the
tunnel probability per mode, $L$ the size of the particle, $\delta$ its level
spacing, $v_{\rm F}$ the Fermi velocity, and $l$ the mean free path.

The magnetoconductance has been calculated in the Hamiltonian approach of Sec.\
\ref{GaussianH}, by replacing the distribution (\ref{GaussEns}) of the $M\times
M$ matrix ${\cal H}$ by the distribution (\ref{PMHdist}) of the Pandey-Mehta
Hamiltonian ${\cal H}={\cal H}_{0}+{\rm i}\alpha{\cal A}$. In the absence of
spin-orbit scattering and for $\Gamma=1$, $N_{1}=N_{2}\gg 1$ the result is
(Pluha\u{r} {\em et al.}, 1994, 1995; Frahm, 1995b):
\begin{mathletters}
\label{magnetoGresult}
\begin{eqnarray}
&&\langle G/G_{0}\rangle=\case{1}{2}N_{1}- \case{1}{4}\bigl[1+(\Phi/\Phi_{\rm
c})^{2}\bigr]^{-1},\label{magnetoGresulta}\\
&&\Phi/\Phi_{\rm c}= 2\alpha\sqrt{M/N_{1}},\label{magnetoGresultb}
\end{eqnarray}
\end{mathletters}%
in agreement with Efetov's (1995) calculation starting from a microscopic
Hamiltonian for a disordered metal grain.\footnote{
Efetov finds for a disordered grain (volume $V$, diffusion coefficient $D$) the
relation \[
(\Phi/\Phi_{\rm c})^{2}=16\pi^{3}(\hbar D/N_{1}\delta)(e/h)^{2} V^{-1}\int_{V}
d\vec{r}\,|\vec{A}|^{2}.
\]
The gauge of the vector potential $\vec{A}$ is such that the normal component
$\hat{n}\cdot\vec{A}$ on the surface of the grain vanishes. This relation
agrees with Eqs.\ (\ref{alphadisordered}) and (\ref{magnetoGresultb}) for a
disordered disk or sphere.}
The Lorentzian flux dependence in Eq.\ (\ref{magnetoGresult}) was first
obtained by Baranger, Jalabert, and Stone (1993a, 1993b), using a semiclassical
theory (reviewed by Baranger, 1996). [They could not derive the prefactor $1/4$
of the Lorentzian, because of an inconsistency in the semiclassical
approximation resolved later by Argaman (1995, 1996).] Unlike random-matrix
theory, the semiclassical theory can also be applied to the case that the
classical motion in the cavity is integrable rather than chaotic. In the
integrable case, Baranger, Jalabert, and Stone find a magnetoconductance which
is {\em linear}, $G(B)-G(0)\propto|B|$, rather than Lorentzian.

\begin{figure}[tb]
\psfig{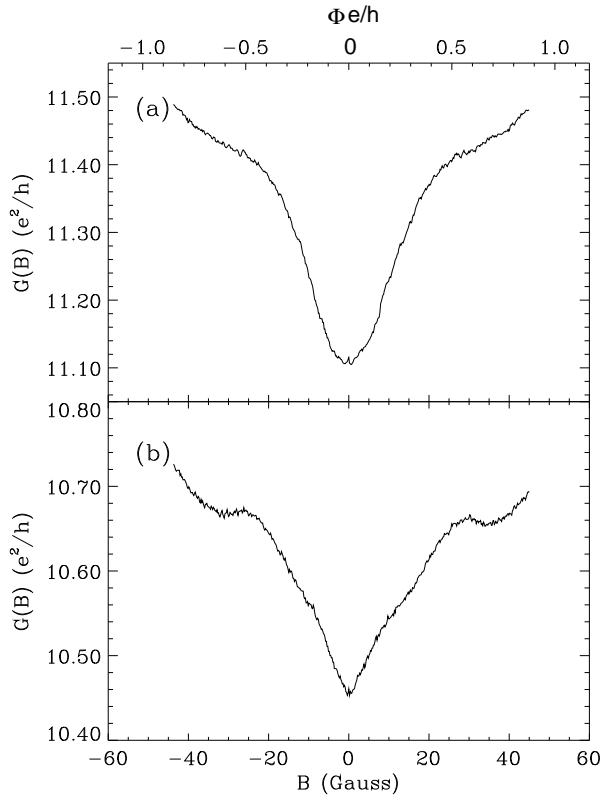}
\medskip
\caption[]{
Magnetoconductance at $50\,{\rm mK}$, averaged over (a) 48 stadium-shaped
cavities and (b) 48 circular-shaped cavities. Insets show the geometry of the
cavities, which are fabricated in the two-dimensional electron gas of a
GaAs/AlGaAs heterostructure. The weak-localization peak has a Lorentzian shape
for the stadium and a triangular shape (linearly decreasing) for the circle, as
expected theoretically for, respectively, chaotic and integrable billiards.
After Chang, Baranger, Pfeiffer, and West (1994).
}\label{fig_Chang1}
\end{figure}

The different flux dependence in the chaotic and integrable cases has been
observed by Chang, Baranger, Pfeiffer, and West (1994). They measured transport
through an array of 48 nominally identical quantum dots connected in series and
parallel. The quantum dots were fabricated in the two-dimensional electron gas
of a GaAs/AlGaAs heterostructure. By measuring on an array, the ensemble
average of the conductance is obtained directly. [Magnetoconductance
experiments on a single quantum dot were done by Marcus {\em et al.} (1992) and
Keller {\em et al.} (1994); Ensemble averaging by varying the shape of the
cavity was done by Chan {\em et al.} (1995), and by varying the Fermi energy by
Keller {\em et al.} (1996).] Two types of arrays were studied, containing
quantum dots in the shape of a stadium or a circle. (Their area was the same,
about $0.8\,\mu{\rm m}^{2}$, and $N_{1}\approx N_{2}\approx 10$ in both cases.)
The classical ballistic motion is chaotic in the stadium and integrable in the
circle. The magnetic-field dependence of the conductance in the two cases is
shown in Fig.\ \ref{fig_Chang1}. (The conductance of the array has been
normalized to that of a single cavity.) The shape of the weak-localization peak
is strikingly different, consistent with the theoretical prediction. Good
agreement could be obtained with a numerical simulation which included
small-angle scattering by a smooth disorder potential. The measured magnitude
$\delta G$ of the peak in the chaotic case is $0.2\,G_{0}$, somewhat smaller
than the theoretical value of $\frac{1}{4}\,G_{0}$. (The discrepancy can be
accounted for by inelastic scattering, {\em cf.} Sec.\ \ref{phasebreaking}.)

\subsection{Universal conductance fluctuations}
\label{UCFdot}

\subsubsection{Conductance}
\label{UCFdotG}

Weak localization is a quantum correction of order $e^{2}/h$ to the
ensemble-averaged conductance. The fluctuations of the conductance from one
member of the ensemble to the other are also of order $e^{2}/h$. These
fluctuations are known as ``universal conductance fluc\-tua\-tions'' ({\em cf.}
Sec.\ \ref{UCF}). The magnitude of the conductance fluctuations in a chaotic
cavity has been calculated in the Hamiltonian approach by Iida,
Weidenm\"{u}ller, and Zuk (1990a, 1990b), and in the scattering-matrix approach
by Baranger and Mello (1994) and Jalabert, Pichard, and Beenakker (1994).

The variance ${\rm Var}\,G=\langle G^{2}\rangle-\langle G\rangle^{2}$ of the
conductance which results from averaging over the circular ensemble of
scattering matrices is
\begin{eqnarray}
&&{\rm Var}\,G/G_{0}=2\beta^{-1}N_{1}N_{2}(N_{1}-1+2/\beta)
(N_{2}-1+2/\beta)\nonumber\\
&&\hspace{1cm}\mbox{}\times(N_{1}+N_{2}-2+2/\beta)^{-1}
(N_{1}+N_{2}-1+4/\beta)^{-1}\nonumber\\
&&\hspace{1cm}\mbox{}\times(N_{1}+N_{2}-1+2/\beta)^{-2}. \label{eq:QDvarC}
\end{eqnarray}
For $N_{1},N_{2}\gg 1$ we may expand
\begin{equation}
{\rm Var}\,G/G_{0}=\frac{2(N_{1}N_{2})^{2}}
{\beta(N_{1}+N_{2})^{4}},\label{VarGdotNbig}
\end{equation}
which for two identical contacts simplifies further to
\begin{equation}
{\rm Var}\,G/G_{0}=\frac{1}{8}\beta^{-1},\;N_{1}=N_{2}\gg
1.\label{VarGdotNequalbig}
\end{equation}

\begin{figure}[tb]
\psfig{figure=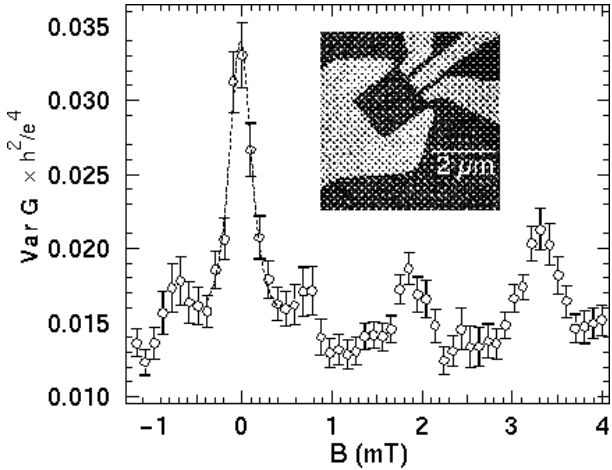,width= 8cm}
\medskip
\caption[]{
Variance of the conductance of a quantum dot at $30\,{\rm mK}$, as a function
of magnetic field. The variance is the mean squared of the fluctuation in the
conductance as the shape of the quantum dot is distorted. The variance
decreases when time-reversal symmetry is broken by a magnetic field. The dashed
curve is a fit to a squared Lorentzian. The inset shows an electron micrograph
of the device, fabricated in the two-dimensional electron gas of a GaAs/AlGaAs
heterostructure. The black rectangle at the center of the inset is the quantum
dot, the gray regions are the gate electrodes on top of the heterostructure.
Electrons can enter and exit the quantum dot through point contacts at the top
and right corner of the rectangle. The side of the rectangle between these two
corners is distorted to generate conductance fluctuations. (The two small
openings in the gate along this side are effectively closed in the electron
gas.) After Chan, Clarke, Marcus, Campman, and Gossard (1995).
}\label{fig_Marcus1}
\end{figure}

In Fig.\ \ref{fig_Marcus1} we show experimental data by Chan {\em et al.}
(1995) for the variance of the conductance of a quantum dot in the
two-dimensional electron gas of a GaAs/AlGaAs heterostructure. An ensemble was
constructed by slightly distorting the shape of the quantum dot by means of a
gate electrode. (The area of the dot varied by less than 5\% around
$2.4\,\mu{\rm m}^{2}$.) The reduction of the conductance fluctuations by a
magnetic field is clearly visible. The two point contacts were adjusted such
that $N_{1}=N_{2}=2$. Equation (\ref{eq:QDvarC}) predicts for this case a
variance of the conductance equal to $72/175\approx 0.41\times (e^{2}/h)^{2}$
for $\beta=1$ and $4/15\approx 0.27\times (e^{2}/h)^{2}$ for $\beta=2$. The
experimental values are considerably smaller, mainly as a result of inelastic
scattering (see Sec.\ \ref{phasebreaking}).

\subsubsection{Other transport properties}
\label{UCFdotother}

To compute the variance of other transport properties than the conductance, one
needs the two-point correlation function of the transmission eigenvalues. In
the limit $N\rightarrow\infty$ we can use the method of functional derivatives
explained in Sec.\ \ref{functionalderivatives}.

The two-point correlation function
$K(\lambda,\lambda')=-\beta^{-1}\delta\rho(\lambda)/\delta V(\lambda')$ is
obtained by variation of the relation (\ref{Mikhlinsol}) between density and
potential:
\begin{eqnarray}
\delta\rho(\lambda)&=&\pi^{-2}[\lambda(\lambda_{\rm
c}-\lambda)]^{-1/2}\int_{0}^{\lambda_{\rm c}}\!\!d\lambda'\nonumber\\
&&\mbox{}\times
\delta V(\lambda')\frac{d}{d\lambda'} \frac{[\lambda'(\lambda_{\rm
c}-\lambda')]^{1/2}}{\lambda-\lambda'}, \label{Mikhlinvary}
\end{eqnarray}
where we used the parameterization $T_{n}=(1+\lambda_{n})^{-1}$. As mentioned
in Sec.\ \ref{functionalderivatives} (footnote), there is no contribution from
variation of the boundary $\lambda_{c}= 4N_{1}N_{2}(N_{1}-N_{2})^{-2}$ of the
spectrum. The variance of the linear statistic $A=\sum_{n}a(\lambda_{n})$
follows upon integration over the two-point correlation function,
\begin{eqnarray}
{\rm Var}\,A&=&\frac{1}{\beta\pi^{2}}\int_{0}^{\lambda_{\rm
c}}\!\!d\lambda\int_{0}^{\lambda_{\rm
c}}\!\!d\lambda'\,\left(\frac{\lambda'(\lambda_{\rm
c}-\lambda')}{\lambda(\lambda_{\rm c}-\lambda)}\right)^{1/2}\nonumber\\
&&\mbox{}\times\frac{a(\lambda)}{\lambda-\lambda'}
\frac{da(\lambda')}{d\lambda'},\label{Mikhlinvaryc}
\end{eqnarray}
where the singular integral is the principal value. For $N_{1}=N_{2}$ one has
$\lambda_{\rm c}\rightarrow\infty$, and Eq.\ (\ref{Mikhlinvaryc}) reduces to
the result (\ref{VarAWDresult2}) for a logarithmic eigenvalue repulsion in the
interval $(0,\infty)$.

\subsubsection{Tunnel barriers}
\label{UCFdottunnel}

We briefly consider the effect of tunnel barriers on the variance of the
conductance. The distribution of the scattering matrix is now the Poisson
kernel instead of the circular ensemble. The parameterization (\ref{eq:Sparam})
reduces the problem to an integration over the unitary group, which can be done
perturbatively for $N\Gamma\gg 1$. The result is  (Efetov, 1995; Brouwer and
Beenakker, 1996a)
\begin{eqnarray}
&&{\rm Var}\,G/G_{0}=2\beta^{-1}
\bigl(g_{1}^{\vphantom{2}}+g_{1}'\bigr)^{-6}
\bigl(2g_{1}^{4}g_{1}'^{2}+4g_{1}^{3}g_{1}'^{3}\nonumber\\
&&\hspace{5mm}\mbox{}-
4g_{1}^{2}g_{2}^{\vphantom{2}}g_{1}'^{3}+
2g_{1}^{2}g_{1}'^{4}-
2g_{1}^{\vphantom{2}}g_{2}^{\vphantom{2}}g_{1}'^{4}+
3g_{2}^{2}g_{1}'^{4}-
2g_{1}^{\vphantom{2}}g_{3}^{\vphantom{2}}g_{1}'^{4}
\nonumber\\
&&\hspace{5mm}\mbox{}+
2g_{2}^{\vphantom{2}}g_{1}'^{5}-
2g_{3}^{\vphantom{2}}g_{1}'^{5}+2g_{1}^{5}g_{2}'-
2g_{1}^4g_{1}'g_{2}'-
4g_{1}^{3}g_{1}'^{2}g_{2}'
\nonumber\\
&&\hspace{5mm}\mbox{}+
6g_{1}^{2}g_{2}^{\vphantom{2}}g_{1}'^{2}g_{2}'+
3g_{1}^{4}g_{2}'^{2}-
2g_{1}^{5}g_{3}'-2g_{1}^{4}g_{1}'g_{3}'\bigr). \label{eq:QDVarDiagr}
\end{eqnarray}
One verifies that Eq.\ (\ref{VarGdotNbig}) is recovered in the absence of
tunnel barriers. For the special case of two identical tunnel barriers
($g_{p}=g_{p}'$), Eq.\ (\ref{eq:QDVarDiagr}) reduces to (Iida,
Weidenm\"{u}ller, and Zuk, 1990a, 1990b)
\begin{equation}
{\rm Var}\,G/G_{0}=(8\beta g_{1}^{2})^{-1}
\left(2g_{1}^{2}-2g_{1}^{\vphantom{2}}g_{2}^{\vphantom{2}}+
3g_{2}^{2}-2g_{1}^{\vphantom{2}}g_{3}^{\vphantom{2}}\right).
\label{eq:QDVarDiagrsimpler}
\end{equation}
Another special case is that of high tunnel barriers, $\Gamma_{n}\ll 1$ for all
$n$, when Eq.\ (\ref{eq:QDVarDiagr}) simplifies to (Zirnbauer, 1993)
\begin{equation}
{\rm Var}\,G/G_{0}=4\beta^{-1}(g_{1}\vphantom{'}+g_{1}')^{-4}
g_{1}^{2}g_{1}'^{2}.\label{highbarrierlimit}
\end{equation}
Finally, if all transmission eigenvalues $\Gamma_{n}\equiv\Gamma$ are equal,
one has ${\rm Var}\,G/G_{0}= (8\beta)^{-1}[1+(1-\Gamma)^{2}]$. A high tunnel
barrier ($1/N\ll\Gamma\ll 1$) doubles the variance.

\subsubsection{Magnetoconductance}
\label{UCFdotmagneto}

A weak magnetic field reduces the variance of the conductance, by a factor of
two if $N\gg 1$. The dashed line in the experimental Figure \ref{fig_Marcus1}
is a fit to a squared Lorentzian, which is the theoretical result of Frahm
(1995b),
\begin{equation}
{\rm Var}\,G/G_{0}= \case{1}{16}+\case{1}{16}\left[1+(\Phi/\Phi_{\rm
c})^{2}\right]^{-2}.\label{Lorentzsquared}
\end{equation}
This result was obtained in the same way as Eq.\ (\ref{magnetoGresult}), for a
system without spin-orbit scattering and assuming $\Gamma=1$, $N_{1}=N_{2}\gg
1$. The characteristic flux $\Phi_{\rm c}$ is related to the parameter $\alpha$
in the Pandey-Mehta Hamiltonian by Eq.\ (\ref{magnetoGresultb}), which in turn
is related to microscopic parameters by Eqs.\ (\ref{alphadisordered}) and
(\ref{alphaballistic}). Up to a numerical coefficient of order unity,
$\Phi_{\rm c}$ is given by Eq.\ (\ref{Phicapproximate}).

Once $\Phi$ is much greater than $\Phi_{\rm c}$, time-reversal symmetry is
effectively broken and the variance of the conductance becomes independent of
the magnetic field. The conductance of a specific sample fluctuates in a random
but reproducible way as a function of magnetic field ({\em cf.} Fig.\
\ref{fig_UCFAu}). These magnetoconductance fluctuations (or
``magnetofingerprints'') are characterized by the correlator
\begin{equation}
C(\Delta\Phi)=\langle G(\Phi)G(\Phi+\Delta \Phi)\rangle-\langle
G(\Phi)\rangle\langle G(\Phi+\Delta\Phi)\rangle,\label{CDeltaPhi}
\end{equation}
where $\langle\cdots\rangle$ represents either an ensemble average or an
average over $\Phi$ ($\gg\Phi_{\rm c}$). This correlator is given by (Efetov,
1995; Frahm, 1995b):
\begin{equation}
C(\Delta\Phi)=\case{1}{16}G_{0}^{2}\left[1+(\Delta\Phi/2\Phi_{\rm
c})^{2}\right]^{-2}.\label{Lorentzsquared2}
\end{equation}
The Lorentzian-squared decay of $C(\Delta\Phi)$ was first derived from
semiclassical theory by Jalabert, Baranger, and Stone (1990; Baranger, 1996).
Experiments by Marcus {\em et al.} (1992; Westervelt, 1996) on the
magnetoconductance fluctuations of a (chaotic) stadium-shaped quantum dot are
in agreement with Eq.\ (\ref{Lorentzsquared2}), and also show the more rapid
decay of the correlator for a (non-chaotic) circular geometry predicted by the
semiclassical theory.

Efetov (1995) has generalized the zero-temperature result
(\ref{Lorentzsquared2}) for the correlator to non-zero temperatures. Thermal
smearing of the Fermi-Dirac distribution function reduces the magnitude of the
magnetoconductance fluctuations, once the thermal energy $k_{\rm B}T$ becomes
greater than the Thouless energy $E_{\rm c,open}\simeq N\Gamma\delta$ of the
open system. (This is in contrast to the weak-localization effect, which is not
influenced by thermal smearing.) In the high-temperature limit $k_{\rm B}T\gg
E_{\rm c,open}$ (and for the case $\Gamma=1$, $N_{1}=N_{2}\gg 1$), the
correlator becomes a Lorentzian,
\begin{equation}
C(\Delta\Phi)=\frac{G_{0}^{2}}{96}\frac{N_{1}\delta}{k_{\rm
B}T}\left[1+(\Delta\Phi/2\Phi_{\rm c})^{2}\right]^{-1},\label{Lorentzian1}
\end{equation}
instead of a squared Lorentzian.

The results in this subsection all follow from the correlator
\begin{eqnarray}
C(\Phi_{1},\Phi_{2},E_{1},E_{2})&=&\langle
G(\Phi_{1},E_{1})G(\Phi_{2},E_{2})\rangle\nonumber\\
&&\mbox{}-\langle G(\Phi_{1},E_{1})\rangle\langle
G(\Phi_{2},E_{2})\rangle\label{CDeltaPhiDeltaEdef}
\end{eqnarray}
of the (zero-temperature) conductance at two different magnetic fluxes
$\Phi_{1},\Phi_{2}$ and two different values $E_{1},E_{2}$ of the Fermi energy.
This correlator has been derived from the Pandey-Mehta Hamiltonian (Frahm
1995b) and from a microscopic Hamiltonian (Efetov, 1995). It also follows from
semiclassical periodic-orbit theory (Bl\"{u}mel and Smilansky, 1988; Jalabert,
Baranger, and Stone, 1990). The result (for $\Gamma=1$, $N_{1}=N_{2}\gg 1$)
\begin{mathletters}
\label{CDeltaPhiDeltaE}
\begin{eqnarray}
C&=&\case{1}{16}G_{0}^{2}[f(\Phi_{\rm tot},\Delta E)+ f(\Delta\Phi,\Delta
E)],\label{CDeltaPhiDeltaEa}\\
f(\Phi,E)&=&\frac{1}{[1+(\Phi/2\Phi_{\rm c})^{2}]^{2}+ (\pi
E/N_{1}\delta)^{2}},\label{CDeltaPhiDeltaEb}
\end{eqnarray}
\end{mathletters}%
is a Lorentzian in energy differences $\Delta E=E_{1}-E_{2}$ and a squared
Lorentzian in magnetic-flux differences $\Delta\Phi=\Phi_{1}-\Phi_{2}$. The
flux dependence contains also a term which depends on the total flux $\Phi_{\rm
tot}=\Phi_{1}+\Phi_{2}$, to ensure that $C$ is an even function of the
individual fluxes $\Phi_{1},\Phi_{2}$.

The correlator (\ref{Lorentzsquared2}) in the absence of time-reversal symmetry
can also be obtained from the Brownian-motion model for the Hamiltonian of
Sec.\ \ref{BrownianE}. An altogether different issue is the question whether a
Brownian-motion model for the {\em scattering matrix\/} can describe the
magnetoconductance of a chaotic cavity. This issue has been addressed by
several authors (Mac\^{e}do, 1994b, 1995; Rau, 1995; Frahm and Pichard, 1995a,
1995b). The answer appears to be negative. Rau and Frahm and Pichard find that
the effect of a magnetic field on the scattering matrix is a Brownian-motion
process only for small flux-increments $\Delta\Phi\ll\Phi_{\rm c}$. Mac\^{e}do
obtains a Lorentzian-squared decay of the magnetocorrelator from a
Brownian-motion model which is not invariant under unitary transformations of
the scattering matrix ($S\rightarrow US$), and therefore seems unjustifiable.

\subsection{Conductance distribution}
\label{Gdistribution}

The conductance of the quantum dot has a Gaussian distribution if the number of
modes in the point contacts is large. Deviations from a Gaussian become
significant when the fluctuations $\simeq e^{2}/h$ of the conductance become
greater than the mean $\simeq N\Gamma e^{2}/h$, {\em i.e.} when $N\lesssim
1/\Gamma$.

\begin{figure}[tb]
\psfig{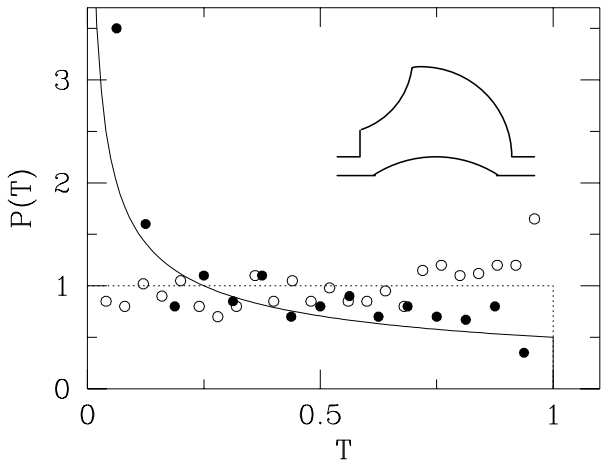}
\medskip
\caption[]{
Distribution of the transmission probability through a chaotic billiard with
two ballistic single-mode point contacts. Data points are numerical results for
the billiard shown in the inset, averaged over a range of Fermi energies and
small variations in shape. Filled data points are for $B=0$, open points for
$B\neq 0$ (a few flux quanta through the billiard). The solid and dotted curves
are the prediction (\ref{PTN1Gamma0}) of the circular ensemble for $\beta=1,2$.
After Baranger and Mello (1994).
}\label{fig_baranger}
\end{figure}

Consider, as an extreme example, the case $\Gamma=1$, $N_{1}=N_{2}=1$ of two
ballistic single-mode point contacts (Baranger and Mello, 1994; Jalabert,
Pichard, and Beenakker, 1994). According to Eq.\ (\ref{PTcircular}), the single
transmission eigenvalue $T$ of the quantum dot has probability distribution
\begin{equation}
P(T)=\case{1}{2}\beta T^{-1+\beta/2},\;\;0<T<1. \label{PTN1Gamma0}
\end{equation}
In the presence of a magnetic field ($\beta=2$), any value of the conductance
$G=G_{0}T$ between $0$ and $G_{0}=2e^{2}/h$ is equally probable. In non-zero
field it is more probable to find a small than a large conductance, provided
that the scattering preserves spin-rotation symmetry ($\beta=1$). In the
presence of spin-orbit scattering ($\beta=4$), a large conductance is more
probable than a small one. In Fig.\ \ref{fig_baranger} we show numerical
calculations of transmission through a chaotic billiard by Baranger and Mello
(1994, see also Ishio, 1995; Yang, Ishio, and Burgd\"{o}rfer, 1995), which
confirm this remarkable sensitivity of the conductance distribution to a
magnetic field.

\begin{figure}[tb]
\psfig{figure=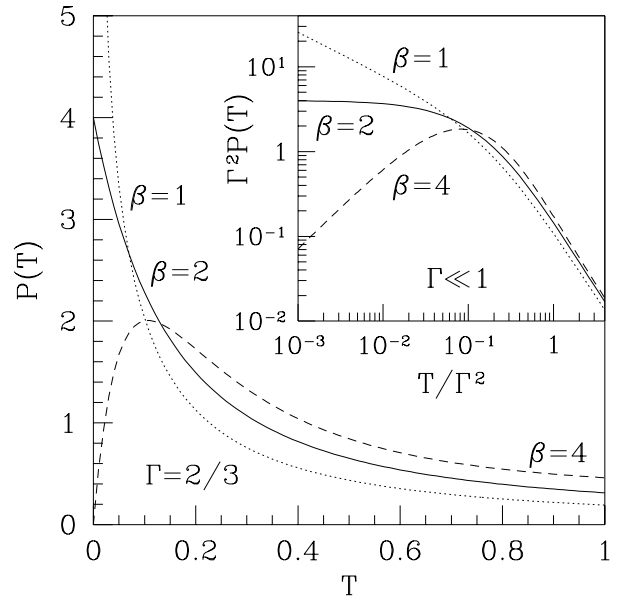,width=
8cm,bbllx=33pt,bblly=166pt,bburx=550pt,bbury=680pt}
\medskip
\caption[]{
Distribution of the transmission probability through a chaotic billiard with
two single-mode point contacts containing a tunnel barrier
($\Gamma_{1}=\Gamma_{2}\equiv\Gamma$). The curves are computed by integrating
over the Poisson kernel, for the three symmetry classes ($\beta=1,2,4$). The
main plot is for $\Gamma=2/3$, the inset shows the asymptotic behavior of
$P(T)$ for $\Gamma\ll 1$ on a log--log scale. Notice that the result $P\propto
T^{-1+\beta/2}$ for ballistic point contacts is recovered if $T\ll\Gamma^{2}$.
After Brouwer and Beenakker (1994).
}\label{fig_nonideal}
\end{figure}

If the point contacts contain a tunnel barrier, the distribution remains
strongly non-Gaussian but becomes less sensitive to a magnetic field. In Fig.\
\ref{fig_nonideal} we show $P(T)$ for the case of two identical single-mode
point contacts with $\Gamma_{1}=\Gamma_{2}\equiv\Gamma=2/3$. The inset shows
the limit $\Gamma\ll 1$. For $T\gtrsim\Gamma^{2}$ the tunnel barriers dominate
the transmission through the entire system, thereby suppressing the
$\beta$-dependence of the distribution. For $T\ll\Gamma^{2}$ the presence of
tunnel barriers is of less importance, and the $\beta$-dependence remains
significant. The curves in Fig.\ \ref{fig_nonideal} were computed using the
Poisson kernel for the distribution of scattering matrices (Brouwer and
Beenakker, 1994; Baranger and Mello, 1996a), and agree with results obtained
from a tunnel Hamiltonian with disorder (Prigodin, Efetov, and Iida, 1993,
1995). Qualitatively similar results have been obtained by Kamenev and Gefen
(1995) for the real part of the frequency-dependent conductance of an isolated
metal ring.

\begin{figure}
\psfig{figure=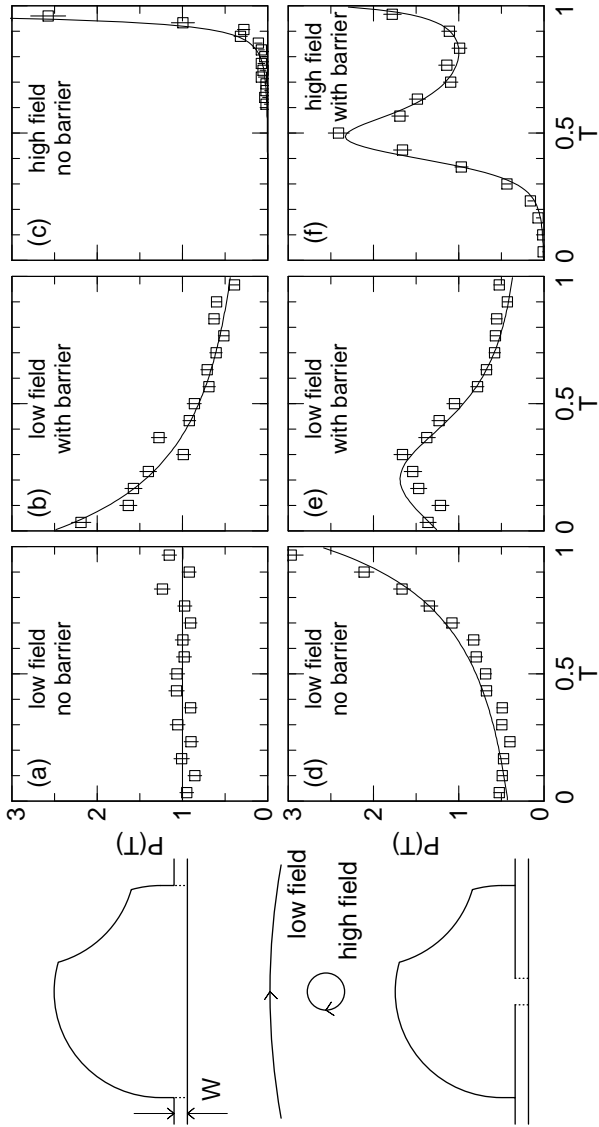,width=
8cm,bbllx=82pt,bblly=55pt,bburx=422pt,bbury=697pt}
\medskip
\caption[]{
Distribution of the transmission probability for the geometry of Fig.\
\ref{fig_baranger} (top row) and for the same geometry with leads extended into
the cavity (bottom row). The magnitude of the magnetic field (``low'' and
``high'' corresponding to 2 and 80 flux quanta through the billiard,
respectively) and the presence or absence of a tunnel barrier at the entrance
to the leads (marked by dotted lines in the sketches of the structures) are
noted in each panel. Cyclotron orbits for both fields, drawn to scale, are
shown on left. The data points with statistical error bars are numerical
results; the curves are the predictions of the Poisson kernel for $\beta=2$,
with $\bar{S}$ extracted from the numerical data. After Baranger and Mello
(1996a).
}\label{fig_mello}
\end{figure}

Tunnel barriers give a non-zero ensemble-averaged scattering matrix $\bar{S}$
in the Poisson kernel because of direct reflection at a point contact
(``direct'' meaning without scattering in the cavity). Baranger and Mello
(1996a) consider also the case that $\bar{S}\neq 0$ because of direct
transmission between the two point contacts. Direct transmission can be
achieved by bringing the point contacts close together, or by increasing the
magnetic field so that electrons can skip along a boundary from one point
contact to the other. Good agreement between the Poisson kernel and numerical
simulations is found in both cases, if the energy-averaged scattering matrix
computed numerically is used as input in the Poisson kernel (see Fig.\
\ref{fig_mello}).

Another generalization, considered by Gopar {\em et al.} (1996) and Baranger
and Mello (1996b), is to quantum dots with a reflection symmetry. (Disordered
conductors with a reflection symmetry had been studied earlier by Hastings,
Stone, and Baranger, 1994.) The scattering matrix for a symmetric geometry
decomposes into blocks in a basis of definite parity with respect to the
symmetry operator. The blocks have independent distributions in the circular
ensemble. Because the conductance couples different blocks, its distribution
differs from the result (\ref{PTN1Gamma0}) for the circular ensemble.

Experiments by Chan {\em et al.} (1995) on quantum dots with ballistic point
contacts ($N_{1}=N_{2}=2$) find a probability distribution for the conductance
which is well described by a Gaussian, presumably as a result of inelastic
scattering ({\em cf.} Sec.\ \ref{phasebreaking}). More recent data by the same
group (Marcus, 1996) shows significant deviations from a Gaussian, in
particular a distribution which is skewed towards small conductance in zero
magnetic field.

\subsection{Phase breaking}
\label{phasebreaking}

Quantum interference effects in the conductance require phase coherence of the
electron wave function to persist on the time scale $\hbar/E_{\rm c,open}$. If
phase coherence is broken after a time $\tau_{\phi}$, then transport becomes
classical if $\hbar/\tau_{\phi}\gtrsim E_{\rm c,open}$. We discuss two
phase-breaking mechanisms: Coupling of the quantum dot to the outside through a
voltage probe, and inelastic scattering inside the quantum dot. In the latter
case, phase breaking occurs uniformly throughout the quantum dot, while in the
former case it occurs locally at the voltage probe.

\subsubsection{Invasive voltage probe}
\label{voltageprobe}

The measurement of a voltage at some point in the sample is an invasive act,
which may destroy the phase coherence throughout the whole sample. The reason
is that electrons which enter the voltage lead are reinjected into the system
without any phase relationship (B\"{u}ttiker, 1986a, 1988a). The phase-breaking
effects of a voltage probe on the conductance of a chaotic cavity have been
investigated by Baranger and Mello (1995) and Brouwer and Beenakker (1995a,
1997a).

The model consists of a quantum dot which is coupled by two leads to source and
drain reservoirs at voltages $V_{1}$ and $V_{2}$. A current $I=I_{1}=-I_{2}$ is
passed from source to drain via leads $1$ and $2$. A third lead is attached to
the quantum dot and connected to a third reservoir at voltage $V_{3}$. This
third lead is a voltage probe, which means that $V_{3}$ is adjusted in such a
way, that no current is drawn ($I_{3}=0$). We denote by $N_{i}$ the number of
modes in lead $i=1,2,3$, and we assume for simplicity that there are no tunnel
barriers in any of the leads.

The scattering-matrix $S$ of the system can be written as
\begin{equation}
S=\left(\begin{array}{ccc}
r_{11}&t_{12}&t_{13}\\
t_{21}&r_{22}&t_{23}\\
t_{31}&t_{32}&r_{33}
\end{array}\right),\label{Svoltageprobe}
\end{equation}
in terms of reflection and transmission matrices $r_{ii}$ and $t_{ij}$ between
leads $i$ and $j$. The currents and voltages satisfy (B\"{u}ttiker, 1986b,
1988b)
\begin{mathletters}
\label{3terminal}
\begin{eqnarray}
&&\frac{h}{2e^{2}}I_{k}=\left(N_{k}-R_{kk}\right)V_{k}-\sum_{l\neq
k}T_{kl}V_{l},\;k=1,2,3,\label{3terminala}\\
&&R_{kk}={\rm Tr}\,r_{kk}^{\vphantom{\dagger}}r_{kk}^{\dagger},\;\;T_{kl}={\rm
Tr}\,t_{kl}^{\vphantom{\dagger}}t_{kl}^{\dagger}.\label{3terminalb}
\end{eqnarray}
\end{mathletters}%
The two-terminal conductance $G=I/(V_{1}-V_{2})$ follows from Eq.\
(\ref{3terminal}) with $I_{1}= -I_{2}=I$, $I_{3}=0$:
\begin{equation}
G=\frac{2e^{2}}{h}\left(T_{12}+\frac{T_{13}T_{32}}{T_{31}+T_{32}}
\right).\label{G3terminal}
\end{equation}

\begin{figure}[tb]
\psfig{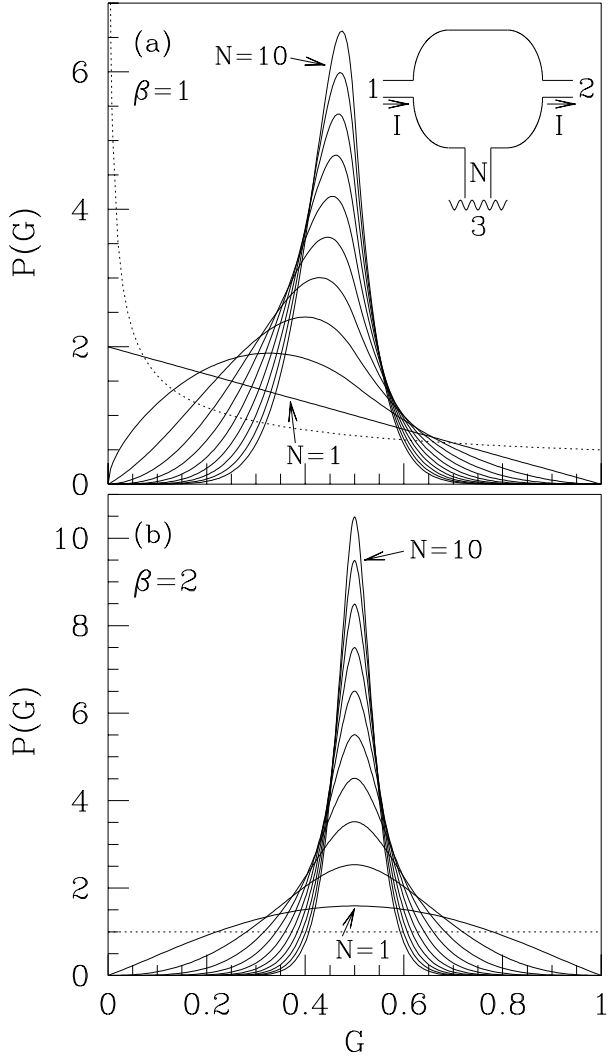}
\medskip
\caption[]{
Effect of an invasive voltage probe on the distribution of the conductance (in
units of $2e^{2}/h$). The current-carrying leads 1 and 2 contain a single mode
each, while the number of modes $N$ in the voltage lead (labeled 3 in the
inset) varies from 1 to 10 with increments of 1 (solid curves). The dotted
curve is the distribution (\ref{PTN1Gamma0}) in the absence of a voltage lead.
The top panel is for $\beta=1$, the bottom panel for $\beta=2$. After Brouwer
and Beenakker (1995a).
}\label{fig_invasive}
\end{figure}

Analytical results for $P(G)$ can be obtained for $N_{1}=N_{2}=1$ and $N_{3}$
arbitrary (Brouwer and Beenakker, 1995a, 1997a). Because of current
conservation,
\begin{mathletters}
\label{currentconserv}
\begin{eqnarray}
&&T_{13}=1-R_{11}-T_{12}=1-|S_{11}|^{2}-|S_{12}|^{2},\label{currentconserva}\\
&&T_{31}=1-R_{11}-T_{21}=1-|S_{11}|^{2}-|S_{21}|^{2},\label{currentconservb}\\
&&T_{32}=1-R_{22}-T_{12}=1-|S_{22}|^{2}-|S_{12}|^{2},\label{currentconservc}
\end{eqnarray}
\end{mathletters}%
so that it suffices to know the marginal distribution of the matrix elements
$S_{kl}$ with $k,l\le 2$. This distribution has been computed by Pereyra and
Mello (1983), and Friedman and Mello (1985b). The resulting $P(G)$ is plotted
in Fig.\ \ref{fig_invasive}, for $\beta=1,2$ and $N_{3}$ ranging from 1 to 10.
Notice the particularly simple result $P(G/G_{0})=2-2G/G_{0}$ in the case
$\beta=1$, $N_{3}=1$. As $N_{3}$ increases, $P(G)$ becomes more and more
sharply peaked around $e^{2}/h$. The limiting distribution as
$N_{3}\rightarrow\infty$ is
\begin{equation}
P(G/G_{0})=\case{1}{2}\beta N_{3}\bigl(1+|y|+(1-2/\beta)y\bigr){\rm e}^{-|y|},
\label{eq:PGdistr}
\end{equation}
where we have abbreviated $y=2\beta N_{3}(G/G_{0}-\frac{1}{2})$. Surprisingly
enough, the distribution remains non-Gaussian for arbitrarily strong dephasing.

The mean and variance for $N_{3}\gg 1$ can be computed analytically for any
$N_{1}=N_{2}$ (Baranger and Mello, 1995),
\begin{eqnarray}
&&\langle G/G_{0}\rangle=\case{1}{2}N_{1}+\case{1}{2}(1-2/\beta)N_{1}/N_{3},
\label{invasivemean}\\
&&{\rm Var}\,G/G_{0}=\frac{2N_{1}+2-\beta}{4\beta N_{1}}(N_{1}/N_{3})^{2}.
\label{invasivevar}
\end{eqnarray}
The variance of $G$ is reduced by a factor of $2+1/N_{1}$ when time-reversal
symmetry is broken in the limit $N_{3}\rightarrow\infty$. The offset of
$\langle G/G_{0}\rangle$ from $\case{1}{2}N_{1}$ when $\beta=1$ is a remnant of
the weak-localization effect.

\subsubsection{Inelastic scattering}
\label{eescattering}

The phase-breaking effects of inelastic scattering (scattering rate
$1/\tau_{\phi}$) can be modeled by an imaginary voltage lead with
$N_{3}=2\pi\hbar/\tau_{\phi}\delta$ (Marcus {\em et al.}, 1993). In this way,
Baranger and Mello (1995) have been able to account for the discrepancies
between the predicted and measured magnitude of quantum corrections mentioned
in Secs.\ \ref{magnetoG}, \ref{UCFdotG}, and \ref{Gdistribution}. There exists
an alternative model, which is to include a (spatially uniform) imaginary
potential in the Hamiltonian, equal to $-\frac{1}{2}{\rm i}\hbar/\tau_{\phi}$.
This second model was used by Efetov (1995) and by McCann and Lerner (1996).
The two models give very different results for the distribution of the
conductance, in particular in the case that the current through the quantum dot
flows through single-mode point contacts. While the distribution $P(G)$ becomes
a delta peak at the classical conductance for very strong dephasing
($\tau_{\phi} \to 0$) in the voltage-probe model, $P(G)$ peaks at zero
conductance in the imaginary potential model.

The origin of the differences lies with certain shortcomings of each model. On
the one hand, the imaginary potential model does not conserve the number of
electrons. On the other hand, the voltage-probe model describes spatially
localized instead of spatially uniform dephasing. There exists a limit of the
voltage-probe model which applies to dephasing processes occurring uniformly in
space (Brouwer and Beenakker, 1997a). This limit is equivalent to a
particle-conserving version of the imaginary potential model. What one needs to
do is to introduce a tunnel barrier (transparency $\Gamma_{3}$) in the voltage
probe and take the limit $N_{3}\rightarrow\infty$, $\Gamma_{3}\rightarrow 0$ at
fixed $N_{3}\Gamma_{3}= 2\pi\hbar/\tau_{\phi}\delta$. The resulting conductance
distribution narrows around the classical series conductance of the two point
contacts when $\tau_{\phi}\rightarrow 0$, in a way which is similar, but not
precisely identical, to the voltage-probe model with $\Gamma_{3}=1$.

Neither the voltage-probe model nor the imaginary potential model provides a
microscopic description of electron-electron scattering, which is the main
source of inelastic scattering at low temperatures. At present there exists a
microscopic theory for dephasing by electron-electron interactions in closed
systems (Sivan, Imry, and Aronov, 1994; Altshuler, Gefen, Kamenev, and Levitov,
1996) --- but not yet in open systems.

\subsection{Coulomb blockade}
\label{Coulombblockade}

So far we have ignored the Coulomb repulsion of electrons in the quantum dot. A
measure of the importance of Coulomb repulsion is the charging energy
$e^{2}/2C$ of a single electron in the quantum dot (capacitance $C$). The
charging energy plays no role if the quantum dot is strongly coupled to the
reservoirs, but it does if the coupling is weak. Strong or weak is determined
by whether the broadening $\gamma$ of the energy levels in the quantum dot is
large or small compared to their spacing $\delta$. The ratio $\gamma/\delta$ is
of the order of the conductance $G$ of the quantum dot in units of $e^{2}/h$,
so that Coulomb repulsion is important if $G\lesssim e^{2}/h$. In addition, the
charging energy should be large compared to the thermal energy $k_{\rm B}T$. If
both these conditions are met, the conductance {\em oscillates\/} as a function
of the Fermi energy, with periodicity $e^{2}/C$ (Shekhter, 1972; Kulik and
Shekhter, 1975). The periodic suppression of the conductance is known as the
Coulomb blockade. There exist several reviews devoted entirely to this
phenomenon (Averin and Likharev, 1991; Van Houten, Beenakker, and Staring,
1992; Meirav and Foxman, 1995). Here we discuss one aspect of it, for which
random-matrix theory is relevant (Jalabert, Stone, and Alhassid, 1992).

If $k_{\rm B}T\ll e^{2}/C$ the oscillations of the conductance develop into a
sequence of well-resolved peaks. If moreover $k_{\rm B}T\ll\delta$, a {\em
single\/} energy level $E_{i}$ in the quantum dot contributes to each peak. The
amplitude of the peaks fluctuates because of fluctuations in the wave functions
of subsequent levels.\footnote{
The amplitude of the minima of the conductance oscillations also fluctuates.
These fluctuations involve virtual transitions to excited states in the quantum
dot (Averin and Nazarov, 1990), and hence depend on the statistics of the
superposition of a large number of wave functions (Aleiner and Glazman, 1996).}
If $k_{\rm B}T\gg\gamma$ the peak amplitude can be calculated using rate
equations (Averin, Korotkov, and Likharev, 1991; Beenakker, 1991). [At lower
temperatures, complications arise because of the Kondo effect (Ng and Lee,
1988; Meir, Wingreen, and Lee, 1991).] The result for the height $G_{\rm max}$
of the $i$-th conductance peak is
\begin{equation}
G_{\rm max}=\frac{e^{2}}{h}\frac{\pi}{2k_{\rm
B}T}\frac{\sum_{n=1}^{N_{1}}\sum_{n'=N_{1}+1}^{N_{1}+N_{2}}
\gamma_{n\vphantom{`}}^{(i)}\gamma_{n'}^{(i)}}
{\sum_{n=1}^{N_{1}+N_{2}}\gamma_{n}^{(i)}}, \label{Gmaxformula}
\end{equation}
where $\gamma_{n}^{(i)}/\hbar$ is the tunnel rate from level $i$ in the quantum
dot to mode $n$ in one of the two leads. In terms of the Hamiltonian
(\ref{HAMHAM}), the tunnel rate $\gamma_{n}^{(i)}$ is determined by the
eigenvalue $w_{n}$ of the coupling-matrix product $WW^{\dagger}$ and by the
matrix $U$ which diagonalizes the Hamiltonian ${\cal H}=U\,{\rm
diag}\,(E_{1},E_{2},\ldots E_{M})U^{\dagger}$ of the isolated quantum dot. In a
basis in which $WW^{\dagger}$ is diagonal, the relation reads
\begin{equation}
\gamma_{n}^{(i)}=2\pi
w_{n}|U_{ni}|^{2}=\frac{\Gamma_{n}M\delta}{2\pi}\,|U_{ni}|^{2}.\label{gammani}
\end{equation}
Equation (\ref{gammani}) follows from the scattering matrix (\ref{SHeq}), under
the assumption that $w_{n}\ll M\delta$, which in view of Eq.\ (\ref{w2Gamma})
implies $\Gamma_{n}\ll 1$. Since $|U_{ni}|^{2}\simeq 1/M$, this also implies
$\gamma_{n}^{(i)}\ll\delta$. Substitution of Eq.\ (\ref{gammani}) into Eq.\
(\ref{Gmaxformula}) gives
\begin{equation}
G_{\rm max}=\frac{e^{2}}{h}\frac{M\delta}{4k_{\rm
B}T}\frac{\sum_{n=1}^{N_{1}}\sum_{n'=N_{1}+1}^{N_{1}+N_{2}}
\Gamma_{n}\Gamma_{n'}|U_{ni}|^{2}|U_{n'i}|^{2}}
{\sum_{n=1}^{N_{1}+N_{2}}\Gamma_{n}|U_{ni}|^{2}}.\label{Gmaxformulab}
\end{equation}

\begin{figure}[tb]
\psfig{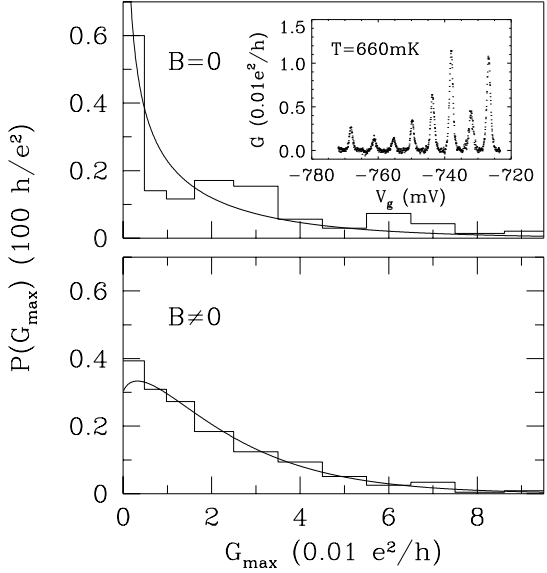}
\medskip
\caption[]{
Inset: Conductance at $660\,{\rm mK}$ of a quantum dot in the two-dimensional
electron gas of a GaAs/AlGaAs heterostructure, as a function of the voltage on
a gate electrode controlling the number of electrons in the dot in equilibrium
(about 100, in an area of $0.25\,\mu{\rm m}\times 0.25\,\mu{\rm m}$; the system
is chaotic because of weak disorder). A conductance peak occurs each time this
number increases by one, because then it costs no charging energy to tunnel
into the dot. The histograms are the measured probability distributions of the
peak heights $G_{\rm max}$ at $75\,{\rm mK}$, with and without a
time-reversal-symmetry breaking magnetic field. The curves are a one-parameter
fit to Eq.\ (\ref{PGmax}) for $\beta=1,2$ (same parameter value
$\Gamma\delta=0.27\,k_{\rm B}T$ in both curves). After Chang, Baranger,
Pfeiffer, West, and Chang (1996).
}\label{fig_Chang2}
\end{figure}

Peak-to-peak fluctuations in $G_{\rm max}$ are due to level-to-level
fluctuations in the eigenfunctions of the quantum dot at the two tunnel
barriers, represented by the vector $\vec{U}\equiv(U_{1i},U_{2i},\ldots
U_{Ni})$ of length $N=N_{1}+N_{2}$. The probability distribution $P(G_{\rm
max})$ of the peak heights follows from the distribution $P(\vec{U})$ in the
limit $M\rightarrow\infty$ at fixed $N$. The distribution $P(\vec{U})$, in
turn, follows from the distribution of the matrix $U$. In zero magnetic field
(without spin-orbit scattering, $\beta=1$), the real matrix $U$ is uniformly
distributed in the orthogonal group. In a magnetic field ($\beta=2$), the
complex matrix $U$ is uniformly distributed in the unitary group. The resulting
distribution of $\vec{U}$ factorizes for $M\gg N$ into independent Gaussian
distributions with zero mean and variance $1/\beta M$ ({\em cf.} Appendix
\ref{UNintegrate}):
\begin{equation}
P(\vec{U})=(\beta M/2\pi)^{\beta N/2}\exp\left(-\case{1}{2}\beta
M\sum_{n=1}^{N}|U_{ni}|^{2}\right).\label{PvecU}
\end{equation}
The distribution $P(G_{\rm max})$ which follows from Eqs.\ (\ref{Gmaxformulab})
and (\ref{PvecU}) takes on a simple form if $N_{1}=N_{2}=1$,
$\Gamma_{1}=\Gamma_{2}\equiv\Gamma$ (Jalabert, Stone, and Alhassid, 1992;
Prigodin, Efetov, and Iida, 1993),
\begin{mathletters}
\label{PGmax}
\begin{eqnarray}
&&P(g)=\left\{\begin{array}{ll}
(\pi g)^{-1/2}\,{\rm e}^{-g},&\beta=1,\\
g[K_{0}(g)+K_{1}(g)]{\rm e}^{-g},&\beta=2,
\end{array}\right.\label{PGmaxa}\\
&&g\equiv G_{\rm max}\frac{h}{e^{2}}\frac{8k_{\rm
B}T}{\Gamma\delta}.\label{PGmaxb}
\end{eqnarray}
\end{mathletters}%
Here $K_{0}$ and $K_{1}$ are Bessel functions. [The case $N_{1},N_{2}>1$ has
been considered by Mucciolo, Prigodin, and Altshuler (1995) and by Alhassid and
Lewenkopf (1995).] The distribution (\ref{PGmax}) has been confirmed
experimentally by Chang {\em et al.} (1996) and by Folk {\em et al.} (1996).
Measurements by Chang {\em et al.} are shown in Fig.\ \ref{fig_Chang2}. Good
agreement is found with Eq.\ (\ref{PGmax}), using a single adjustable
parameter.

Folk {\em et al.} also measured the correlator
\begin{equation}
C_{\rm max}(\Delta B)=\left\langle G_{\rm max}(B)G_{\rm max}(B+\Delta
B)\right\rangle-\left\langle G_{\rm max}(B)\right\rangle^{2}\label{CDeltaB}
\end{equation}
of the height of a given peak at different magnetic fields. To obtain this
correlator theoretically, one can use the Brownian-motion model described in
Sec.\ \ref{BrownianE}. If the field $B$ is sufficiently large to break
time-reversal symmetry, the matrix elements ${\cal H}_{nm}$ of the Hamiltonian
of the closed system execute a Brownian motion in the Gaussian unitary
ensemble, in the fictitious time $\tau\propto(\Delta B)^{2}$. The problem is to
extract the evolution of the matrix of eigenfunctions $U$ from the Brownian
motion of ${\cal H}$. This problem has been studied by Alhassid and Attias
(1996) and by Bruus, Lewenkopf, and Mucciolo (1996). An analytical solution
exists only for $\Delta B$ small compared to the correlation field $B_{\rm c}$.
Numerical calculations suggest $C_{\rm max}(\Delta B)\propto[1+(\Delta B/B_{\rm
c})^{2}]^{-2}$, which is roughly in agreement with experiment.

\subsection{Frequency dependence}
\label{frequency}

Throughout this review we focus on zero-frequency (DC) transport properties.
The generalization to non-zero frequencies $\omega$ in the case of a quantum
dot is briefly discussed in this subsection, following Gopar, Mello, and
B\"{u}ttiker (1996), and Brouwer and B\"{u}ttiker (1997). Variations in the
currents $I_{i}(\omega)$ and voltages $V_{j}(\omega)$ in the two leads
($i,j=1,2$) are related by the conductance coefficients
$G_{ij}(\omega)=\partial I_{i}/\partial V_{j}$. At zero frequency, current
conservation implies that $G_{11}=G_{22}=-G_{12}=-G_{21}$ equals the DC
conductance $G$. At non-zero frequency all four conductance coefficients are
different in general (B\"{u}ttiker, 1993; B\"{u}ttiker, Pr\^{e}tre, and Thomas,
1993; B\"{u}ttiker and Christen, 1996). If we ignore the screening of charges
accumulated temporarily in the system, the conductance coefficients are related
to the scattering matrix by
\begin{eqnarray}
G^{\infty}_{ij}&=&\frac{2e^{2}}{h}\int_{-\infty}^{\infty}
\frac{d\varepsilon}{\hbar\omega}
\bigl[f(\varepsilon-\case{1}{2}\hbar\omega)-f(\varepsilon+
\case{1}{2}\hbar\omega)\bigr]
\nonumber\\
&&\mbox{}\times{\rm Tr}\,\left[\delta_{ij}-
S^{\dagger}_{ij}(\varepsilon-\case{1}{2}\hbar\omega)
S^{\vphantom{\dagger}}_{ij}(\varepsilon+\case{1}{2}\hbar\omega)\right],
\label{Gomega}
\end{eqnarray}
where $f(\varepsilon)=[1+\exp(\varepsilon/k_{\rm B}T)]^{-1}$ is the Fermi
function. The $N_{i}\times N_{j}$ matrix $S_{ij}(\varepsilon)$ contains the
scattering amplitudes from lead $j$ into lead $i$ at energy $\varepsilon$,
measured relative to the Fermi energy. Neglecting screening amounts to putting
the capacitance $C$ of the system equal to infinity, hence the superscript
$\infty$ on $G_{ij}$.

Prigodin {\em et al.} (1994, 1995) have computed the average $\langle
G^{\infty}_{12}\rangle$ for the single-mode case $N_{1}=N_{2}=1$. The case
$N_{1},N_{2}\gg 1$ was considered by Brouwer and B\"{u}ttiker (1997), who found
\begin{eqnarray}
\frac{h}{2e^{2}}\langle
G^{\infty}_{ij}\rangle&=&\delta_{ij}N_{i}-\frac{N_{i}N_{j}}{N(1-{\rm i}\omega
\tau_{\rm dwell})}-\frac{(1-2/\beta)N_{i}}{N(1-{\rm i}\omega\tau_{\rm
dwell})}\nonumber\\
&&\mbox{}\times\left(\frac{N_{j}(1-2{\rm i}\omega\tau_{\rm dwell})}{N(1-{\rm
i}\omega\tau_{\rm dwell})^2}-\delta_{ij}\right),\label{eq:Tabavg}
\end{eqnarray}
where $N=N_{1}+N_{2}$ and $\tau_{\rm dwell}\equiv 2\pi\hbar/N\delta$ is the
mean dwell time of an electron in the quantum dot. One can check that the
result (\ref{GSnmaverageb}) is recovered in the limit $\omega\rightarrow 0$.

Screening is irrelevant for the DC conductance, but has an essential effect on
the frequency dependence. If the potential inside the quantum dot can be
assumed to be spatially uniform, the conductance coefficients take the form
(B\"{u}ttiker, Pr\^{e}tre, and Thomas, 1993)
\begin{equation}
G_{ij}=G_{ij}^{\infty}+\frac{\sum_{k,l=1}^{2}G_{ik}^{\infty}G_{lj}^{\infty}}
{{\rm i}\omega C-\sum_{k,l=1}^{2}G_{kl}^{\infty}}, \label{eq:GIab1a}
\end{equation}
so that the expression (\ref{Gomega}) is recovered in the limit
$C\rightarrow\infty$. Since fluctuations in $G_{ij}^{\infty}$ are of relative
order $N^{-2}$, we may directly substitute the result (\ref{eq:Tabavg}) into
Eq.\ (\ref{eq:GIab1a}), to obtain the average
\begin{eqnarray}
\frac{h}{2e^{2}}\langle
G_{ij}\rangle&=&\delta_{ij}N_{i}-\frac{N_{i}N_{j}}{N(1-{\rm i}\omega
\tau_{C})}-\frac{(1-2/\beta)N_{i}}{N(1-{\rm i}\omega\tau_{\rm
dwell})}\nonumber\\
&&\mbox{}\times\left(\frac{N_{j}(1-2{\rm i}\omega\tau_{C})}{N(1-{\rm
i}\omega\tau_{C})^2}-\delta_{ij}\right),\label{eq:Gadmavg}
\end{eqnarray}
with $1/\tau_{C}=1/\tau_{\rm dwell}+2e^{2}N/hC$. The ${\cal O}(N)$ term in Eq.\
(\ref{eq:Gadmavg}) is the classical Drude conductance, with $\tau_{C}$ playing
the role of the $RC$-time of the circuit. The $\beta$-dependent ${\cal O}(1)$
term is the frequency dependent weak-localization correction. For $C\rightarrow
0$, the $RC$-time $\tau_{C}$ vanishes. In this limit all four conductance
coefficients are the same, $G_{11}=G_{22}=-G_{12}=-G_{21}\equiv G$, with
average
\begin{equation}
\frac{h}{2e^{2}}\langle
G\rangle=\frac{N_{1}N_{2}}{N}+\frac{(1-2/\beta)N_{1}N_{2}}{N^{2}(1-{\rm
i}\omega\tau_{\rm dwell})}.\label{Gomega0}
\end{equation}
The frequency dependence of the conductance is now due entirely to the
weak-localization effect.

\section{Disordered wires}
\label{wires}

\subsection{DMPK equation}
\label{DMPKeq}

\subsubsection{Scaling approach to localization}
\label{scaling}

The scaling approach to localization (Abrahams {\em et al.}, 1979) studies the
limiting behavior of the conductance as one or more of the dimensions of the
system tend towards infinity. Classically, Ohm's law tells us that the
conductance $G\propto L^{d-2}$ in an $L\times L\times L$ cube ($d=3)$, an
$L\times L$ square ($d=2$), or a chain of length $L$ ($d=1$). The fundamental
result of Abrahams, Anderson, Licciar\-dello, and Ramakrishan is that this
classical scaling is only valid in three dimensions and for sufficiently weak
disorder. For $d=3$ and strong disorder, or for $d=1$ and any disorder
strength, the conductance $G\propto\exp(-L/\xi)$ decays exponentially for large
$L$. Two dimensions is the marginal case ($G\rightarrow 0$ as
$L\rightarrow\infty$ for $d=2$ in the absence of spin-orbit scattering, but the
decay is not necessarily exponential). The localization length $\xi$ depends on
the mean free path $l$. For a chain, $\xi\lesssim l$. For a cube, $\xi$
diverges with some power of $1/(l_{\rm c}-l)$ as $l$ increases towards a
critical value $l_{\rm c}$, which is of the order of the Fermi wavelength
$\lambda_{\rm F}$. At $l=l_{\rm c}$ a transition occurs from a metal ($l>l_{\rm
c}$) to an insulator ($l<l_{\rm c}$). This disorder-induced metal-insulator
transition is known as the Anderson transition (for reviews, see: Lee and
Ramakrishnan, 1985; Vollhardt and W\"{o}lfle, 1992; Brezini and Zekri, 1992;
Kramer and MacKinnon, 1993). No metal-insulator transition occurs for $d=1$ (or
for $d=2$ in the absence of spin-orbit scattering). In one dimension the system
scales towards an insulator even in the case $l\gg\lambda_{\rm F}$ of weak
disorder.

The 1979 paper of Abrahams {\em et al.} was based on a qualitative relationship
(Edwards and Thouless, 1972; Thouless, 1977) between the conductance of an open
system and the response to a change in boundary conditions of eigenstates of
the corresponding closed system. In 1980, Anderson, Thouless, Abrahams, and
Fisher proposed a {\em ``New method for a scaling theory of localization''},
based on the more precise relationship (Landauer, 1957, 1970) between the
conductance and the scattering states of the open system. They considered a
one-dimensional (1D) chain with weak scattering ($l\gg\lambda_{\rm F}$), and
computed how the transmission probability $T$ (and hence the conductance
$G=T\times 2e^{2}/h$) scales with the chain length $L$. For $L>l$ an
exponential decay was obtained, demonstrating localization. In the following
decade the scaling theory of 1D localization was developed in great detail
(Abrikosov, 1981; Mel'nikov, 1981; Kirkman and Pendry, 1984; Kumar, 1985;
Mello, 1986) and the complete distribution $P(T,L)$ of the transmission
probability was found. This solved the problem of 1D localization due to weak
disorder (for the opposite regime of strong disorder, see the reviews by
Erd\"{o}s and Herndon, 1982, and by Pendry, 1994).

A real metal wire is not one-dimensional. Typically, the width $W$ is much
greater than $\lambda_{\rm F}$, so that the number $N$ of transverse modes at
the Fermi level is much greater than one. Instead of a single transmission
probability $T$, one now has the $N$ eigenvalues $T_{n}$ of the transmission
matrix product $tt^{\dagger}$. To obtain the distribution of the conductance
$G=(2e^{2}/h)\sum_{n}T_{n}$ one now needs the joint probability distribution
$P(T_{1},T_{2},\ldots T_{N},L)$. This distribution differs essentially from the
distribution in the 1D chain, because of correlations induced by the repulsion
of nearby eigenvalues. As a consequence of the eigenvalue repulsion, the
localization length $\xi\simeq Nl$ is increased by a factor of $N$ in
comparison to the 1D case (Thouless, 1977). One can therefore distinguish a
metallic and an insulating regime. On length scales $l\ll L\ll Nl$ the
conductance decreases linearly rather than exponentially with $L$. This is the
(diffusive) metallic regime, where mesoscopic effects as weak localization and
universal conductance fluctuations occur. The insulating regime of
exponentially small conductance is entered for wire lengths $L\gtrsim Nl$.

A scaling theory of localization in multi-mode wires was pioneered by Dorokhov
(1982) and independently by Mello, Pereyra, and Kumar (1988). The DMPK equation
\begin{mathletters}
\label{DMPK}
\begin{eqnarray}
l\frac{\partial P}{\partial L}&=&
\frac{2}{\beta N+2-\beta}\sum_{n=1}^{N}
\frac{\partial}{\partial\lambda_{n}}\lambda_{n}(1+\lambda_{n})
J\frac{\partial}{\partial\lambda_{n}}\frac{P}{J},\label{DMPKa}\\
J&=&\prod_{i=1}^{N}\prod_{j=i+1}^{N}|\lambda_{j}-\lambda_{i}|^{\beta},
\label{DMPKb}
\end{eqnarray}
\end{mathletters}%
describes the evolution with increasing wire length of the distribution
function $P(\lambda_{1},\lambda_{2},\ldots \lambda_{N},L)$. [We recall the
definition of the variables $\lambda_{n}= (1-T_{n})/T_{n}$.] Equation
(\ref{DMPK}) is known as a Fokker-Planck equation (or generalized diffusion
equation) in the theory of Brownian motion (Van Kampen, 1981). One can say that
the DMPK equation is a description of scaling in a multi-mode wire as the
Brownian motion of transmission eigenvalues.

For a 1D chain ($N=1$) the Jacobian $J\equiv 1$, and  Eq.\ (\ref{DMPK})
simplifies to
\begin{equation}
l\frac{\partial}{\partial L}P(\lambda,L)=
\frac{\partial}{\partial\lambda}\lambda(1+\lambda)
\frac{\partial}{\partial\lambda}P(\lambda,L),\label{DMPK1D}
\end{equation}
independent of the symmetry index $\beta$. The diffusion equation
(\ref{DMPK1D}) was derived and solved as early as 1959 by Gertsenshtein and
Vasil'ev, in an article entitled {\em ``Wave\-guides with random
inhomogeneities and Brownian motion in the Lobachevsky plane''}. This
remarkable paper on the exponential decay of radio-waves due to weak disorder
contains many of the results which were rediscovered in the eighties for the
problem of 1D localization of electrons (see the references listed above). The
paper was noticed in the literature on classical wave propagation (Gazaryan,
1969; Papanicolaou, 1971), but apparently not among solid-state physicists.

\subsubsection{Brownian motion of transmission eigenvalues}
\label{Brownian}

Equation (\ref{DMPK}) was derived by Dorokhov (1982, for $\beta=2$), and by
Mello, Pereyra, and Kumar (1988, for $\beta=1$, with generalizations to
$\beta=2,4$ by Mello and Stone, 1991, and by Mac\^{e}do and Chalker, 1992), by
computing the incremental change of the transmission eigenvalues upon
attachment of a thin slice to the wire. It is assumed that the conductor is
weakly disordered ($l\gg\lambda_{\rm F}$), so that the scattering in the thin
slice can be treated perturbatively. A key simplification is the {\em isotropy
assumption\/} that the flux incident in one scattering channel is, on average,
equally distributed among all outgoing channels. This assumption restricts the
applicability of the DMPK equation to a wire geometry ($L\gg W$), since it
ignores the finite time scale for transverse diffusion. The derivation of the
DMPK equation given in this subsection emphasizes the fact that it holds on
length scales $\gg l$ regardless of the microscopic scattering properties of
the conductor. It is similar in spirit to the derivation given by Mello and
Shapiro (1988). An altogether different derivation has been given by
Tartakovski (1995).

\begin{figure}[tb]
\centerline{
\psfig{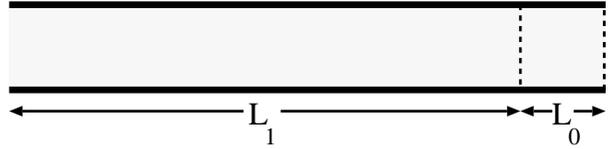}
}%
\medskip
\caption[]{
Disordered wire of length $L_{1}$ to which a segment of length $L_{0}$ is
attached. This scaling operation leads to a Brownian motion of the transmission
eigenvalues.
}\label{fig_segment}
\end{figure}

We consider a wire of length $L_{1}$, to which we attach a segment of length
$L_{0}$ (see Fig.\ \ref{fig_segment}). The combined system, of length
$L_{2}=L_{1}+L_{0}$, has transmission matrix
\begin{equation}
t_{2}=t_{1}(1-r_{0}r_{1})^{-1}t_{0},\label{t2def}
\end{equation}
where $t_{i}$ and $r_{i}$ are $N\times N$ transmission and reflection matrices
of the segment of length $L_{i}$. Equation (\ref{t2def}) ignores the
propagation of evanescent modes, which is justified if $L_{0}\gg\lambda_{\rm
F}$.  We denote by $T_{n}$ ($n=1,2,\ldots N$) the eigenvalues of the
transmission matrix product $t_{1}^{\vphantom{\dagger}}t_{1}^{\dagger}$, and by
$T_{n}+\delta T_{n}$ the eigenvalues of
$t_{2}^{\vphantom{\dagger}}t_{2}^{\dagger}$. If $L_{0}\ll l$, the change
$\delta T_{n}$ of the transmission eigenvalues can be computed by perturbation
theory. In view of the earlier requirement $L_{0}\gg\lambda_{\rm F}$, this
implies a restriction to weak scattering, $l\gg\lambda_{\rm F}$.

To second order in perturbation theory one has
\begin{equation}
\delta T_{n}=w_{nn}+\sum_{m\,(\neq n)}\frac{|w_{nm}|^{2}}{T_{n}-T_{m}}+{\cal
O}(L_{0}/l)^{3/2}.\label{deltaT}
\end{equation}
The matrix element $w_{nm}$ [of order $(L_{0}/l)^{1/2}$] is an element of the
Hermitian matrix $w=t_{2}^{\vphantom{\dagger}}t_{2}^{\dagger}-
t_{1}^{\vphantom{\dagger}}t_{1}^{\dagger}$ in the basis where
$t_{1}^{\vphantom{\dagger}}t_{1}^{\dagger}$ is diagonal. To determine $w$ in
this basis we use a polar decomposition of the transmission and reflection
matrices [cf.\ Eq.\ (\ref{polarS})],
\begin{mathletters}
\label{polar}
\begin{eqnarray}
&&t_{0}=U_{0}\sqrt{{\cal T}_{0}}V'_{0},\;\; r_{0}=U_{0}\sqrt{1-{\cal
T}_{0}}U'_{0},\label{polara}\\
&&t_{1}=U_{1}\sqrt{{\cal T}_{1}}V'_{1},\;\; r_{1}=V_{1}\sqrt{1-{\cal
T}_{1}}V'_{1}.\label{polarb}
\end{eqnarray}
\end{mathletters}%
The $U$'s and $V$'s are $N\times N$ unitary matrices, and the diagonal matrices
${\cal T}_{i}$ contain the $N$ transmission eigenvalues of segment $i$.
Combining Eqs.\ (\ref{t2def}) and (\ref{polar}), and using that $1-{\cal
T}_{0}={\cal O}(L_{0}/l)$, one obtains the expansion
\begin{eqnarray}
w&=&
\sqrt{{\cal T}_{1}}V\sqrt{1-{\cal T}_{0}}U\sqrt{1-{\cal T}_{1}}\sqrt{{\cal
T}_{1}}+{\rm H.c.}\nonumber\\
&&\mbox{}+\sqrt{{\cal T}_{1}}\left(V\sqrt{1-{\cal T}_{0}}U\sqrt{1-{\cal
T}_{1}}\right)^{2}\sqrt{{\cal T}_{1}}+{\rm H.c.}\nonumber\\
&&\mbox{}+\sqrt{{\cal T}_{1}}V\sqrt{1-{\cal T}_{0}}U(1-{\cal
T}_{1})U^{\dagger}\sqrt{1-{\cal T}_{0}}V^{\dagger}\sqrt{{\cal
T}_{1}}\nonumber\\
&&\mbox{}-\sqrt{{\cal T}_{1}}V(1-{\cal T}_{0})V^{\dagger}\sqrt{{\cal
T}_{1}}+{\cal O}(L_{0}/l)^{3/2},\label{wpolar}
\end{eqnarray}
with the definitions $U= U'_{0}V_{1}$, $V= V'_{1}U_{0}$. The abbreviation H.c.\
stands for Hermitian conjugate.

We now assume that the segment of length $L_{0}$ is taken from an ensemble with
an isotropically distributed scattering matrix. This means that the unitary
matrices in the polar decomposition (\ref{polara}) are uniformly distributed in
the unitary group. The matrix of transmission eigenvalues ${\cal T}_{0}$ may
have an arbitrary distribution. The mean free path $l$ is defined in terms of
its first moment,
\begin{equation}
\langle{\rm Tr}\,{\cal T}_{0}\rangle= N(1-L_{0}/l).\label{TrT0def}
\end{equation}
We will see in Sect.\ \ref{classicalG} that this definition differs by a
numerical coefficient (dependent on the dimensionality of the Fermi surface)
from that of the transport mean free path $l_{\rm tr}$ of kinetic theory:
\begin{equation}
\frac{l}{l_{\rm tr}}=\left\{\begin{array}{ll}
2 & (\mbox{1D chain}),\\
\pi/2 & (\mbox{Fermi circle}),\\
4/3 & (\mbox{Fermi sphere}).
\end{array}\right.\label{ltrdef}
\end{equation}
The ensemble average $\langle\cdots\rangle$ can be performed in two steps,
averaging first over the unitary matrices and then over the transmission
eigenvalues. In the absence of time-reversal symmetry ($\beta=2$), the matrices
$U$ and $V$ are independent, hence $\langle f(U,V)\rangle=\int d\mu(U)\int
d\mu(V)\,f(U,V)$. In the presence of time-reversal symmetry ($\beta=1$; the
case $\beta=4$ requires separate treatment\footnote{
If $\beta=4$ the quaternion matrices $U$ and $V$ are each others dual,
$V=U^{\rm R}$. To apply the formulas of Appendix \ref{UNintegrate}, we
represent the $N\times N$ quaternion matrix $U$ by a $2N\times 2N$ complex
matrix ${\cal U}$. Then $\langle f(U,V)\rangle=\int d\mu({\cal U})\,f({\cal
U},{\cal C}^{\rm T}{\cal U}^{\rm T}{\cal C})$, where the matrix ${\cal C}$ is
defined in Eq.\ (\ref{calCdef}).}), the matrices $U$ and $V$ are each others
transpose, hence $\langle f(U,V)\rangle=\int d\mu(U)\,f(U,U^{\rm T})$. The
integrals over the unitary group [with invariant measure $d\mu(U)$] can be
performed with the help of the formulas in Appendix \ref{UNintegrate}.

{}From Eqs.\ (\ref{deltaT}) and (\ref{wpolar}) we compute the moments of
$\delta T_{n}$ to first order in $\delta s= L_{0}/l$,
\begin{mathletters}
\label{moments}
\begin{eqnarray}
\frac{1}{\delta s}\langle\delta T_{n}\rangle&=&-T_{n}+\frac{2T_{n}}{\beta
N+2-\beta}\Bigl(1-T_{n}\nonumber\\
&&\mbox{}+\frac{\beta}{2}\sum_{m\,(\neq
n)}\frac{T_{n}+T_{m}-2T_{n}T_{m}}{T_{n}-T_{m}}\Bigr),\label{momentsa}\\
\frac{1}{\delta s}\langle\delta T_{n}\delta
T_{m}\rangle&=&\delta_{nm}\frac{4T_{n}^{2}(1-T_{n})}{\beta
N+2-\beta}.\label{momentsb}
\end{eqnarray}
\end{mathletters}%
The third and higher moments vanish to first order in $\delta s$. It follows
from the theory of Brownian motion (Van Kampen, 1981) that the probability
distribution $P(T_{1},T_{2},\ldots T_{N},s)$ of the transmission eigenvalues
evolves with increasing $s= L/l$ according to the Fokker-Planck
equation\footnote{The It\^{o}-Stratonovich ambiguity (Van Kampen, 1981) of
Brownian motion with a position-dependent diffusion coefficient does not arise,
because Eq.\ (\ref{moments}) explicitly relates the change in the variables
$T_{n}$ to their value {\em prior\/} to the change.}
\begin{equation}
\frac{\partial P}{\partial s}=\frac{1}{\delta
s}\sum_{n=1}^{N}\frac{\partial}{\partial T_{n}}\Bigl(-\langle\delta
T_{n}\rangle P+\frac{1}{2}\sum_{m=1}^{N}\frac{\partial}{\partial
T_{m}}\langle\delta T_{n}\delta T_{m}\rangle P\Bigr).\label{FokkerPlanckT}
\end{equation}
Equation (\ref{FokkerPlanckT}) becomes the DMPK equation (\ref{DMPK}) upon a
change of variables from $T_{n}$ to $\lambda_{n}=(1-T_{n})/T_{n}$.

The derivation of the DMPK equation given here rests on the assumption of {\em
isotropy\/} of the distribution of scattering matrices. It is possible to
replace the isotropy assumption by the weaker assumption of {\em equivalent
scattering channels\/} (Mello and Tomsovic, 1991, 1992). This is the assumption
that the first two moments of the reflection matrix $r_{0}$ of the thin slice
are the same as one would obtain for an isotropic distribution. For example,
for $\beta=2$ the requirement of equivalent channels is
\begin{mathletters}
\label{equichannels}
\begin{eqnarray}
&&\langle (r_{0})_{ij}\rangle=0,\;\;\langle
(r_{0})_{ij}(r_{0})_{kl}\rangle=0,\label{equichannelsa}\\
&&\langle (r_{0}^{\vphantom{\ast}})_{ij}(r_{0}^{\ast})_{kl}\rangle=
N^{-2}\langle{\rm Tr}(1-{\cal
T}_{0})\rangle\delta_{ik}\delta_{jl}.\label{equichannelsb}
\end{eqnarray}
\end{mathletters}%
One can see that Eq.\ (\ref{equichannels}) is a weaker assumption than the
isotropy assumption, by considering the case of a thin slice without
scattering. Then $r_{0}=0$ and ${\cal T}_{0}=1$, so that Eq.\
(\ref{equichannels}) is trivially satisfied, but the scattering matrix of the
thin slice has a delta-function rather than an isotropic distribution. Dorokhov
(1988) has constructed a model of $N$ weakly coupled chains for which the
equivalent-channel assumption is exact. This is a special model. More
generally, neither the isotropy nor the equivalent-channel assumption hold
exactly. For example, the transmission probability which follows from the
Boltzmann equation with isotropic impurity scattering is about twice as large
for normal incidence than it is for grazing incidence (Nieuwenhuizen and Luck,
1993), simply because about half of the electrons at grazing incidence are
scattered back before penetrating a mean-free-path deep into the disordered
region. [More subtle, quantum mechanical deviations have been noticed in
simulations of the Anderson model by Jalabert and Pichard (1995).] Still, as we
will see, the DMPK equation provides a remarkably accurate description of the
distribution of the transmission eigenvalues of a disordered wire, on length
scales ranging from below the mean free path to above the localization length.
Moreover, in the metallic regime the restriction $L\gg W$ to a wire geometry
can be relaxed considerably. It is only when the transverse dimension $W$
becomes comparable to the localization length that the DMPK equation breaks
down completely.

\subsubsection{Mapping to a free-fermion model}
\label{freefermion}

In the absence of time-reversal symmetry, the DMPK equation is equivalent to a
Schr\"{o}dinger equation for $N$ non-interacting fermions in one dimension.
This formal correspondence permits an exact solution of the DMPK equation for
$\beta=2$ (Beenakker and Rejaei, 1993, 1994a).

To carry out the mapping it is convenient to first write the DMPK equation in
terms of a new set of variables $x_{n}$, related to $\lambda_{n}$ and $T_{n}$
by
\begin{equation}
\lambda_{n}=\sinh^{2}x_{n},\;\;T_{n}=1/\cosh^{2}x_{n},\;\;x_{n}\geq
0.\label{xlambdaTrel}
\end{equation}
As discussed in Sec.\ \ref{SandM}, $\exp(\pm 2x_{n})$ is an eigenvalue of the
transfer-matrix product $MM^{\dagger}$. By applying this change of variables to
Eq.\ (\ref{DMPK}), one finds that the distribution $P(x_{1},x_{2},\ldots
x_{N},s)$ of the $x_{n}$'s evolves with increasing $s=L/l$ according to a
Fokker-Planck equation with a {\em constant\/} diffusion coefficient,
\begin{mathletters}
\label{FokkerPlanckx}
\begin{eqnarray}
\frac{\partial P}{\partial s}&=&
\frac{1}{2\gamma}\sum_{n=1}^{N}\frac{\partial}{\partial x_{n}}
\Bigl(\frac{\partial P}{\partial x_{n}}+
\beta P\frac{\partial\Omega}{\partial x_{n}}\Bigr),
\label{FPxa}\\
\Omega&=&-\sum_{i=1}^{N}\sum_{j=i+1}^{N}
\ln|\sinh^{2}x_{j}-\sinh^{2}x_{i}|\nonumber\\
&&\mbox{}-\frac{1}{\beta}\sum_{i=1}^{N}\ln|\sinh 2x_{i}|.\label{FPxb}
\end{eqnarray}
\end{mathletters}%
We have abbreviated
\begin{equation}
\gamma=\beta N+2-\beta.\label{gammadef}
\end{equation}

The probability distribution $P(\{x_{n}\},s)$ is related to a wave function
$\Psi(\{x_{n}\},s)$ by the transformation
\begin{equation}
P=\Psi\,{\rm e}^{-\beta\Omega/2},\label{PPsirelation}
\end{equation}
originally introduced by Sutherland (1972) to solve the Fokker-Planck equation
of Dyson's Brownian-motion model (given by Eq.\ (\ref{PEtauFP}) for $\beta=2$).
Substitution into Eq.\ (\ref{FokkerPlanckx}) yields for $\Psi$ a
Schr\"{o}dinger equation in imaginary time,
\begin{mathletters}
\label{Schrodinger}
\begin{eqnarray}
-\frac{\partial\Psi}{\partial s}&=&{\cal H}\Psi,
\label{Schrodingera}\\
{\cal H}&=&-\frac{1}{2\gamma}\sum_{i}\left(\frac{\partial^{2}}
{\partial x_{i}^{2}}+\frac{1}{\sinh^{2}2x_{i}}\right)\nonumber\\
&&\mbox{}+\frac{\beta(\beta-2)}{2\gamma}\sum_{i<j}
\frac{\sinh^{2}2x_{j}+\sinh^{2}2x_{i}}
{(\cosh 2x_{j}-\cosh 2x_{i})^{2}}.\label{Schrodingerb}
\end{eqnarray}
\end{mathletters}%
The Fokker-Planck equation considered by Sutherland maps onto a Hamiltonian
with a translationally invariant interaction potential $(x-x')^{-2}$ (the
Calogero-Sutherland Hamiltonian: Calogero, 1969; Sutherland, 1971). The
interaction potential in Eq.\ (\ref{Schrodinger}) is not translationally
invariant. Using a trigonometric identity it can be rewritten as
$\sinh^{-2}(x-x')+\sinh^{-2}(x+x')$, to show that the breaking of translational
invariance is due to the interaction between $x$ and an ``image charge'' at
$-x'$. Caselle (1995) has pointed out that the Hamiltonian (\ref{Schrodinger})
belongs to the same family as the Calogero-Sutherland Hamiltonian, in the sense
that both represent the Laplacian on a certain curved space (first identified
by H\"{u}ffmann, 1990).

For $\beta=2$ the interaction vanishes identically, reducing ${\cal H}$ to a
sum of single-particle Hamiltonians ${\cal H}_{0}$,
\begin{equation}
{\cal H}_{0}=-\frac{1}{4N}\frac{\partial^{2}}
{\partial x^{2}}-\frac{1}{4N\sinh^{2}2x}.\label{H0def}
\end{equation}
The spectrum of ${\cal H}_{0}$ is continuous, with eigenvalues
$\varepsilon=\frac{1}{4}k^{2}/N$ and eigenfunctions (scattering states)
\begin{equation}
\psi_{k}(x)=[\pi k\tanh(\case{1}{2}\pi k)\sinh(2x)]^{1/2}\,{\rm
P}_{\frac{1}{2}({\rm i}k-1)}(\cosh 2x)\label{psikdef}
\end{equation}
labeled by a wave number $k>0$. (The Legendre functions ${\rm
P}_{\frac{1}{2}({\rm i}k-1)}$ are known as ``toroidal functions'', because they
appear as solutions to the Laplace equation in toroidal coordinates.) An
antisymmetric $N$-fermion wave function $\Psi$ can be constructed from a Slater
determinant of single-particle scattering states. The transformation
\begin{equation}
P=\Psi\times\prod_{i<j}(\sinh^{2}x_{j}-\sinh^{2}x_{i})\prod_{i}(\sinh
2x_{i})^{1/2}\label{PPsirelation2}
\end{equation}
then yields a symmetric probability distribution $P$.

We conclude that the Brownian motion of $N$ transmission eigenvalues in the
absence of time-reversal symmetry is equivalent to a scattering problem of $N$
non-interacting fermions in one dimension. The correlations due to eigenvalue
repulsion are fully accounted for by the requirement of an antisymmetric
$N$-fermion wave function. An exact solution for $P$ can be written down for
arbitrary initial conditions. For the application to an ensemble of disordered
wires we need the ballistic initial condition\footnote{
Other initial conditions have been considered by Beenakker and Melsen (1994)
[see Sec.\ \ref{obstacle_initial}] and by Frahm and M\"{u}ller-Groeling (1996)
[to compute the correlator $\langle G(s)G(s+\Delta s)\rangle$].}
\begin{equation}
\lim_{s\rightarrow
0}P(\{x_{n}\},s)=\prod_{i=1}^{N}\delta(x_{i}-0^{+}),\label{ballistic}
\end{equation}
which says that all $T_{n}$'s are equal to 1 if $L\ll l$. The exact solution of
the DMPK equation (\ref{DMPK}) with $\beta=2$ for the initial condition
(\ref{ballistic}) is (Beenakker and Rejaei, 1993, 1994a)
\begin{eqnarray}
P=C(s)\prod_{i<j}(\sinh^{2}x_{j}-\sinh^{2}x_{i})
\prod_{i}\sinh 2x_{i}\nonumber\\
\mbox{}\times {\rm Det}\,\Bigl[\int_{0}^{\infty}\!\!dk\,
{\rm e}^{-k^{2}s/4N}\tanh(\case{1}{2}\pi k)k^{2m-1}\nonumber\\
\mbox{}\times{\rm P}_{\frac{1}{2}({\rm i}k-1)}(\cosh 2x_{n})\Bigr].
\label{exactsolution}
\end{eqnarray}
Here $C(s)$ is an $x$-independent normalization factor and ${\rm Det}\,a_{nm}$
denotes the determinant of the matrix with elements $a_{nm}$ ($1\le n,m\le N$).
For $N=1$, Eq.\ (\ref{exactsolution}) reduces to the $\beta$-independent
solution of the scaling equation (\ref{DMPK1D}) for a 1D chain (first obtained
by Gertsenshtein and Vasil'ev, 1959). Equation (\ref{exactsolution})
generalizes the 1D-chain solution to arbitrary $N$, for the case $\beta=2$.

In the presence of time-reversal symmetry, for $\beta=1$ or $4$, there exists
no exact solution of the DMPK equation in terms of functions of a single
variable [such as the Legendre functions in Eq.\ (\ref{exactsolution})]. The
interaction potential in the Hamiltonian (\ref{Schrodinger}) leads to
correlations between the $x_{n}$'s which can not be represented by a Slater
determinant. Caselle (1995) has shown that these correlations are described by
functions of $N$ variables known in the mathematical literature as ``zonal
spherical functions''. It is only for $\beta=2$ that these functions factorize
into functions of a single variable. For $\beta=1$ or $4$, an expression for
$P$ in terms of functions of a single variable can be obtained in the metallic
regime $s\ll N$ (Sec.\ \ref{Trepulsion}) and also in the insulating regime
$s\gg N$ (Sec.\ \ref{lognormal}) --- but not in the crossover regime.

\subsubsection{Equivalence to a supersymmetric field theory}
\label{fieldtheory}

A field theory for localization in disordered wires has been developed by
Efetov and Larkin (1983), building on work by Wegner (1979). (For reviews, see
Efetov, 1983, 1996.) The diffusion modes are represented by matrix fields $Q$
containing an equal number of commuting and anti-commuting elements. By analogy
with the supersymmetry between bosons and fermions in particle physics, such
matrices are called supersymmetric matrices, or ``supermatrices''. The
interaction between the diffusion modes is described by a model known in
quantum field theory as the non-linear $\sigma$ model (Itzykson and Zuber,
1980). The adjective ``non-linear'' refers to the constraint $Q^{2}=1$, and the
letter $\sigma$ originates from an early notation for the fields. The
restriction to a wire geometry makes the fields one-dimensional (1D) in the
spatial coordinate. It has been demonstrated by Brouwer and Frahm (1996)
(following up on a paper by Rejaei, 1996) that the 1D $\sigma$ model is
equivalent to the thick-wire limit of the DMPK equation. The thick-wire limit
is defined by $N\rightarrow\infty$, $L/l\rightarrow\infty$ at constant ratio
$Nl/L$. The equivalence holds for an arbitrary $p$-point correlation function
of the transmission eigenvalues. Here we give an outline of the equivalence
proof for the simplest case $p=1$. It is then sufficient to consider $8\times
8$ supermatrices in the $\sigma$ model.

The quantity which relates the 1D $\sigma$ model to the DMPK equation is the
generating function $F(\theta_{1},\theta_{2},\theta_{3},\theta_{4};
T_{1},T_{2},\ldots T_{N})$, defined by
\begin{mathletters}
\label{Fgeneratingdef}
\begin{eqnarray}
F(\{\theta_{i}\};\{T_{n}\})= \prod_{n=1}^{N}\left(\frac{1-T_{n}\sin^{2}\bigl[
\frac{1}{2}(\theta_{3}+\theta_{4})\bigr]} {1+T_{n}\sinh^{2}\bigl[
\frac{1}{2}(\theta_{1}+\theta_{2})\bigr]}\right. \nonumber\\
\mbox{}\times\left. \frac{1-T_{n}\sin^{2}\bigl[
\frac{1}{2}(\theta_{3}-\theta_{4})\bigr]} {1+T_{n}\sinh^{2}\bigl[
\frac{1}{2}(\theta_{1}-\theta_{2})
\bigr]}\right)^{\textstyle\frac{1}{2}(1+\delta_{\beta,4})},
\label{Fgeneratingdefa}\\
\theta_{2}\equiv 0 \;{\rm if}\;\beta\in\{2,4\};\;\theta_{4}\equiv 0\;{\rm
if}\;\beta\in\{1,2\}. \label{Fgeneratingdefb}
\end{eqnarray}
\end{mathletters}%
The doubling of the exponent in Eq.\ (\ref{Fgeneratingdefa}) for $\beta=4$
originates from Kramers' degeneracy of the transmission eigenvalues $T_{n}$.
(The product over $n$ runs only over the $N$ distinct eigenvalues.) The
function $F$ is called a generating function because its ensemble average
yields the density $\rho(T)$ of the transmission eigenvalues,
\begin{equation}
\rho(T)=-\bigl(\pi T\sqrt{1-T}\bigr)^{-1}\,{\rm
Re}\,\frac{\partial}{\partial\theta_{3}}\langle
F\rangle,\label{rhogeneratingdef}
\end{equation}
evaluated at $\theta_{2}=\theta_{4}=0$, $\sin^{2}\frac{1}{2}\theta_{3}=
-\sinh^{2}\frac{1}{2}\theta_{1}=(T+{\rm i}0^{+})^{-1}$. The angles $\theta_{i}$
parameterize the supermatrices of the non-linear $\sigma$ model. There are
three independent $\theta_{i}$'s if $\beta=1,4$ and two if $\beta=2$. For
example, if $\beta=2$ the parameterization is (Efetov, 1983, 1996)
\begin{equation}
Q=\left(\begin{array}{cc}u&0\\0&v\end{array}\right)
\left(\begin{array}{cc}\cos\hat{\theta}&{\rm i}\sin\hat{\theta}\\ -{\rm
i}\sin\hat{\theta}&-\cos\hat{\theta}\end{array}\right)
\left(\begin{array}{cc}u^{-1}&0\\0&v^{-1}\end{array}\right). \label{Qparam}
\end{equation}
Here $u$ and $v$ are $4\times 4$ unitary supermatrices and $\hat{\theta}$ is a
$4\times 4$ diagonal matrix with elements $\theta_{3},\theta_{3},{\rm
i}\theta_{1},{\rm i}\theta_{1}$ on the diagonal. The variable
$\theta_{1}\in(0,\infty)$ is called a non-compact angle, and
$\theta_{3}\in(0,\pi)$ is called a compact angle. A similar parameterization of
$Q$ exists for $\beta=1$ (with two non-compact angles $\theta_{1},\theta_{2}$
and one compact angle $\theta_{3}$), and for $\beta=4$ (with two compact angles
$\theta_{3},\theta_{4}$ and one non-compact angle $\theta_{1}$).

A remarkable property of the function $F$ is that the DMPK equation yields a
{\em closed\/} evolution equation for its ensemble average (Brouwer and Frahm,
1996):
\begin{equation}
l\frac{\partial\langle F\rangle}{\partial
L}=\frac{\beta(1+\delta_{\beta,4})^{-1}}{\beta N+2-\beta}\sum_{i}
\frac{1}{J_{\theta}}\frac{\partial}{\partial\theta_{i}}J_{\theta}
\frac{\partial}{\partial\theta_{i}}\langle F\rangle.\label{Fevolution}
\end{equation}
The sum over $i$ runs over the two ($\beta=2$) or three ($\beta=1,4$)
independent angles. The factor $J_{\theta}$ is the Jacobian from the space of
supermatrices $Q$ to the space of angles $\theta_{i}$, given by
\begin{equation}
J_{\theta}=\left\{
\begin{array}{cl}
\frac{\displaystyle\sinh\theta_{1}\sinh\theta_{2}\sin^{3}\theta_{3}}
{\displaystyle\prod_{\sigma_{1},\sigma_{2}=\pm 1}\sinh^{2}[\case{1}{2}
(\theta_{1}+\sigma_{1}\theta_{2}+{\rm i}\sigma_{2}\theta_{3})]}
&{\rm if}\;\beta=1,\\
\frac{\displaystyle\sinh\theta_{1}\sin\theta_{3}}
{\displaystyle\prod_{\sigma_{1}=\pm 1}\sinh^{2}[\case{1}{2} (\theta_{1}+{\rm
i}\sigma_{1}\theta_{3})]}
&{\rm if}\;\beta=2,\\
\frac{\displaystyle\sinh^{3}\theta_{1}\sin\theta_{3}\sin\theta_{4}}
{\displaystyle\prod_{\sigma_{1},\sigma_{2}=\pm 1}\sinh^{2}[\case{1}{2}
(\theta_{1}+{\rm i}\sigma_{1}\theta_{3}+{\rm i}\sigma_{2}\theta_{4})]}
&{\rm if}\;\beta=4.
\end{array}\right.
\label{Jacobiantheta}
\end{equation}
The practical importance of Eq.\ (\ref{Fevolution}) is that it allows one to
compute the ensemble average of $F$ (and hence the eigenvalue density) by
solving a partial differential equation involving only two or three variables
--- in contrast to the $N$ variables in the DMPK equation (\ref{DMPK}). [A
similar method exists for Dyson's Brownian-motion model (Guhr, 1996).] The
conceptual importance of Eq.\ (\ref{Fevolution}) is that (for $N\gg 1$) the
same evolution equation is obtained if one computes $\langle F\rangle$ from the
$1D$ $\sigma$ model. This was shown by Rejaei (1996; for $\beta=2$) and by
Brouwer and Frahm (1996; for $\beta=1,4$). The conclusion is that $\rho(T)$ is
the same whether computed from the 1D $\sigma$ model or from the DMPK equation.
This equivalence can be generalized to all $p$-point correlation functions. It
holds for arbitrary $Nl/L$ if the same initial conditions are chosen in both
descriptions, but requires the thick-wire limit $N\gg 1$ (since the $\sigma$
model can only be formulated in this limit).

The 1D $\sigma$ model has been derived from three microscopic descriptions of
the conductor. Efetov and Larkin (1983) started from a homogeneous wire with a
white-noise potential. Iida, Weidenm\"{u}ller, and Zuk (1990a, 1990b) studied a
chain of disordered grains, each grain having a random Hamiltonian drawn from
the Gaussian ensemble. Fyodorov and Mirlin (1991, 1994) considered a
tight-binding Hamiltonian, whose non-zero elements lie in a band around the
diagonal (a so-called banded random matrix). Because of the equivalence
discussed above, each of these three models can also be considered as being a
microscopic model for the DMPK equation.

\subsection{Metallic regime}
\label{metallic}

\subsubsection{Conductance}
\label{classicalG}

In the metallic regime, for wire lengths $L$ much less than the localization
length $Nl$, the conductance is known to decrease linearly with $L$ (Ohm's
law). Let us verify that the DMPK equation correctly describes this classical
scaling for $Nl/L\gg 1$.

We use the method of moments of Mello and Stone (Mello, 1988; Mello and Stone,
1991). This is a method for computing the moments of
\begin{equation}
M_{q}=\sum_{n=1}^{N}T_{n}^{q},\;\;q=1,2,\ldots,\label{Mqdef}
\end{equation}
as an expansion in inverse powers of $N$. From the DMPK equation (\ref{DMPK})
one derives a hierarchy of coupled evolution equations for moments of $M_{q}$.
For example, the evolution of $M_{1}$, $M_{2}$, and $M_{3}$ is coupled by the
equation
\begin{eqnarray}
\frac{\partial}{\partial s}\langle M_{1}^{p}\rangle&=&\frac{-p\beta}{\beta
N+2-\beta}\left\langle
M_{1}^{p+1}-(1-2/\beta)M_{1}^{p-1}M_{2}^{\vphantom{p}}\right.\nonumber\\
&&\left.\mbox{}-2(p-1)\beta^{-1}M_{1}^{p-2}(M_{2}^{\vphantom{p}}-
M_{3}^{\vphantom{p}})\right\rangle. \label{M1equation}
\end{eqnarray}
The hierarchy closes order by order in the large-$N$ expansion. Indeed, since
$M_{q}^{p}={\cal O}(N^{p})$, Eq.\ (\ref{M1equation}) reduces to leading order
in $N$ to
\begin{equation}
\frac{\partial}{\partial s}\langle M_{1}^{p}\rangle=-pN^{-1}\langle
M_{1}^{p+1}\rangle+{\cal O}(N^{p-1}).\label{M1leadingorder}
\end{equation}
Notice that the symmetry index $\beta$ has dropped out in this order. The
ballistic initial condition ($T_{n}=1$ for all $n$ if $s=0$) implies
\begin{equation}
\lim_{s\rightarrow 0}\langle M_{1}^{p}\rangle=N^{p}.\label{M1initial}
\end{equation}
Equation (\ref{M1leadingorder}), with $p=1,2,\ldots$, forms a recursive set of
differential equations. The solution with initial condition (\ref{M1initial})
is
\begin{equation}
\langle M_{1}^{p}\rangle=N^{p}(1+s)^{-p}+{\cal O}(N^{p-1}).\label{M1solution}
\end{equation}

In view of the Landauer formula (\ref{Landauer}), the average conductance is
$\langle G\rangle=G_{0}\langle M_{1}\rangle$ (with $G_{0}=2e^{2}/h$), hence
\begin{equation}
\langle G/G_{0}\rangle=N(1+s)^{-1}+{\cal O}(N^{0}).\label{GDrude}
\end{equation}
In the diffusive limit ($s= L/l\gg 1$) the conductance decreases linearly with
$L$, $\langle G/G_{0}\rangle\rightarrow Nl/L$, as expected. Furthermore,
comparison with the Drude formula (Ashcroft and Mermin, 1976) shows that the
mean free path $l$ of the scaling theory [defined in Eq.\ (\ref{TrT0def})] is
related to the transport mean free path $l_{\rm tr}$ of kinetic theory by the
numerical coefficient given in Eq.\ (\ref{ltrdef}). In the ballistic limit
($s\ll 1$) the conductance $\langle G/G_{0}\rangle\rightarrow N$ reaches the
contact conductance of an $N$-mode wire between wide reservoirs. The crossover
from the ballistic to the diffusive limit, for $s\simeq 1$, is not described
exactly by Eq.\ (\ref{GDrude}) --- but the error is small [about 3\% for
isotropic impurity scattering (De Jong, 1994)].

By carrying out the expansion of the moments to order $N^{0}$, one can compute
the weak-localization correction $\delta G= \langle G\rangle-G_{0}(1+s)^{-1}$
to the average conductance (\ref{GDrude}), as well as the variance ${\rm
Var}\,G=\langle G^{2}\rangle-\langle G\rangle^{2}$. The results are (Mello,
1988)
\begin{eqnarray}
\delta G/G_{0}&=&\frac{1}{3}(1-2/\beta)\frac{s^{3}}{(1+s)^{3}}+{\cal
O}(N^{-1}),\label{deltaGDMPK}\\
{\rm Var}\,G/G_{0}&=&\frac{2}{15}\beta^{-1}
\left(1-\frac{1+6s}{(1+s)^{6}}\right)
+{\cal O}(N^{-1}).\label{VarGDMPK}
\end{eqnarray}
The diffusive limits $s\rightarrow\infty$ of Eqs.\ (\ref{deltaGDMPK}) and
(\ref{VarGDMPK}),
\begin{equation}
\delta G/G_{0}\rightarrow\case{1}{3}(1-2/\beta),\;\;{\rm
Var}\,G/G_{0}\rightarrow\case{2}{15}\beta^{-1},\label{deltaGVarGDMPK}
\end{equation}
agree precisely with diagrammatic perturbation theory (Lee and Stone, 1985;
Mello and Stone, 1991).

The method of moments can in principle be applied to all polynomial linear
statistics, {\em i.e.} transport properties of the form $A=\sum_{n}a(T_{n})$
with $a(T)$ a polynomial in $T$. It is an efficient way to compute the mean and
variance of the conductance (for which $a(T)=T$), since only a few levels of
the hierarchy of evolution equations have to be considered. With a great deal
of effort it is possible to apply the method of moments to the shot-noise power
(De Jong and Beenakker, 1992), for which $a(T)=T-T^{2}$. Other transport
properties, for which $a(T)$ is not a polynomial, require the more general
method discussed in the next subsection.

\begin{figure}[tb]
\centerline{
\psfig{figure=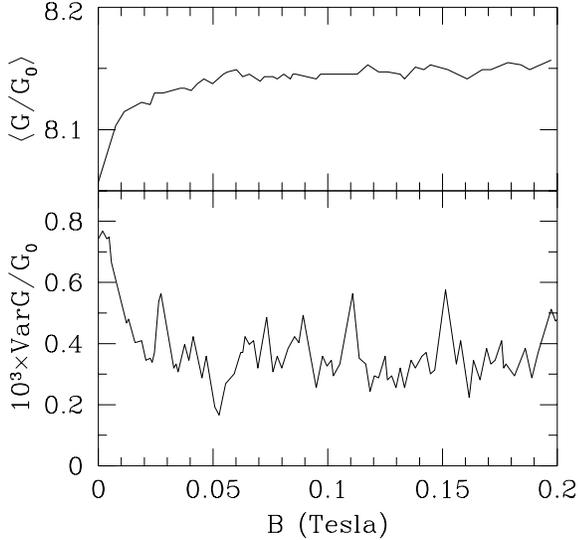,width= 8cm}
}%
\caption[]{
Mean and variance of the conductance as a function of magnetic field, measured
by averaging over 50 impurity configurations in a single Si-doped GaAs wire
($T=45\,{\rm mK}$, $W=0.09\,\mu{\rm m}$, $L=10\,\mu{\rm m}$). Uncorrelated
impurity configurations were generated by thermal cycling to room temperature.
A short phase-coherence length $l_{\phi}\ll L$ reduces the zero-field
weak-localization correction and variance to $\delta G/G_{0}=-l_{\phi}/L$,
${\rm Var}\,G/G_{0}=3(l_{\phi}/L)^{3}$. Comparing with the data we estimate
$l_{\phi}\approx 0.65\,\mu{\rm m}$. The $\beta=1\rightarrow\beta=2$ transition
leads to an increase of the average conductance and to a halving of the
variance, as observed in the experiment. After Mailly and Sanquer (1992).
}\label{fig_UCF}
\end{figure}

In experiments on disordered wires, phase coherence is in general not
maintained throughout the whole wire length. The numerical coefficients in Eq.\
(\ref{deltaGVarGDMPK}) are therefore much larger than measured (for a review,
see Beenakker and Van Houten, 1991). The $\beta$-dependence is insensitive,
however, to phase-breaking processes. Moreover, while the numerical
coefficients are specific for a wire geometry, the $\beta$-dependence is the
same in wires, thin films and bulk samples. In the absence of spin-orbit
scattering, application of a magnetic field induces a
$\beta=1\rightarrow\beta=2$ transition, leading to an increase of the average
conductance and a reduction of the variance by a factor of two. Measurements by
Mailly and Sanquer (1992) of this symmetry-class transition are shown in Fig.\
\ref{fig_UCF}. The typical field scale for the transition is one flux quantum
through a phase coherent region.\footnote{At much higher fields (when the
Zeeman energy becomes larger than the Thouless energy), a further reduction of
the variance by a factor of two takes place, associated with the removal of
spin degeneracy (measured by Debray {\em et al.}, 1989).} In the presence of
strong spin-orbit scattering, a magnetic field induces a
$\beta=4\rightarrow\beta=2$ transition, leading to a decrease of the average
conductance. The change in $\beta$ is accompanied by a removal of Kramers'
degeneracy of the transmission eigenvalues. The net result is that ${\rm
Var}\,G$ is reduced by a factor of two, just as in the absence of spin-orbit
scattering (Altshuler and Shklovski\u{\i}, 1986; measured by Birge, Golding,
and Haemmerle, 1989; Millo {\em et al.}, 1990).

\subsubsection{Other transport properties}
\label{generalA}

To compute the mean and variance of arbitrary linear statistics
$A=\sum_{n}a(x_{n})$, one needs the density of transmission eigenvalues
\begin{equation}
\rho(x)=\left\langle\sum_{i}\delta(x-x_{i})\right\rangle\label{rhoxdef}
\end{equation}
and the two-point correlation function
\begin{equation}
K(x,x')=\left\langle\sum_{i,j}\delta(x-x_{i})
\delta(x'-x_{j})\right\rangle-\rho(x)\rho(x').\label{Kxdef}
\end{equation}
(We recall the parameterization $T_{n}= 1/\cosh^{2}x_{n}$, $x_{n}\geq 0$.) In
the metallic regime it is sufficient to know $\rho$ and $K$ to order $N^{0}$.
Dorokhov (1984) and Mello and Pichard (1989) computed the leading order term in
$\rho$ (which is of order $N$), while Beenakker (1994b) and Mac\^{e}do and
Chalker (1994) computed the next term (of order $N^{0}$). The leading order
term in $K$ (which is of order $N^{0}$) was computed by Chalker and Mac\^{e}do
(1993) and by Beenakker and Rejaei (1993). The derivations are given in the
following subsections. The results are (in the diffusive limit $L\gg l$):
\begin{equation}
\rho(x)=\frac{Nl}{L}+\left(1-\frac{2}{\beta}\right)\left[\frac{1}{4}\,
\delta(x-0^{+})+\frac{1}{4x^{2}+\pi^{2}}\right],\label{RHOXRESULT}
\end{equation}
\begin{mathletters}
\label{Kxresult}
\begin{eqnarray}
K(x,x')&=&{\cal K}(x-x')+{\cal K}(x+x'),\label{Kxresulta}\\
{\cal
K}(x)&=&-\frac{1}{2\beta\pi^{2}}\,\frac{d^{\,2}}{dx^{2}}\ln[1+(\pi/x)^{2}]
.\label{Kxresultb}
\end{eqnarray}
\end{mathletters}%

The density $\rho(x)$ has a cutoff at $x\simeq L/l$, such that
$\int_{0}^{\infty}\!dx\,\rho(x)=N$. Since only the range $x\lesssim 1$
contributes to transport properties, this large-$x$ cutoff need not be
specified more accurately. The integrable singularities at $x=0$ in both $\rho$
and ${\cal K}$ are in reality smeared out over a few eigenvalue spacings.
Equations (\ref{RHOXRESULT}) and (\ref{Kxresult}) are sufficiently accurate if
$a(x)$ is smooth on the scale $\delta x$ of the eigenvalue spacing. Since
$\delta x\simeq L/Nl\ll 1$ in the metallic regime, this is not a strong
requirement.

The mean and variance of $A=\sum_{n}a(x_{n})$ follow upon integration,
\begin{eqnarray}
\langle A\rangle&=&\frac{Nl}{L}\int_{0}^{\infty}\!\!dx\,a(x)\nonumber\\
&&\mbox{}+\left(1-\frac{2}{\beta}\right)\left[\frac{1}{4}a(0)+
\int_{0}^{\infty}\!\!dx\,\frac{a(x)}
{4x^{2}+\pi^{2}}\right],\label{meanAresult}
\end{eqnarray}
\begin{mathletters}
\label{VarAresult}
\begin{eqnarray}
{\rm Var}\,A&=&\frac{1}{2\beta\pi^{2}}\int_{0}^{\infty}
\!\!dx\int_{0}^{\infty}
\!\!dx'\left(\frac{da(x)}{dx}\right)
\left(\frac{da(x')}{dx'}\right)\nonumber\\
&&\mbox{}\times\ln
\left(\frac{1+\pi^{2}(x-x')^{-2}}{1+\pi^{2}(x+x')^{-2}}\right).
\label{VarAresulta}
\end{eqnarray}
The double integral in Eq.\ (\ref{VarAresulta}) reduces to a single integral if
the Fourier transform $a(k)=2\int_{0}^{\infty}\!\!dx\,a(x)\cos kx$ is known,
\begin{equation}
{\rm Var\,}A=\frac{1}{2\beta\pi^{2}}\int_{0}^{\infty}\!\!dk\, \left(1-{\rm
e}^{-\pi k}\right)k\,|a(k)|^{2}.\label{VarAresultb}
\end{equation}
\end{mathletters}%
The first term  in Eq.\ (\ref{meanAresult}) is the semiclassical value of $A$,
which is of order $N$ and $\beta$-independent. The second term is the
weak-localization correction, which is of order $N^{0}$ and has a $1-2/\beta$
dependence on the symmetry index. The variance (\ref{VarAresult}) has no order
$N$ contribution ({\em universality}). The leading order term is of order
$N^{0}$ and is inversely proportional to $\beta$.

Equations (\ref{meanAresult}) and (\ref{VarAresult}) reduce the computation of
$\langle A\rangle$ and ${\rm Var}\,A$ to a quadrature, regardless of the
complexity of the function $a(x)$. Let us check these formulas for the case
that $A$ is the conductance, $G/G_{0}=\sum_{n}1/\cosh^{2}x_{n}$. Substitution
of $a(x)=1/\cosh^{2}x$ into Eq.\ (\ref{meanAresult}) yields $\langle
G/G_{0}\rangle=Nl/L+\frac{1}{3}(1-2/\beta)$, in agreement with Eqs.\
(\ref{GDrude}) and (\ref{deltaGDMPK}) in the diffusive limit ($s\gg 1$).
Similarly, substitution of the Fourier transform $a(k)=\pi
k/\sinh(\frac{1}{2}\pi k)$ into Eq.\ (\ref{VarAresultb}) yields ${\rm
Var\,}G/G_{0}=\frac{2}{15}\beta^{-1}$, in agreement with Eq.\ (\ref{VarGDMPK}).

\subsubsection{Transmission eigenvalue density}
\label{Tdensity}

The derivation which we present of the eigenvalue density in the metallic
regime is based on Mello and Pichard (1989) for the ${\cal O}(N)$ term and
Beenakker (1994b) for the ${\cal O}(N^{0})$ correction. Starting point is the
DMPK equation (\ref{DMPK}) for the probability distribution
$P(\{\lambda_{n}\},s)$ of the $\lambda$-variables. We seek to reduce it to an
equation for the density $\rho(\lambda,s)=
\langle\sum_{n}\delta(\lambda-\lambda_{n})\rangle$. Multiplying both sides of
Eq.\ (\ref{DMPK}) by $\sum_{n}\delta(\lambda-\lambda_{n})$ and integrating over
$\lambda_{1},\lambda_{2},\ldots\lambda_{N}$, one obtains the equation
\begin{mathletters}
\label{rho1}
\begin{eqnarray}
&&\frac{\partial\rho}{\partial s}=\frac{2}{\gamma}\,
\frac{\partial}{\partial\lambda}\lambda(1+\lambda)
\left(\frac{\partial\rho}{\partial\lambda}-\beta I\right),\label{rho1a}\\
&&I(\lambda,s)=\int_{0}^{\infty}\!\!
\frac{d\lambda'}{\lambda-\lambda'}\left\langle\sum_{i\neq
j}\delta(\lambda-\lambda_{i})\delta(\lambda'-\lambda_{j})
\right\rangle.\label{rho1b}
\end{eqnarray}
\end{mathletters}%
(Recall the definition $\gamma=\beta N+2-\beta$.) The integral over the pair
distribution function has the large-$N$ expansion (Dyson, 1972; see Appendix
\ref{Dysonexpansion} for a derivation)
\begin{equation}
\frac{I(\lambda,s)}{\rho(\lambda,s)}=\int_{0}^{\infty}\!\!
d\lambda'\,\frac{\rho(\lambda',s)}{\lambda-\lambda'}
+\frac{1}{2}\,\frac{\partial}{\partial\lambda}\ln\rho(\lambda,s)
+{\cal O}(N^{-1}).\label{Dyson}
\end{equation}
Substitution into Eq.\ (\ref{rho1}) gives a non-linear evolution equation for
the eigenvalue density,
\begin{eqnarray}
\frac{\partial\rho}{\partial s}=\frac{1}{\gamma}\,
\frac{\partial}{\partial\lambda}\lambda(1+\lambda)\rho
\frac{\partial}{\partial\lambda}\left(
\vphantom{\int_{0}^{\infty}}(2-\beta)\ln\rho
\right.\nonumber\\
\left.\mbox{}-2\beta\int_{0}^{\infty}\!\!
d\lambda'\,\rho(\lambda',s)\ln|\lambda-\lambda'|\right).\label{rho2}
\end{eqnarray}

At this point it is convenient to switch from the $\lambda$ to the
$x$-variables (defined by $\lambda_{n}=\sinh^{2}x_{n}$). The densities are
related by $\rho(x,s)=\rho(\lambda,s)\,d\lambda/dx$. In terms of the
$x$-variables, Eq.\ (\ref{rho2}) takes the form
\begin{eqnarray}
\frac{\partial\rho}{\partial s}=\frac{1}{4\gamma}\,
\frac{\partial}{\partial x}\rho
\frac{\partial}{\partial x}\left(\vphantom{\int_{0}^{\infty}}
(2-\beta)\bigl(\ln\rho-\ln|\sinh 2x|\bigr)\right.\nonumber\\
\left.\mbox{}-2\beta\int_{0}^{\infty}\!\!
dx'\,\rho(x',s)\ln |\sinh^{2}x-\sinh^{2}x'|\right).\label{rho3}
\end{eqnarray}
We need to solve Eq.\ (\ref{rho3}) to the same order in $N$ as the expansion
(\ref{Dyson}), {\em i.e.} neglecting terms of order $N^{-1}$. To this end we
decompose $\rho=\rho_{0}+\delta\rho$, with $\rho_{0}$ of order $N$ and
$\delta\rho$ of order $N^{0}$. Substitution into Eq.\ (\ref{rho3}) yields to
order $N$ an equation for $\rho_{0}$,
\begin{eqnarray}
\frac{\partial\rho_{0}}{\partial s}=-\frac{1}{2N}\,
\frac{\partial}{\partial x}\rho_{0}
\frac{\partial}{\partial x}\int_{0}^{\infty}\!\!
dx'\,\rho_{0}(x',s)\nonumber\\
\mbox{}\times\ln |\sinh^{2}x-\sinh^{2}x'|.\label{rho0eq}
\end{eqnarray}
It is possible to solve this evolution equation for all $s$, as we will discuss
in the next subsection. Here we only need the solution in the diffusive limit
$s\gg 1$. Then the $x$-variables have the uniform density
\begin{equation}
\rho_{0}(x,s)=
\left\{\begin{array}{cl}
N/s & {\rm if}\;\; x\lesssim s,\\
0 & {\rm if}\;\; x\gtrsim s,
\end{array}\right.\label{rho0result}
\end{equation}
as one can verify by substitution into Eq.\ (\ref{rho0eq}).

Linearization of Eq.\ (\ref{rho3}) around $\rho_{0}$ yields an equation for
$\delta\rho$,
\begin{eqnarray}
&&\frac{1}{2}\,\frac{d^{\,2}}{dx^{2}}\int_{0}^{\infty}\!\!
dx'\,\delta\rho(x')\ln|\sinh^{2}x-\sinh^{2}x'|
+\frac{d}{dx}(x\delta\rho)\nonumber\\
&&\hspace{2cm}\mbox{}
=\frac{1}{4}\left(1-\frac{2}{\beta}\right)\,\frac{d^{\,2}}{dx^{2}}\ln|\sinh
2x|.\label{rho5}
\end{eqnarray}
The $s$-dependence has dropped out in the limit $s\gg 1$. The
integro-differential equation (\ref{rho5}) can be solved by means of the
identity
\begin{eqnarray}
&&\int_{0}^{\infty}\!\!
dx'\,f(x')\ln|\sinh^{2}x-\sinh^{2}x'|\nonumber\\
&&\hspace{2cm}\mbox{}=\int_{-\infty}^{\infty}\!\!
dx'\,f(|x'|)\ln|\sinh(x-x')|,\label{identity}
\end{eqnarray}
which simplifies the integration to a convolution. The Fourier transform of
$\delta\rho(x)$ then satisfies an ordinary differential equation, which is
easily solved. The result is
\begin{equation}
\delta\rho(x)=(1-2/\beta)[\case{1}{4}\delta(x-0^{+})+
(4x^{2}+\pi^{2})^{-1}].\label{DELTARHO}
\end{equation}
The correction (\ref{DELTARHO}) to the uniform density (\ref{rho0result}) takes
the form of a deficit (for $\beta=1$) or an excess (for $\beta=4$),
concentrated in the region $x\lesssim 1$. Equations (\ref{rho0result}) and
(\ref{DELTARHO}) together form the result (\ref{RHOXRESULT}) used in the
previous subsection.

\subsubsection{Scaling as a hydrodynamic flow}
\label{hydrodynamic}

The uniform eigenvalue density (\ref{rho0result}) is the large-$s$ limit of the
solution $\rho_{0}(x,s)$ of the evolution equation (\ref{rho0eq}). Let us
investigate how this limit is reached starting from an initially non-uniform
density. It turns out that the non-linear integro-differential equation
(\ref{rho0eq}) can be solved exactly, for arbitrary initial condition
(Beenakker, Rejaei, and Melsen, 1994). The solution is based on a mapping of
Eq.\ (\ref{rho0eq}) onto Euler's equation of hydrodynamics. A similar mapping
exists for Dyson's Brownian-motion model (Pandey and Shukla, 1991). For
notational simplicity we will write $\rho$ instead of $\rho_{0}$ in this
subsection, being only concerned here with the leading order contribution in
powers of $N$.

We begin by rewriting Eq.\ (\ref{rho0eq}) in terms of the $\lambda$-variables,
\begin{equation}
\frac{\partial\rho}{\partial s}=-\frac{2}{N}
\frac{\partial}{\partial\lambda}\lambda(1+\lambda)\rho
\frac{\partial}{\partial\lambda}
\int_{0}^{\infty}\!\!d\lambda'\,
\rho(\lambda',s)\ln|\lambda-\lambda'|.\label{rho0eqlambda}
\end{equation}
The density $\rho(\lambda,s)$ has the Stieltjes transform
\begin{equation}
F(z,s)=\int_{0}^{\infty}\!\!d\lambda'\,
\frac{\rho(\lambda',s)}{z-\lambda'}.\label{Fdef}
\end{equation}
The function $F(z,s)$ is an analytic function of $z$ in the complex plane cut
by the positive real axis, which vanishes for large $|z|$ as
\begin{equation}
\lim_{|z|\rightarrow\infty}F(z,s)=N/z.\label{Flimit}
\end{equation}
It has a discontinuity for $z=\lambda\pm{\rm i}0^{+}$ ($\lambda>0$). The
limiting values $F_{\pm}(\lambda,s)\equiv F(\lambda\pm{\rm i}0^{+},s)$ are
\begin{equation}
F_{\pm}=\pm\frac{\pi}{\rm i}\rho(\lambda,s)+\frac{\partial}{\partial\lambda}
\int_{0}^{\infty}\!\!d\lambda'\,\rho(\lambda',s)\ln|\lambda-\lambda'|.
\label{Fdisc}
\end{equation}
Combination of Eqs.\ (\ref{rho0eqlambda}) and (\ref{Fdisc}) gives
\begin{equation}
N\frac{\partial}{\partial s}(F_{+}-F_{-})=-\frac{\partial}
{\partial\lambda}\lambda(1+\lambda) (F_{+}^{2}-F_{-}^{2}),\label{scaling3}
\end{equation}
which implies that the function
\begin{equation}
{\cal F}(z,s)=N\frac{\partial}{\partial s}F(z,s)+\frac{\partial} {\partial
z}z(1+z)F^{2}(z,s)\label{calFdef}
\end{equation}
is analytic in the whole complex plane, including the real axis. Moreover,
${\cal F}\rightarrow 0$ for $|z|\rightarrow\infty$, in view of Eq.\
(\ref{Flimit}). We conclude that ${\cal F}\equiv 0$, since the only analytic
function which vanishes at infinity is identically zero.

We now return from the $\lambda$ to the $x$-variables. The mapping
$z=\sinh^{2}\zeta$ maps the $z$-plane onto the strip in the $\zeta$-plane
between the lines $y=0$ and $y=-\pi/2$, where $\zeta=x+{\rm i}y$. The mapping
is conformal if we cut the $z$-plane by the two half{}lines $\lambda>0$ and
$\lambda<-1$ on the real axis. On this strip we define the auxiliary function
$U=U_{x}+{\rm i}U_{y}$ by
\begin{equation}
U(\zeta,s)\equiv\frac{F}{2N}\,\frac{dz}{d\zeta}
=\frac{\sinh 2\zeta}{2N} \int_{0}^{\infty}\!\!dx'\,\frac{\rho(x',s)}
{\sinh^{2}\zeta-\sinh^{2}x'}.\label{Udef}
\end{equation}
The equation ${\cal F}\equiv 0$ then takes the form
\begin{equation}
\frac{\partial}{\partial s}U(\zeta,s)+U(\zeta,s)\frac{\partial} {\partial
\zeta}U(\zeta,s)=0,\label{Euler}
\end{equation}
which we recognize as {\em Euler's equation\/} of hydrodynamics:
$(U_{x},U_{y})$ is the velocity field in the $(x,y)$ plane of a two-dimensional
ideal fluid at constant pressure. Euler's equation is easily solved. For
initial condition $U(\zeta,0)=U_{0}(\zeta)$ the solution to Eq.\ (\ref{Euler})
is
\begin{equation}
U(\zeta,s)=U_{0}\biglb(\zeta-sU(\zeta,s)\bigrb).\label{Eulersol}
\end{equation}
{}From $U$ we obtain the eigenvalue density
\begin{equation}
\rho(x,s)=(2N/\pi)\,{\rm Im}\,U(x-{\rm i}0^{+},s).\label{rhoxU}
\end{equation}

\begin{figure}[tb]
\centerline{
\psfig{figure=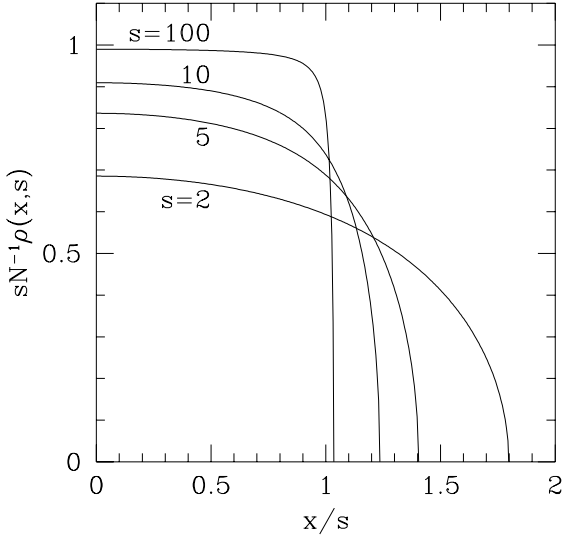,width= 8cm}
}%
\caption[]{
Eigenvalue density in the metallic regime. The variable $x$ is related to the
transmission eigenvalue by $T=1/\cosh^{2}x$. Curves are computed from Eq.\
(\protect\ref{Eulersol}) for four values of $s=L/l$, with the ballistic initial
condition $\rho(x,0)=N\delta(x-0^{+})$. For $s\gg 1$ the density tends to the
limit (\protect\ref{rho0result}) of a step-function profile.
}\label{rho_ballistic}
\end{figure}

The ballistic initial condition $\rho(x,0)=N\delta(x-0^{+})$ corresponds to
$U_{0}(\zeta)={\rm cotanh}\,\zeta$. The solution of the implicit equation
(\ref{Eulersol}) is plotted in Fig.\ \ref{rho_ballistic} for several values of
$s=L/l$. With increasing disorder the eigenvalue density spreads along the
$x$-axis, such that $0<\rho\leq N/s$ for $x<x_{\rm max}$ and $\rho\equiv 0$ for
$x\geq x_{\rm max}$. The edge $x_{\rm max}$ of the density profile is located
at
\begin{eqnarray}
x_{\rm max}&=&\case{1}{2}\,{\rm
arcosh}\,(1+2s)+\case{1}{2}\sqrt{(1+2s)^2-1}\nonumber\\
&=&s+\case{1}{2}\ln 4s+\case{1}{2}+{\cal O}(1/s),\label{xmax}
\end{eqnarray}
with $\rho\propto\sqrt{x_{\rm max}-x}$ for $x$ close to $x_{\rm max}$. For
$s\gg 1$ the density tends to the diffusive limit (\ref{rho0result}) of a
step-function profile: $\rho(x,s)\rightarrow N/s$ for $s\rightarrow\infty$ at
fixed $x/s<1$. In fact, this limit is reached regardless of the particular
initial condition, because of the fixed point $\zeta=-\frac{1}{2}{\rm i}\pi$ at
which $U=0$ [{\em cf.} Eq.\ (\ref{Udef})]. To see this, define
$\zeta_{0}=\zeta-sU(\zeta,s)$ and write Eq.\ (\ref{Eulersol}) in the form
$\zeta-\zeta_{0}=sU_{0}(\zeta_{0})$. For $s\rightarrow\infty$,
$\zeta_{0}\rightarrow-\frac{1}{2}{\rm i}\pi$ so that the product
$sU_{0}(\zeta_{0})$ remains finite. It follows that
$U(\zeta,s)\rightarrow(\zeta+\frac{1}{2}{\rm i}\pi)/s$, hence
$\rho(x,s)\rightarrow N/s$.

\subsubsection{Non-logarithmic eigenvalue repulsion}
\label{Trepulsion}

The tiny difference between the variance of the conductance in a quantum dot
and in a disordered wire has a fundamental implication for the repulsion of the
transmission eigenvalues ({\em cf.} Sec.\ \ref{UCF}). Since a logarithmic
repulsion of the $\lambda$-variables implies that $\beta\,{\rm
Var}\,G/G_{0}=\frac{1}{8}$, while in a wire one has $\beta\,{\rm
Var}\,G/G_{0}=\frac{2}{15}$, it follows that the repulsion of the $\lambda$'s
can not be precisely logarithmic (Beenakker, 1993a, 1993c). Recall that a
logarithmic repulsion follows from the Jacobian from matrix to eigenvalue
space. A non-logarithmic repulsion means that there exist correlations between
the eigenvalues which do not have a geometric origin. To determine the
eigenvalue repulsion in a disordered wire in the metallic regime, we consider
the exact solution (\ref{exactsolution}) of the DMPK equation for $\beta=2$.

If $1\ll s\ll N$ the dominant contribution to the integral over $k$ in Eq.\
(\ref{exactsolution}) comes from the range $k\gtrsim (N/s)^{1/2}\gg 1$. In this
range $\tanh(\case{1}{2}\pi k)\rightarrow 1$ and the Legendre function
simplifies to a Bessel function,
\begin{equation}
{\rm P}_{\frac{1}{2}({\rm i}k-1)}(\cosh 2x)\rightarrow J_{0}(kx)(2x/\sinh
2x)^{1/2}.
\label{PJrelation}
\end{equation}
The $k$-integration can now be carried out analytically,
\begin{eqnarray}
&&\int_{0}^{\infty}\!\!dk\,{\rm e}^{-k^{2}s/4N}\,
k^{2m-1}\,J_{0}(kx_{n})=\case{1}{2}(m-1)!\,(4N/s)^{m}\nonumber\\
&&\hspace{2cm}\mbox{}\times{\rm
e}^{-x_{n}^{2}N/s}\,L_{m-1}(x_{n}^{2}N/s),\label{integral}
\end{eqnarray}
with $L_{m-1}$ a Laguerre polynomial. We then apply the determinantal identity
\begin{equation}
{\rm Det\,}L_{m-1}(x_{n}^{2}N/s)=c\,{\rm
Det\,}(x_{n}^{2})^{m-1}=c\prod_{i<j}(x_{j}^{2}-x_{i}^{2}),
\label{Ldet}
\end{equation}
with $c$ an $x$-independent number [which can be absorbed in $C(s)$].  The
first equality in Eq.\ (\ref{Ldet}) holds because the determinant of a matrix
is unchanged if any one column of the matrix is added to any other column, so
that we can reduce the Laguerre polynomial in $x^{2}$ of degree $m-1$ to just
its highest order term $x^{2(m-1)}$ times a numerical coefficient. The second
equality expands the Vandermonde determinant.

\begin{figure}[tb]
\centerline{
\psfig{figure=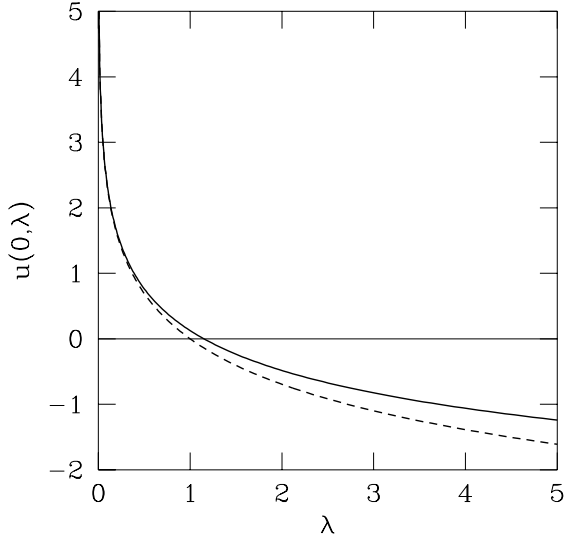,width= 8cm}
}%
\caption[]{
Eigenvalue interaction potential $u(\lambda_{i},\lambda_{j})$ for
$\lambda_{i}=0$ as a function of $\lambda_{j}\equiv\lambda$. The solid curve is
the result (\protect\ref{ulambda1}) from the DMPK equation in the metallic
regime. The dashed curve is the logarithmic repulsion $-\ln|\lambda|$ dictated
by the Jacobian from matrix to eigenvalue space. For $\lambda\ll 1$ the two
curves coincide. For $\lambda\rightarrow\infty$ their ratio approaches a factor
of two. After Beenakker and Rejaei (1994a).
}\label{fig_interaction}
\end{figure}

Collecting results, we find that the solution (\ref{exactsolution}) of the DMPK
equation for $\beta=2$ simplifies in the metallic regime to (Beenakker and
Rejaei, 1993, 1994a)
\begin{eqnarray}
P=C(s)\prod_{i<j}\left[(\sinh^{2}x_{j}-
\sinh^{2}x_{i})(x_{j}^{2}-x_{i}^{2})\right]\nonumber\\
\mbox{}\times\prod_{i}\left[\exp(-x_{i}^{2}N/s)(x_{i}\sinh
2x_{i})^{1/2}\right].\label{exactsolution2}
\end{eqnarray}
Caselle (1995) has generalized Eq.\ (\ref{exactsolution2}) to $\beta=1$ and 4.
The result for the three values of $\beta$ can be written in the form of a
Gibbs distribution,
\begin{mathletters}
\label{EXACTSOLUTION3X}
\begin{eqnarray}
&&P=C(s)\exp\Bigl[-\beta\Bigl(\sum_{i<j}u(x_{i},x_{j})+
\sum_{i}V(x_{i})\Bigr)\Bigr],\label{final3Px}\\
&&u(x_{i},x_{j})=-\case{1}{2}\ln|\sinh^{2}x_{j}-\sinh^{2}x_{i}|\nonumber\\
&&\hspace{4cm}\mbox{}-\case{1}{2}\ln|x_{j}^{2}-x_{i}^{2}|,
\label{EXACTSOLUTION3Xa}\\
&&V(x)=\case{1}{2}(\gamma/s)\beta^{-1}x^{2}-\case{1}{2}\beta^{-1}\ln|x\sinh
2x|.
\label{EXACTSOLUTION3Xb}
\end{eqnarray}
\end{mathletters}%
In terms of the $\lambda$-variables ($\lambda=\sinh^{2}x$), the interaction
takes the form
\begin{eqnarray}
u(\lambda_{i},\lambda_{j})=-\case{1}{2}\ln|{\rm
arsinh}^{2}\lambda_{j}^{1/2}-{\rm arsinh}^{2}\lambda_{i}^{1/2}|\nonumber\\
\mbox{}-\case{1}{2}\ln|\lambda_{j}-\lambda_{i}|.\label{ulambda1}
\end{eqnarray}
For $\lambda\ll 1$ ({\em i.e.} for $T=(1+\lambda)^{-1}$ close to unity)
$u(\lambda_{i},\lambda_{j})\rightarrow -\ln|\lambda_{j}-\lambda_{i}|$, so we
obtain a logarithmic repulsion for the strongly transmitting scattering
channels. However, for $\lambda\approx 1$ the interaction (\ref{ulambda1}) is
non-logarithmic. For fixed $\lambda_{i}\ll 1$, $u(\lambda_{i},\lambda_{j})$ as
a function of $\lambda_{j}$ crosses over from $-\ln|\lambda_{j}-\lambda_{i}|$
to $-\frac{1}{2}\ln|\lambda_{j}-\lambda_{i}|$ as $\lambda_{j}\rightarrow\infty$
(see Fig.\ \ref{fig_interaction}). We conclude that, for weakly transmitting
channels, the interaction is twice as small as predicted by considerations
based solely on the Jacobian.

The reduced level repulsion for weakly transmitting channels enhances the
variance of the conductance above the result $\beta\,{\rm
Var}\,G/G_{0}=\frac{1}{8}$ for a purely logarithmic repulsion. To see this, we
compute the two-point correlation function $K(x,x')$ from the interaction
potential $u(x,x')$, using the property that $K$ and $u$ are each others
functional inverse in the large-$N$ limit (Beenakker, 1993a, 1993c; {\em cf.}
Sec.\ \ref{functionalderivatives}):
\begin{equation}
\beta\int_{0}^{\infty}\!\!
dx''\,u(x,x'')K(x'',x')=\delta(x-x'),\;\;x,x'>0.\label{uKrelation}
\end{equation}
To solve this integral equation, we note that the decomposition of the
interaction potential (\ref{EXACTSOLUTION3Xa}) into
\begin{mathletters}
\label{decomposeu}
\begin{eqnarray}
u(x,x')&=&{\cal U}(x-x')+{\cal U}(x+x')+{\rm constant},\label{decomposeua}\\
{\cal U}(x)&=&-\case{1}{2}\ln|2x\sinh x|,\label{decomposeub}
\end{eqnarray}
\end{mathletters}%
implies for the two-point correlation function the decomposition
\begin{equation}
K(x,x')={\cal K}(x-x')+{\cal K}(x+x').\label{decomposeK}
\end{equation}
(The additive constant in Eq.\ (\ref{decomposeua}) is irrelevant, because of
the sum rule $\int_{0}^{\infty}dx''\,K(x'',x')=0$ implied by the definition
(\ref{Kxdef}) of $K$.) The functions ${\cal U}(x)$ and ${\cal K}(x)$, defined
for both positive and negative $x$, are related by the convolution
\begin{equation}
\beta\int_{-\infty}^{\infty}\!\! dx''\,{\cal U}(x-x''){\cal K}(x''-x')
=\delta(x-x')+{\rm constant},\label{uKrelation2}
\end{equation}
which is readily inverted by Fourier transformation:
\begin{equation}
{\cal K}(k)=
\frac{|k|}{\beta\pi}\left(1-{\rm e}^{-\pi|k|}\right).\label{KUk}
\end{equation}
Transforming back from $k$ to $x$ one finds the two-point correlation function
(\ref{Kxresult}) used in Sec.\ \ref{generalA} to compute the variance of the
conductance. The result $\beta\,{\rm Var}\,G/G_{0}=\frac{2}{15}$ is only
slightly larger than the value $\frac{1}{8}$ for a logarithmic repulsion,
because only the weakly transmitting channels (which contribute little to the
conductance) are affected by the non-logarithmic interaction.

The derivation of the two-point correlation function presented here emphasizes
the relationship with the eigenvalue interaction (Beenakker and Rejaei, 1993,
1994a; Caselle, 1995). There exists an alternative derivation (Chalker and
Mac\^{e}do, 1993; Mac\^{e}do and Chalker, 1994), which starts directly from the
DMPK equation and reduces it to an evolution equation for $K$ in the metallic
regime. The diffusive limit $s\rightarrow\infty$ then leads to the result
(\ref{Kxresult}). We discussed a similar approach in Sec.\ \ref{Tdensity}, in
connection with the eigenvalue density. It is worthwhile to check that the
result (\ref{DELTARHO}) for $\delta\rho$ obtained there agrees with the density
implied by the distribution (\ref{EXACTSOLUTION3X}). This alternative route to
the eigenvalue density is described in Appendix \ref{Tdensity2}.

\subsection{Localized regime}
\label{localized}

\subsubsection{Log-normal distribution of the conductance}
\label{lognormal}

In the metallic regime, the root-mean-square fluctuations of the conductance
are a factor $L/Nl$ smaller than the average conductance. The sample-to-sample
fluctuations are therefore relatively unimportant, since $L\ll Nl$. As the
length $L$ of the wire increases beyond the localization length $\xi\simeq Nl$,
the localized regime is entered. Then fluctuations become as large as the
average, which is no longer representative for the conductance of a single
sample. The conductance distribution $P(G)$, which was well approximated by a
Gaussian in the metallic regime,\footnote{The third cumulant of $G$ is of order
$(L/Nl)^{2}(e^{2}/h)^{3}$ (Mac\^{e}do, 1994a; Gopar, Mart\'{\i}nez, and Mello,
1995).} becomes very broad and asymmetric, with a peak at small $G$ and a long
tail towards large $G$.

It follows from general properties of products of random matrices that $P(G)$
is log-normal in the limit $L/Nl\rightarrow\infty$, that is to say, $\ln G$ has
a Gaussian distribution (Imry, 1986a). To see this, note that the scaling
operation of Fig.\ \ref{fig_segment} corresponds to the multiplication of
transfer matrices: $M=\prod_{i}M_{i}$, where $M_{i}$ is the transfer matrix of
segment $i$ and $M$ is the transfer matrix of the entire wire. The limit
$L\rightarrow\infty$, at fixed $N$ and $l$, corresponds to the multiplication
of an infinite number of random matrices, drawn independently from the same
ensemble. In this limit, the $2N$ random eigenvalues $\exp(\pm 2x_{n})$ of
$MM^{\dagger}$ tend to the non-random values $\exp(\pm 2L/\xi_{n})$, with
$\xi_{n}$ independent of $L$ (Pichard and Sarma, 1981; Pichard and Andr\'{e},
1986). This is known as the ``multiplicative ergodic theorem'' (Oseledec, 1968;
Crisanti, Paladin, and Vulpiani, 1993). The inverse localization lengths
$1/\xi_{n}$ are referred to as the Lyapunov exponents of the random matrix
product. For large, but finite $L$, the $x_{n}$'s have small Gaussian
fluctuations around their asymptotic limit $L/\xi_{n}$. The conductance
$G=G_{0}\sum_{n}\cosh^{-2}x_{n}\approx 4G_{0}\exp(-2x_{1})$ is dominated by the
smallest $x_{n}$, say $x_{1}$. The conclusion is that
$-\frac{1}{2}\ln(G/4G_{0})$ has the same Gaussian distribution as $x_{1}$.

The mean and variance of the log-normal distribution of the conductance follow
directly from the DMPK equation (\ref{FokkerPlanckx}) for the probability
distribution $P(x_{1},x_{2},\ldots x_{N},s)$ (Dorokhov, 1982, 1983; Pichard,
1991). In the limit $L/Nl\equiv s/N\rightarrow\infty$ the variables $x_{1}\ll
x_{2}\ll\cdots\ll x_{N}$ become widely separated and $\gg 1$, so that the term
$\Omega$ in Eq.\ (\ref{FokkerPlanckx}) may be approximated by
\begin{equation}
\Omega\approx -2\beta^{-1}\sum_{n=1}^{N}(1+\beta n-\beta)x_{n}+{\rm
constant}.\label{Omegadecoupled}
\end{equation}
The solution of Eq.\ (\ref{FokkerPlanckx}) then factorizes into a product of
Gaussians,
\begin{mathletters}
\label{Pxdecoupled}
\begin{eqnarray}
&&P\approx\left(\frac{\gamma l}{2\pi
L}\right)^{N/2}\prod_{n=1}^{N}\exp\left[-\frac{\gamma
l}{2L}\,(x_{n}-L/\xi_{n})^{2}\right],\label{Pxdecoupleda}\\
&&\xi_{n}=\gamma l(1+\beta n-\beta)^{-1}.\label{Pxdecoupledb}
\end{eqnarray}
\end{mathletters}%
The root-mean-square fluctuation of the $x_{n}$'s equals $\sqrt{L/\gamma l}$,
which is indeed much smaller than their spacing $\beta L/\gamma l$. The
conductance is dominated by $x_{1}$, which has a mean $L/\gamma l$ equal to its
variance. The Gaussian distribution of $-\ln G/G_{0}\approx 2x_{1}+{\cal O}(1)$
therefore has a mean which is half the variance,
\begin{equation}
-\langle \ln G/G_{0}\rangle=\case{1}{2}\,{\rm Var}\,(\ln G/G_{0})=2L/\gamma
l.\label{meanvariancelnG}
\end{equation}
The localization length $\xi$ is obtained from the exponential decay of the
typical conductance, by identifying $\exp\langle\ln
G/G_{0}\rangle\equiv\exp(-2L/\xi)$. Hence
\begin{equation}
\xi=\gamma l=(\beta N+2-\beta)l.\label{localizationlength}
\end{equation}
The average conductance $\langle G\rangle$ decays more slowly than the typical
conductance $\exp\langle\ln G\rangle$:
\begin{eqnarray}
\langle G/G_{0}\rangle&\propto&\int_{0}^{\infty}\!dx\,{\rm
e}^{-2x}\,\exp\left[-\frac{\gamma l}{2L}\,(x-L/\gamma l)^{2}\right]\nonumber\\
&\propto&\exp(-L/2\xi).\label{Gaveragedecay}
\end{eqnarray}

For $N\gg 1$ the localization length $\xi\approx\beta Nl$ becomes proportional
to the symmetry index $\beta$. This $\beta$-dependence can be measured by
studying the effect of a magnetic field on the conductance (Pichard, Sanquer,
Slevin, and Debray, 1990). In the absence of spin-orbit scattering, a
time-reversal-symmetry breaking magnetic field induces a transition from
$\beta=1$ to $\beta=2$, and hence a doubling of $\xi$. The spin degeneracy of
the $N$ scattering channels is not broken. In the case of strong spin-orbit
scattering, breaking of time-reversal symmetry induces a transition from
$\beta=4$ to $\beta=2$ and also breaks Kramers' degeneracy of the scattering
channels. The combined result of $\beta=4\rightarrow\beta=2$ and $N\rightarrow
2N$ is that $\xi$ remains unchanged (Efetov and Larkin, 1983; the role of
Kramers' degeneracy has been emphasized by Mirlin, 1994). To observe the
doubling of $\xi$ induced by a magnetic field in the absence of spin-orbit
scattering requires field strengths $B\gtrsim h/e\xi^{2}$. In weaker fields the
magnetoconductance is dominated by thermally activated processes (Mott
hopping), which lead to an increase of $G$ with $B$ both in the absence and
presence of spin-orbit scattering (Nguyen, Spivak, and Shklovski\u{\i}, 1985;
Meir, Wingreen, Entin-Wohlman, and Altshuler, 1991: Meir and Entin-Wohlman,
1993; for reviews, see: Shklovski\u{\i} and Spivak, 1990; Imry, 1995). It is
this positive magnetoconductance which is usually measured in
insulators.\footnote{An exception is formed by the experiments reported by
Pichard, Sanquer, Slevin, and Debray (1990) on an insulating amorphous Y/Si
alloy with strong spin-orbit scattering, which show a negative
magnetoconductance ($G$ decreases with $B$). It is not clear how to reconcile
this with a theory which properly accounts for Kramers' degeneracy.} The simple
and universal $\beta$-dependence of the localization length
(\ref{localizationlength}) is special for a wire geometry
(quasi-one-dimensional sample) and has not yet been observed experimentally. In
two- and three-dimensional samples the $B$-dependence of $\xi$ is more
complicated and not universal (Lerner and Imry, 1995).

\begin{figure}[tb]
\psfig{figure=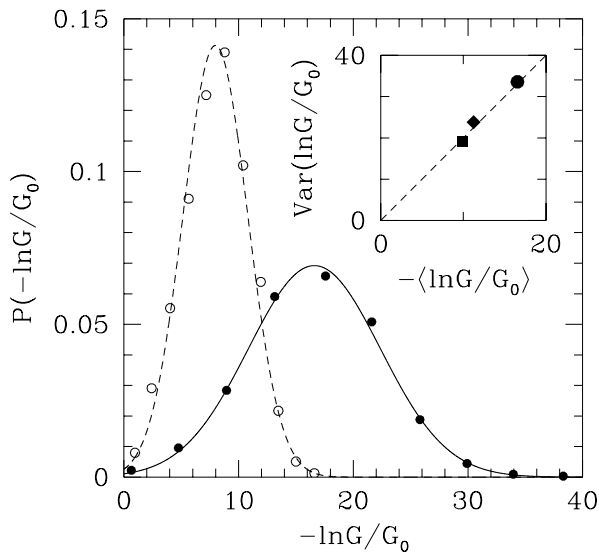,width= 8cm}
\caption[]{
Distribution of $-\ln G/G_{0}$ from a numerical simulation of the Anderson
model on a $10\times 250$ strip (filled dots) and a $10\times 10$ square (open
dots), in zero magnetic field. The dashed and solid curves are Gaussians with
variance equal to the mean and to twice the mean, respectively. The results for
a strip are as expected from the DMPK equation, those for a square have no
known explanation. The inset shows the mean and variance for the strip for
three values of the magnetic field: circle, diamond, and square correspond,
respectively, to $B=0$, $3\cdot 10^{-4}$, and $3\cdot 10^{-2}\times h/ea^{2}$.
The localization length $\xi=-2L/\langle\ln G/G_{0}\rangle$ increases by a
factor close to the value $20/11$ predicted by Eq.\
(\protect\ref{localizationlength}) for the $\beta=1\rightarrow\beta=2$
transition with $N=10$. The dashed line indicates the predicted factor of two
between mean and variance. After Pichard (1991).
}\label{fig_lognormal}
\end{figure}

The log-normal distribution of the conductance in the localized regime has been
verified in numerical simulations of the Anderson model (Pichard, 1991). The
disordered region is modeled by a tight-binding Hamiltonian on a
two-dimensional square lattice (lattice constant $a$, width $W$, length $L$),
with a constant hopping term $U_{0}=\hbar^{2}/2ma^{2}$ between neighboring
sites and with a random impurity potential at each site (uniformly distributed
between $\pm\frac{1}{2}U_{\rm d}$). The Fermi level is chosen at the center of
the tight-binding band ($4U_{0}$ from the band bottom), at which the number $N$
of propagating modes equals the number $W/a$ of sites in a row (for hard-wall
boundary conditions at the two ends of the row). Results for a $10\times 250$
strip are shown in Fig.\ \ref{fig_lognormal} (filled dots). The disorder is
sufficiently strong ($U_{\rm d}=3U_{0}$) that the wire is deep in the localized
regime ($L\approx 8\xi$). The distribution of $-\ln G/G_{0}$ is well fitted by
a Gaussian (solid curve), with a variance equal to twice the mean. The inset
shows that a magnetic field significantly increases the localization length
(there is no spin-orbit scattering in the simulation). At $B=0.03\,h/ea^{2}$
the increase by a factor of 1.7 is close to the factor $20/11\approx 1.8$
predicted by Eq.\ (\ref{localizationlength}) for $N=10$. Note also that the
factor-of-two between mean and variance is observed to be $B$-independent, as
expected. The open dots and dashed curve in Fig.\ \ref{fig_lognormal} show a
still unexplained feature of two-dimensional insulators. The simulation of a
$10\times 10$ square with $U_{\rm d}=12\,U_{0}$ yields a log-normal
distribution of the conductance with a variance equal to the mean --- not twice
the mean as in a quasi-one-dimensional insulator.

\subsubsection{Crystallization of transmission eigenvalues}
\label{crystallization}

The exponential decay with increasing wire length of the conductance in the
localized regime is associated with a ``crystallization'' of the transmission
eigenvalues (Pichard, Zanon, Imry, and Stone, 1990; Muttalib, 1990; Stone,
Mello, Muttalib, and Pichard, 1991). In the limit $L/Nl\rightarrow\infty$, the
$x_{n}$'s form a one-dimensional lattice with spacing $\delta x=L/Nl$ (for
$N\gg 1$). The fluctuations of the $x_{n}$'s around their lattice positions
grow as $L$ decreases, and become comparable to the lattice spacing when
$L\simeq Nl$. If $L\ll Nl$ the density is nearly constant, with small ripples
of periodicity $\delta x$, reminiscent of a liquid. This is the metallic
regime, in which the conductance scales linearly with $L$.

The transition from a liquid-like to a crystal-like eigenvalue density can be
obtained from the exact solution (\ref{exactsolution}) of the DMPK equation for
$\beta=2$, by integrating out $N-1$ of the $x_{n}$'s:
\FL
\begin{equation}
\rho(x,s)=N\int_{0}^{\infty}\!\!dx_{2}\cdots\!\int_{0}^{\infty}\!\!dx_{N}\,
P(x,x_{2},\ldots x_{N},s).\label{rhodef2}
\end{equation}
The calculation was carried out by Frahm (1995a), using a generalization of the
method of orthogonal polynomials suggested by Muttalib (1995). (The same result
was obtained from the 1D $\sigma$ model by Rejaei, 1996.) The idea is to write
the probability distribution as the product of two determinants,
\begin{equation}
P=\biglb({\rm Det}\,a_{m}(x_{n},s)\bigrb)\biglb({\rm
Det}\,b_{m}(x_{n},s)\bigrb),\label{Pdetsquared}
\end{equation}
in such a way that the functions $a$ and $b$ are bi-orthogonal:
\begin{equation}
\int_{0}^{\infty}\!\!dx\,a_{n}(x,s)b_{m}(x,s)=\delta_{nm},\;\;1\le n,m\le N.
\label{biorthogonal}
\end{equation}
Then the integrals in Eq.\ (\ref{rhodef2}) reduce to a finite series,
\begin{equation}
\rho(x,s)=\sum_{n=1}^{N}a_{n}(x,s)b_{n}(x,s),\label{rhoab}
\end{equation}
and similar series exist for the correlation functions of the $x_{n}$'s.

\begin{figure}[tb]
\hspace*{-1.3cm}
\psfig{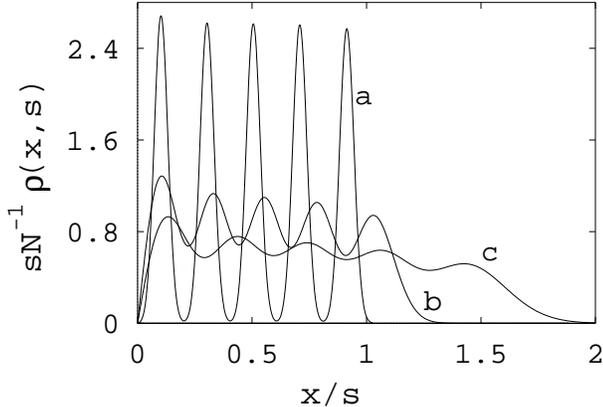}
\caption[]{
Exact eigenvalue density for $\beta=2$, computed from Eqs.\
(\protect\ref{rhoab}) and (\protect\ref{abdef}) for $N=5$ and three values of
$s=L/l$ ($s=100$, 10, and 2, respectively, for curves a, b, and c). The density
of the $x_{n}$'s is not quite uniform in the metallic regime (curve c), because
of the relatively small value of $N$ (compare with Fig.\
\protect\ref{rho_ballistic} for the large-$N$ limit). In the insulating regime
the eigenvalues ``crystallize'' at equally spaced average positions, with small
Gaussian fluctuations around the average (curve a). After Frahm (1995a).
}\label{fig_frahm}
\end{figure}

The exact solution (\ref{exactsolution}) is of the form (\ref{Pdetsquared})
(since the term $\prod_{i<j}$ in Eq.\ (\ref{exactsolution}) is a Vandermonde
determinant), but the functions in the determinants are not bi-orthogonal.
Frahm (1995a) constructed a linear combination of these functions such that
Eq.\ (\ref{biorthogonal}) is realized:
\begin{mathletters}
\label{abdef}
\begin{eqnarray}
a_{n}(x,s)&=&\exp[-(2n-1)^{2}s/4N]\,{\rm P}_{n-1}(\cosh 2x),\label{adef}\\
b_{n}(x,s)&=&\case{1}{2}\sinh{2x}\int_{0}^{\infty}\!\!dk\,
{\rm e}^{-k^{2}s/4N}k\tanh(\case{1}{2}\pi k)\nonumber\\
&&\mbox{}\times Q_{n}(k^{2})\,{\rm P}_{\frac{1}{2}({\rm i}k-1)}(\cosh
2x),\label{bdef}
\end{eqnarray}
\end{mathletters}%
where ${\rm P}_{n}$ is a Legendre polynomial and $Q_{n}$ is the interpolation
polynomial of Lagrange,
\begin{equation}
Q_{n}(k^{2})=\prod_{p=1\,(p\neq n)}^{N}\frac{k^{2}+(2p-1)^{2}}
{(2p-1)^{2}-(2n-1)^{2}}.\label{Ldef}
\end{equation}
The resulting eigenvalue density is plotted in Fig.\ \ref{fig_frahm}, for $N=5$
and values of $s$ in the metallic, crossover, and insulating regime. One
recognizes the appearance of deep minima in the density due to eigenvalue
repulsion, upon entering the localized regime.

This non-perturbative result for the density of transmission eigenvalues is for
the case $\beta=2$ of broken time-reversal symmetry (Frahm, 1995a; Rejaei,
1996). The result for $\beta=1,4$ is not known. What is known
non-perturbatively for any $\beta$ is the first and second moment of the
conductance. See Zirnbauer (1992), Mirlin, M\"{u}ller-Groeling, and Zirnbauer
(1994) for the cases $\beta=1,2$, and Brouwer and Frahm (1996) for the case
$\beta=4$. Non-perturbative results for entire distributions (rather than
moments) exist for the distributions of $|t_{nm}|^{2}$ and
$\sum_{n}|t_{nm}|^{2}$, in the case $\beta=2$ (Van Langen, Brouwer, and
Beenakker, 1996). The distribution of the conductance
$G/G_{0}=\sum_{n,m}|t_{nm}|^{2}$ is not known exactly for any $\beta$.

\subsection{Disordered wire with obstacles}
\label{obstacles}

\subsubsection{Obstacle as initial condition for scaling}
\label{obstacle_initial}

So far we have concentrated on the DMPK equation with the ballistic initial
condition (\ref{ballistic}). This means that all scattering in the wire is due
to disorder, so that for $L/l\rightarrow 0$ all transmission eigenvalues
$T_{n}$ are equal to 1. In this subsection we consider the case that the
disordered wire contains obstacles, such as point contacts or tunnel barriers,
which provide additional scattering. As we will discuss below, if the
scattering matrix of each obstacle has an isotropic distribution, then the
presence of the obstacles can be accounted for by a non-ballistic initial
condition on the DMPK equation (Beenakker and Melsen, 1994).

\begin{figure}[tb]
\centerline{
\psfig{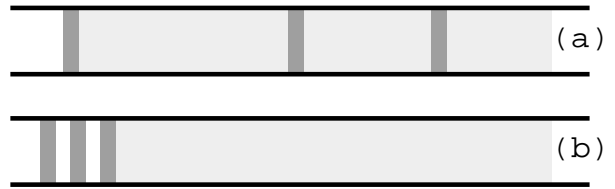}
}%
\vspace*{3mm}
\caption[]{
Two wires containing obstacles (shaded) in series with disordered segments
(dotted). For isotropic probability distributions the geometries of (a) and (b)
are equivalent. This permits one to treat the effect of the obstacles as an
initial condition on the DMPK equation.
}\label{fig_obstacles}
\end{figure}

The wire geometry we have in mind is sketched in Fig.\ \ref{fig_obstacles}a.
Disordered segments (dotted) alternate with obstacles (shaded). The disordered
segments have $N$ propagating modes, mean free path $l$, and a total length
$L$. We model the scattering by the impurities and by the obstacles by {\em
independent\/} and {\em isotropic\/} transfer matrices. That is to say, we
write the transfer matrix $M$ of the whole system as the product
$M=\prod_{i}M_{i}$ of the transfer matrices $M_{i}$ of its segments, and then
we assume that the $M_{i}$'s are distributed according to independent and
isotropic distributions $p_{i}(M_{i})$. A distribution $p(M)$ is called
isotropic if it is only a function of the eigenvalues of $MM^{\dagger}$. Under
these assumptions, the geometry of Fig.\ \ref{fig_obstacles}a is equivalent to
that of Fig.\ \ref{fig_obstacles}b, where the obstacles are in series with a
disordered segment of length $L$. To see this, note that the transfer matrices
$M$ and $M'$ of Figs.\ \ref{fig_obstacles}a and \ref{fig_obstacles}b differ by
a permutation of the $M_{i}$'s. The probability distribution $p(M)$ is given by
\begin{equation}
p=p_{1}\ast p_{2}\ast p_{3}\ast\cdots,\label{pdef}
\end{equation}
where the symbol $\ast$ denotes a convolution,
\begin{equation}
p_{i}\ast p_{j}(M)=\int\!d\mu(M_{j})\,p_{i}(MM_{j}^{-1})p_{j}(M_{j}),
\label{convolution}
\end{equation}
and $d\mu(M)$ is the invariant measure on the group of transfer matrices
(Mello, Pereyra, and Kumar, 1988). The probability distribution $p'(M')$ is
also given by a convolution of the $p_{i}$'s, but in a different order. It is a
property of isotropic distributions that their convolution does not depend on
the order: $p_{i}\ast p_{j}=p_{j}\ast p_{i}$ if both $p_{i}$ and $p_{j}$ are
isotropic. It follows that $p=p'$, and hence that the geometries of Figs.\
\ref{fig_obstacles}a and \ref{fig_obstacles}b are equivalent.

The obstacles in Fig.\ \ref{fig_obstacles}b form an initial condition on the
DMPK equation,
\begin{equation}
\lim_{L\rightarrow 0}P(\lambda_{1},\lambda_{2},\ldots
\lambda_{N},L)=P_{0}(\lambda_{1},\lambda_{2},\ldots
\lambda_{N}).\label{P0initial}
\end{equation}
Here $P_{0}$ is the probability distribution of the obstacles in the absence of
disordered segments. By solving the DMPK equation with initial condition
(\ref{P0initial}), one determines how disordered segments between the obstacles
affect the distribution of the transmission eigenvalues. In the following
subsections we will apply this general method to point contacts and tunnel
barriers.

\subsubsection{Point contact}
\label{obstacle_pointcontact}

\begin{figure}[tb]
\centerline{
\psfig{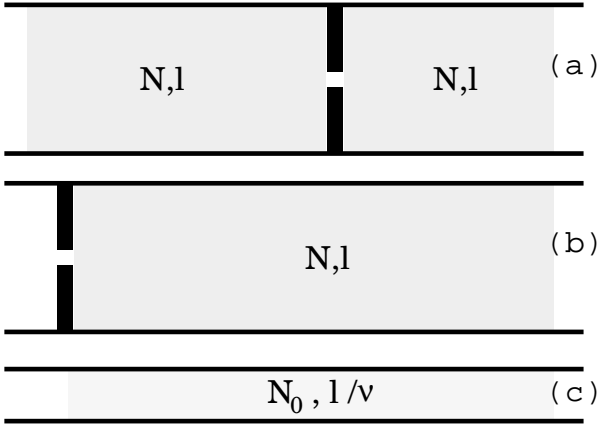}
}%
\vspace*{3mm}
\caption[]{
Equivalence of constricted and unconstricted geometries. (a) Sketch of a
ballistic constriction (with conductance $N_{0}G_{0}$) in a disordered wire
(with mean free path $l$ and $N$ transverse modes). (b) Constricted geometry
with all disorder at one side of the point contact, equivalent to (a) for
isotropic transfer matrices. (c) Unconstricted geometry, with mean free path
$l/\nu$ and $N_{0}$ transverse modes, equivalent to (a) and (b) for $\nu$ given
by Eq.\ (\protect\ref{nudef}). After Beenakker and Melsen (1994).
}\label{fig_constricted}
\end{figure}

As a first application, we consider a disordered wire containing a constriction
much narrower than the mean free path (Beenakker and Melsen, 1994). This is
known as a ballistic point contact, or a ``Sharvin'' point contact (Sharvin,
1965). [The opposite limit is known as a diffusive or ``Maxwell'' point contact
(Maxwell, 1891).] The assumption of independent transfer matrices for the
constriction and disordered regions requires a spatial separation of scattering
by the impurities and by the point contact. This is justified if the mean
separation $d_{\rm imp}$ of the impurities is much greater than the width
$W_{0}$ of the constriction. Since $d_{\rm imp}$ is much smaller than the mean
free path $l$, the condition $d_{\rm imp}\gg W_{0}$ is stronger than the
condition $l\gg W_{0}$ for a ballistic point contact. The isotropy assumption
for the transfer matrix of the constriction is a simple but realistic model of
the coupling between wide and narrow regions, which implies that all $N$
transverse modes in the wide regions (of width $W$) to the left and right of
the constriction (of width $W_{0}$) are equally coupled to each other (Szafer
and Stone, 1989). The isotropy assumption for the transfer matrices of the
disordered regions (of length $L_{1}$ and $L_{2}$) requires aspect ratios
$L_{1}/W,L_{2}/W\gg 1$ corresponding to a wire geometry. As argued in the
previous subsection, the geometry of Fig.\ \ref{fig_constricted}a, with lengths
$L_{1}$ and $L_{2}$ of disordered wire to the left and right of the point
contact, is equivalent to the geometry of Fig.\ \ref{fig_constricted}b, with a
length $L=L_{1}+L_{2}$ of disordered wire to one side only. We will now argue
that the {\em constricted\/} geometry of Fig.\ \ref{fig_constricted}b is, in
turn, equivalent to the {\em unconstricted\/} geometry of Fig.\
\ref{fig_constricted}c, consisting of a disordered wire with $N_{0}$ transverse
modes and mean free path $l/\nu$. The number $N_{0}$ is determined by the
conductance $N_{0}(2e^{2}/h)$ of the point contact. The fraction $\nu$ is
defined by
\begin{equation}
\nu=\frac{\beta N_{0}+2-\beta}{\beta N+2-\beta}.\label{nudef}
\end{equation}

The argument goes as follows. A ballistic point contact has to a good
approximation $T_{n}=1$ ($\lambda_{n}=0$) for $1\leq n\leq N_{0}$, and
$T_{n}=0$ ($\lambda_{n}\rightarrow\infty$) for $N_{0}+1\leq n\leq N$. (This is
a statement about transmission {\em eigenvalues}, not about the transmission
probabilities of individual modes, which are all of order $N_{0}/N$.) The
initial condition (\ref{P0initial}) becomes
\begin{equation}
\lim_{L\rightarrow 0}P=\lim_{\Lambda\rightarrow\infty}
\prod_{n=1}^{N_{0}}\delta(\lambda_{n})\,\prod_{n=N_{0}+1}^{N}
\delta(\lambda_{n}-\Lambda).\label{PinitialQPC}
\end{equation}
The closed channels $N_{0}+1\leq n\leq N$ are irrelevant for conduction and can
be integrated out. The reduced distribution function
$\tilde{P}(\lambda_{1},\lambda_{2},\ldots \lambda_{N_{0}},L)$ is defined by
\begin{equation}
\tilde{P}=\int_{0}^{\infty}\!d\lambda_{N_{0}+1}
\int_{0}^{\infty}\!d\lambda_{N_{0}+2}\cdots
\int_{0}^{\infty}\!d\lambda_{N}\,P,\label{Ptildedef}
\end{equation}
and satisfies the following evolution equation plus initial condition [{\em
cf.} Eqs.\ (\ref{DMPK}) and (\ref{PinitialQPC})]:
\begin{mathletters}
\label{FPtilde}
\begin{eqnarray}
&&\frac{l}{2}(\beta N+2-\beta)\frac{\partial\tilde{P}}{\partial L}=
\sum_{n=1}^{N_{0}}
\frac{\partial}{\partial\lambda_{n}}\lambda_{n}(1+\lambda_{n})
\tilde{J}\frac{\partial}{\partial\lambda_{n}}
\frac{\tilde{P}}{\tilde{J}},\nonumber\\
\label{FPtildea}\\
&&\tilde{J}=\prod_{i=1}^{N_{0}}\prod_{j=i+1}^{N_{0}}
|\lambda_{j}-\lambda_{i}|^{\beta},
\label{FPtildeb}\\
&&\tilde{P}(\lambda_{1},\lambda_{2},\ldots
\lambda_{N_{0}},0)=\prod_{n=1}^{N_{0}}\delta(\lambda_{n}).\label{FPtildec}
\end{eqnarray}
\end{mathletters}%
We now compare with the unconstricted geometry of Fig.\ \ref{fig_constricted}c,
with $N_{0}$ modes and mean free path $l/\nu$. The probability distribution
$P_{\nu}(\lambda_{1},\lambda_{2},\ldots \lambda_{N_{0}},L)$ for this geometry
is determined by
\begin{mathletters}
\label{FPN0}
\begin{eqnarray}
&&\frac{l}{2\nu}(\beta N_{0}+2-\beta) \frac{\partial P_{\nu}}{\partial L}=
\sum_{n=1}^{N_{0}}
\frac{\partial}{\partial\lambda_{n}}\lambda_{n}(1+\lambda_{n})
\tilde{J}\frac{\partial}{\partial\lambda_{n}}
\frac{P_{\nu}}{\tilde{J}},\nonumber\\
\label{FPN0a}\\
&&P_{\nu}(\lambda_{1},\lambda_{2},\ldots
\lambda_{N_{0}},0)=\prod_{n=1}^{N_{0}}\delta(\lambda_{n}).\label{FPN0b}
\end{eqnarray}
\end{mathletters}%
Comparison of Eqs.\ (\ref{FPtilde}) and (\ref{FPN0}) shows that
$\tilde{P}=P_{\nu}$ if $\nu$ is given by Eq.\ (\ref{nudef}).

\begin{figure}[tb]
\centerline{
\psfig{figure=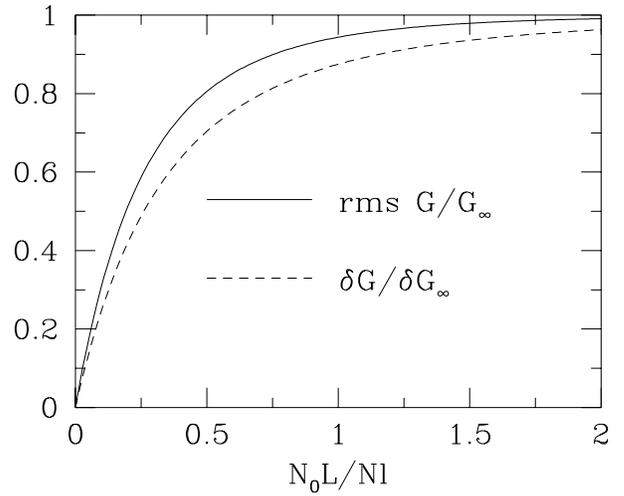,width=
8cm,bbllx=51pt,bblly=157pt,bburx=554pt,bbury=571pt}
}%
\medskip
\caption[]{
Suppression by the point contact of the weak-localization correction $\delta G$
and the root-mean-square conductance fluctuations ${\rm rms}\,G=({\rm
Var}\,G)^{1/2}$. The dashed and solid curves are from Eqs.\
(\protect\ref{deltaGconstricted}) and (\protect\ref{VarGconstricted}),
respectively. [The limit $N_{0}/N\rightarrow 0$ is taken in Eq.\
(\ref{deltaGconstricted}).] For $N_{0}L/Nl\gg 1$ the curves approach the values
$\delta G_{\infty}$ and ${\rm rms}\,G_{\infty}$ of an unconstricted disordered
wire (normalized to unity in the plot). After Beenakker and Melsen (1994).
}\label{fig_vargplot}
\end{figure}

The mapping between constricted and unconstricted geometries allows us to
obtain the effect of the point contact on the conductance directly from the
results for disordered wires in Sec.\ \ref{classicalG}. Let us first assume
that the individual conductances of the point contact and of the disordered
region are both $\gg G_{0}$, or in other words, that $N_{0}$ and $N/s$ are both
$\gg 1$. (Recall the definitions $G_{0}=2e^{2}/h$, $s=L/l$.) Substitution of
$N\rightarrow N_{0}$ and $s\rightarrow\nu s$ in Eqs.\ (\ref{GDrude}) and
(\ref{deltaGDMPK}) yields the average conductance $\langle
G/G_{0}\rangle=G_{\rm series}+\delta G$, with $G_{\rm
series}=G_{0}(N_{0}^{-1}+s/N)^{-1}$ and
\begin{eqnarray}
\delta G/G_{0}&=&(1-2/\beta)\left[
\frac{1}{3}\left(\frac{N_{0}s/N}{1+N_{0}s/N}\right)^{3}\right.\nonumber\\
&&\mbox{}+\left.\left(
\vphantom{\left(\frac{N_{0}s/N}{1+N_{0}s/N}\right)^{3}}
1-\frac{N_{0}}{N}\right)
\frac{N_{0}s/N}{(1+N_{0}s/N)^{2}}\right].\label{deltaGconstricted}
\end{eqnarray}
Similarly, from Eq.\ (\ref{VarGDMPK}) one finds the variance
\begin{equation}
{\rm Var}\,G/G_{0}=\frac{2}{15}\beta^{-1}\left(1-
\frac{1+6N_{0}s/N}{(1+N_{0}s/N)^{6}}\right).\label{VarGconstricted}
\end{equation}
The term $G_{\rm series}$ is the series addition of the Sharvin conductance
$G_{0}N_{0}$ of the ballistic point contact and the Drude conductance
$G_{0}N/s$ of the disordered region. The term $\delta G$ is the
weak-localization correction to the classical series conductance. This term
depends on the ratio $N_{0}s/N$ of the Sharvin and Drude conductances as well
as on the ratio $N_{0}/N$ of the width of the point contact and the wide
regions. The variance ${\rm Var}\,G$ of the conductance depends only on
$N_{0}s/N$.
In Fig.\ \ref{fig_vargplot} we have plotted $\delta G$ and $({\rm
Var}\,G)^{1/2}$ as a function of $N_{0}s/N$. (The limit $N_{0}/N\rightarrow 0$
is assumed for $\delta G$.) For large $N_{0}s/N$ the curves tend to $\delta
G_{\infty}=\frac{1}{3}(1-2/\beta)G_{0}$ and ${\rm
Var}\,G_{\infty}=\frac{2}{15}\beta^{-1}G_{0}^{2}$, which are the results
(\ref{deltaGVarGDMPK}) for weak localization and universal conductance
fluctuations in a wire geometry without a point contact. These values are
universal to the extent that they are independent of wire length and mean free
path. The presence of a point contact breaks this universality, but only if the
Sharvin conductance is smaller than the Drude conductance. For $N_{0}>N/s$ the
universality is quickly restored. For $N_{0}<N/s$ both $\delta G$ and ${\rm
Var}\,G$ are suppressed by the presence of the point contact,
\begin{eqnarray}
&&\delta G/G_{0}=(1-2/\beta)N_{0}s/N+{\cal
O}(N_{0}s/N)^{2},\label{deltaGsmall}\\
&&{\rm Var}\,G/G_{0}=2\beta^{-1}(N_{0}s/N)^{2}+{\cal
O}(N_{0}s/N)^{3},\label{VarGsmall}
\end{eqnarray}
as first noticed by Maslov, Barnes, and Kirczenow (1993a, 1993b) in a study of
the quasi-ballistic regime $l\gg L$. For a $d$-dimensional point contact,
$N_{0}\propto W_{0}^{d-1}$, so that Eq.\ (\ref{VarGsmall}) implies that the
root-mean-square fluctuations scale as ${\rm rms}\,G\propto W_{0}^{d-1}$ with
the point contact width. A classical argument of series addition of a
non-fluctuating contact resistance with a fluctuating background would instead
imply the much stronger suppression ${\rm rms}\,G\propto W_{0}^{2d-2}$.

Holweg {\em et al.} (1991) have measured magnetoconductance fluctuations of a
three-dimensional point contact between two Ag films. The mean free path
$l\simeq 200-240\,{\rm nm}$ is comparable to the thickness of the metal films,
and much larger than the diameter $W_{0}\simeq 10-30\,{\rm nm}$ of the point
contact. The mean separation of the impurities $d_{\rm imp}\simeq(l\lambda_{\rm
F})^{1/3}\simeq 3\,{\rm nm}$ is, however, considerably smaller than $W_{0}$.
The root-mean-square conductance fluctuations in the experiment are much larger
than predicted by Eq.\ (\ref{VarGsmall}), and moreover scale linearly rather
than quadratically with $W_{0}$. Kozub, Caro, and Holweg (1996) have argued
that the presence of impurities near the opening of the constriction leads to a
substantial enhancement of the conductance fluctuations and to a linear $W_{0}$
dependence, in agreement with their experiment. Possibly, experiments on point
contacts in a two-dimensional electron gas (with much larger $\lambda_{\rm F}$
and $l$) can reach the regime $d_{\rm imp}\gg W_{0}$ where Eq.\
(\ref{VarGsmall}) is expected to apply. In a numerical simulation of such a
system, Maslov, Barnes, and Kirczenow (1993a, 1993b) have indeed obtained
results consistent with Eq.\ (\ref{VarGsmall}).\footnote{Maslov, Barnes, and
Kirczenow (1993a, 1993b) consider a geometry as in Fig.\
\protect\ref{fig_constricted}a, with $L_{1}=L_{2}=\frac{1}{2}L$, and relate the
variance ${\rm Var}\,G$ of the whole system to the variance ${\rm Var}\,G_{1}$
of one of the two disordered segments of length $\frac{1}{2}L$. Their result is
${\rm Var}\,G=(N_{0}s/N)^{2}(l/L_{1})^{2}{\rm Var}\,G_{1}$, in agreement with
Eq.\ (\protect\ref{VarGsmall}) [since ${\rm
Var}\,G_{1}=2\beta^{-1}(L_{1}/l)^{2}$ for $L_{1}\ll l$].}

\subsubsection{Single-channel limit}
\label{singlechannel}

The results (\ref{deltaGconstricted}) and (\ref{VarGconstricted}) require
$N_{0}\gg 1$, which means that the width $W_{0}$ of the point contact should be
much greater than the Fermi wavelength $\lambda_{\rm F}$. Such a point contact
is called ``classical''. A ``quantum'' point contact has $W_{0}$ comparable to
$\lambda_{\rm F}$, so that $N_{0}$ is a small integer. Let us consider the
single-channel limit $N_{0}=1$. We assume $N\gg 1$, hence $\nu=2/\beta N$. The
evolution equation for the distribution $\tilde{P}(\lambda_{1},L)$ of the
single transmitted channel is given by Eq.\ (\ref{DMPK1D}) with a rescaled mean
free path ($l\rightarrow\frac{1}{2}\beta Nl$) and a ballistic initial
condition,
\begin{mathletters}
\label{DMPK1Dnu}
\begin{eqnarray}
&&\case{1}{2}\beta Nl\frac{\partial\tilde{P}}{\partial L}=
\frac{\partial}{\partial\lambda_{1}}\lambda_{1}(1+\lambda_{1})
\frac{\partial}{\partial\lambda_{1}}\tilde{P},\label{DMPK1Dnua}\\
&&\tilde{P}(\lambda_{1},0)=\delta(\lambda_{1}).\label{DMPK1Dnub}
\end{eqnarray}
\end{mathletters}%
The solution is (Gertsenshtein and Vasil'ev, 1959; Abrikosov, 1981)
\begin{eqnarray}
&&\tilde{P}(\lambda_{1},L)=(2\pi)^{-1/2}(\beta Nl/2L)^{3/2}\exp(-L/2\beta Nl)
\nonumber\\
&&\hspace{1cm}\mbox{}\times\int_{{\rm
arcosh}(1+2\lambda_{1})}^{\infty}du\frac{u\exp(-u^{2}\beta Nl/8L)}{(\cosh
u-1-2\lambda_{1})^{1/2}}.\label{DMPK1Dsol}
\end{eqnarray}
(This solution is the single-channel limit of Eq.\ (\ref{exactsolution}), with
an integral representation for the Legendre function.) From Eq.\
(\ref{DMPK1Dsol}) we obtain immediately the distribution $P(\delta R)$ of the
excess resistance $\delta R=1/G-1/G_{0}=\lambda_{1}/G_{0}$. In the metallic
regime $Nl/L\gg 1$ the integral over $u$ can be carried out analytically, with
the result (Beenakker and Melsen, 1994)
\begin{equation}
P(\delta R)=\frac{G_{0}\beta Nl}{2L}\exp\left(-\frac{G_{0}\beta Nl}{2L}\delta
R\right),\;\delta R\geq 0.\label{Pexponential}
\end{equation}
The width $2L/\beta Nl$ of this exponential distribution decreases by a factor
of two upon breaking time-reversal symmetry in the absence of spin-orbit
scattering ($\beta=1\rightarrow\beta=2$).

\begin{figure}[tb]
\centerline{
\psfig{figure=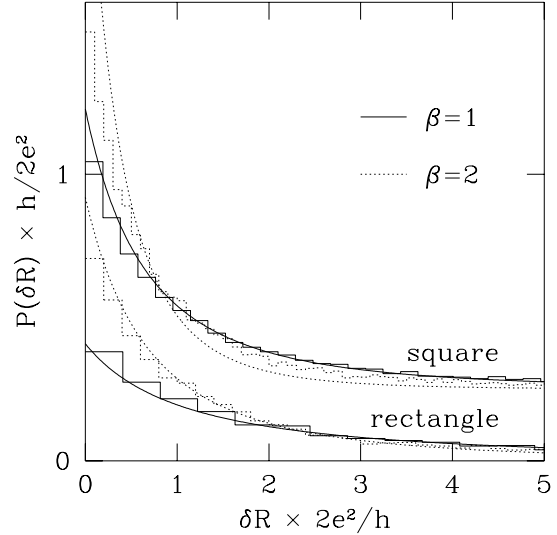,width= 8cm}
}%
\caption[]{
Distribution of the excess resistance $\delta R=R-h/2e^{2}$ of a single-channel
point contact in series with a disordered region (square: $N=20$, $L=W=8.3\,l$;
rectangle: $N=10$, $L=2W=8.3\,l$). The histograms are the numerical data
(averaged over $10^{4}$ impurity configurations), the smooth curves are
computed from Eq.\ (\ref{DMPK1Dsol}) (with $\delta R=\lambda_{1}h/2e^{2}$).
Solid curves are for zero magnetic field ($\beta=1$), dotted curves for a flux
of $50\,h/e$ through the disordered region ($\beta=2$). For clarity, the curves
for the square geometry are offset vertically by $0.25$. After Beenakker and
Melsen (1994).
}\label{fig_exponential}
\end{figure}

A comparison with numerical simulations is shown in Fig.\
\ref{fig_exponential}. The model is the Anderson model on a two-dimensional
square lattice described in Sec.\ \ref{lognormal}. The single-channel point
contact is introduced by assigning a large potential energy to sites at one end
of the lattice, so as to create a nearly impenetrable barrier with a narrow
opening in the center. The Fermi energy is chosen at $E_{\rm F}=1.5\,U_{0}$
from the band bottom. Two geometries are considered for the wide disordered
region: A square geometry ($L=W=47\,a$, corresponding to $N=20$), and a
rectangular geometry ($L=47\,a$, $W=23\,a$, corresponding to $N=10$). The mean
free path $l$ which appears in the DMPK equation is computed numerically from
${\rm Tr}\,t_{\rm d}^{\vphantom{\dagger}}t_{\rm d}^{\dagger}=N(1+L/l)^{-1}$,
with $t_{\rm d}$ the transmission matrix of the wide disordered region without
the constriction. [We recall that this mean free path differs by a numerical
coefficient from the transport mean free path of kinetic theory, {\em cf.}
Eqs.\ (\ref{ltrdef}) and (\ref{GDrude}).] The results for $P(\delta R)$ plotted
in Fig.\ \ref{fig_exponential} are for $L/l=8.3$ (disorder strength $U_{\rm
d}=3U_{0}$). To compare the cases $\beta=1$ and $\beta=2$, the simulations are
repeated in the presence of a magnetic flux of $50\,h/e$ through the disordered
region. The numerical results (histograms) are seen to be in good agreement
with the theoretical predictions (smooth curves), without any adjustable
parameters. The theory agrees comparably well with the simulations for the
square and rectangular geometries, which shows that the condition $L\gg W$ for
the validity of the DMPK equation can be relaxed to a considerable extent.

An experimental observation of the exponential distribution of the excess
resistance in a quantum point contact is still lacking.

\subsubsection{Double-barrier junction}
\label{obstacle_tunnel}

\begin{figure}[tb]
\centerline{
\psfig{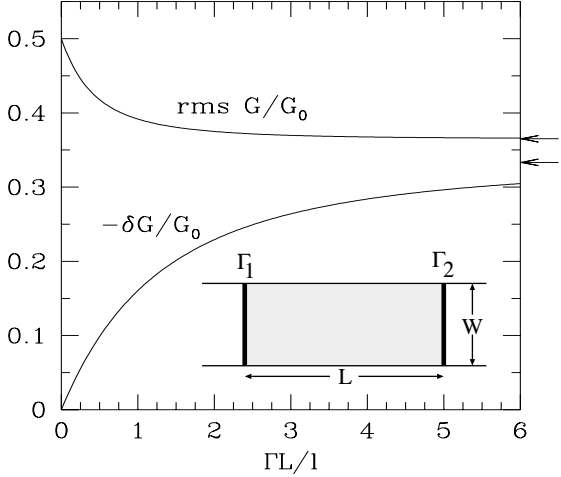}
}%
\caption[]{
Weak-localization correction $\delta G$ to the average conductance and
root-mean-square fluctuations ${\rm rms}\,G$ (in units of $G_{0}=2e^{2}/h$),
computed from Eqs.\ (\ref{deltaGGammas}) and (\ref{VarGGammas}) for $\beta =
1$. The arrows give the limit $\Gamma L/l\gg 1$. The inset shows the geometry
of the double-barrier junction (the disordered region is dotted). The curves
plotted in the figure are for a symmetric junction,
$\Gamma_{1}=\Gamma_{2}\equiv\Gamma\ll 1$. After Melsen and Beenakker (1995).
}\label{fig_doublebarrier}
\end{figure}

As a second application, we consider the case that the obstacles in the
disordered wire are formed by two tunnel barriers (Melsen and Beenakker, 1995).
The geometry is shown in the inset of Fig.\ \ref{fig_doublebarrier}. A
disordered region (length $L$, mean free path $l$, width $W$) is separated from
ideal $N$-mode leads by two tunnel barriers, with conductances
$N\Gamma_{i}G_{0}$ ($i=1,2$). We assume $N\Gamma_{i}\gg 1$, so that the
transmission in the absence of disorder occurs via a large number of
overlapping resonances. (For the opposite regime of isolated transmission
resonances, see Fertig and Das Sarma, 1989; Leo and MacDonald, 1990; Berkovits
and Feng, 1992; Lerner and Raikh, 1992.) Two types of disorder can play a role
in a double-barrier junction, interface roughness at the barriers and
impurities between the barriers. Interface roughness leads to mesoscopic
(sample-to-sample) fluctuations in the conductance even in the absence of  any
phase coherence, because the tunnel probability $\Gamma$ of a single barrier
depends strongly on its thickness. Conductance fluctuations for a single rough
tunnel barrier have been studied by Raikh and Ruzin (1991). Here we discuss the
case of impurity scattering in the absence of interface roughness. Phase
coherence is then essential.

We first assume that the disorder is weak enough that its effect on the average
conductance is negligibly small ($l\gg\Gamma L$), but strong enough to fully
mix the transverse modes in the inter-barrier region ($l,W\ll L/\Gamma$). We
may then describe the disorder-induced mode-mixing by a random $N\times N$
unitary matrix $\Omega$, distributed according to the circular ensemble. The
transmission eigenvalues $T_{n}$ are related to the eigenvalues $\exp({\rm
i}\phi_{n})$ of $\Omega$ by
\begin{mathletters}
\label{Tnab}
\begin{eqnarray}
&&T_{n}=(a+b\cos\phi_{n})^{-1},\label{Tnaba}\\
&&a=[1+(1-\Gamma_1)(1-\Gamma_2)]/\Gamma_1\Gamma_2,\label{Tnabab}\\
&&b=2\sqrt{(1-\Gamma_1)(1-\Gamma_2)}/\Gamma_1\Gamma_2\label{Tnabc}.
\end{eqnarray}
\end{mathletters}%
The statistics of the conductance $G=G_{0}\sum_{n}T_{n}$ follows from the
probability distribution (\ref{Pphicircular}) of the $\phi_{n}$'s in the
circular ensemble.

We seek the average $\langle A\rangle$ and variance ${\rm Var}\,A$ of linear
statistics $A=\sum_{n=1}^{N}a(\phi_{n})$ on the eigenphases $\phi_{n}$. Since
in the circular ensemble the $\phi_{n}$'s are uniformly distributed in
$(0,2\pi)$, the average is exactly equal to
\begin{equation}
\langle
A\rangle=\frac{N}{2\pi}\int_{0}^{2\pi}\!d\phi\,a(\phi).\label{Abarcircular}
\end{equation}
An exact expression for the variance can also be given (Mehta, 1991) but is
cumbersome to evaluate. For $N\gg 1$ we can use a formula analogous to Eq.\
(\ref{VarAWDresult}) (Forrester, 1995; Melsen and Beenakker, 1995):
\begin{mathletters}
\label{VarAcircular}
\begin{eqnarray}
&&{\rm Var}\,A=\frac{1}{\beta\pi^{2}}\sum_{n=1}^{\infty}n|a_{n}|^{2}+{\cal
O}(N^{-1}),\label{VarAcirculara}\\
&&a_{n}=\int_{0}^{2\pi}\!d\phi\,{\rm e}^{{\rm
i}n\phi}a(\phi).\label{VarAcircularb}
\end{eqnarray}
\end{mathletters}%
For the conductance we substitute $a(\phi)=(a + b\cos\phi)^{-1}$, with Fourier
coefficients $a_{n}={2\pi}(a^{2}-b^{2})^{-1/2}b^{-n}
[(a^{2}-b^{2})^{1/2}-a]^{n}$. The results are (assuming $\Gamma_{i}\ll 1$)
\begin{eqnarray}
\langle
G/G_{0}\rangle&=&N(1/\Gamma_{1}+1/\Gamma_{2})^{-1},\label{Gbarcircular}\\
{\rm Var}\,G/G_{0}&=&\frac{4}{\beta}\frac{\Gamma_{1}^{2}\Gamma_{2}^{2}}
{(\Gamma_{1}+\Gamma_{2})^{4}}.\label{VarGcircular}
\end{eqnarray}

Equation (\ref{Gbarcircular}) for the average conductance is what one would
expect from classical addition of the resistances $(N\Gamma_{i}G_{0})^{-1}$ of
the individual barriers. Each member of the ensemble contains a different set
of overlapping transmission resonances, and the ensemble average removes any
trace of resonant tunneling in $\langle G\rangle$. Equation
(\ref{VarGcircular}) for the conductance fluctuations tells us that ${\rm
Var}\,G$ becomes completely independent of $N$ in the limit
$N\rightarrow\infty$. (More precisely, corrections to Eq.\ (\ref{VarGcircular})
are of order $\langle G/G_{0}\rangle^{-1}$, which is $\ll 1$ if $N\Gamma_{i}\gg
1$.) The variance reaches a $\Gamma$-independent maximum for two equal
barriers: ${\rm Var}\,G/G_{0}=\frac{1}{4}\beta^{-1}$ if
$\Gamma_{1}=\Gamma_{2}$. A smaller numerical coefficient ($\frac{3}{16}$
instead of $\frac{1}{4}$) has been obtained by Fal'ko (1995), using a different
method. The origin of the difference is not yet understood.

We now relax the assumption $l\gg\Gamma L$ to include the case that the
impurity scattering is sufficiently strong to affect the average conductance.
The $L$-dependence of the distribution of the transmission eigenvalues is
governed by the DMPK equation, with the circular ensemble as initial condition.
The mean and variance of the conductance can be computed using the method of
moments described in Sec.\ \ref{classicalG}. The results for
$\Gamma_{1}=\Gamma_{2}\equiv\Gamma\ll 1$ are
\begin{eqnarray}
&&\langle G\rangle=NG_{0}(s+1/\Gamma_{1}+1/\Gamma_{2})^{-1}+\delta
G,\label{GbarGammas}\\
&&\delta G/G_{0}=\frac{1}{3}(1-2/\beta)-\frac{1-2/\beta}{(2+\Gamma
s)^{3}}\left(\frac{8}{3}+2\Gamma s\right),\label{deltaGGammas}\\
&&{\rm Var}\,G/G_{0}=\frac{2}{15\beta}+\frac{4}{\beta(2+\Gamma s)^{6}}\left(
\Gamma^{2}s^{2}+\frac{8}{5}\Gamma s+\frac{28}{15}\right).\nonumber\\
\label{VarGGammas}
\end{eqnarray}
Equations (\ref{deltaGGammas}) and (\ref{VarGGammas}) are plotted in Fig.\
\ref{fig_doublebarrier}, for the case $\beta=1$. We see that impurity
scattering leads to the appearance of a weak-localization effect on the average
conductance. The conductance fluctuations become universal ({\em i.e.},
independent of $\Gamma$) if $L$ exceeds a length $l/\Gamma$ which is
parametrically greater than the mean free path. A similar conclusion has been
reached by Iida, Weidenm\"{u}ller, and Zuk (1990a, 1990b), who used the
supersymmetry technique to study the conductance fluctuations of a chain of
disordered grains as a function of the coupling strength to two electron
reservoirs. Their model (which has also been studied by Argaman (1995, 1996)
using a semiclassical method) is qualitatively similar but different in detail
from the homogeneously disordered conductor considered here ({\em cf.} Sec.\
\ref{overview}).

Experimentally, the effects of disorder on tunneling through double-barrier
junctions have been studied mainly in semiconductor quantum wells, where the
resonances are widely separated because of the small barrier separation $L$
relative to the Fermi wave length $\lambda_{\rm F}$. Conductance fluctuations
of order $e^{2}/h$ in such a structure have been observed by Ghenim {\em et
al.} (1996). The results presented above apply to the opposite regime of
strongly overlapping resonances, relevant to metal structures (where
$\lambda_{\rm F}$ is very short, comparable to the inter-atomic separation), or
to tunneling in the plane of a two-dimensional electron gas (where $L$ can be
quite long, because of the large phase-coherence length).

\subsection{Shot noise}
\label{shotnoiseN}

The shot-noise power $P$, defined in Eq.\ (\ref{Pdef}), contains information on
temporal correlations in the current which is not contained in the conductance.
A familiar example is a tunnel  diode, where
$P=2e\bar{I}\equiv P_{\rm Poisson}$, with $\bar{I}$ the time-averaged current.
This tells us that the electrons traverse the conductor in completely
uncorrelated fashion, as in a Poisson process. In a degenerate electron gas the
shot noise can be smaller than $P_{\rm Poisson}$, due to correlations in the
electron transmission imposed by the Pauli principle (Kulik and Omel'yanchuk,
1984; Khlus, 1987; Lesovik, 1989; Yurke and Kochanski, 1990; B\"{u}ttiker,
1990, 1992; Martin and Landauer, 1992). Here we consider the sub-Poissonian
shot noise in a metallic diffusive conductor, relevant for random-matrix
theory. For reviews specifically devoted to shot noise, we refer to Martin
(1994) and De Jong and Beenakker (1997).

Starting point is the relationship (\ref{PTrelation}) between the
zero-temperature, zero-frequency shot-noise power $P$ and the transmission
eigenvalues $T_{n}$. One sees that $P=2eVG=P_{\rm Poisson}$ for a conductor
where all $T_{n}\ll 1$ (such as a high tunnel barrier). However, if some
$T_{n}$ are near 1 (open channels), then the shot noise is reduced below
$P_{\rm Poisson}$. In the metallic diffusive regime ($l\ll L\ll Nl$) the
variables $x_{n}$ have the uniform density (\ref{rho0result}). This means that
the transmission eigenvalues $T_{n}=1/\cosh^{2}x_{n}$ have a {\em bimodal\/}
distribution,
\begin{equation}
\rho_{0}(T)=\frac{Nl}{2L}\frac{1}{T\sqrt{1-T}},\;{\rm e}^{-2L/l}\lesssim T\leq
1,\label{rhoTresultP}
\end{equation}
with a peak at unit transmission and a peak at exponentially small transmission
(Dorokhov, 1984; Imry, 1986a; Pendry, MacKinnon, and Roberts, 1992). Averaging
of Eq.\ (\ref{PTrelation}) with the density (\ref{rhoTresultP}) yields
\begin{equation}
\langle P\rangle=P_{0}\int_{0}^{1}\!dT\,T(1-T)\rho_{0}(T)=
P_{0}\frac{Nl}{3L}=\frac{1}{3}P_{\rm Poisson},\label{Ponethird}
\end{equation}
where we have used that $P_{0}Nl/L=2eV\langle G\rangle=P_{\rm Poisson}$. The
bimodal distribution of the transmission eigenvalues causes a {\em one-third
suppression\/} of the shot noise (Beenakker and B\"{u}ttiker, 1992).

Although the derivation of the eigenvalue density in Sec.\ \ref{Tdensity} is
based on the DMPK equation, and hence requires a wire geometry, its validity is
independent of the dimensionality of the conductor (Nazarov, 1994a; Altshuler,
Levitov, and Yakovets, 1994).\footnote{
Corrections to Eq.\ (\ref{Ponethird}) from the weak-localization effect do
depend on the dimensionality. In a wire geometry the weak-localization
correction is $\delta P=-\frac{4}{45}P_{0}$ in zero magnetic field (De Jong and
Beenakker, 1992).}
Furthermore, although the concept of a transmission eigenvalue requires
phase-coherence, this is not required for the one-third suppression. An
alternative derivation exists that starts from a semiclassical kinetic
equation, in which the Pauli principle is accounted for but the electron motion
is treated classically (Nagaev, 1992; De Jong and Beenakker, 1995, 1996). The
one-third suppression thus applies regardless of whether $L$ is long or short
compared to the phase-coherence length $l_{\phi}$.

Loss of phase coherence is one consequence of electron-electron interactions.
Another consequence is thermalization of the distribution of the electrons
among the available energy levels. The thermalization length $l_{\rm th}$ is
generally much greater than $l_{\phi}$. [Thermalization requires interactions
with large transfer of energy, while a small transfer of energy is sufficient
to destroy phase coherence (Altshuler and Aronov, 1985; Imry, 1996).] If $L$
becomes greater than $l_{\rm th}$ the shot-noise power increases slightly, from
$P=\frac{1}{3}P_{\rm Poisson}=0.67\,e\bar{I}$ to $P=\frac{1}{4}\sqrt{3}\,P_{\rm
Poisson}=0.87\,e\bar{I}$ (Nagaev, 1995; Kozub and Rudin, 1995; Steinbach,
Martinis, and Devoret, 1995; De Jong and Beenakker, 1996). On longer length
scales $L>l_{\rm in}$, inelastic electron-phonon scattering equilibrates the
electron gas with the lattice, thereby averaging the shot noise out to zero.

\begin{figure}[tb]
\centerline{
\psfig{figure=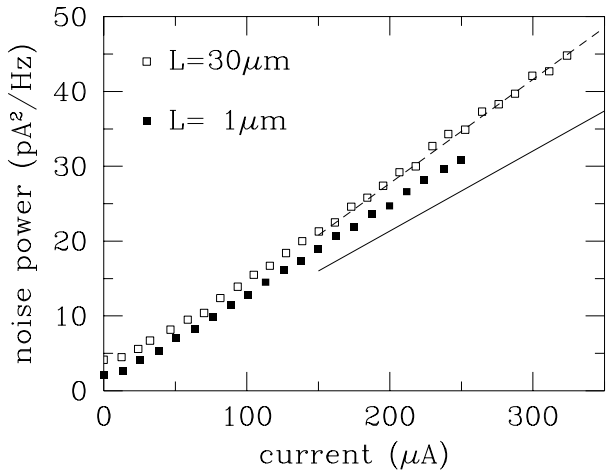,width=
8cm,bbllx=19pt,bblly=158pt,bburx=547pt,bbury=571pt}
}%
\medskip
\caption[]{
Current dependence of the noise power at 50~mK in two thin-film Ag wires of
different length $L$. The noise power is linear in the current, indicating shot
noise, except at the lowest currents, where thermal noise $P_{\rm
thermal}=4k_{\rm B}TG$ takes over. (The conductance $G=0.76\,\Omega^{-1}$ for
$L=1\,\mu{\rm m}$ and $1.47\,\Omega^{-1}$ for $L=30\,\mu{\rm m}$.) A current-
and temperature-independent background noise has been subtracted from the data.
The lines indicate the two theoretical predictions $P/P_{\rm
Poisson}=\frac{1}{3}$ for short wires (solid) and $P/P_{\rm
Poisson}=\frac{1}{4}\sqrt{3}$ for long wires (dashed). After Steinbach,
Martinis, and Devoret (1996).
}\label{fig_steinbach}
\end{figure}

The length-scale dependence of the shot-noise power has been studied
experimentally by Steinbach, Martinis, and Devoret (1996) on Ag thin-film wires
of different lengths. (Sub-Poissonian shot noise had earlier been measured by
Liefrink {\em et al.} (1994), on a narrow two-dimensional electron gas in a
GaAs/AlGaAs heterostructure.) The data for $L=1\,\mu{\rm m}$ and $30\,\mu{\rm
m}$ at temperature $T=50\,{\rm mK}$ is shown in Fig.\ \ref{fig_steinbach}. The
noise-power is linear in the average current for $\bar{I}\gtrsim 50\,\mu{\rm
A}$. (The saturation at smaller currents is due to the residual thermal noise
$P_{\rm thermal}=4k_{\rm B}TG$.) The ratio $P/2e\bar{I}$ is close to
$\frac{1}{4}\sqrt{3}$ (dashed line) for the $30\,\mu{\rm m}$ wire, indicating
that this length is in the range $l_{\rm th}\ll L\ll l_{\rm in}$. For the
$1\,\mu{\rm m}$ wire the slope is clearly smaller than in the longer wire, but
still above the $\frac{1}{3}$ prediction for $L\ll l_{\rm th}$ --- presumably
because $L$ is not quite small enough in the experiment.

\section{Normal-metal--superconductor junctions}
\label{junctions}

\subsection{Scattering theory}
\label{NSscatteringtheory}

\subsubsection{Andreev reflection}
\label{Andreevreflection}

\begin{figure}[tb]
\centerline{
\psfig{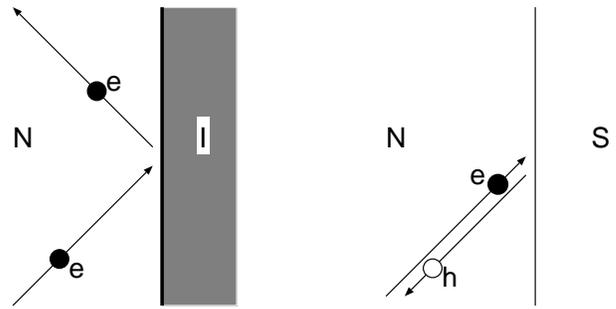}
}%
\medskip
\caption[]{
Normal reflection by an insulator (I) versus Andreev reflection by a
superconductor (S) of an electron excitation in a normal metal (N) near the
Fermi level. Normal reflection (left) conserves charge but does not conserve
momentum. Andreev reflection (right) conserves momentum but does not conserve
charge: The electron (e) is reflected as a hole (h) with the same momentum and
opposite velocity. The missing charge of $2e$ is absorbed as a Cooper pair by
the superconducting condensate.
}\label{reflection}
\end{figure}

At the interface between a normal metal and a superconductor, dissipative
electrical current is converted into dissipationless supercurrent. The
mechanism for this conversion was discovered in 1964 by Andreev: An electron
excitation slightly above the Fermi level in the normal metal is reflected at
the interface as a hole excitation slightly below the Fermi level (see Fig.\
\ref{reflection}). The missing charge of $2e$ is removed as a Cooper pair. The
reflected hole has (approximately) the same momentum as the incident electron.
(The two momenta are precisely equal at the Fermi level.) The velocity of the
hole is minus the velocity of the electron ({\em cf.} the notion of a hole as a
``time-reversed'' electron). This curious scattering process is known as
retro-reflection or {\em Andreev reflection}.

The early theoretical work on the conductance of a normal-metal--superconductor
(NS) junction treats the dynamics of the quasiparticle excitations
semiclassically, as is appropriate for macroscopic junctions. Phase coherence
of the electrons and the Andreev-reflected holes is ignored. Interest in
``mesoscopic'' NS junctions, where phase coherence plays an important role, is
a recent development. Significant advances have been made in our understanding
of quantum interference effects due to phase-coherent Andreev reflection. Much
of the motivation has come from the technological advances in the fabrication
of a highly transparent contact between a superconducting film and the
two-dimensional electron gas in a semiconductor heterostructure. The advantages
of a two-dimensional electron gas over a metal are the large Fermi wavelength,
large mean free path, and the possibility to confine the electrons
electrostatically by means of gate electrodes. Andreev reflection requires
relatively transparent NS interfaces. Semiconductor --- superconductor
junctions are convenient, since the Schottky barrier at the interface is much
more transparent than a typical dielectric tunnel barrier. The technological
effort is directed towards making the interface as transparent as possible.

The random-matrix theory of phase-coherent Andreev reflection is based on a
scattering formulation, in which the conductance $G_{\rm NS}$ of the NS
junction is related to the transmission matrix $t$ in the normal state. In the
limit of zero temperature, zero voltage, and zero magnetic field, the
relationship is (Beenakker, 1992a)
\begin{equation}
G_{\rm NS}=\frac{4e^{2}}{h}\sum_{n=1}^{N}
\frac{T_{n}^{2}}{(2-T_{n})^{2}},\label{keyzero}
\end{equation}
where the transmission eigenvalue $T_{n}$ is an eigenvalue of the matrix
product $tt^{\dagger}$. The same numbers $T_{n}$ ($n=1,2,\ldots N$) determine
the conductance $G_{\rm N}$ in the normal state, according to the Landauer
formula (\ref{Landauer}). (In this section we append the subscript N to the
normal-state conductance, to help distinguish it from $G_{\rm NS}$.) The fact
that the same eigenvalues determine both $G_{\rm N}$ and $G_{\rm NS}$ means
that one can use the same random-matrix ensembles as in the normal state. This
is a substantial technical and conceptual simplification.

Let us discuss how Eq.\ (\ref{keyzero}) is obtained.

\subsubsection{Bogoliubov-De Gennes equation}
\label{BdGequation}

\begin{figure}[tb]
\centerline{
\psfig{figure=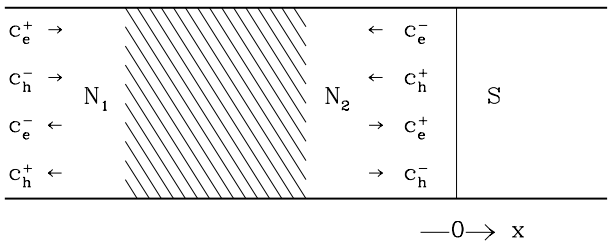,width=
8cm,bbllx=106pt,bblly=320pt,bburx=545pt,bbury=497pt}
}%
\medskip
\caption[]{
Normal-metal--superconductor junction containing a disordered normal region
(shaded). Scattering states in the two normal leads ${\rm N}_{1}$ and ${\rm
N}_{2}$ are indicated schematically.
}\label{diagram}
\end{figure}

The model considered is illustrated in Fig.\ \ref{diagram}. It consists of a
disordered normal region (shaded) adjacent to a superconductor (S). The
disordered region may also contain a geometrical constriction or a tunnel
barrier. To obtain a well-defined scattering problem we insert ideal
(impurity-free) normal leads ${\rm N}_{1}$ and ${\rm N}_{2}$ to the left and
right of the disordered region. The NS interface is located at $x=0$. We assume
that the only scattering in the superconductor consists of Andreev reflection
at the NS interface, {\em i.e.} we consider the case that the disorder is
contained entirely within the normal region. The spatial separation of Andreev
and normal scattering is the key simplification which allows us to relate the
conductance directly to the normal-state scattering matrix. The model is
directly applicable to a superconductor in the clean limit (mean free path in S
large compared to the superconducting coherence length $\xi$), or to a
point-contact junction (formed by a constriction which is narrow compared to
$\xi$). In both cases the contribution of scattering within the superconductor
to the junction resistance can be neglected.

The scattering states at energy $\varepsilon$ are eigenfunctions of an equation
called the Bogoliubov-De Gennes equation (De Gennes, 1966) --- although
historically it made its first appearance in a paper by Andreev (1964). This
equation has the form of two Schr\"{o}dinger equations for electron and hole
wavefunctions ${\rm u}(\vec{r})$ and ${\rm v}(\vec{r})$, coupled by the pair
potential ${\Delta}(\vec{r})$:
\begin{equation}
\left(\begin{array}{cc} {\cal H}_{0}&{\Delta}\\ {\Delta}^{\ast}&-{\cal
H}_{0}^{\ast} \end{array}\right) \left(\begin{array}{c}{\rm u}\\{\rm
v}\end{array}\right)=\varepsilon \left(\begin{array}{c}{\rm u}\\{\rm
v}\end{array}\right) .\label{BdG1}
\end{equation}
Here ${\cal H}_{0}$ is the single-electron Hamiltonian, which we assume to be
independent of the electron spin. (We will include spin-orbit scattering
later.) The excitation energy $\varepsilon$ is measured relative to the Fermi
energy $E_{\rm F}$. To simplify construction of the scattering basis we assume
that the magnetic field $\vec{B}$ (in the $z$-direction) vanishes outside the
disordered region.

The pair potential in the bulk of the superconductor ($x\gg\xi$) has amplitude
${\Delta_{0}}$ and phase $\phi$. The spatial dependence of ${\Delta}(\vec{r})$
near the NS interface is determined by the self-consistency relation
\begin{equation}
{\Delta}(\vec{r})=g(\vec{r})\sum_{\varepsilon>0}{\rm v}^{\ast}(\vec{r}){\rm
u}(\vec{r})[1-2f(\varepsilon)],\label{selfconsist}
\end{equation}
where the sum is over all states with positive eigenvalue,\footnote{
A cutoff at $\hbar\omega_{\rm D}$, with $\omega_{\rm D}$ the Debije frequency,
has to be introduced as usual in the BCS theory.}
and $f(\varepsilon)=[1+\exp(\varepsilon/k_{\rm B}T)]^{-1}$ is the Fermi
function. The coefficient $g$ is the interaction constant of the BCS theory of
superconductivity. At an NS interface, $g$ drops abruptly (over atomic
distances) to zero, in the assumed absence of any pairing interaction in the
normal region. Therefore, ${\Delta}(\vec{r})\equiv 0$ for $x<0$. At the
superconducting side of the NS interface, ${\Delta}(\vec{r})$ recovers its bulk
value $\Delta_{0}{\rm e}^{{\rm i}\phi}$ only at some distance from the
interface. We will neglect the suppression of ${\Delta}(\vec{r})$ on
approaching the NS interface, and use the step-function model
\begin{equation}
{\Delta}(\vec{r})={\Delta_{0}}{\rm e}^{{\rm
i}\phi}\theta(x).\label{stepfunction}
\end{equation}
This model is also referred to in the literature as a ``rigid
boundary-condition''. Likharev (1979) discusses in detail the conditions for
its validity: If the width $W$ of the NS junction is small compared to $\xi$,
the non-uniformities in ${\Delta}(\vec{r})$ extend only over a distance of
order $W$ from the junction (because of ``geometrical dilution'' of the
influence of the narrow junction in the wide superconductor). Since
non-uniformities on length scales $\ll\xi$ do not affect the dynamics of the
quasiparticles, these can be neglected and the step-function model holds. A
point contact or microbridge belongs in general to this class of junctions.
Alternatively, the step-function model holds also for a wide junction if the
resistivity of the junction region is much bigger than the resistivity of the
bulk superconductor. A semiconductor --- superconductor junction is typically
in this second category. Note that both cases are consistent with our
assumption that the disorder is contained entirely within the normal region.

It is worth emphasizing that the absence of a pairing interaction in the normal
region ($g(\vec{r})\equiv 0$ for $x<0$) implies a vanishing pair potential
${\Delta}(\vec{r})$, according to Eq.\ (\ref{selfconsist}), but does not imply
a vanishing order parameter $\Psi(\vec{r})$, which is given by
\begin{equation}
\Psi(\vec{r})=\sum_{\varepsilon>0}{\rm v}^{\ast}(\vec{r}){\rm
u}(\vec{r})[1-2f(\varepsilon)].\label{orderparam}
\end{equation}
Phase coherence between the electron and hole wave functions u and v leads to
$\Psi(\vec{r})\neq 0$ for $x<0$. The term ``proximity effect'' can therefore
mean two different things: One is the suppression of the pair potential
${\Delta}$ at the superconducting side of the NS interface. This is a small
effect which we neglect. The other is the induction of a non-zero order
parameter $\Psi$ at the normal side of the NS interface. This effect is fully
included, even though $\Psi$ does not appear explicitly in the expressions
which follow. The reason is that the order parameter quantifies the degree of
phase coherence between electrons and holes, but does not itself affect the
dynamics of the quasiparticles. (The Bogoliubov-De Gennes equation (\ref{BdG1})
contains ${\Delta}$ not $\Psi$.)

\subsubsection{Scattering formula for the conductance}

We now construct a basis for the scattering matrix. In the normal lead ${\rm
N}_{2}$ the eigenfunctions of the Bogoliubov-De Gennes equation (\ref{BdG1})
can be written in the form
\begin{mathletters}
\label{PsiN}
\begin{eqnarray}
&&\psi_{n,{\rm e}}^{\pm}({\rm N}_{2})=
{\renewcommand{\arraystretch}{0.6}
\left(\begin{array}{c}1\\ 0\end{array}\right)}\Phi_{n}(y,z)\exp(\pm{\rm
i}k_{n}^{\rm e}x),\label{PsiNa}\\
&&\psi_{n,{\rm h}}^{\pm}({\rm N}_{2})=
{\renewcommand{\arraystretch}{0.6}
\left(\begin{array}{c}0\\ 1\end{array}\right)}\Phi_{n}(y,z)\exp(\pm{\rm
i}k_{n}^{\rm h}x),\label{PsiNb}
\end{eqnarray}
\end{mathletters}%
where the wavenumbers $k_{n}^{\rm e}$ and $k_{n}^{\rm h}$ are given by
\begin{equation}
k_{n}^{\rm e,h}\equiv (2m/\hbar^{2})^{1/2}(E_{\rm F}-E_{n}+\sigma^{\rm
e,h}\varepsilon)^{1/2}, \label{keh}
\end{equation}
and we have defined $\sigma^{\rm e}\equiv 1$, $\sigma^{\rm h}\equiv -1$. The
labels e and h indicate the electron or hole character of the wavefunction. The
index $n$ labels the modes, $\Phi_{n}(y,z)$ is the transverse wavefunction of
the $n$-th mode, and $E_{n}$ its threshold energy. The $\Phi_{n}$'s are
normalized such that each wavefunction in the basis (\ref{PsiN}) carries the
same amount of quasiparticle current. The eigenfunctions in lead ${\rm N}_{1}$
are chosen similarly.

A wave incident on the disordered normal region is described in the basis
(\ref{PsiN}) by a vector of coefficients
\begin{equation}
c_{\rm N}^{\rm in}\equiv\bigl(
c_{\rm e}^{+}({\rm N}_{1}), c_{\rm e}^{-}({\rm N}_{2}),
c_{\rm h}^{-}({\rm N}_{1}), c_{\rm h}^{+}({\rm N}_{2})\bigr).\label{cNin}
\end{equation}
(The mode-index $n$ has been suppressed for simplicity of notation.) The
reflected and transmitted wave has vector of coefficients
\begin{equation}
c_{\rm N}^{\rm out}\equiv\bigl(
c_{\rm e}^{-}({\rm N}_{1}), c_{\rm e}^{+}({\rm N}_{2}),
c_{\rm h}^{+}({\rm N}_{1}), c_{\rm h}^{-}({\rm N}_{2})\bigr).\label{cNout}
\end{equation}
The scattering matrix $s_{\rm N}$ of the normal region relates these two
vectors,
\begin{equation}
c_{\rm N}^{\rm out}=s_{\rm N}^{\vphantom{{\rm in}}}c_{\rm N}^{\rm
in}.\label{sNdef}
\end{equation}
Because the normal region does not couple electrons and holes, this matrix has
the block-diagonal form
\begin{equation}
s_{\rm N}(\varepsilon)=
{\renewcommand{\arraystretch}{0.8}
\left(\begin{array}{cc}
s_{0}(\varepsilon)&0\\
\!0&s_{0}(-\varepsilon)^{\ast}
\end{array}\right)},\,
s_{\rm 0}\equiv{\renewcommand{\arraystretch}{0.6}
\left(\begin{array}{cc}
r_{11}&t_{12}\\t_{21}&r_{22}
\end{array}\right)}.
\label{sN}
\end{equation}
Here $s_{0}$ is the unitary scattering matrix associated with the
single-electron Hamiltonian ${\cal H}_{0}$. The reflection and transmission
matrices $r(\varepsilon)$ and $t(\varepsilon)$ are $N\times N$ matrices,
$N(\varepsilon)$ being the number of propagating modes at energy $\varepsilon$.
(We assume for simplicity that the number of modes in leads ${\rm N}_{1}$ and
${\rm N}_{2}$ is the same.) The matrix $s_{0}$ is unitary
($s_{0}^{\vphantom{\dagger}}s_{0}^{\dagger}=1$) and satisfies the symmetry
relation $s_{0}(\varepsilon,B)_{ij}=s_{0}(\varepsilon,-B)_{ji}$.

For energies $0<\varepsilon<{\Delta_{0}}$ there are no propagating modes in the
superconductor. We can then define a scattering matrix for Andreev reflection
at the NS interface which relates the vector of coefficients $\bigl( c_{\rm
e}^{-}({\rm N}_{2}), c_{\rm h}^{+}({\rm N}_{2})\bigr)$ to $\bigl( c_{\rm
e}^{+}({\rm N}_{2}), c_{\rm h}^{-}({\rm N}_{2})\bigr)$. The elements of this
scattering matrix can be obtained by matching the wavefunctions (\ref{PsiN}) at
$x=0$ to the decaying solutions in S of the Bogoliubov-De Gennes equation. If
terms of order ${\Delta_{0}}/E_{\rm F}$ are neglected (the socalled Andreev
approximation), the result is simply
\begin{mathletters}
\label{sA}
\begin{eqnarray}
&&c_{\rm e}^{-}({\rm N}_{2})= \alpha\,{\rm e}^{{\rm i}\phi}c_{\rm h}^{-}({\rm
N}_{2}),\label{sAa}\\
&&c_{\rm h}^{+}({\rm N}_{2})= \alpha\,{\rm e}^{-{\rm i}\phi}c_{\rm e}^{+}({\rm
N}_{2}),\label{sAb}
\end{eqnarray}
\end{mathletters}%
where $\alpha\equiv\exp[-{\rm i}\arccos(\varepsilon/{\Delta_{0}})]$. Andreev
reflection transforms an electron mode into a hole mode, without change of mode
index. The transformation is accompanied by a phase shift, which consists of
two parts: (1) A phase shift $-\arccos(\varepsilon/{\Delta_{0}})$ due to the
penetration of the wavefunction into the superconductor; (2) A phase shift
equal to plus or minus the phase $\phi$ of the pair potential in the
superconductor ({\em plus\/} for reflection from hole to electron, {\em
minus\/} for the reverse process).

We can combine the $2N$ linear relations (\ref{sA}) with the $4N$ relations
(\ref{sNdef}) to obtain a set of $2N$ linear relations between the incident
wave in lead ${\rm N}_{1}$ and the reflected wave in the same lead:
\begin{mathletters}
\label{sdef}
\begin{eqnarray}
&&c_{\rm e}^{-}({\rm N}_{1})= s_{\rm ee}^{\vphantom{+}}c_{\rm e}^{+}({\rm
N}_{1})+s_{\rm eh}^{\vphantom{+}}c_{\rm h}^{-}({\rm N}_{1}),\label{sdefa}\\
&&c_{\rm h}^{+}({\rm N}_{1})= s_{\rm he}^{\vphantom{+}}c_{\rm e}^{+}({\rm
N}_{1})+s_{\rm hh}^{\vphantom{+}}c_{\rm h}^{-}({\rm N}_{1}).\label{sdefb}
\end{eqnarray}
\end{mathletters}%
The four $N\times N$ matrices $s_{\rm ee}$, $s_{\rm hh}$, $s_{\rm eh}$, and
$s_{\rm he}$ form together the scattering matrix $s$ of the whole system for
energies $0<\varepsilon<{\Delta_{0}}$. An electron incident in lead ${\rm
N}_{1}$ is reflected either as an electron (with scattering amplitudes $s_{\rm
ee}$) or as a hole (with scattering amplitudes $s_{\rm he}$). Similarly, the
matrices $s_{\rm hh}$ and $s_{\rm eh}$ contain the scattering amplitudes for
reflection of a hole as a hole or as an electron. After some algebra we find
for these matrices the expressions
\begin{mathletters}
\label{sehheeehh}
\begin{eqnarray}
s_{\rm
ee}^{\vphantom{\ast}}(\varepsilon)&=&r_{11}^{\vphantom{\ast}}(\varepsilon)+
\alpha^{2}t_{12}^{\vphantom{\ast}}(\varepsilon)r_{22}^{\ast}
(-\varepsilon)M_{\rm e}^{\vphantom{\ast}}t_{21}^{\vphantom{\ast}}(\varepsilon),
\label{see}\\
s_{\rm hh}^{\vphantom{\ast}}(\varepsilon)&=&r_{11}^{\ast}(-\varepsilon)+
\alpha^{2}t_{12}^{\ast}(-\varepsilon)r_{22}^{\vphantom{\ast}}
(\varepsilon)M_{\rm
h}^{\vphantom{\ast}}t_{21}^{\ast}(-\varepsilon),\label{shh}\\
s_{\rm eh}^{\vphantom{\ast}}(\varepsilon)&=&\alpha\,{\rm e}^{{\rm
i}\phi}t_{12}^{\vphantom{\ast}}(\varepsilon)M_{\rm
h}^{\vphantom{\ast}}t_{21}^{\ast}(-\varepsilon),\label{seh}\\
s_{\rm he}^{\vphantom{\ast}}(\varepsilon)&=&\alpha\,{\rm e}^{-{\rm
i}\phi}t_{12}^{\ast}(-\varepsilon)M_{\rm
e}^{\vphantom{\ast}}t_{21}^{\vphantom{\ast}}(\varepsilon), \label{she}
\end{eqnarray}
\end{mathletters}%
where we have defined the matrices
\begin{mathletters}
\label{MeMh}
\begin{eqnarray}
&&M_{\rm e}^{\vphantom{\ast}}\equiv[1-\alpha^{2}
r_{22}^{\vphantom{\ast}}(\varepsilon)r_{22}^{\ast}(-\varepsilon)]^{-1},
\label{MeMha}\\
&&M_{\rm h}^{\vphantom{\ast}}\equiv[1-\alpha^{2}
r_{22}^{\ast}(-\varepsilon)r_{22}(^{\vphantom{\ast}}\varepsilon)]^{-1}.
\label{MeMhb}
\end{eqnarray}
\end{mathletters}%
One can verify that the scattering matrix constructed from these four
sub-matrices satisfies unitarity ($ss^{\dagger}=1$) and the symmetry relation
$s(\varepsilon,B,\phi)_{ij}=s(\varepsilon,-B,-\phi)_{ji}$, as required by
quasiparticle-current conservation and by time-reversal invariance,
respectively.

The differential conductance $G_{\rm NS}$ of the NS junction at zero
temperature and subgap voltage $V\leq{\Delta_{0}}/e$ is given by (Blonder,
Tinkham, and Klapwijk, 1982; Lambert, 1991; Takane and Ebisawa, 1992a)
\begin{eqnarray}
G_{\rm NS}&=&G_{0}{\rm Tr}\,(1-s_{\rm ee}^{\vphantom{\dagger}}s_{\rm
ee}^{\dagger}+s_{\rm he}^{\vphantom{\dagger}}s_{\rm he}^{\dagger})\nonumber\\
&=&2G_{0}{\rm Tr}\,s_{\rm he}^{\vphantom{\dagger}}s_{\rm he}^{\dagger}=
2G_{0}{\rm Tr}\,s_{\rm eh}^{\vphantom{\dagger}}s_{\rm
eh}^{\dagger}.\label{Gdef}
\end{eqnarray}
(The second and third equalities follow from unitarity of $s$.) The conductance
quantum $G_{0}=2e^{2}/h$, the factor of two being due to spin degeneracy. The
scattering matrix elements are to be evaluated at energy $\varepsilon=eV$. We
now substitute Eq.\ (\ref{seh}) into Eq.\ (\ref{Gdef}), and obtain the
expression (Beenakker, 1992a)
\begin{mathletters}
\label{key}
\begin{eqnarray}
&&G_{\rm NS}=2G_{0}{\rm Tr}\,m(eV)m^{\dagger}(eV).\label{keya}\\
&&m(\varepsilon)=t_{12}^{\vphantom{\dagger}}(\varepsilon)
[1-\alpha^{2}r_{22}^{\ast}(-\varepsilon)
r_{22}^{\vphantom{\ast}}(\varepsilon)]^{-1}t_{21}^{\ast}(-\varepsilon).
\label{keyb}
\end{eqnarray}
\end{mathletters}%
The advantage over Eq.\ (\ref{Gdef}) is that Eq.\ (\ref{key}) can be evaluated
with the same techniques developed for quantum transport in the normal state,
since the only input is the normal-state scattering matrix. The effects of
multiple Andreev reflections are fully incorporated by the matrix inversion in
Eq.\ (\ref{keyb}).

In the limit $V\rightarrow 0$ of linear response we only need the scattering
matrix elements at the Fermi level, {\em i.e.} at $\varepsilon=0$. We will
restrict ourselves to this limit in most of what follows, and omit the argument
$\varepsilon$. Note that $\alpha=-{\rm i}$ for $\varepsilon=0$. In the absence
of a magnetic field, the general formula (\ref{key}) simplifies considerably.
Since the scattering matrix $s_{0}$ of the normal region is symmetric for
$B=0$, one has $r_{22}^{\ast}=r_{22}^{\dagger}$ and
$t_{21}^{\ast}=t_{12}^{\dagger}$. Equation (\ref{key}) then takes the form
\begin{eqnarray}
G_{\rm NS}&=&2G_{0}{\rm Tr}\,t_{12}^{\vphantom{\dagger}}
(1+r_{22}^{\dagger}r_{22}^{\vphantom{\ast}})^{-1}
t_{12}^{\dagger}t_{12}^{\vphantom{\rm T}}
(1+r_{22}^{\dagger}r_{22}^{\vphantom{\rm T}})^{-1}t_{12}^{\dagger}\nonumber\\
&=&2G_{0}{\rm Tr}\left(t_{12}^{\dagger}t_{12}^{\vphantom{\dagger}}
(2-t_{12}^{\dagger}t_{12}^{\vphantom{\dagger}})^{-1}\right)^{2}.\label{GBzero}
\end{eqnarray}
In the second equality we have used the unitarity relation
$r_{22}^{\dagger}r_{22}^{\vphantom{\ast}}+
t_{12}^{\dagger}t_{12}^{\vphantom{\dagger}}=1$. The trace
(\ref{GBzero}) depends only on the eigenvalues $T_{n}$ ($n=1,2,\ldots
N$) of the transmission matrix product
$t_{12}^{\dagger}t_{12}^{\vphantom{\dagger}}$. We thus obtain the
relation (\ref{keyzero}) between the conductance and the transmission
eigenvalues. Equation (\ref{keyzero}) holds for an arbitrary
transmission matrix $t$, {\em i.e.} for arbitrary disorder potential.
It is the multi-channel generalization of a formula first obtained by
Blonder, Tinkham, and Klapwijk (1982) (and subsequently by Shelankov,
1984, and Za\u{\i}tsev, 1984) for the single-channel case (appropriate
for a geometry such as a planar tunnel barrier, where the different
scattering channels are uncoupled).

Slevin, Pichard, and Mello (1996; Altland and Zirnbauer, 1996b; Brouwer and
Beenakker, 1996a) have considered the modifications required by the inclusion
of spin-orbit scattering. The scattering matrix elements are then quaternion
numbers. The complex conjugate $Q^{\ast}$ and the Hermitian conjugate
$Q^{\dagger}$ of a matrix $Q$ with quaternion elements
$Q_{nm}=a_{nm}\openone+{\rm i}b_{nm}\sigma_{x}+{\rm i}c_{nm}\sigma_{y}+{\rm
i}d_{nm}\sigma_{z}$ have matrix elements
\begin{eqnarray}
Q_{nm}^{\ast}&=&a_{nm}^{\ast}\openone+{\rm i}b_{nm}^{\ast}\sigma_{x}+{\rm
i}c_{nm}^{\ast}\sigma_{y}+{\rm i}d_{nm}^{\ast}\sigma_{z},\label{qastdef}\\
Q_{nm}^{\dagger}&=&a_{mn}^{\ast}\openone-{\rm i}b_{mn}^{\ast}\sigma_{x}-{\rm
i}c_{mn}^{\ast}\sigma_{y}-{\rm i}d_{mn}^{\ast}\sigma_{z}.\label{qdaggerdef}
\end{eqnarray}
Notice that the definition of the Hermitian conjugate of the $N\times N$
quaternion matrix $Q$ is the same as for the corresponding $2N\times 2N$
complex matrix, while the definition of the complex conjugate is different. The
dual $Q^{\rm R}$ of a quaternion matrix is defined by $Q^{\rm
R}=(Q^{\ast})^{\dagger}$, which differs from the transpose of a complex matrix.
The trace of $Q$ is defined by ${\rm Tr}\,Q=\sum_{n}a_{nn}$, which is half the
trace of the corresponding $2N\times 2N$ matrix. With these definitions Eq.\
(\ref{key}) remains valid in the presence of spin-orbit scattering. What about
Eq.\ (\ref{keyzero})? In zero magnetic field, $s_{0}$ is a self-dual matrix:
$s_{0}^{\vphantom{\rm R}}=s_{0}^{\rm R}$. Hence
$r_{22}^{\ast}=r_{22}^{\dagger}$ and $t_{21}^{\ast}=t_{12}^{\dagger}$, which
combined with unitarity implies Eq.\ (\ref{GBzero}). The linear-response
conductance in zero magnetic field is therefore still given by Eq.\
(\ref{keyzero}).

In summary, the linear-response conductance of an NS junction is a linear
statistic on the transmission eigenvalues for $\beta=1$ or 4, given by Eq.\
(\ref{keyzero}). For $\beta=2$, or for finite voltage, the more general
expression (\ref{key}) is required, which is not a linear statistic.

\subsection{Ideal NS interface}
\label{NSideal}

In this subsection we investigate the case of an ideal ({\em i.e.} perfectly
transparent) interface between the normal metal and the superconductor. The
effect of a tunnel barrier at the NS interface will be considered in the next
subsection. The disordered normal region is supposed to have a length $L$ much
greater than its width $W$ (see Fig.\ \ref{fig_reservoirs}). We concentrate on
the metallic diffusive regime, in which $L$ is greater than the mean free path
$l$ for elastic impurity scattering, but smaller than the localization length
$Nl$.

\subsubsection{Average conductance}
\label{GNSmean}

Let us begin by calculating the average conductance of the junction, averaged
over an ensemble of impurity configurations (Beenakker, 1992a). According to
Eq.\ (\ref{keyzero}), this average is given in zero magnetic field by
\begin{eqnarray}
\langle G_{\rm NS}\rangle&=&\frac{4e^{2}}{h}\int_{0}^{1}\!dT\,\rho(T)
\frac{T^{2}}{(2-T)^{2}}\nonumber\\
&=&\frac{4e^{2}}{h}\int_{0}^{\infty}\!dx\,\rho(x)
\frac{1}{\cosh^{2}2x},\label{GNSaverage}
\end{eqnarray}
where in the second equality we have substituted the parameterization
$T=1/\cosh^{2}x$ introduced in Sec.\ \ref{SandM}. Equation (\ref{GNSaverage})
is to be compared with the equation for the average conductance in the normal
state,
\begin{equation}
\langle G_{\rm N}\rangle=\frac{2e^{2}}{h}\int_{0}^{\infty}\!dx\,\rho(x)
\frac{1}{\cosh^{2}x},\label{GNaverage}
\end{equation}
which follows from the Landauer formula (\ref{Landauer}). As we discussed in
Sec.\ \ref{Tdensity}, the density $\rho(x)=Nl/L+{\cal O}(N^{0})$ is uniform in
the metallic diffusive regime, up to weak-localization corrections. Ignoring
these corrections for the moment, we find that
\begin{equation}
\langle G_{\rm NS}\rangle=\frac{2e^{2}}{h}\,\frac{Nl}{L}+{\cal
O}(N^{0})=\langle G_{\rm N}\rangle.\label{GNSNaverage}
\end{equation}
We conclude that --- although $G_{\rm NS}$ according to Eq.\ (\ref{keyzero}) is
of {\em second\/} order in the transmission eigenvalues $T_{n}$ --- the
ensemble average $\langle G_{\rm NS}\rangle$ is of {\em first\/} order in
$l/L$. The resolution of this paradox is that the $T_{n}$'s are not distributed
uniformly, but are either exponentially small or of order unity ({\em cf.}
Sec.\ \ref{shotnoiseN}). Hence the average of $T_{n}^{2}$ is of the same order
as the average of $T_{n}$.

Differences between $\langle G_{\rm NS}\rangle$ and $\langle G_{\rm N}\rangle$
may appear because of several effects. One effect is that of a finite
temperature. Equation (\ref{GNSaverage}) holds if the thermal energy $k_{\rm
B}T$ is much smaller than the Thouless energy $E_{\rm c}=\hbar D/L^{2}$ (with
$D$ the diffusion coefficient). Nazarov and Stoof (1996; see also Stoof and
Nazarov, 1996a; Golubov, Wilhelm, and Zaikin, 1996) have calculated that
$\langle G_{\rm NS}\rangle$ increases by $10\,\%$ as $k_{\rm B}T$ is raised to
$E_{\rm c}$, and then drops back to $\langle G_{\rm N}\rangle$ at higher
temperatures. Such a non-monotonic temperature dependence was first predicted
for a point-contact geometry by Artemenko, Volkov, and Za\u{\i}tsev (1979).
Experimental confirmation has been provided by Charlat {\em et al.} (1996).
Nazarov and Stoof have also shown that at zero temperature, attractive
(repulsive) interactions between the quasiparticles in the normal metal lead to
an increase (decrease) of $\langle G_{\rm NS}\rangle$ relative to $\langle
G_{\rm N}\rangle$.

\begin{figure}[tb]
\centerline{
\psfig{figure=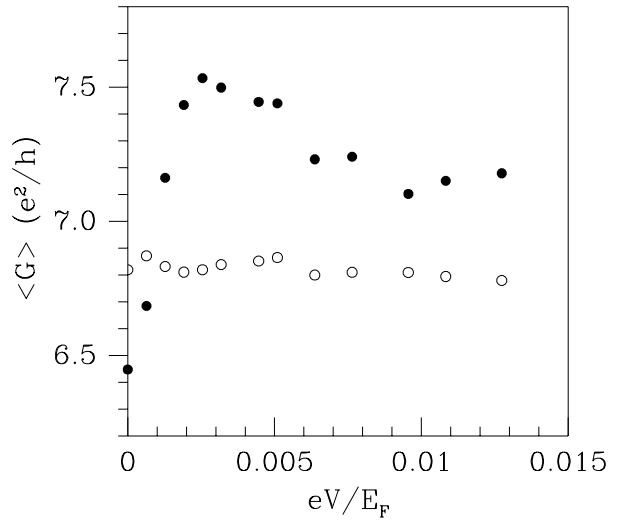,width=
8cm,bbllx=27pt,bblly=182pt,bburx=555pt,bbury=637pt}
}%
\medskip
\caption[]{
Numerical simulation of the voltage dependence of the average differential
conductance of a two-dimensional wire ($L/W=4.8$, $N=15$, $l/L=0.31$), in zero
magnetic field. The filled circles represent $\langle G_{\rm NS}\rangle$, for
the case that the wire is connected to a superconducting reservoir; the open
circles represent the $V$-independent conductance $\langle G_{\rm N}\rangle$ in
the normal state. At zero voltage, $\langle G_{\rm NS}\rangle$ is smaller than
$\langle G_{\rm N}\rangle$ because of the weak-localization effect. At high
voltage, $\langle G_{\rm NS}\rangle$ is larger than $\langle G_{\rm N}\rangle$
because of the contact-resistance effect. After Brouwer and Beenakker (1995b).
}\label{fig_contactNS}
\end{figure}

Contact resistances are yet another effect. As discussed in Sec.\
\ref{classicalG}, the contact resistance in the normal state is $h/2Ne^{2}$, so
that
\begin{equation}
\frac{2Ne^{2}}{h}\langle G_{\rm N}\rangle^{-1}=\frac{L}{l}+1.\label{RNaverage}
\end{equation}
In an NS junction, the contact resistance is voltage and magnetic-field
dependent (Brouwer and Beenakker, 1995b):
\begin{equation}
\frac{2Ne^{2}}{h}\langle G_{\rm NS}\rangle^{-1}=\left\{\begin{array}{c}
\frac{L}{l}+1\;{\rm if}\;B=0\;{\rm and}\;V=0,\\
\frac{L}{l}+\frac{1}{2}\;{\rm if}\;B\gg B_{\rm c}\;{\rm or}\;V\gg E_{\rm c}/e,
\end{array}\right.\label{RNSaverage}
\end{equation}
where $B_{\rm c}=h/eLW$. At zero voltage and zero magnetic field the contact
resistance of the NS junction is the same as in the normal state. Application
of either a voltage or a magnetic field reduces the contact resistance in the
NS junction by a factor of two. This leads to an increase of $\langle G_{\rm
NS}\rangle$ by approximately $(l/L)^{2}Ne^{2}/h$. A numerical simulation of
this contact-resistance effect is shown in Fig.\ \ref{fig_contactNS}, where the
differential conductance at zero magnetic field is plotted as a function of
voltage. For $V\gg E_{\rm c}/e$, $\langle G_{\rm NS}\rangle$ is larger than
$\langle G_{\rm N}\rangle$ because of the difference in contact resistance. The
non-monotonic $V$-dependence at intermediate voltages, observed in the
simulation, has been studied theoretically by Yip (1995), Volkov, Allsopp, and
Lambert (1996), and Lesovik, Fauch\`{e}re, and Blatter (1997), and
experimentally by Poirier, Mailly, and Sanquer (1996), and Charlat {\em et al.}
(1996). (It is closely related to the non-monotonic temperature dependence
mentioned above.) The difference between $\langle G_{\rm NS}\rangle$ and
$\langle G_{\rm N}\rangle$ at $V=0$ is due to the term of order $N^{0}$ in Eq.\
(\ref{GNSNaverage}). This represents the weak-localization effect, which we
discuss in the following subsection.

\subsubsection{Weak localization}
\label{NSWL}

In the presence of time-reversal symmetry, {\em i.e.} for $\beta=1$ or 4, the
weak-localization correction to the average conductance of the NS junction can
be computed from the ${\cal O}(N^{0})$ correction to the transmission
eigenvalue density. We write $\langle G_{\rm NS}\rangle=Nl/L+\delta G_{\rm NS}$
and $\rho(x)=Nl/L+\delta\rho(x)$. For $\beta=1$ or 4, the ${\cal O}(N^{0})$
corrections $\delta G_{\rm NS}$ and $\delta\rho(x)$ are related by
\begin{equation}
\delta G_{\rm NS}=\frac{4e^{2}}{h}\int_{0}^{\infty}\!dx\,\delta\rho(x)
\frac{1}{\cosh^{2}2x},\label{deltaGNSaverage}
\end{equation}
in view of Eq.\ (\ref{keyzero}) and the definition $T=1/\cosh^{2}x$. The
function $\delta\rho(x)$ is given by Eq.\ (\ref{DELTARHO}). Substitution into
Eq.\ (\ref{deltaGNSaverage}) yields the result (Beenakker, 1994b; see also
Mac\^{e}do and Chalker, 1994; Takane and Otani, 1994; Nazarov, 1995b)
\begin{equation}
\delta G_{\rm
NS}=(1-2/\beta)\left(2-\frac{8}{\pi^{2}}\right)\frac{e^{2}}{h},\;\;{\rm
if}\;\;\beta=1,4.\label{deltaGNSresult}
\end{equation}

Equation (\ref{deltaGNSresult}) does not apply to $\beta=2$, because then Eq.\
(\ref{keyzero}) for $G_{\rm NS}$ on which it is based does not hold. Instead,
one should start from the more general expression (\ref{key}). It turns out
(Brouwer and Beenakker, 1995b) that breaking time-reversal symmetry is not
sufficient to suppress the weak-localization correction in an NS junction, but
only reduces $\delta G_{\rm NS}$ by about a factor of two (see Table
\ref{TABLEDELTAGNS}). To achieve $\delta G_{\rm NS}=0$ requires, in addition to
a magnetic field, a sufficiently large voltage to break the degeneracy in
energy between the electrons (occupied states at energy $eV$ above the Fermi
level) and the holes (empty states at energy $eV$ below the Fermi level). The
electron-hole degeneracy is effectively broken when $eV$ exceeds the Thouless
energy $E_{\rm c}$. Weak localization in an NS junction {\em coexists with a
magnetic field}, as long as $eV \ll E_{c}$.

\begin{table}[tb]
\caption[]{
Dependence of the weak-localization correction $\delta G_{\rm NS}$ of a
normal-metal wire attached to a superconductor on the presence or
absence of time-reversal symmetry (TRS) and electron-hole degeneracy
(ehD). The results  are for a metal without spin-orbit scattering
(Brouwer and Beenakker, 1995b). In the presence of strong spin-orbit
scattering each entry is to be multiplied by $-1/2$ (Slevin, Pichard,
and Mello, 1996). For comparison, the corresponding result in the
normal state is listed between $\{\cdots\}$.
}\label{TABLEDELTAGNS}
\begin{tabular}{c|cc} $\delta G_{\rm NS}\,[e^2/h]$ & TRS & no TRS\\
\hline\\
ehD & $-2+8/\pi^{2}$ $\{-2/3\}$ & $-2/3$ $\{0\}$\\
no ehD & $-4/3$ $\{-2/3\}$ & $0$ $\{0\}$
\end{tabular}
\end{table}

\begin{figure}[tb]
\centerline{
\psfig{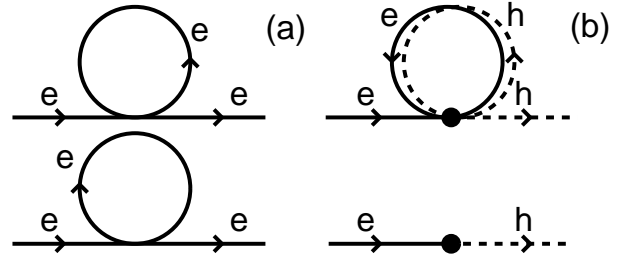}
}%
\bigskip
\caption[]{
Interfering Feynman paths. (a) Two paths interfering constructively in the
presence of time-reversal symmetry. (b) Two paths involving Andreev reflection
(solid dot), which interfere destructively both in the presence and absence of
time-reversal symmetry. In the normal state, weak localization (meaning a
negative correction to the conductance of order $e^{2}/h$) arises from the
paths in (a). In an NS junction, weak localization coexists in the presence of
a magnetic field because of the paths in (b). After Brouwer and Beenakker
(1995b).
}\label{fig_feynman}
\end{figure}

All this is in marked contrast with weak localization in the normal state,
where $\delta G$ vanishes in a magnetic field regardless of the voltage. (In
fact, $\delta G$ is independent of $V$ on the scale of $E_{\rm c}$.) In normal
metals, weak localization is understood (Khmel'nitski\u{\i}, 1984; Bergmann,
1984) as constructive interference of pairs of time-reversed Feynman paths
(Fig.\ \ref{fig_feynman}a). This interference is destroyed by a magnetic field.
What kind of interfering paths are responsible for weak localization in an NS
junction without time-reversal symmetry? The two simplest interfering paths are
shown in Fig.\ \ref{fig_feynman}b. Regardless of whether time-reversal symmetry
is broken or not, there is an exact cancellation of the phase shifts
accumulated by the electron and the hole which traverse the loop in the same
direction. What remains is a phase shift of $\pi$ due to the double Andreev
reflection. As a consequence, the path with the double loop interferes
destructively with the path without a loop, giving rise to a negative $\delta
G_{\rm NS}$. In the diagrammatic perturbation theory of weak localization
(Gor'kov, Larkin, and Khmel'nitski\u{\i}, 1979; Anderson, Abrahams, and
Rama\-krishnan, 1979), the two interfering time-reversed paths of Fig.\
\ref{fig_feynman}a correspond to a diagram known as the cooperon. The two paths
involving Andreev reflection of Fig.\ \ref{fig_feynman}b correspond to a new
type of diagram, first identified by Altland and Zirnbauer (1996a).

\begin{figure}[tb]
\centerline{
\psfig{figure=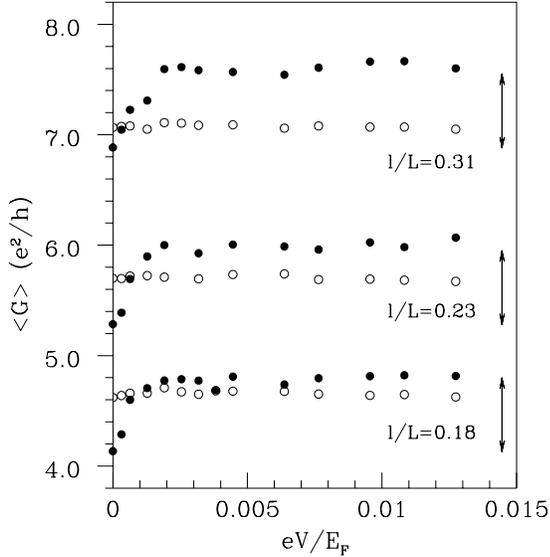,width= 8cm}
}%
\medskip
\caption[]{
Numerical simulation of the voltage dependence of the average differential
conductance of a two-dimensional wire ($L/W=4.8$, $N=15$), for a magnetic flux
of $6\,h/e$ through the disordered region. The filled circles represent
$\langle G_{\rm NS}\rangle$, the open circles $\langle G_{\rm N}\rangle$. The
three sets of data points are for three different values of the ratio $l/L$.
The arrows indicate the theoretically predicted increase of $G_{\rm NS}$ by
$\frac{2}{3}e^{2}/h$, due to the weak-localization effect. The
contact-resistance effect is suppressed by the magnetic field. After Brouwer
and Beenakker (1995b).
}\label{fig_deltaGNS}
\end{figure}

The interested reader is referred to Appendix \ref{deltaGNSdetails} for the
calculation of $\delta G_{\rm NS}$. The results, summarized in Table
\ref{TABLEDELTAGNS}, imply a universal $B$ and $V$-dependence of the
conductance of an NS microbridge. Raising first $B$- and then $V$ leads to two
subsequent increases of the conductance, while raising first $V$ and then $B$
leads first to a decrease and then to an increase. The $V$-dependence of the
differential conductance in a time-reversal-symmetry breaking magnetic field is
shown in Fig.\ \ref{fig_deltaGNS}. The dots are numerical simulations of the
Anderson model, while the arrows indicate the increase of $\langle G_{\rm
NS}\rangle$ by $\frac{2}{3}e^{2}/h$ predicted by Table \ref{TABLEDELTAGNS}. The
agreement is quite satisfactory. The $V$-dependence of the weak-localization
correction at zero $B$ (or the $B$-dependence at zero $V$) is obscured by the
$B$- and $V$-dependent contact resistance of the previous subsection, which can
only be neglected if $N(l/L)^{2}\ll 1$. This condition is difficult to meet in
numerical simulations, and possibly also in experiments. This complication was
not understood in earlier simulations by Marmorkos, Beenakker, and Jalabert
(1993) and experiments by Lenssen {\em et al.} (1994).

\subsubsection{Universal conductance fluctuations}
\label{NSUCF}

\begin{figure}[tb]
\centerline{
\psfig{figure=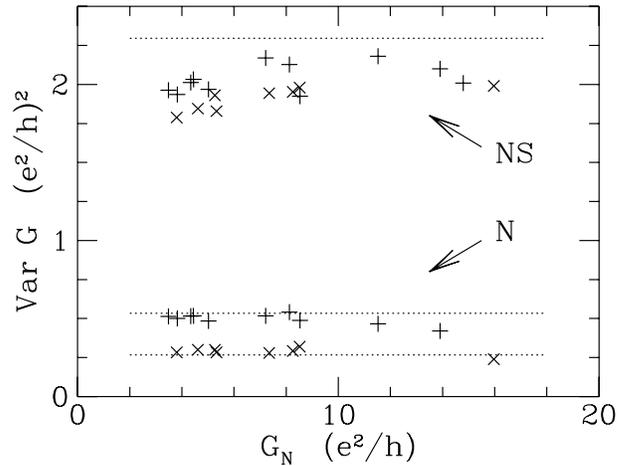,width=
8cm,bbllx=53pt,bblly=171pt,bburx=560pt,bbury=562pt}
}%
\medskip
\caption[]{
Numerical simulation of the variance of the conductance of a two-dimensional
wire, for different values of the average conductance ($+$ for $B=0$; $\times$
for a flux of $10\,h/e$ through the disordered region). The labels N and NS
indicate the case that the wire connects two normal reservoirs or one normal
and one superconducting reservoir, respectively. Dotted lines are the
analytical results from Eqs.\ (\protect\ref{VarG215}) and
(\protect\ref{VarGNSB0}). Note the absence of a factor-of-two reduction in
${\rm Var}\,G_{\rm NS}$ on applying a magnetic field. After Marmorkos,
Beenakker, and Jalabert (1993).
}\label{fig_variance}
\end{figure}

So far we have considered the ensemble average $\langle G_{\rm NS}\rangle$ of
the  conductance of the NS junction. In Fig.\ \ref{fig_variance} we show
results of numerical simulations by Marmorkos, Beenakker, and Jalabert (1993)
for the sample-to-sample fluctuations. A range of parameters $L,W,l,N$ was used
to collect this data, in the quasi-one-dimensional, metallic, diffusive regime
$l<W<L<Nl$. The normal-state results are in accord with the prediction
(\ref{VarG215}) of the Altshuler--Lee--Stone theory of ``universal conductance
fluctuations''. As implied by the $1/\beta$ dependence of ${\rm Var}\,G_{\rm
N}$, the variance is reduced by a factor of two upon application of a
time-reversal-symmetry breaking magnetic field (see the two dotted lines in the
lower part of Fig.\ \ref{fig_variance}). The data for ${\rm Var\,}G_{\rm NS}$
at $B=0$ shows approximately a four-fold increase over the normal state. For
$B\neq 0$, the simulation shows that ${\rm Var\,}G_{\rm NS}$ is {\em
insensitive to a magnetic field}. In contrast to the situation in the normal
state, the theory for universal conductance fluctuations in an NS junction is
quite different for zero and for non-zero magnetic field, as we now discuss.

In zero magnetic field, the conductance of the NS junction is a linear
statistic on the transmission eigenvalues, according to Eq.\ (\ref{keyzero}).
The variance then follows immediately from the general formula
(\ref{VarAresult}) for the variance of an arbitrary linear statistic in a wire
geometry (Beenakker and Rejaei, 1993; Chalker and Mac\^{e}do, 1993).
Substitution of $a(x)=(4e^{2}/h)\cosh^{-2}2x$ into Eq.\ (\ref{VarAresult})
yields
\begin{eqnarray}
{\rm Var\,}G_{\rm
NS}&=&\frac{64}{15}\left(1-\frac{45}{\pi^{4}}\right)\beta^{-1}
\left(\frac{e^{2}}{h}\right)^{2}\nonumber\\
&=&4.30\,\beta^{-1}{\rm Var\,}G_{\rm N},\;{\rm if}\;\beta=1,4.\label{VarGNSB0}
\end{eqnarray}
A factor of four between ${\rm Var\,}G_{\rm NS}$ and ${\rm Var\,}G_{\rm N}$ was
estimated by Takane and Ebisawa (1992b). (A diagrammatic calculation by the
same authors (Takane and Ebisawa, 1991) gave a factor of six, presumably
because only the dominant diagram was included.) The numerical data in Fig.\
\ref{fig_variance} is within 10~\% of the theoretical prediction
(\ref{VarGNSB0}) for $\beta=1$ (upper dotted line). Similar numerical results
for ${\rm Var\,}G_{\rm NS}$ in zero magnetic field were obtained by Takane and
Ebisawa (1992b), and by Bruun, Hui, and Lambert (1994).

We conclude that the phenomenon of universal conductance fluctuations in zero
magnetic field is basically the same for $G_{\rm N}$ and $G_{\rm NS}$, because
both quantities are linear statistics for $\beta=1,4$. If time-reversal
symmetry is broken by a magnetic field, the situation is qualitatively
different. For $G_{\rm N}$, breaking time-reversal symmetry does not affect the
universality of the fluctuations, but merely reduces the variance by a factor
of two. No such simple behavior is to be expected for $G_{\rm NS}$, since it is
no longer a linear statistic for $\beta=2$. Indeed, the numerical data of Fig.\
\ref{fig_variance} demonstrates that ${\rm Var}\,G_{\rm NS}$ is unaffected by a
magnetic field, within the 10\% statistical uncertainty of the simulations. Is
there some symmetry principle hidden behind these findings?

Motivated by this question, a calculation of conductance fluctuations in an NS
junction for $\beta=2$ was carried out by Brouwer and Beenakker (1995c). The
result is that ${\rm Var}\, G_{\rm NS}$ for a disordered wire attached to a
superconductor is reduced by $(2-90/\pi^4)^{-1}=0.929$ upon breaking
time-reversal symmetry (see Table \ref{TABLEVARGNS}). This number is
sufficiently close to 1 to be consistent with the numerical simulations, but
not precisely equal to 1 --- so that we can be sure that no rigorous symmetry
principle exists. Still, an approximate symmetry argument could be found, as we
now discuss. For simplicity, we first assume zero voltage and no spin-orbit
scattering.

\begin{table}[tb]
\caption[]{
Dependence of the variance ${\rm Var}\,G_{\rm NS}$ of the conductance of a
normal-metal wire attached to a superconductor on the presence or absence of
time-reversal symmetry (TRS) and electron-hole degeneracy (ehD). The results
are for a metal without spin-orbit scattering (Brouwer and Beenakker, 1995b).
In the presence of strong spin-orbit scattering each entry is to be multiplied
by $1/4$ (Brouwer and Beenakker, 1996a). For comparison, the corresponding
result in the normal state is listed between $\{\cdots\}$.
}\label{TABLEVARGNS}
\begin{tabular}{c|cc} ${\rm Var}\,G_{\rm NS}\,[e^{4}/h^{2}]$ & TRS & no TRS\\
\hline\\
ehD & $64/15-192/\pi^{4}$ $\{8/15\}$ & $32/15$ $\{4/15\}$\\
no ehD & $32/15$ $\{8/15\}$ & $16/15$ $\{4/15\}$
\end{tabular}
\end{table}

\begin{figure}[tb]
\centerline{
\psfig{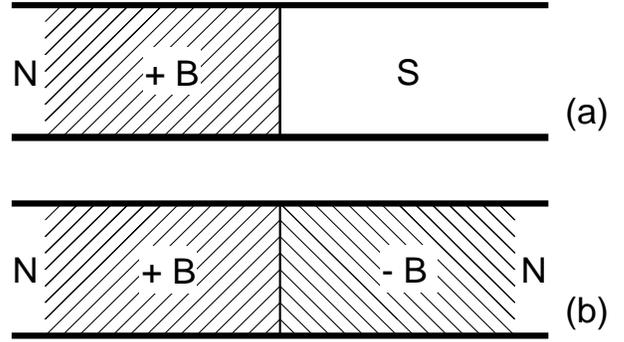}
}%
\bigskip
\caption[]{
Exchange of symmetries. (a): Schematic drawing of a disordered normal metal (N)
connected to a superconductor (S), in a time-reversal-symmetry breaking
magnetic field $B$. In (b) the normal region is connected in series with its
mirror image. As indicated, the magnetic field $B$ changes sign upon
reflection. The variance of the conductance fluctuations in (a) is exactly four
times the variance in (b). The variance in (b) is exactly two times the
variance in the absence of the reflection symmetry. The exchange of
time-reversal symmetry for reflection symmetry explains the insensitivity to a
magnetic field of the conductance fluctuations in an NS junction. After Brouwer
and Beenakker (1995c).
}\label{fig_NNNS}
\end{figure}

The argument is based on the general expression (\ref{key}) for the conductance
$G_{\rm NS}$ of an NS junction, in terms of the scattering matrix $s_{0}$ of
the normal region. We compare $G_{\rm NS}$ with the conductance $G_{\rm NN}$ of
an entirely normal metal consisting of two segments in series (see Fig.\
\ref{fig_NNNS}). The first segment has scattering matrix $s_{0}$, the second
segment is the mirror image of the first. That is to say, the disorder
potential is specularly reflected and the sign of the magnetic field is
reversed. The system NN thus has a reflection symmetry (RS), both in the
presence and absence of time-reversal symmetry (TRS). The scattering matrix of
the second segment is $Xs_{0}X$, where $X$ is a $2N \times 2N$ matrix with zero
elements, except for $X_{i,N+i}=X_{N+i,i}=1$ ($i=1,2,\ldots N$). (The matrix
$X$ exchanges scattering states incident from left and right.) The conductance
$G_{\rm NN}$ follows from the transmission matrix through the two segments in
series by means of the Landauer formula,
\begin{equation}
G_{\rm NN}({\rm RS})=G_{0}\,{\rm Tr}\,m'm'^{\dagger},\;\;
m'=t_{12}[1-(r_{22})^{2}]^{-1}t_{21}.\label{SNN}
\end{equation}
The difference between $r^{\ast}r$ in Eq.\ (\ref{key}) and $r^{2}$ in Eq.\
(\ref{SNN}) is crucial in the presence of time-reversal symmetry, but not in
its absence. Indeed, an explicit calculation shows that, for broken
time-reversal symmetry, the variance of ${\rm Tr}\,mm^{\dagger}$ equals that of
${\rm Tr}\,m'm'^{\dagger}$, hence
\begin{equation}
{\rm Var}\,G_{\rm NS}(\mbox{no TRS})=4\,{\rm Var}\,G_{\rm NN}(\mbox{RS, no
TRS}). \label{VarNSNN}
\end{equation}

The system NN is special because it possesses a reflection symmetry. Breaking
reflection symmetry amounts to the replacement of the mirror-imaged segment by
a different segment, with scattering matrix $s_{0}'$ which is independent of
$s_{0}$ but drawn from the same ensemble. This reduces the variance of the
conductance fluctuations by a factor of two, regardless of whether
time-reversal symmetry is present or not,
\begin{equation}
{\rm Var}\,G_{\rm NN}({\rm RS})=2\,{\rm Var}\,G_{\rm NN}(\mbox{no RS}).
\label{VarNNNN}
\end{equation}
One may check this relation by an explicit calculation, but it is intuitively
obvious if one considers that the eigenstates separate into even and odd states
which fluctuate independently. Since breaking time-reversal symmetry by itself
reduces the variance of $G_{\rm NN}$ by a factor of two, we may write
\begin{equation}
{\rm Var}\,G_{\rm NN}(\mbox{RS, no TRS})={\rm Var}\,G_{\rm NN}(\mbox{TRS, no
RS}). \label{VarTRSSRS}
\end{equation}
Equations (\ref{VarNSNN})--(\ref{VarTRSSRS}) are exact, for any distribution of
the scattering matrix which depends only on the transmission eigenvalues. We
need one more relationship, which is approximate and holds only for the case of
a disordered wire:
\begin{equation}
{\rm Var}\,G_{\rm NS}({\rm TRS})\approx 4\,{\rm Var}\, G_{\rm NN}(\mbox{TRS, no
RS}).\label{VarNSNNTRS}
\end{equation}
Equation (\ref{VarNSNNTRS}) is approximate, because the correct coefficient
according to Eq.\ (\ref{VarGNSB0}) is $4.3$ and not $4$. Taken together, Eqs.\
(\ref{VarNSNN})--(\ref{VarNSNNTRS}) imply the approximate relationship ${\rm
Var}\,G_{\rm NS}({\rm TRS})\approx{\rm Var}\,G_{\rm NS}(\mbox{no TRS})$.

One can thus understand the insensitivity of the conductance fluctuations to a
magnetic field as an {\em exchange of symmetries}: Breaking time-reversal
symmetry introduces an approximate reflection symmetry into the structure of
the scattering matrix. This reflection symmetry compensates the reduction of
the conductance fluctuations due to breaking of time-reversal symmetry, and
explains the anomalous insensitivity of the fluctuations in a magnetic field
discovered in computer simulations.

We conclude this subsection by mentioning the effects of a voltage and of
spin-orbit scattering (Brouwer and Beenakker, 1996a). If electron-hole
degeneracy (ehD) is broken by a voltage $V\gg E_{\rm c}/e$, then the NS
junction is equivalent to the system NN {\em without\/} reflection symmetry:
\begin{equation}
{\rm Var}\,G_{\rm NS}(\mbox{no ehD})=4\,{\rm Var}\,G_{\rm NN}(\mbox{no RS}).
\label{VarNSNNnoehD}
\end{equation}
This relationship holds regardless of whether time-reversal symmetry is broken
or not (see Table \ref{TABLEVARGNS}). Concerning spin-orbit scattering, we know
that ${\rm Var}\,G_{\rm N}$ is four times smaller with spin-orbit scattering
than without, either because $1/\beta=1/4$ instead of 1, or because
$G_{0}^{2}=(e^{2}/h)^{2}$ instead of $(2e^{2}/h)^{2}$ ({\em cf.} Sec.\
\ref{classicalG}). The same factor of four applies to ${\rm Var}\,G_{\rm NS}$,
because the relationships (\ref{VarNSNN})--(\ref{VarNSNNnoehD}) between the
systems NS and NN hold both with and without spin-orbit scattering.

\subsection{NS junction containing a tunnel barrier}
\label{NStunnelbarrier}

\begin{figure}[tb]
\centerline{
\psfig{figure=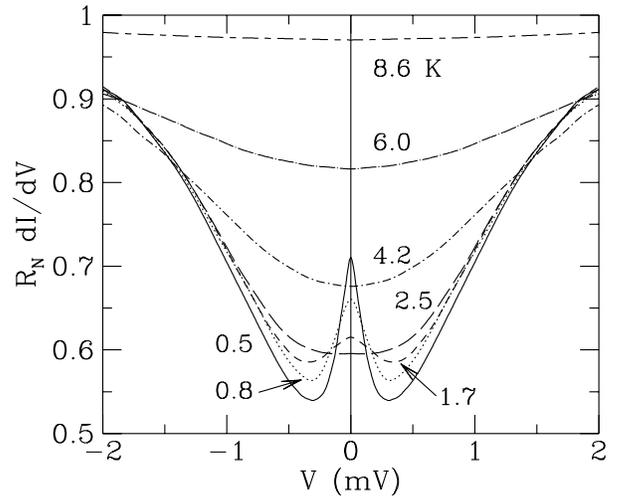,width=
8cm,bbllx=19pt,bblly=190pt,bburx=552pt,bbury=612pt}
}%
\bigskip
\caption[]{
Experimental data of the differential conductance (normalized by the
normal-state resistance $R_{\rm N}=0.27\,\Omega$) of a Nb--InGaAs junction, as
a function of applied voltage at seven different temperatures. After Kastalsky
{\em et al.} (1991).
}\label{fig_Kleinsasser}
\end{figure}

Although the case of an ideal NS interface, considered in the previous
subsection, is of considerable conceptual importance, it is more common in
experiments to have a high potential barrier at the interface between the
normal metal and the superconductor. The resulting interplay between normal and
Andreev reflections causes a new quantum interference effect on the
conductance, as was first appreciated by Van Wees {\em et al.} (1992). The
effect, now known as {\em reflectionless tunneling}, was discovered in 1991 by
Kastalsky {\em et al.}, as a large and narrow peak in the differential
conductance of a Nb--InGaAs junction. We reproduce their data in Fig.\
\ref{fig_Kleinsasser}. There exists similar data from many other groups
(Nguyen, Kroemer, and Hu, 1992; Mani, Ghenim, and Theis, 1992; Agra\"{\i}t,
Rodrigo, and Vieira, 1992; Xiong, Xiao, and Laibowitz, 1993; Lenssen {\em et
al.}, 1994; Bakker {\em et al.}, 1994; Magn\'{e}e {\em et al.}, 1994). The
effect can be explained in terms of the disorder-induced opening of tunneling
channels (Nazarov, 1994a; Beenakker, Rejaei, and Melsen, 1994), or equivalently
as a non-equilibrium proximity effect (Volkov, Za\u{\i}tsev, and Klapwijk,
1993; Nazarov, 1994b). To set the stage we begin by discussing the
phenomenology of the effect, which gave it its name (Marmorkos, Beenakker, and
Jalabert, 1993).

\subsubsection{Reflectionless tunneling}
\label{reflectionless}

It is instructive to first discuss the classical resistance $R_{\rm NS}^{\rm
class}$ of the NS junction. The basic approximation in $R_{\rm NS}^{\rm class}$
is that currents rather than amplitudes are matched at the NS interface
(Andreev, 1966). The result is
\begin{equation}
R_{\rm NS}^{\rm
class}=\frac{h}{2Ne^{2}}\bigl(L/l+2\Gamma^{-2}\bigr),\label{RNSclass}
\end{equation}
where $L$ is the length of the disordered region, $l$ the mean free path, $N$
the number of transverse modes, and $\Gamma\ll 1$ the tunnel probability per
mode through the barrier. The contribution from the barrier to the resistance
is $\propto\Gamma^{-2}$ because tunneling into a superconductor is a
two-particle process (Shelankov, 1980): Both the incident electron and the
Andreev-reflected hole have to tunnel through the barrier (the net result being
the addition of a Cooper pair to the superconducting condensate). Equation
(\ref{RNSclass}) is to be contrasted with the classical resistance $R_{\rm
N}^{\rm class}$ in the normal state,
\begin{equation}
R_{\rm N}^{\rm
class}=\frac{h}{2Ne^{2}}\bigl(L/l+\Gamma^{-1}\bigr),\label{GNclass}
\end{equation}
where the contribution of a resistive barrier is $\propto\Gamma^{-1}$. Let us
now see how these classical results compare with the results of numerical
simulations (Marmorkos, Beenakker, and Jalabert, 1993; Takane and Ebisawa,
1993).

\begin{figure}[tb]
\centerline{
\psfig{figure=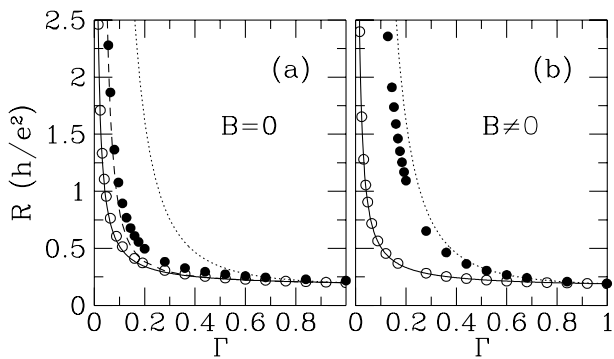,width=
8cm,bbllx=38pt,bblly=183pt,bburx=584pt,bbury=505pt}
\medskip
}%
\caption[]{
Reflectionless tunneling. Filled circles: Numerically calculated resistance
$R_{\rm NS}$ of a disordered NS junction, versus the transmission probability
per mode $\Gamma$ of the tunnel barrier at the NS interface; Open circles:
Resistance $R_{\rm N}$ of the same junction in the normal state; (a) is for
zero magnetic field, (b) is for a flux of $10\,h/e$ through the disordered
region. The dotted and solid curves are the classical Eqs.\
(\protect\ref{RNSclass}) and (\protect\ref{GNclass}). The dashed curve is the
theory of Volkov, Za\u{\i}tsev, and Klapwijk (1993), which for $\Gamma\gg
l/L\approx 0.12$ coincides with Eq.\ (\protect\ref{RNSzeroB}). After Marmorkos,
Beenakker, and Jalabert (1993).
}\label{marmo_fig1}
\end{figure}

In Fig.\ \ref{marmo_fig1} we show the resistance (at $V=0$) as a function of
$\Gamma$ in the absence and presence of a magnetic field. There is good
agreement with the classical Eqs.\ (\ref{RNSclass}) and (\ref{GNclass}) for a
magnetic field corresponding to 10 flux quanta through the disordered segment
(Fig.\ \ref{marmo_fig1}b). For $B=0$, however, the situation is different
(Fig.\ \ref{marmo_fig1}a). The normal-state resistance (open circles) still
follows approximately the classical formula (solid curve). (Deviations due to
weak localization are noticeable, but small on the scale of the figure.) In
contrast, the resistance of the NS junction (filled circles) lies much below
the classical prediction (dotted curve). The numerical data shows that for
$\Gamma\gg l/L$ one has approximately
\begin{equation}
R_{\rm NS}(B=0,V=0)\approx R_{\rm N}^{\rm class},\label{RNSzeroB}
\end{equation}
which for $\Gamma\ll 1$ is much smaller than $R_{\rm NS}^{\rm class}$. This is
the phenomenon of {\em reflectionless tunneling}: In Fig.\ \ref{marmo_fig1}a
the barrier contributes to $R_{\rm NS}$ in order $\Gamma^{-1}$, just as for
single-particle tunneling, and not in order $\Gamma^{-2}$, as expected for
two-particle tunneling. It is as if the Andreev-reflected hole is not reflected
by the barrier.

The numerical data of Fig.\ \ref{marmo_fig1}a is in good agreement with the
Green function calculation of Volkov, Za\u{\i}tsev, and Klapwijk (1993) (dashed
curve). In the next subsection we discuss a scaling theory of reflectionless
tunneling, based on the DMPK equation (Beenakker, Rejaei, and Melsen, 1994).
This theory is equivalent to the Green function calculation, but has the
advantage of explicitly demonstrating how the opening of tunneling channels on
increasing the length $L$ of the disordered region induces a transition from a
$\Gamma^{-2}$ dependence to a $\Gamma^{-1}$ dependence when $L\simeq l/\Gamma$.

\subsubsection{Scaling theory}
\label{scalingtheory}

We use the parameterization $T_{n}=1/\cosh^{2}x_{n}$ of the transmission
eigenvalues, and consider the density $\rho(x,s)$ of the $x_{n}$'s for a
(dimensionless) length $s=L/l$ of disordered region. For $s=0$, {\em i.e.} in
the absence of disorder, we have the initial condition imposed by the barrier,
\begin{equation}
\rho(x,0)=N\delta(x-x_{0}),\label{rhox0}
\end{equation}
with $\Gamma=1/\cosh^{2}x_{0}$. The DMPK equation (\ref{FokkerPlanckx})
describes how the entire distribution of the $x_{n}$'s evolves with increasing
$s$. In the large-$N$ limit, this equation reduces to the non-linear diffusion
equation (\ref{rho0eq}) for the eigenvalue density (Mello and Pichard, 1989).
In Sec.\ \ref{hydrodynamic} we showed how Eq.\ (\ref{rho0eq}) can be solved
exactly, by a mapping onto Euler's equation of hydrodynamics. The solution is
\begin{equation}
\rho(x,s)=(2N/\pi)\,{\rm Im}\,U(x-{\rm i}0^{+},s),\label{rhoxUsec4}
\end{equation}
where the complex function $U(\zeta,s)$ is determined by
\begin{equation}
U(\zeta,s)=U_{0}\biglb(\zeta-sU(\zeta,s)\bigrb).\label{Udefsec4}
\end{equation}
The function $U_{0}(\zeta)$ is fixed by the initial condition (\ref{rhox0}) on
$\rho$,
\begin{eqnarray}
U_{0}(\zeta)&=&\frac{\sinh 2\zeta}{2N} \int_{0}^{\infty}\!\!{\rm
d}x'\,\frac{\rho(x',0)}{\sinh^{2}\zeta-\sinh^{2}x'}\nonumber\\
&=&\case{1}{2}\sinh 2\zeta\,(\cosh^{2}\zeta
-\Gamma^{-1})^{-1}.\label{U0defsec4}
\end{eqnarray}
The implicit equation (\ref{Udefsec4}) has multiple solutions in the entire
complex plane; We need the solution for which both $\zeta$ and
$\zeta-sU(\zeta,s)$ lie in the strip between the lines $y=0$ and $y=-\pi/2$,
where $\zeta=x+{\rm i}y$.

\begin{figure}[tb]
\centerline{
\psfig{figure=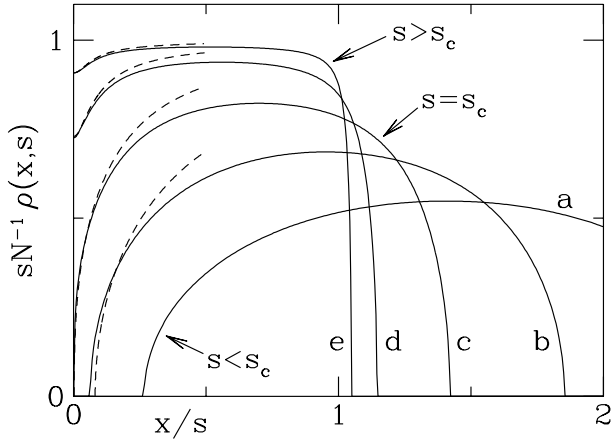,width=
8cm,bbllx=53pt,bblly=193pt,bburx=552pt,bbury=562pt}
\medskip
}%
\caption[]{
Eigenvalue density $\rho(x,s)$ as a function of $x$ (in units of $s=L/l$) for
$\Gamma=0.1$. Curves a,b,c,d,e are for $s=2,4,9,30,100$, respectively. The
solid curves are from Eq.\ (\protect\ref{rhoxUsec4}), the dashed curves from
Eq.\ (\protect\ref{rhoxapprox}). The collision of the density profile with the
boundary at $x=0$, for $s=s_{\rm c}=(1-\Gamma)/\Gamma$, signals the
disorder-induced opening of tunneling channels responsible for the
reflectionless-tunneling effect. After Beenakker, Rejaei, and Melsen (1994).
}\label{rhoplot}
\end{figure}

The resulting density (\ref{rhoxUsec4}) is plotted in Fig.\ \ref{rhoplot}
(solid curves), for $\Gamma=0.1$ and several values of $s$. For $s\gg 1$ and
$x\ll s$ it simplifies to
\begin{mathletters}
\label{rhoxapprox}
\begin{eqnarray}
&&x=\case{1}{2}{\rm arcosh}\,\tau-\case{1}{2}\Gamma
s(\tau^{2}-1)^{1/2}\cos\sigma,\label{rhoxapproxa}\\
&&\sigma\equiv \pi sN^{-1}\rho(x,s),\;\;\tau\equiv\sigma(\Gamma
s\sin\sigma)^{-1},\label{rhoxapproxb}
\end{eqnarray}
\end{mathletters}%
shown dashed in Fig.\ \ref{rhoplot}. Equation (\ref{rhoxapprox}) agrees with
the result of a Green function calculation by Nazarov (1994a). For $s=0$ (no
disorder), $\rho$ is a delta function at $x_{0}$. On adding disorder the
eigenvalue density rapidly spreads along the $x$-axis (curve a), such that
$\rho\leq N/s$ for $s>0$. The sharp edges of the density profile, so
uncharacteristic for a diffusion profile, reveal the hydrodynamic nature of the
scaling equation (\ref{rho0eq}). The upper edge is at
\begin{equation}
x_{\rm max}=s+\case{1}{2}\ln(s/\Gamma)+{\cal O}(1).\label{xmaxNS}
\end{equation}
Since $L/x$ has the physical significance of a localization length ({\em cf.}
Sec.\ \ref{lognormal}), this upper edge corresponds to a minimum localization
length $\xi_{\rm min}=L/x_{\rm max}$ of order $l$. The lower edge at $x_{\rm
min}$ propagates from $x_{0}$ to $0$ in a ``time'' $s_{\rm
c}=(1-\Gamma)/\Gamma$. For $1\ll s\leq s_{\rm c}$ one has
\begin{equation}
x_{\rm min}=\case{1}{2}{\rm arcosh}\,(s_{\rm c}/s) -\case{1}{2}[1-(s/s_{\rm
c})^{2}]^{1/2}.\label{xmin}
\end{equation}
It follows that the maximum localization length $\xi_{\rm max}=L/x_{\rm min}$
{\em increases\/} if disorder is added to a tunnel junction. This paradoxical
result, that disorder enhances transmission, becomes intuitively obvious from
the hydrodynamic correspondence, which implies that $\rho(x,s)$ spreads both to
larger {\em and\/} smaller $x$ as the fictitious time $s$ progresses. When
$s=s_{\rm c}$ the diffusion profile hits the boundary at $x=0$ (curve c), so
that $x_{\rm min}=0$. This implies that for $s>s_{\rm c}$ there exist
scattering states (eigenfunctions of $tt^{\dagger}$) which tunnel through the
barrier with near-unit transmission probability, even if $\Gamma\ll 1$. The
number $N_{\rm open}$ of transmission eigenvalues close to one (open channels)
is of the order of the number of $x_{n}$'s in the range $0$ to $1$ (since
$T_{n}\equiv 1/\cosh^{2}x_{n}$ vanishes exponentially if $x_{n}>1$). For $s\gg
s_{\rm c}$ (curve e) we estimate
\begin{equation}
N_{\rm open}\simeq\rho(0,s)=N(s+\Gamma^{-1})^{-1},\label{Nopen}
\end{equation}
where we have used Eq.\ (\ref{rhoxapprox}). The disorder-induced opening of
tunneling channels was discovered by Nazarov (1994a). It is the fundamental
mechanism for the $\Gamma^{-2}$ to $\Gamma^{-1}$ transition in the conductance
of an NS junction, as we now discuss.

We compare the integral expressions (\ref{GNSaverage}) and (\ref{GNaverage})
for the average conductances $\langle G_{\rm NS}\rangle$ and $\langle G_{\rm
N}\rangle$. For $\Gamma\gg l/L$ one is in the regime $s\gg s_{\rm c}$ of curve
e in Fig.\ \ref{rhoplot}. Then the dominant contribution to the integrals over
$x$ comes from the range $x/s\ll 1$ where
$\rho(x,s)\approx\rho(0,s)=N(s+\Gamma^{-1})^{-1}$ is approximately independent
of $x$. Substitution of $\rho(x,s)$ by $\rho(0,s)$ in Eqs.\ (\ref{GNSaverage})
and (\ref{GNaverage}) yields directly
\begin{equation}
\langle G_{\rm NS}\rangle\approx\langle G_{\rm N}\rangle\approx 1/R_{\rm
N}^{\rm class},\label{GNSGN}
\end{equation}
in agreement with the result (\ref{RNSzeroB}) of the numerical simulations.

\begin{figure}[tb]
\centerline{
\psfig{figure=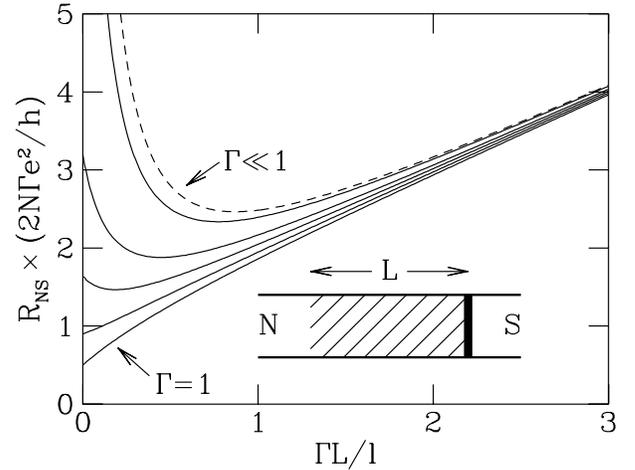,width=
8cm,bbllx=49pt,bblly=174pt,bburx=552pt,bbury=570pt}
\medskip
}%
\caption[]{
Dependence of the resistance $R_{\rm NS}$ on the length $L$ of the disordered
normal region (shaded in the inset), for different values of the transmittance
$\Gamma$ of the NS interface. Solid curves are computed from Eq.\
(\protect\ref{GNSresult}), for $\Gamma= 1,0.8,0.6,0.4,0.1$ from bottom to top.
For $\Gamma\ll 1$ the dashed curve is approached. After Beenakker, Rejaei, and
Melsen (1994).
}\label{GNSplot}
\end{figure}

Equation (\ref{GNSGN}) has the linear $\Gamma$ dependence characteristic for
reflectionless tunneling. The crossover to the quadratic $\Gamma$ dependence
when $\Gamma\lesssim l/L$ is obtained by evaluating the integrals
(\ref{GNSaverage}) and (\ref{GNaverage}) with the density $\rho(x,s)$ given by
Eq.\ (\ref{rhoxUsec4}). The result is
\begin{eqnarray}
\langle G_{\rm NS}\rangle&=&(2Ne^{2}/h)(s+Q^{-1})^{-1},\label{GNSresult}\\
\langle G_{\rm N}\rangle&=&(2Ne^{2}/h)(s+\Gamma^{-1})^{-1}.\label{GNresult}
\end{eqnarray}
The ``effective'' tunnel probability $Q$ is defined by
\begin{equation}
Q=\frac{\theta}{s\cos\theta}\left(\frac{\theta} {\Gamma
s\cos\theta}(1+\sin\theta)-1\right),\label{Qdef}
\end{equation}
where $\theta\in(0,\pi/2)$ is the solution of the transcendental equation
\begin{equation}
\theta[1-\case{1}{2}\Gamma(1-\sin\theta)]=\Gamma s\cos\theta. \label{phidef}
\end{equation}
For $\Gamma\ll 1$ (or $s\gg 1$) Eqs.\ (\ref{Qdef}) and (\ref{phidef}) simplify
to $Q=\Gamma\sin\theta$, $\theta=\Gamma s\cos\theta$, in precise agreement with
the Green function calculation of Volkov, Za\u{\i}tsev, and Klapwijk (1993).
According to Eq.\ (\ref{GNresult}), the normal-state resistance increases {\em
linearly\/} with the length $L$ of the disordered region, as expected from
Ohm's law. This classical reasoning fails if one of the contacts is in the
superconducting state. The scaling of the resistance $R_{\rm NS}\equiv
1/\langle G_{\rm NS}\rangle$ with length, computed from Eq.\ (\ref{GNSresult}),
is plotted in Fig.\ \ref{GNSplot}. For $\Gamma=1$ the resistance increases
monotonically with $L$. The ballistic limit $L\rightarrow 0$ equals
$h/4Ne^{2}$, half the contact resistance of a normal junction because of
Andreev reflection. For $\Gamma\lesssim 0.5$ a {\em resistance minimum\/}
develops, somewhat below $L=l/\Gamma$. The resistance minimum is associated
with the crossover from a quadratic to a linear dependence of $R_{\rm NS}$ on
$1/\Gamma$.

If $\Gamma s\gg 1$ one has $\theta\rightarrow\pi/2$, hence
$Q\rightarrow\Gamma$. In the opposite regime $\Gamma s\ll 1$ one has
$\theta\rightarrow\Gamma s$, hence $Q\rightarrow\Gamma^{2}s$. The corresponding
asymptotic expressions for $\langle G_{\rm NS}\rangle$ are (assuming $\Gamma\ll
1$ and $s\gg 1$):
\begin{mathletters}
\label{asymptab}
\begin{eqnarray}
\langle G_{\rm NS}\rangle&=&(2Ne^{2}/h)(s+\Gamma^{-1})^{-1},\;\;{\rm if}\;\;
\Gamma s\gg 1,\label{asympta}\\
\langle G_{\rm NS}\rangle&=&(2Ne^{2}/h)\Gamma^{2}s,\;\;{\rm if}\;\; \Gamma s\ll
1.\label{asymptb}
\end{eqnarray}
\end{mathletters}%
In either limit the conductance is greater than the classical result
\begin{equation}
G_{\rm NS}^{\rm class}=(2Ne^{2}/h)(s+2\Gamma^{-2})^{-1}, \label{GNSclass}
\end{equation}
which holds if phase coherence between electrons and holes is destroyed by a
voltage or magnetic field. The peak in the conductance around $V,B=0$ is of
order $\Delta G_{\rm NS}=\langle G_{\rm NS}\rangle-G_{\rm NS}^{\rm class}$,
which has the relative magnitude
\begin{equation}
\frac{\Delta G_{\rm NS}}{\langle G_{\rm
NS}\rangle}\approx\frac{2}{2+\Gamma^{2}s}.\label{peakheight}
\end{equation}

The scaling theory assumes zero temperature. Hekking and Nazarov (1993, 1994),
and Zhou, Spivak, and Zyuzin (1995) have studied the conductance of a resistive
NS interface at finite temperatures, when $L$ is greater than the correlation
length $L_{\rm c}=(\hbar D/k_{\rm B}T)^{1/2}$. (This is the length scale at
which the Thouless energy $E_{\rm c}$ equals the thermal energy $k_{\rm B}T$.)
Their result is consistent with the limiting expression (\ref{asymptb}), if
$s=L/l$ is replaced by $L_{\rm c}/l$. The implication is that, if $L>L_{\rm
c}$, the non-linear scaling of the resistance shown in Fig.\ \ref{GNSplot} only
applies to a disordered segment of length $L_{\rm c}$ adjacent to the
superconductor. For the total resistance one should add the Ohmic contribution
of order $(h/e^{2})(L-L_{\rm c})/l$ from the rest of the wire.

\subsubsection{Double-barrier junction}
\label{doublebarrierNS}

In the previous subsection we have discussed how the opening of tunneling
channels ({\em i.e.} the appearance of transmission eigenvalues close to one)
by disorder leads to a minimum in the resistance when $L\simeq l/\Gamma$. The
minimum separates a $\Gamma^{-1}$ from a $\Gamma^{-2}$ dependence of the
resistance on the transparency of the interface. We referred to the
$\Gamma^{-1}$ dependence as ``reflectionless tunneling'', since it is as if one
of the two quasiparticles which form the Cooper pair can tunnel through the
barrier with probability one. In the present subsection we will show, following
Melsen and Beenakker (1994), that a qualitatively similar effect occurs if the
disorder in the normal region is replaced by a second tunnel barrier (tunnel
probability $\Gamma'$). The resistance at fixed $\Gamma$ shows a minimum as a
function of $\Gamma'$ when $\Gamma'\simeq\Gamma$.  For $\Gamma'\lesssim\Gamma$
the resistance has a $\Gamma^{-1}$ dependence, so that we can speak again of
reflectionless tunneling.

We consider an ${\rm NI}_{1}{\rm NI}_{2}{\rm S}$ junction, where N = normal
metal, S = superconductor, and ${\rm I}_i$ = insulator or tunnel barrier
(transmission probability per mode $\Gamma_{i}\equiv 1/\cosh^{2}\alpha_{i}$).
We assume ballistic motion between the barriers. (The effect of disorder is
discussed later.) A straightforward calculation yields the transmission
probabilities $T_{n}$ of the two barriers in series [{\em cf.} Eq.\
(\ref{Tnab})],
\begin{mathletters}
\label{TnabNS}
\begin{eqnarray}
&&T_{n}=(a+b\cos\phi_{n})^{-1},\label{eq:tnphin}\\
&&a=\case{1}{2}+\case{1}{2}\cosh 2\alpha_{1}\cosh
2\alpha_{2},\label{coeffalphaa}\\
&&b=\case{1}{2}\sinh 2\alpha_{1}\sinh 2\alpha_{2},\label{coeffalphab}
\end{eqnarray}
\end{mathletters}%
where $\phi_{n}$ is the phase accumulated between the barriers by mode $n$. We
assume that $L\gg\lambda_{\rm F}$ and $N\Gamma_{i}\gg 1$, so that the
conductance is not dominated by a single resonance. In this case, the phases
$\phi_{n}$ are distributed uniformly in the interval $(0,2\pi)$ and we may
replace the sum over the transmission eigenvalues in Eqs.\ (\ref{Landauer}) and
(\ref{keyzero}) by integrals over $\phi$:
$\sum_{n=1}^{N}f(\phi_{n})\rightarrow(N/2\pi)\int_{0}^{2\pi}d\phi\,f(\phi)$.
The result is
\begin{eqnarray}
G_{\rm NS}&=&\frac{4Ne^2}{h}\frac{\cosh 2\alpha_{1}\cosh 2\alpha_{2}} {\left(
\cosh^{2}2\alpha_{1}+\cosh^{2}2\alpha_{2}-1 \right)^{3/2}},\label{gnsintphi}\\
G_{\rm N}&=&\frac{4Ne^{2}}{h}(\cosh 2\alpha_{1}+\cosh
2\alpha_{2})^{-1}\label{gnintphi}.
\end{eqnarray}
These expressions are symmetric in the indices 1 and 2: It does not matter
which of the two barriers is closest to the superconductor. In the same way we
can compute the entire distribution of the transmission eigenvalues,
$\rho(T)\equiv\sum_{n}\delta(T-T_{n})
\rightarrow(N/2\pi)\int_{0}^{2\pi}d\phi\,\delta\biglb(T-T(\phi)\bigrb)$.
Substituting $T(\phi)=(a+b\cos\phi)^{-1}$ from Eq.\ (\ref{TnabNS}), one finds
\begin{equation}
\rho(T)=\frac{N}{\pi T}\left(b^{2}T^{2}-(aT-1)^{2}\right)^{-1/2}.\label{rhoT}
\end{equation}

\begin{figure}[tb]
\centerline{
\psfig{figure=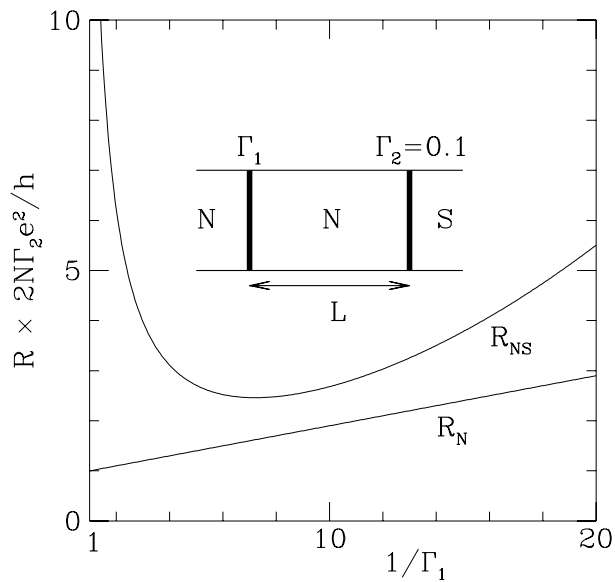,width=
8cm,bbllx=19pt,bblly=160pt,bburx=541pt,bbury=661pt}
\medskip
}%
\caption[]{
Dependence of the resistances $R_{\rm N}$ and $R_{\rm NS}$ of ballistic NININ
and NINIS structures, respectively, on barrier transparency $\Gamma_{1}$, while
transparency $\Gamma_{2}=0.1$ is kept fixed [computed from Eqs.\
(\protect\ref{gnsintphi}) and (\protect\ref{gnintphi})]. The inset shows the
NINIS structure considered. After Melsen and Beenakker (1994).
}\label{NINISfig}
\end{figure}

In Fig.\ \ref{NINISfig} we plot the resistances following from Eqs.\
(\ref{gnsintphi}) and (\ref{gnintphi}). Notice that $R_{\rm N}$ follows Ohm's
law,
\begin{equation}
R_{\rm N}=\frac{h}{2Ne^2}(1/\Gamma_{1}+1/\Gamma_{2}-1),\label{Ohmslaw}
\end{equation}
as expected from classical considerations. In contrast, the resistance $R_{\rm
NS}$ has a {\em minimum\/} if one of the $\Gamma$'s is varied while keeping the
other fixed. This resistance minimum cannot be explained by classical series
addition of barrier resistances. If $\Gamma_{2}\ll 1$ is fixed and $\Gamma_{1}$
is varied, as in Fig.\ \ref{NINISfig}, the minimum occurs when
$\Gamma_{1}=\sqrt{2}\,\Gamma_{2}$. The minimal resistance $R_{\rm NS}^{\rm
min}$ is of the same order of magnitude as the resistance $R_{\rm N}$ in the
normal state at the same value of $\Gamma_{1}$ and $\Gamma_{2}$. In particular,
we find that $R_{\rm NS}^{\rm min}$ depends linearly on $1/\Gamma_{i}$, whereas
for a single barrier $R_{\rm NS}\propto 1/\Gamma^{2}$.

\begin{figure}[tb]
\centerline{
\psfig{figure=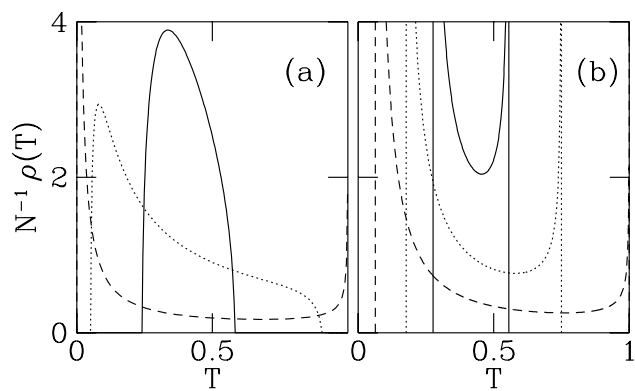,width=
8cm,bbllx=76pt,bblly=183pt,bburx=583pt,bbury=504pt}
}%
\medskip
\caption[]{
Density of transmission eigenvalues through a normal region containing a
potential barrier (transmission probability $\Gamma=0.4$). The left panel (a)
shows the disorder-induced opening of tunneling channels (solid curve:
$s=0.04$; dotted: $s=0.4$; dashed: $s=5$; where $s\equiv L/l$). The right panel
(b) shows the opening of channels by a second tunnel barrier (transparency
$\Gamma'$; solid curve: $\Gamma'=0.95$; dotted: $\Gamma'=0.8$; dashed:
$\Gamma'=0.4$). The curves in (a) are computed from Eq.\
(\protect\ref{rhoxUsec4}), the curves in (b) from Eq.\ (\protect\ref{rhoT}).
After Melsen and Beenakker (1994).
}\label{rhotplot}
\end{figure}

The linear dependence on the barrier transparency shows the qualitative
similarity of a ballistic \mbox{NINIS} junction to the disordered NIS junction
considered in the previous subsection. To illustrate the similarity, we compare
in Fig.\ \ref{rhotplot} the densities of normal-state transmission eigenvalues.
The left panel is for an NIS junction [computed using Eq.\ (\ref{rhoxUsec4})],
the right panel is for an \mbox{NINIS} junction [computed from Eq.\
(\ref{rhoT})]. In the NIS junction, disorder leads to a bimodal distribution
$\rho(T)$, with a peak near zero transmission and another peak near unit
transmisssion (dashed curve). A similar bimodal distribution appears in the
ballistic \mbox{NINIS} junction, for approximately equal transmission
probabilities of the two barriers. There are also differences between the two
cases: The NIS junction has a uni-modal $\rho(T)$ if $L/l<1/\Gamma$, while the
\mbox{NINIS} junction has a bimodal $\rho(T)$ for any ratio of $\Gamma_{1}$ and
$\Gamma_{2}$. In both cases, the opening of tunneling channels, {\em i.e.} the
appearance of a peak in $\rho(T)$ near $T=1$, is the origin for the $1/\Gamma$
dependence of the resistance.

The DMPK scaling equation can be used to investigate what happens to the
resistance minimum if the region of length $L$ between the tunnel barriers
contains impurities, with elastic mean free path $l$ (Melsen and Beenakker,
1994). In the diffusive regime ($l\ll L$) the scaling theory is found to agree
with the Green function calculation by Volkov, Za\u{\i}tsev, and Klapwijk
(1993) for a disordered \mbox{NINIS} junction. For strong barriers
($\Gamma_{1},\Gamma_{2}\ll 1$) and strong disorder ($L\gg l$), one has the two
asymptotic formulas
\begin{mathletters}
\label{eq:glimitlss}
\begin{eqnarray}
\langle G_{\rm NS}\rangle&=& \frac{2Ne^2}{h}\frac{\Gamma_{1}^{2}\Gamma_{2}^{2}}
{\left(\Gamma_{1}^{2}+\Gamma_{2}^{2}\right)^{3/2}},\;\;{\rm if}\;\;
\Gamma_{1},\Gamma_{2}\ll l/L,\label{eq:glimitls}\\
\langle G_{\rm
NS}\rangle&=&\frac{2Ne^2}{h}(L/l+1/\Gamma_{1}+1/\Gamma_{2})^{-1},\;\;{\rm
if}\;\;\Gamma_{1},\Gamma_{2}\gg l/L.\nonumber\\
\label{eq:glimitss}
\end{eqnarray}
\end{mathletters}%
Equation (\ref{eq:glimitls}) coincides with Eq.\ (\ref{gnsintphi}) in the limit
$\alpha_{1},\alpha_{2}\gg 1$ (recall that $\Gamma_{i}\equiv
1/\cosh^2\alpha_{i}$). This shows that the effect of disorder on the resistance
minimum can be neglected as long as the resistance of the junction is dominated
by the barriers. In this case $\langle G_{\rm NS}\rangle$ depends linearly on
$\Gamma_{1}$ and $\Gamma_{2}$ only if $\Gamma_{1}\approx\Gamma_{2}$. Equation
(\ref{eq:glimitss}) shows that if the disorder dominates, $\langle G_{\rm
NS}\rangle$ has a linear $\Gamma$-dependence regardless of the relative
magnitude of $\Gamma_{1}$ and $\Gamma_{2}$.

\begin{figure}[tb]
\centerline{
\psfig{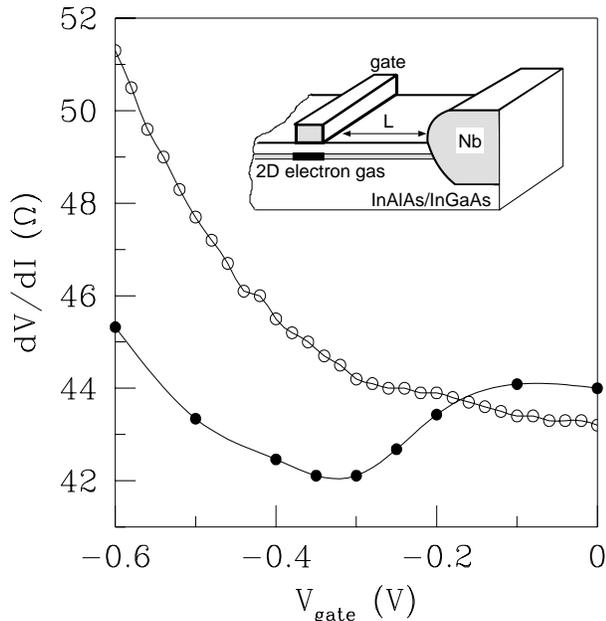}
}%
\caption[]{
Differential resistance $dV/dI$ of a gated Nb--InAlAs/InGaAs junction as a
function of the gate voltage $V_{\rm gate}$. The more negative $V_{\rm gate}$,
the higher the tunnel barrier in the two-dimensional electron gas below the
gate. The gate is at a separation $L=0.5\,\mu{\rm m}$ from the Nb contact (see
the inset). A second Nb contact (not shown) is at $4\,\mu{\rm m}$ from the
first. The filled circles are for $V=0$, the open circles are for $V>3\,{\rm
mV}$, where $V$ is the voltage between the two Nb contacts. (The curve through
the data points is a guide to the eye.) The data for $V=0$ shows the resistance
minimum expected for a ballistic double-barrier NS junction ({\em cf.} Fig.\
\protect\ref{NINISfig}). After Takayanagi, Toyoda, and Akazaki (1996).
}\label{fig_takayanagi}
\end{figure}

The resistance minimum predicted by Eq.\ (\ref{gnsintphi}) has been observed by
Takayanagi, Toyoda, and Akazaki (1996), in a junction between Nb and the
two-dimensional electron gas in a InAlAs/InGaAs heterostructure. [Similar
experiments using doped GaAs instead of a heterostructure have been performed
by Poirier, Mailly, and Sanquer (1996).] One of the two barriers is present
naturally at the interface between Nb and the heterostructure. The other
barrier is created electrostatically by means of a gate on top of the
heterostructure, at a separation $L=0.5\,\mu{\rm m}$ from the Nb interface.
This separation is much less than the mean free path $l=2.8\,\mu{\rm m}$ in the
electron gas. By making the voltage $V_{\rm gate}$ on the gate more negative,
the transparency $\Gamma_{1}$ of the tunnel barrier below the gate is reduced.
The transparency of the tunnel barrier at the Nb-interface is fixed, and
estimated at $\Gamma_{2}=0.7$ from the high-temperature resistance. The
low-temperature resistance is plotted as a function of the gate voltage in
Fig.\ \ref{fig_takayanagi} (filled circles). Also shown is the normal-state
resistance (open circles), obtained by applying a voltage greater than
$2{\Delta}/e=3\,{\rm mV}$ over the junction. The former has a minimum, while
the latter decreases monotonically with gate voltage. A quantitative comparison
with the theory needs to take into account the series resistance from a second
Nb contact, at a distance of $3.5\,\mu{\rm m}$ from the gate. Takayanagi {\em
et al.} have found that quite a good agreement with the theoretical result
(\ref{gnsintphi}) can be obtained.

\subsubsection{Circuit theory}
\label{circuit}

The scaling theory of reflectionless tunneling, which was the subject of Sec.\
\ref{scalingtheory}, describes the transition from the ballistic to the
diffusive regime. In the diffusive regime it is equivalent to the Green
function theory of Volkov, Za\u{\i}tsev, and Klapwijk (1993). A convenient
formulation of the Green function theory has been presented by Nazarov (1994b).
Starting from a continuity equation for the non-equilibrium Green function
(Keldysh, 1964; Larkin and Ovchinnikov, 1975, 1977), and applying the
appropriate boundary conditions, Nazarov was able to formulate a set of rules
which reduce the problem of computing the resistance of an NS junction to a
simple exercise in circuit theory. [An alternative derivation of these rules,
using scattering matrices, has been given recently by Argaman (1997).] The
approach can be applied without further complications to multi-terminal
networks involving several normal and superconducting reservoirs. In this
subsection we describe Nazarov's circuit theory and compare it with the results
obtained from the DMPK equation.

Zero temperature is assumed, as well as infinitesimal voltage differences
between the normal reservoirs (linear response). The superconducting reservoirs
$S_{i}$ are all at the same voltage, because they are effectively
short-circuited by the supercurrent. The pair potential in $S_{i}$ has phase
$\phi_{i}$. The reservoirs are connected by a set of diffusive normal-state
conductors (length $L_{i}$, mean free path $l_{i}$; $s_{i}\equiv L_{i}/l_{i}\gg
1$). Between the conductors there may be tunnel barriers (tunnel probability
$\Gamma_{i}$). The presence of superconducting reservoirs has no effect on the
resistance $(h/2Ne^{2})s_{i}$ of the diffusive conductors, but affects only the
resistance $h/2Ne^{2}\Gamma_{i}^{\rm eff}$ of the tunnel barriers. The tunnel
probability $\Gamma_{i}$ of barrier $i$ is renormalized to an effective tunnel
probability $\Gamma_{i}^{\rm eff}$, which depends on the entire circuit.

Nazarov's rules to compute the effective tunnel probabilities are as follows.
To each node and to each terminal of the circuit one assigns a vector
$\vec{n}_{i}$ of unit length. For a normal reservoir, $\vec{n}_{i}=(0,0,1)$ is
at the north pole, for a superconducting reservoir,
$\vec{n}_{i}=(\cos\phi_{i},\sin\phi_{i},0)$ is at the equator. For a node,
$\vec{n}_{i}$ is somewhere on the northern hemisphere. The vector $\vec{n}_{i}$
is called a ``spectral vector'' because it determines the density of states.
(The $z$-component of the spectral vector is the local density of states at the
Fermi energy divided by the density of states in the normal reservoirs.) If the
tunnel barrier is located between spectral vectors $\vec{n}_{1}$ and
$\vec{n}_{2}$, its effective tunnel probability is
\begin{equation}
\Gamma^{\rm eff}=(\vec{n}_{1}\cdot\vec{n}_{2})\Gamma=
\Gamma\cos\theta_{12},\label{Gamma-eff}
\end{equation}
where $\theta_{12}$ is the angle between $\vec{n}_{1}$ and $\vec{n}_{2}$. The
rule to compute the spectral vector of node $i$ follows from the continuity
equation for the Green function. Let the index $k$ label the nodes or terminals
connected to node $i$ by a single tunnel barrier (with tunnel probability
$\Gamma_{k}$). Let the index $q$ label the nodes or terminals connected to $i$
by a diffusive conductor (with $L/l\equiv s_{q}$). The spectral vectors then
satisfy the sum rule
\begin{equation}
\sum_{k}(\vec{n}_{i}\times\vec{n}_{k})\Gamma_{k}+
\sum_{q}(\vec{n}_{i}\times\vec{n}_{q})\frac{{\rm
arccos}(\vec{n}_{i}\cdot\vec{n}_{q})}
{s_{q}\sqrt{1-(\vec{n}_{i}\cdot\vec{n}_{q})^{2}}}=0.\label{sumrule}
\end{equation}
This is a sum rule for a set of vectors perpendicular to $\vec{n}_{i}$ of
magnitude $\Gamma_{k}\sin\theta_{ik}$ or $\theta_{iq}/s_{q}$, depending on
whether the element connected to node $i$ is a tunnel barrier or a diffusive
conductor. There is a sum rule for each node, and together the sum rules
determine the spectral vectors of the nodes.

These rules can be readily generalized (Nazarov, 1995a) to include the case
that a tunnel barrier is replaced by a ballistic point contact (conductance
$2Ne^{2}/h$). Instead of the effective tunnel probability (\ref{Gamma-eff}) one
then has the effective number of modes
\begin{equation}
N^{\rm eff}=\frac{2N}{1+(\vec{n}_{1}\cdot\vec{n}_{2})}=
N\cos^{-2}\case{1}{2}\theta_{12}. \label{N-eff}
\end{equation}
The corresponding replacement in the sum rule (\ref{sumrule}) is
\begin{equation}
(\vec{n}_{i}\times\vec{n}_{k})\Gamma_{k}\rightarrow
\frac{2(\vec{n}_{i}\times\vec{n}_{k})N_{k}}{1+(\vec{n}_{i}\cdot\vec{n}_{k})}.
\label{sumruleN}
\end{equation}
A further generalization, to include the effect of an Aharonov-Bohm ring, has
been given by Stoof and Nazarov (1996b).

\begin{figure}[tb]
\centerline{
\psfig{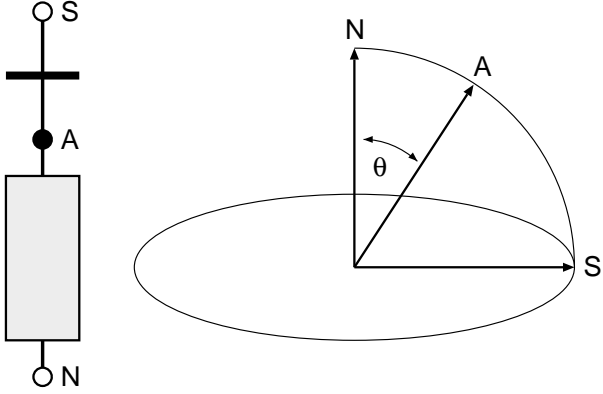}
}%
\medskip
\caption[]{
Representation of a circuit by spectral vectors. At left: Circuit containing
two terminals (open circles), one node (filled circle), and two elements: A
diffusive conductor (dotted) and a tunnel barrier (black). At right: Spectral
vectors associated with the terminals N,S and with the node A.
}\label{NScircuit}
\end{figure}

As a simple example, let us consider the system of Sec.\ \ref{scalingtheory},
consisting of one normal terminal (N), one superconducting terminal (S), one
node (labeled A), and two elements: A diffusive conductor (with $L/l\equiv s$)
between N and A, and a tunnel barrier (tunnel probability $\Gamma$) between A
and S (see Fig.\ \ref{NScircuit}). There are three spectral vectors,
$\vec{n}_{\rm N}$, $\vec{n}_{\rm S}$, and $\vec{n}_{\rm A}$. All spectral
vectors lie in one plane. (This holds for any network with a single
superconducting terminal.) The resistance of the circuit is given by
$R=(h/2Ne^{2})(s+1/\Gamma^{\rm eff})$, with the effective tunnel probability
\begin{equation}
\Gamma^{\rm eff}=\Gamma\cos\theta_{\rm AS}=\Gamma\sin\theta.\label{thetaAS}
\end{equation}
Here $\theta\in[0,\pi/2]$ is the polar angle of $\vec{n}_{\rm A}$. This angle
is determined by the sum rule (\ref{sumrule}), which in this case takes the
form
\begin{equation}
\Gamma\cos\theta-\theta/s=0.\label{sumruleAS}
\end{equation}
Comparison with Sec.\ \ref{scalingtheory} shows that $\Gamma^{\rm eff}$
coincides with the effective tunnel probability $Q$ of Eq.\ (\ref{Qdef}) in the
limit $s\gg 1$, {\em i.e.} if one restricts oneself to the diffusive regime.
That is the basic requirement for the application of the circuit theory.

\begin{figure}[tb]
\centerline{
\psfig{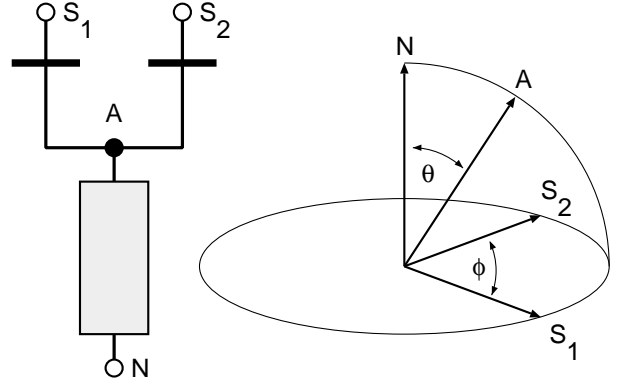}
}%
\medskip
\caption[]{
Circuit diagram and spectral vectors for a structure containing one normal and
two superconducting terminals (phase difference $\phi$).
}\label{forkcircuit}
\end{figure}

Let us now consider the ``fork junction'' of Fig.\ \ref{forkcircuit}, with one
normal terminal (N) and two superconducting terminals ${\rm S}_{1}$ and ${\rm
S}_{2}$ (phases $\phi_{1}\equiv-\phi/2$ and $\phi_{2}\equiv\phi/2$). There is
one node (A), which is connected to N by a diffusive conductor ($L/l\equiv s$),
and to ${\rm S}_{1}$ and ${\rm S}_{2}$ by tunnel barriers ($\Gamma_{1}$ and
$\Gamma_{2}$). This structure was studied theoretically by Hekking and Nazarov
(1993) and experimentally by Pothier {\em et al.} (1994) and Dimoulas {\em et
al.} (1995). For simplicity, let us assume two identical tunnel barriers
$\Gamma_{1}=\Gamma_{2}\equiv\Gamma$. Then the spectral vector $\vec{n}_{\rm
A}=(\sin\theta,0,\cos\theta)$ of node A lies symmetrically between the spectral
vectors of terminals ${\rm S}_{1}$ and ${\rm S}_{2}$. The sum rule
(\ref{sumrule}) now takes the form
\begin{equation}
2\Gamma|\cos\case{1}{2}\phi|\cos\theta-\theta/s=0.\label{sumrulefork}
\end{equation}
Its solution determines the effective tunnel rate $\Gamma^{\rm
eff}=\Gamma|\cos\case{1}{2}\phi|\sin\theta$ of each of the two barriers in
parallel, and hence the conductance of the fork junction,
\begin{equation}
G=\frac{2Ne^{2}}{h}[s+\case{1}{2}(\Gamma
|\cos\case{1}{2}\phi|\sin\theta)^{-1}]^{-1}.\label{Gfork}
\end{equation}
Two limiting cases of Eqs.\ (\ref{sumrulefork}) and (\ref{Gfork}) are
\begin{mathletters}
\label{Gforkab}
\begin{eqnarray}
G&=&\frac{2Ne^{2}}{h}(s+\case{1}{2}\Gamma^{-1}
|\cos\case{1}{2}\phi|^{-1})^{-1},\;\;{\rm
if}\;\;s\Gamma|\cos\case{1}{2}\phi|\gg 1,\nonumber\\
\label{Gforka}\\
G&=&\frac{4Ne^{2}}{h}\,s\Gamma^{2}(1+\cos\phi),\;\;{\rm
if}\;\;s\Gamma|\cos\case{1}{2}\phi|\ll 1.\label{Gforkb}
\end{eqnarray}
\end{mathletters}%
For $\phi=0$ (and $2\Gamma\rightarrow\Gamma$) these expressions reduce to the
result (\ref{asymptab}) for an NS junction with a single superconducting
reservoir. The limit (\ref{Gforkb}) agrees with the finite-temperature result
of Hekking and Nazarov (1993), if $s$ is replaced by $L_{\rm c}/l$ and a series
resistance is added due to the normal segment which is further than a
correlation length from the NS interfaces.

\begin{figure}[tb]
\centerline{
\psfig{figure=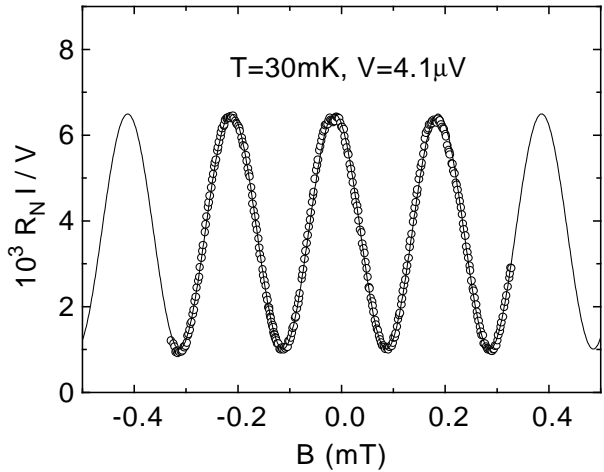,width=
8cm,bbllx=188pt,bblly=445pt,bburx=559pt,bbury=736pt}
}%
\medskip
\caption[]{
Conductance of a fork junction as a function of magnetic field, showing the
dependence on the phase difference $\phi$ of the superconductor at two tunnel
barriers. The circles are measurements by Pothier {\em et al.} (1994) of the
current $I$ through a Cu wire connected to an oxidized Al fork (normal-state
resistance $R_{\rm N}=1.56\,{\rm k}\Omega$). The applied voltage $V$ is
sufficiently low that $I/V$ is close to the linear-response conductance. (The
amplitude of the oscillations at $V=0$ is $3.94\cdot 10^{-6}\,\Omega^{-1}$,
somewhat larger than in the figure.) The solid curve is a cosine fit to the
data. The offset of maximum conductance from $B=0$ is attributed to a small
residual field in the cryostat. After Pothier, Gu\'{e}ron, Esteve, and Devoret
(1994).
}\label{Pothier}
\end{figure}

Experimental data by Pothier {\em et al.} (1994) for the $\phi$-dependence of
the conductance of a fork junction is shown in Fig.\ \ref{Pothier}. The
conductance of a Cu wire attached to an oxidized Al fork oscillates as a
function of the applied magnetic field. The period corresponds to a flux
increment of $h/2e$ through the area enclosed by the fork and the wire, and
thus to $\Delta\phi=2\pi$. The experiment is in the regime where the junction
resistance is dominated by the tunnel barriers, as in Eq.\ (\ref{Gforkb}). (The
metal-oxide tunnel barriers in these structures have typically very small
transmission probabilities $\Gamma\simeq 10^{-5}$, so that the regime of Eq.\
(\ref{Gforka}) is not easily accessible.) Equation (\protect\ref{Gforkb})
provides only a qualitative description of the experiment, mainly because the
motion in the arms of the fork is diffusive rather than ballistic. This is why
the conductance minima in Fig.\ \protect\ref{Pothier} do not go to zero. A
solution of the diffusion equation in the actual experimental geometry is
required for a quantitative comparison with the theory.

\subsection{NS junction containing a point contact}
\label{NSpointcontact}

\begin{figure}[tb]
\centerline{
\psfig{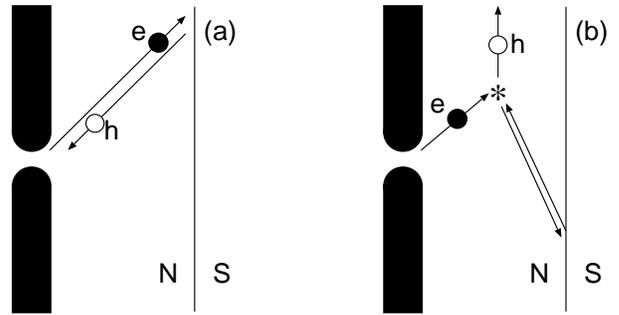}
}%
\bigskip
\caption[]{
Classical trajectories of an electron injected through a point contact towards
a superconductor, where it is Andreev reflected as a hole. In (a) the
trajectory is ballistic, in (b) the electron and hole are scattered in random
directions by an impurity (asterisk). }\label{fig_NSpc}
\end{figure}

Andreev reflection doubles the conductance of a point contact (Za\u{\i}tsev,
1980; Blonder, Tinkham, and Klapwijk, 1982; Shelankov, 1984). As illustrated in
Fig.\ \ref{fig_NSpc}a, an electron injected through the point contact is
reflected back as a hole. Because electron and hole carry the same current
(both in magnitude and in direction), the current through the point contact,
and hence its conductance, are doubled. If $N_{0}$ is the number of transverse
modes in the cross-sectional area of the point contact, then its conductance is
given by
\begin{equation}
G_{\rm NS}=N_{0}\frac{4e^{2}}{h},\label{GNSpc}
\end{equation}
which is twice the conductance
\begin{equation}
G_{\rm N}=N_{0}\frac{2e^{2}}{h}\label{GNpc}
\end{equation}
in the normal state (Sharvin, 1965). Equations (\ref{GNSpc}) and (\ref{GNpc})
apply to ballistic transport, without scattering of the electrons by
impurities. What is the effect of impurities in the region between the point
contact and the superconductor? Classically, one would expect these to destroy
the conductance doubling from Andreev reflection, because the hole no longer
retraces the path of the electron (Fig.\ \ref{fig_NSpc}b). It is therefore
unlikely to find its way back through the point contact, so that no current
doubling is to be expected. This classical picture is correct if the separation
$L$ between the point contact and the NS interface is greater than the
correlation length $L_{\rm c}=(\hbar D/k_{\rm B}T)^{1/2}$. At sufficiently low
temperatures that $L\ll L_{\rm c}$, however, the conductance doubling may
persist in the presence of impurity scattering (Golubov and Kupriyanov, 1995;
Beenakker, Melsen, and Brouwer, 1995). To explain this effect, we first
consider the angular distribution of the holes which are reflected by a
disordered NS junction.

\subsubsection{Giant backscattering peak}
\label{giantpeak}

The angular distribution of electrons reflected by a disordered normal metal
has a narrow peak at the angle of incidence. This peak has the same origin as
the weak-localization correction to the average conductance, namely the
constructive interference of time-reversed sequences of multiple scattering
events (Berkovits and Feng, 1994). The peak is at most twice as high as the
background. In this subsection we discuss the giant enhancement of the
backscattering peak, which occurs if the normal metal is in contact with a
superconductor (Beenakker, Melsen, and Brouwer, 1995). At the interface with
the superconductor an electron incident from the normal metal is reflected
either as an electron (normal reflection) or as a hole (Andreev reflection).
Both scattering processes contribute to the backscattering peak. Normal
reflection contributes a factor of two. In contrast, we will see that Andreev
reflection contributes a factor $G/G_{0}$, which is $\gg 1$.

\begin{figure}[tb]
\vspace*{-2cm}
\centerline{
\psfig{figure=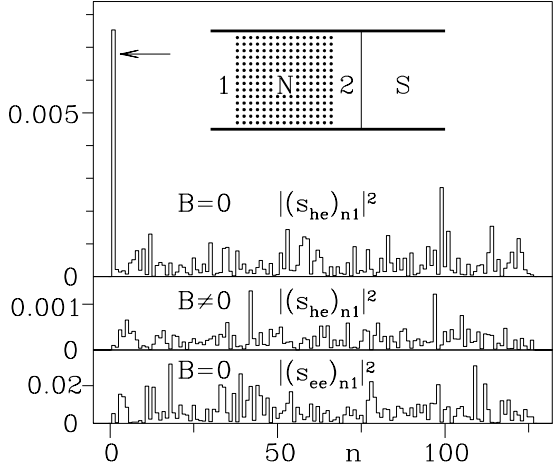,width= 8cm}
}%
\medskip
\caption[]{
Numerical simulation of a $300\times 300$ tight-binding model for a disordered
normal metal ($L=9.5\,l$), in series with a superconductor (inset). The
histograms give the modal distribution for reflection of an electron at normal
incidence (mode number 1). The top two panels give the distribution of
reflected holes (for $B=0$ and $B=10\,h/eL^{2}$), the bottom panel of reflected
electrons (for $B=0$). The arrow indicates the ensemble-averaged height of the
backscattering peak for Andreev reflection, predicted from Eq.\
(\protect\ref{Andreev_giant}). After Beenakker, Melsen, and Brouwer (1995).
(The original figure has a mislabeled vertical axis.)
}\label{fig_giant}
\end{figure}

We consider a disordered normal-metal conductor which is connected at one end
to a superconductor (see inset of Fig.\ \ref{fig_giant}). An electron at the
Fermi level incident from the opposite end in mode $m$ is reflected into some
other mode $n$, either as an electron or as a hole, with probability amplitudes
$(s_{\rm ee})_{nm}$ and $(s_{\rm he})_{nm}$, respectively. The $N\times N$
matrices $s_{\rm ee}$ and $s_{\rm he}$ are given by Eqs.\ (\ref{sehheeehh}) and
(\ref{MeMh}) (with $\varepsilon=0$, $\alpha=-{\rm i}$). In terms of the polar
decomposition (\ref{polarS}) of the transmission and reflection matrices, we
can write
\begin{equation}
s_{\rm ee}=-2U\frac{\sqrt{1-{\cal T}}}{2-{\cal T}}U',\;\;
s_{\rm he}=-{\rm i}U^{\ast}\frac{\cal T}{2-{\cal T}}U'.
\label{rUrelation}
\end{equation}

We first consider zero magnetic field ($B=0$). Time-reversal symmetry then
requires that $U'=U^{\rm T}$. We make the isotropy assumption that $U$ is
uniformly distributed over the unitary group. The average over $U$ (using the
formulas of Appendix \ref{UNintegrate}) yields
\begin{mathletters}
\label{rnmT}
\begin{eqnarray}
\left\langle|(s_{\rm
ee})_{nm}|^{2}\right\rangle&=&\frac{\delta_{nm}+1}{N^{2}+N}\left(N-
\left\langle\sum_{k}\sigma_{k}^{2}\right\rangle\right),\label{rnmTee}\\
\left\langle|(s_{\rm he})_{nm}|^{2}\right\rangle&=&
\frac{\delta_{nm}+1}{N^{2}+N}\left\langle\sum_{k}\sigma_{k}^{2}\right\rangle
\nonumber\\
&&\mbox{}+\frac{N\delta_{nm}-1}{N^{3}-N}\left\langle\sum_{k\neq
k'}\sigma_{k}\sigma_{k'}\right\rangle,\label{rnmThe}
\end{eqnarray}
\end{mathletters}%
where we have defined $\sigma_{k}\equiv T_{k}(2-T_{k})^{-1}$. In the metallic
regime $N\gg L/l\gg 1$. In this large-$N$ limit we may factorize
$\langle\sum_{k\neq k'}\sigma_{k}\sigma_{k'}\rangle$ into
$\langle\sum_{k}\sigma_{k}\rangle^{2}$, which can be evaluated using Eq.\
(\ref{meanAresult}):
\begin{equation}
\left\langle\sum_{k}f(T_{k})\right\rangle=\frac{Nl}{L}\int_{0}^{\infty}
dx\,f(1/\cosh^{2}x).\label{fTaverage}
\end{equation}
The result for normal reflection is
\begin{equation}
\left\langle|(s_{\rm ee})_{nm}|^{2}\right\rangle=(1+\delta_{nm})\frac{1}{N}
\left(1-\frac{l}{2L}\right). \label{normal_giant}
\end{equation}
Off-diagonal ($n\neq m$) and diagonal ($n=m$) reflection differ by precisely a
factor of two, just as in the normal state (Mello, Akkermans, and Shapiro,
1988). In contrast, for Andreev reflection we find
\begin{equation}
\left\langle|(s_{\rm he})_{nm}|^{2}\right\rangle=\frac{l}{2NL} \;\;(n\neq
m),\;\; \left\langle|(s_{\rm he})_{nn}|^{2}\right\rangle=\left(\frac{\pi
l}{4L}\right)^{2}.\label{Andreev_giant}
\end{equation}
Off-diagonal and diagonal reflection now differ by an order of magnitude
$Nl/L\simeq G/G_{0}\gg 1$.

Eqs.\ (\ref{normal_giant}) and (\ref{Andreev_giant}) hold for $B=0$. If
time-reversal symmetry is broken (by a magnetic field $B\gtrsim B_{\rm c}\equiv
h/eLW$), then the matrices $U$ and $U'$ are independent. Carrying out the
average over the unitary group in the large-$N$ limit, we find
\begin{equation}
\left\langle|(s_{\rm
ee})_{nm}|^{2}\right\rangle=\frac{1}{N}\left(1-\frac{l}{2L}\right),\;\;
\left\langle|(s_{\rm
he})_{nm}|^{2}\right\rangle=\frac{l}{2NL}.\label{noTRS_giant}
\end{equation}
Diagonal and off-diagonal reflection now occur with the same probability.

In Fig.\ \ref{fig_giant} we compare this theoretical prediction of a giant
backscattering peak with a numerical simulation of the Anderson model. The
results shown are raw data from a single sample. For normal reflection (bottom
panel) the backscattering peak is not visible due to statistical fluctuations
in the reflection probabilities (speckle noise). The backscattering peak for
Andreev reflection is much larger than the fluctuations and is clearly visible
(top panel). A magnetic flux of $10\,h/e$ through the disordered region
completely destroys the peak (middle panel). The arrow in the top panel
indicates the ensemble-averaged peak height from Eq.\ (\ref{Andreev_giant}),
consistent with the simulation within the statistical fluctuations. The peak is
just one mode wide, as predicted by Eq.\ (\ref{Andreev_giant}).\footnote{
The angular reflection distribution follows from the modal distribution in the
large-$N$ limit. For example, in a two-dimensional conductor each transverse
mode $n$ is associated to angles $\pm\theta$ with the normal such that $k_{\rm
F}\sin|\theta|=n\pi/W$. One can distinguish between reflection at $\theta$ and
at $-\theta$ by choosing as a new basis the sum and difference of two adjacent
modes. The enhanced backscattering occurs within $\Delta\theta\simeq 2\pi
/k_{\rm F}W{\rm cos}\,\theta$ around the angle of incidence $\theta$. The
decrease of $\Delta\theta$ with increasing $W$ stops when $W\simeq L$, due to
the breakdown of the isotropy assumption.}

Coherent backscattering in the normal state is intimately related to the
weak-localization correction to the average conductance. We have seen that the
backscattering peak for Andreev reflection is increased by a factor $G/G_{0}$.
However, as was discussed in Sec.\ \ref{NSWL}, the weak-localization correction
in an NS junction remains of order $G_{0}$. The reason is that, according to
Eq.\ (\ref{Gdef}), the conductance
\begin{equation}
G_{\rm NS}=2G_{0}\sum_{n,m}|(s_{\rm he})_{nm}|^{2}\label{GNS_giant}
\end{equation}
contains the sum over all Andreev reflection probabilities, so that the
backscattering peak is averaged out. Indeed, Eqs.\ (\ref{Andreev_giant}) and
(\ref{noTRS_giant}) give the same $\langle G_{\rm NS}\rangle$, up to
corrections smaller by factors $1/N$ and $l/L$. In order to observe the
enhanced backscattering in a transport experiment one has to increase the
sensitivity to Andreev reflection at the angle of incidence. This can be done
by injecting the electrons through a point contact, as we discuss next.

\subsubsection{Conductance doubling}
\label{doubleG}

\begin{figure}[tb]
\centerline{
\psfig{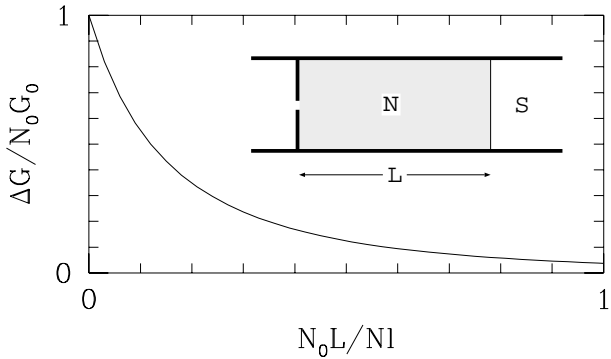}
}%
\medskip
\caption[]{
Excess conductance $\Delta G=\langle G(B=0)\rangle -\langle G(B\gtrsim B_{\rm
c})\rangle$ of a point contact in series with a disordered NS junction (inset),
computed from Eqs.\ (\protect\ref{GBzeroab}) and (\protect\ref{GBc}). At $B=0$
the contact conductance is twice the Sharvin conductance $N_{0}G_{0}$, provided
$N_{0}L/Nl\ll 1$. After Beenakker, Melsen, and Brouwer (1995).
}\label{fig_doubleG}
\end{figure}

The point-contact geometry is shown in the inset of Fig.\ \ref{fig_doubleG}.
The point contact contains $N_{0}$ transverse modes, and the disordered region
between point contact and superconductor contains $N$ transverse modes. The
disordered region has length $L$ and mean free path $l$. We assume ballistic
motion through the point contact, which requires that its width is much smaller
than $l$. (For the opposite regime of diffusive motion through the point
contact, see Volkov, 1994.) Furthermore, we assume that both $N_{0}$ and $Nl/L$
are $\gg 1$, so that the conductance $G_{\rm NS}$ of the system is much greater
than the conductance quantum $G_{0}\equiv 2e^{2}/h$.

In zero magnetic field we can compute the average conductance from Eq.\
(\ref{keyzero}),
\begin{equation}
\langle G_{\rm NS}\rangle=2G_{0}\int_{0}^{\infty}\!dx\,\rho(x,L)
\frac{1}{\cosh^{2}2x},
\end{equation}
in the parameterization $T=1/\cosh^{2}x$. The $L$-dependence of the density
$\rho(x,L)$ follows from Eqs.\ (\ref{Eulersol}) and (\ref{rhoxU}), with the
initial condition $U_{0}(\zeta)=(N_{0}/N)\,{\rm cotanh}\,\zeta$ corresponding
to a point contact. The resulting conductance is
\begin{mathletters}
\label{GBzeroab}
\begin{eqnarray}
&&\langle G_{\rm NS}\rangle=G_{0}[\case{1}{2}(1+\sin\theta)/N_{0}+L/Nl\,]^{-1},
\label{GBzeroa}\\
&&\case{1}{2}\theta(1+\sin\theta)=(N_{0}L/Nl)\cos\theta,\;\;
\theta\in(0,\pi/2).\label{GBzerob}
\end{eqnarray}
\end{mathletters}%
The implicit equation (\ref{GBzeroab}) has the two limiting solutions
\begin{equation}
\langle G_{\rm NS}\rangle=\left\{
\begin{array}{ll}
G_{0}(1/2N_{0}+L/Nl)^{-1}&{\rm if}\;\;N_{0}L/Nl\ll 1,\\
G_{0}(1/N_{0}+L/Nl)^{-1}&{\rm if}\;\;N_{0}L/Nl\gg 1.
\end{array}\right.\label{GBzeroablimits}
\end{equation}
The contribution from disorder remains the same in the two limits, but the
contribution from the point contact differs by a factor of two.

These results hold in zero magnetic field. A magnetic field $B$ greater than
$B_{\rm c}=h/eLW$ effectively breaks time-reversal symmetry. Instead of Eqs.\
(\ref{GBzeroab}) and (\ref{GBzeroablimits}) one then has
\begin{equation}
\langle G_{\rm NS}\rangle=G_{0}(1/N_{0}+L/Nl)^{-1},\label{GBc}
\end{equation}
for $1\ll N_{0}\ll N$ and $l\ll L\ll Nl$ --- but regardless of the ratio
$N_{0}L/Nl$. Equation (\ref{GBc}) is just the classical addition in series of
the Sharvin conductance $N_{0}G_{0}$ of the point contact and the Drude
conductance $(Nl/L)G_{0}$ of the disordered region. The same equation
(\ref{GBc}) applies if a voltage $V$ greater than the Thouless energy $E_{\rm
c}=\hbar D/L^{2}$ breaks the electron-hole degeneracy. If both $B\ll B_{\rm c}$
and $eV\ll E_{\rm c}$, in contrast, the contribution from the point contact
depends on the ratio $N_{0}L/Nl$, according to Eq.\ (\ref{GBzeroablimits}).

In Fig.\ \ref{fig_doubleG} we show the difference $\Delta G=\langle G_{\rm
NS}(B=0)\rangle-\langle G_{\rm NS}(B\gtrsim B_{\rm c})\rangle$ of Eqs.\
(\ref{GBzeroab}) and (\ref{GBc}). If $N_{0}/N\ll l/L\ll 1$ the conductance
drops from $2N_{0}G_{0}$ to $N_{0}G_{0}$ upon breaking time-reversal symmetry.
As discussed at the beginning of this subsection, a doubling of the contact
conductance at $B=0$ is a classical effect in {\em ballistic\/} NS junctions
($l\gg L$): An electron injected towards the superconductor is reflected back
as a hole, doubling the current through the point contact. We now understand
that the conductance doubling can survive multiple scattering in a {\em
diffusive\/} junction ($l\ll L$), because of the enhanced backscattering at the
angle of incidence. The difference between ballistic and diffusive junctions
appears in the width of the conductance peak around $B,V=0$. For a ballistic
junction the width in magnetic field is $mv_{\rm F}/eL$ (determined by the
curvature of the electron trajectories), and the width in voltage is the
superconducting energy gap ${\Delta}$. These values are much greater than the
values $B_{\rm c}$ and $E_{\rm c}$ for a diffusive junction. Experiments on the
conductance doubling have been done by Van Son, Van Kempen, and Wyder (1987,
1988). The anomalously narrow conductance peak reported in their 1988 paper may
well be due to the effects of disorder discussed here.

\subsection{Chaotic Josephson junction}
\label{Jjunction}

A Josephson junction is a weak link between two superconductors. The weak link
could be a tunnel barrier, a point contact, or a piece of normal metal. (For a
review of Josephson junctions, see Likharev, 1979.) In this subsection we will
consider the special case that the weak link consists of a small
(phase-coherent) metal grain (a ``quantum dot''). A random-matrix theory of
induced superconductivity (``proximity effect'') in such a system can be
constructed, based on the assumption that the classical motion in the quantum
dot is chaotic (Altland and Zirnbauer, 1996a, 1996b; Frahm {\em et al.}, 1996;
Melsen {\em et al.}, 1996). A phase difference $\phi$ between the
superconductors induces a current through the junction. This current flows in
equilibrium; It is a thermodynamic property of the system, and as such falls
outside the scope of this review (we refer to Brouwer and Beenakker, 1997b).

\begin{figure}[tb]
\centerline{
\psfig{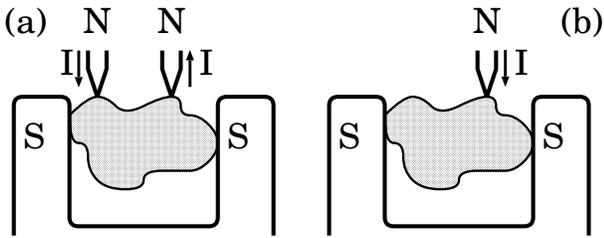}
}%
\bigskip
\caption[]{
Chaotic Josephson junction in a four-terminal (a) and three-terminal (b)
configuration. The three-terminal configuration is equivalent as a circuit to
the fork junction of Fig.\ \protect\ref{forkcircuit}. After Brouwer and
Beenakker (1996b).
}\label{fig_chaoticJJ}
\end{figure}

The problem considered here is the injection of non-equilibrium quasiparticles
into the Josephson junction. The system is shown schematically in Fig.\
\ref{fig_chaoticJJ}a. A quantum dot is contacted by four ballistic point
contacts (with $N_{i}$ modes transmitted through contact $i=1,2,3,4$). The
classical motion in the quantum dot should be chaotic on time scales greater
than $\tau_{\rm ergodic}$ and the point contacts should be sufficiently small
that the dwell time $\tau_{\rm dwell}\gg\tau_{\rm ergodic}$ ({\em cf.} Sec.\
\ref{dots}). The quantum dot forms a Josephson junction in a superconducting
ring. Coupling to the two superconducting banks is via point contacts 3 and 4
(phase difference $\phi$, same voltage). Contacts 1 and 2 are connected to
normal metals (voltage difference $V$). A current $I$ is passed between
contacts 1 and 2 and one measures the conductance $G=I/V$ as a function of
$\phi$.

Spivak and Khmel'nitski\u{\i} (1982) and Altshuler, Khmel'nitski\u{\i}, and
Spivak (1983) computed the ensemble average $\langle G(\phi)\rangle$ in the
high-temperature regime $k_{\rm B}T\gg\hbar/\tau_{\rm dwell}$. (For more recent
work in this regime, see Zhou and Spivak, 1996.) They obtained $\pi$-periodic
oscillations with an amplitude of order $G_{0}$. In experiments ({\em e.g.} by
De Vegvar {\em et al.}, 1994) these are obscured by $2\pi$-periodic
sample-specific fluctuations of the same order of magnitude (Altshuler and
Spivak, 1987). At low temperatures the fundamental periodicity of the
oscillations in $\langle G(\phi)\rangle$ doubles, and their amplitude increases
to become much greater than $G_{0}$ (Za\u{\i}tsev, 1994; Beenakker, Melsen, and
Brouwer, 1995; Kadigrobov {\em et al.}, 1995; Allsopp {\em et al.}, 1996;
Volkov and Za\u{\i}tsev, 1996; Claughton, Raimondi, and Lambert, 1996).
Experiments by Petrashov {\em et al.} (1995) (and similar measurements by
Courtois {\em et al.}, 1996) showed such giant conductance oscillations, but
these are now believed to have been caused by the thermal effect of Sec.\
\ref{GNSmean} (Nazarov and Stoof, 1996). The sample-specific fluctuations
remain of order $G_{0}$ at low temperatures, and have been studied
experimentally (Den Hartog {\em et al.}, 1996) and theoretically (Brouwer and
Beenakker, 1996b).

In the first part of this subsection we review the theory of the
low-temperature oscillations in the ensemble-averaged conductance.
Sample-specific fluctuations at low temperatures are discussed in the second
part.

\subsubsection{Average conductance}
\label{GaverageJJ}

We have discussed conductance oscillations before, in the three-terminal fork
junction of Sec.\ \ref{circuit}. The four-terminal Josephson junction
considered here differs from the three-terminal configuration (shown in Fig.\
\ref{fig_chaoticJJ}b) in the following respect: In the three-terminal
configuration the current flows from a normal metal reservoir into a
superconducting reservoir, whereas in the four-terminal configuration the
current flows between two normal metal reservoirs. The four-terminal
configuration shows the phase-coherent effects in a ``cleaner'' way, because
without phase coherence in the normal metal the superconductor would have no
effect at all on the conductance. In the three-terminal configuration, in
contrast, there is an effect on the conductance because of the excitation gap
in the bulk superconductor --- even in the absence of any phase coherence
between electrons and holes in the normal metal.

The matrices $s_{\rm ee}$ and $s_{\rm he}$ [with elements $(s_{\rm
ee})_{ij,nm}$ and $(s_{\rm he})_{ij,nm}$] contain the combined effect of
scattering in the quantum dot (described by the matrix $S$) and Andreev
reflection at the two contacts with the superconductor. The scattering matrix
$S$ of the quantum dot has submatrices $s_{ij}$, the matrix element $s_{ij,nm}$
being the scattering amplitude from mode $m$ in contact $j$ to mode $n$ in
contact $i$. By summing a series of multiple Andreev reflections we obtain for
$s_{\rm ee}$ and $s_{\rm he}$ expressions analogous to Eq.\ (\ref{sehheeehh}),
\begin{mathletters}
\label{rsrelation2}
\begin{eqnarray}
&&s_{\rm ee}=a-b\,\Omega c^{\ast}\Omega^{\ast}(1+c\,\Omega
c^{\ast}\Omega^{\ast})^{-1}d,\label{ree2}\\
&&s_{\rm he}=-{\rm i}b^{\ast}\Omega^{\ast}(1+c\,\Omega
c^{\ast}\Omega^{\ast})^{-1}d,\label{rhe2}
\end{eqnarray}
\end{mathletters}%
where we have abbreviated
\begin{eqnarray*}
&&a=
{\renewcommand{\arraystretch}{0.6}
\left(\begin{array}{cc}
s_{11}&s_{12}\\s_{21}&s_{22}
\end{array}\right)},\;\;
b=
{\renewcommand{\arraystretch}{0.6}
\left(\begin{array}{cc}
s_{13}&s_{14}\\s_{23}&s_{24}
\end{array}\right)},\;\;
c=
{\renewcommand{\arraystretch}{0.6}
\left(\begin{array}{cc}
s_{33}&s_{34}\\s_{43}&s_{44}
\end{array}\right)},\nonumber\\
&&d=
{\renewcommand{\arraystretch}{0.6}
\left(\begin{array}{cc}
s_{31}&s_{32}\\s_{41}&s_{42}
\end{array}\right)},\;\;
\Omega=
{\renewcommand{\arraystretch}{0.6}
\left(\begin{array}{ll}
{\rm e}^{{\rm i}\phi/2}&0\\0&{\rm e}^{-{\rm i}\phi/2}
\end{array}\right)}.
\end{eqnarray*}
The four-terminal generalization of Eq.\ (\ref{Gdef}) is (Lambert, 1991, 1993;
Lambert, Hui, and Robinson, 1993)
\begin{mathletters}
\label{GRrelation}
\begin{eqnarray}
&&G/G_{0}=R_{21}^{\rm ee}+R_{21}^{\rm he}+\frac{2(R_{11}^{\rm he}R_{22}^{\rm
he}-R_{12}^{\rm he}R_{21}^{\rm he})}{R_{11}^{\rm he}+R_{22}^{\rm
he}+R_{12}^{\rm he}+R_{21}^{\rm he}},\label{GRrelationa}\\
&&R_{ij}^{\rm ee}=\sum_{n,m}|(s_{\rm ee})_{ij,nm}|^{2},\;\; R_{ij}^{\rm
he}=\sum_{n,m}|(s_{\rm he})_{ij,nm}|^{2}.\label{GRrelationb}
\end{eqnarray}
\end{mathletters}%

\begin{figure}[tb]
\centerline{
\psfig{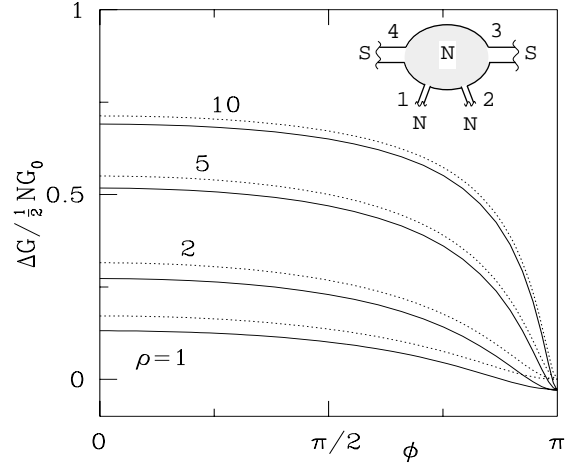}
}%
\medskip
\caption[]{
Excess conductance $\Delta G=\langle G\rangle_{\rm COE}$ $\mbox{}-\langle
G\rangle_{\rm CUE}$ of a chaotic four-terminal Josephson junction (inset). The
solid curves are computed from Eqs.\ (\protect\ref{rsrelation2}) and
(\protect\ref{GRrelation}) for $N_{1}=N_{2}\equiv N$, $N_{3}=N_{4}\equiv\rho
N$, with $N=10$. The dotted curves are the large-$N$ limit
(\protect\ref{GCOEab}). The excess conductance at $\phi=0$ is a factor
$G/G_{0}={\cal O}(N)$ larger than the negative weak-localization correction at
$\phi=\pi$. After Beenakker, Melsen, and Brouwer (1995).
}\label{fig_giantosc}
\end{figure}

We evaluate the average conductance $\langle G\rangle$ by averaging $S$ over
the circular ensemble ({\em cf.} Sec.\ \ref{circular}). At $B=0$ this means
that $S=UU^{\rm T}$ with $U$ uniformly distributed in the group ${\cal U}(N)$
of $N\times N$ unitary matrices ($N=\sum_{i=1}^{4}N_{i}$). This is the circular
orthogonal ensemble (COE). If time-reversal symmetry is broken, then $S$ itself
is uniformly distributed in ${\cal U}(N)$. This is the circular unitary
ensemble (CUE). In the CUE the average can be done analytically for any $N_{i}$
and $\phi$. The result is
\begin{equation}
\langle G\rangle_{\rm CUE}=G_{0}\,\frac{N_{1}N_{2}}{N_{1}+N_{2}},\label{GCUE}
\end{equation}
independent of $\phi$. In the COE one can do the average numerically, by
generating a large number of random matrices in ${\cal U}(N)$. An analytical
result can be obtained for $N\gg 1$. The easiest way to do this is to use
Nazarov's circuit theory, described in Sec.\ \ref{circuit}. The result for the
symmetric case $N_{1}=N_{2}\gg 1$, $N_{3}=N_{4}\equiv\rho N_{1}$ is given by
\begin{mathletters}
\label{GCOEab}
\begin{eqnarray}
&&\langle G\rangle_{\rm COE}=\frac{N_{1}G_{0}}{1+\cos\theta},\label{GCOEa}\\
&&\frac{\sin\theta+\sin^{2}\theta\,\cos\case{1}{2}\phi}
{\cos\theta+\cos^{2}\theta}=\rho\cos\case{1}{2}\phi,\;\;\theta\in(0,\pi/2).
\label{GCOEb}
\end{eqnarray}
\end{mathletters}%
In Fig.\ \ref{fig_giantosc} we show the excess conductance $\Delta G=\langle
G\rangle_{\rm COE}-\langle G\rangle_{\rm CUE}$ as a function of $\phi$. For
$N_{i}\gtrsim 10$ the numerical finite-$N$ curves (solid) are close to the
analytical large-$N$ limit (\ref{GCOEab}) (dotted). The excess conductance
$\Delta G$ is of order $N_{1}G_{0}$ and positive --- except for $\phi$ close to
$\pi$, where a small negative weak-localization correction of order $G_{0}$
appears.

The excess conductance is a manifestation of the giant backscattering peak of
Sec.\ \ref{giantpeak}. To see this, note that $\langle R_{12}^{\rm
he}\rangle=\langle R_{21}^{\rm he}\rangle$, $\langle R_{11}^{\rm
he}\rangle=\langle R_{22}^{\rm he}\rangle$. (For simplicity, we assume again
the symmetric case $N_{1}=N_{2}$.) Current conservation requires $R_{11}^{\rm
he}+R_{21}^{\rm he}+R_{11}^{\rm ee}+R_{21}^{\rm ee}=N_{1}$. For $N\gg 1$ we may
replace $\langle f(R_{ij})\rangle$ by $f(\langle R_{ij}\rangle)$. The average
of Eq.\ (\ref{GRrelation}) then becomes
\begin{equation}
\langle G/G_{0}\rangle=\case{1}{2}N_{1}-\case{1}{2}\langle R_{11}^{\rm
ee}-R_{21}^{\rm ee}\rangle+\case{1}{2}\langle R_{11}^{\rm he}-R_{21}^{\rm
he}\rangle.\label{Gsimplified}
\end{equation}
The first term $\case{1}{2}N_{1}$ is the classical series conductance. The
second term is the weak-localization correction due to enhanced backscattering
for normal reflection. Since $\langle R_{11}^{\rm ee}-R_{21}^{\rm
ee}\rangle={\cal O}(1)$ this negative correction to $\case{1}{2}N_{1}$ can be
neglected if $N\gg 1$. The third term gives the excess conductance due to
enhanced backscattering for Andreev reflection. Since $\langle R_{11}^{\rm
he}-R_{21}^{\rm he}\rangle={\cal O}(N)$ this positive contribution is a factor
$G/G_{0}={\cal O}(N)$ greater than the negative weak-localization correction.

\subsubsection{Conductance fluctuations}
\label{GfluctuationsJJ}

\begin{figure}[tb]
\centerline{
\psfig{figure=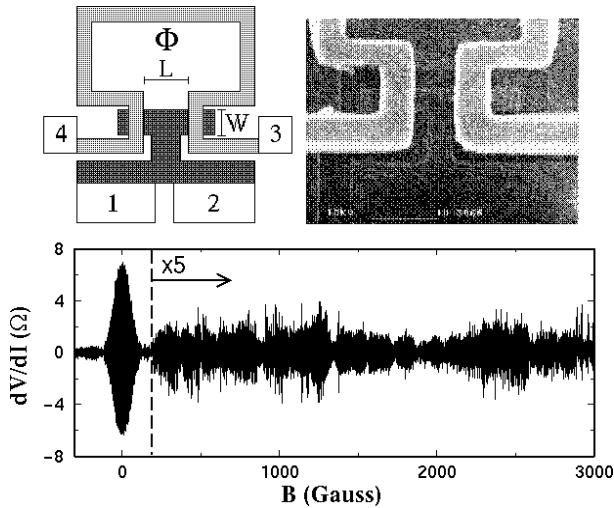,width= 8cm}
}%
\medskip
\caption[]{
Three-terminal Josephson junction. The upper left panel shows a schematic
picture of a T-shaped 2D electron gas beneath a Nb loop. The upper right panel
shows a scanning electron micrograph of the actual device. The dimensions are
$L=0.7\,\mu{\rm m}$, $W=0.3\,\mu{\rm m}$. The current $I$ flows from contacts
1,2 (connected to the 2D electron gas) to contacts 3,4 (connected to the
superconducting Nb). The voltage $V$ is measured between contacts 1,2 and 3,4.
The flux $\Phi\approx B\times 10.3\,\mu{\rm m}^{2}$ through the loop determines
the phase difference $\phi=(4\pi e/h)\Phi$ of contacts 3 and 4. The
magneto-resistance oscillations at $50\,{\rm mK}$ are plotted in the lower
panel, after subtraction of a $\phi$-independent background. The amplitude near
$B=0$ corresponds to a conductance of $0.1\,e^{2}/h$. The amplitude for
$B\gtrsim h/eLW\equiv B_{\rm c}$ is strongly suppressed, but small oscillations
remain (amplitude $\approx 0.005\,e^{2}/h$). The envelope of these small
oscillations fluctuates randomly on the scale of $B_{\rm c}$. After Den Hartog
{\em et al.} (1996).
}\label{fig_hartog}
\end{figure}

The conductance of the Josephson junction contains two types of sample-specific
fluctuations: Aperiodic fluctuations as a function of the magnetic field $B$
and $2\pi$-periodic fluctuations as a function of the superconducting phase
difference $\phi$. To observe the fluctuations in $G(B,\phi)$, the magnetic
field should be sufficiently large to break time-reversal symmetry, otherwise
the fluctuations will be obscured by the much stronger $B$- and
$\phi$-dependence of the ensemble average. Den Hartog {\em et al.} (1996) have
reported the experimental observation of phase-dependent magnetoconductance
fluctuations in a T-shaped two-dimensional electron gas (see Fig.\
\ref{fig_hartog}). The horizontal arm of the T is connected to two
superconductors, the vertical arm to a normal metal reservoir. The observed
magnitude of the fluctuations was much smaller than $e^2/h$, presumably because
the motion in the T-junction was nearly ballistic. Larger fluctuations are
expected if the arms of the T are closed, leaving only a small opening (a point
contact) for electrons to enter or leave the junction. Motion in the junction
can be ballistic or diffusive, as long as it is chaotic the statistics of the
conductance fluctuations will only depend on the number of modes in the point
contacts and not on the microscopic details of the junction. We review the
theory of universal conductance fluctuations in a chaotic Josephson junction,
following Brouwer and Beenakker (1996b).

For the conductance fluctuations there is no essential difference between the
three- and four-terminal configurations of Fig.\ \ref{fig_chaoticJJ}. We focus
on the three-terminal configuration, because it corresponds to the experiment
of Den Hartog {\em et al.} We assume that the two point contacts to the
superconductor contain $N_{3}=N_{4}$ modes each, and denote by $N_{1}$ the
number of modes in the contact to the normal metal. (There is only one such
contact, so $N_{2}\equiv 0$.) The total number of modes in the three point
contacts is $N=N_{1}+2N_{3}$. There are two regimes, depending on the relative
magnitude of $N_{1}$ and $N_{3}$. For $N_{1}\ll N_{3}$ the $\phi$-dependence of
the conductance is strongly anharmonic. (This is the regime studied by
Altshuler and Spivak, 1987.) For $N_{1}\gtrsim N_{3}$ the oscillations are
nearly sinusoidal, as observed by Den Hartog {\em et al.} The difference
between the two regimes can be understood qualitatively in terms of interfering
Feynman paths. In the second regime, only paths with a single Andreev
reflection contribute to the conductance. Each such path depends on $\phi$ with
a phase factor ${\rm e}^{\pm{\rm i}\phi/2}$. Interference of these paths yields
a sinusoidal $\phi$-dependence of the conductance. In the first regime,
quasiparticles undergo many Andreev reflections before leaving the junction.
Hence higher harmonics appear, and the conductance becomes a random
$2\pi$-periodic function of $\phi$.

The conductance $G(B,\phi)=G_{0}(B)+G_{\phi}(B,\phi)$ consists of a
$\phi$-independent background
\begin{equation}
G_{0}(B)=\int_{0}^{2\pi}\frac{d\phi}{2\pi}\,G(B,\phi),\label{Gbackground}
\end{equation}
plus $2\pi$-periodic fluctuations $G_{\phi}$. In the absence of time-reversal
symmetry, the ensemble average $\langle G(B,\phi)\rangle\equiv\langle G\rangle$
is independent of $B$ and $\phi$. Hence $\langle G_0(B)\rangle=\langle
G\rangle$ and $\langle G_{\phi}(B,\phi)\rangle= 0$. The correlator of $G$ is
\begin{equation}
C(\delta B,\delta \phi)=\langle G(B,\phi)G(B+\delta
B,\phi+\delta\phi)\rangle-\langle G\rangle^{2}.\label{Gcorrelator1}
\end{equation}
Fluctuations of the background conductance are described by the correlator of
$G_{0}$,
\begin{eqnarray}
C_{0}(\delta B)&=&\langle G_{0}(B)G_{0}(B+\delta B)\rangle -\langle
G\rangle^{2}\nonumber\\
&=&\int_{0}^{2\pi}\frac{d\delta\phi}{2\pi}\,C(\delta B,\delta
\phi).\label{Gcorrelator2}
\end{eqnarray}
(In the second equality we used that $\langle G_{\phi}G_{0}\rangle=0$.) The
difference $C_{\phi}=C-C_{0}$ is the correlator of $G_{\phi}$,
\begin{equation}
C_{\phi}(\delta B,\delta\phi)=\langle G_{\phi}(B,\phi)G_{\phi}(B+\delta
B,\phi+\delta\phi)\rangle.\label{Gcorrelator3}
\end{equation}

\begin{figure}[tb]
\centerline{
\psfig{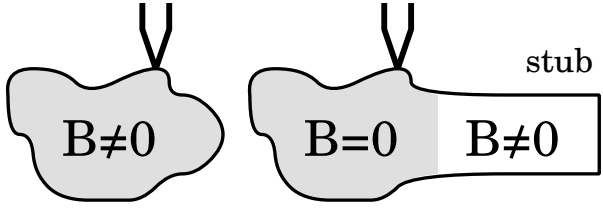}
}%
\bigskip
\caption[]{
Schematic picture, showing how a magnetic field can be included in the
scattering-matrix ensemble. A chaotic cavity with a spatially homogeneous
magnetic field (left diagram) is statistically equivalent to a chaotic cavity
in zero magnetic field (right diagram), which is coupled to a closed lead (a
stub) having a non-symmetric reflection matrix. After Brouwer and Beenakker
(1996b).
}\label{fig_stub}
\end{figure}

For chaotic scattering without time-reversal symmetry, the scattering matrix
$S$ is distributed according to the circular unitary ensemble (CUE). The CUE
does not specify how $S$ at different values of $B$ is correlated. There exists
a method to extend the CUE, such that it includes the parametric dependence of
the scattering matrix on the magnetic field (Brouwer, 1997). The method
consists in replacing the magnetic field by a time-reversal-symmetry breaking
stub (see Fig.\ \ref{fig_stub}). This idea is similar in spirit to
B\"{u}ttiker's method (1986a, 1988a) of modeling inelastic scattering by a
phase-breaking lead. The stub contains $N_{\rm stub}$ modes. The end of the
stub is closed, so that it conserves the number of particles without breaking
phase coherence. (B\"{u}ttiker's lead, in contrast, is attached to a reservoir,
which conserves the number of particles by matching currents, not amplitudes,
and therefore breaks phase coherence.) We choose the scattering basis such that
the $N_{\rm stub} \times N_{\rm stub}$ reflection matrix $r_{\rm stub}(B)$ of
the stub equals the unit matrix at $B=0$. For non-zero magnetic fields we take
\begin{equation}
r_{\rm stub}(B)={\rm e}^{BA},\;a^{2}\equiv\sum_{n< m}A_{nm}^{2},\label{eq:VB}
\end{equation}
where the matrix $A$ is real and antisymmetric:
$A_{nm}^{\vphantom{*}}=A_{nm}^{*} =-A_{mn}^{\vphantom{*}}$. Particle-number is
conserved by the stub because $r_{\rm stub}$ is unitary, but time-reversal
symmetry is broken, because $r_{\rm stub}$ is not symmetric if $B\neq 0$. In
order to model a spatially homogeneous magnetic field, it is essential that
$N_{\rm stub} \gg N$. The value of $N_{\rm stub}$ and the precise choice of $A$
are irrelevant, all results depending only on the single parameter $a$.

The magnetic-field dependent scattering matrix $S(B)$ in this model takes the
form
\begin{equation}
S(B)=U_{11}+U_{12}\left[1-r_{\rm stub}(B)U_{22}\right]^{-1}r_{\rm
stub}(B)U_{21}.\label{eq:CUEB}
\end{equation}
The matrices $U_{ij}$ are the four blocks of a matrix $U$ representing the
scattering matrix of the quantum dot at $B=0$, with the stub replaced by a
regular lead. The distribution of $U$ is the circular orthogonal ensemble
(COE). The distribution of $S(B)$ resulting from Eqs.\ (\ref{eq:VB}) and
(\ref{eq:CUEB}) crosses over from the COE for $B=0$ to the CUE for $B\gg B_{\rm
c}$. It is equivalent to the distribution of scattering matrices following from
the Pandey-Mehta Hamiltonian (\ref{PMH}). The parameter $a$ is related to the
parameters $\alpha$ and $M$ in Eq.\ (\ref{PMH}) by $Ba=\alpha\sqrt{2M}$. (The
relationship between $\alpha$,$M$ and microscopic properties of the quantum dot
was discussed in Sec.\ \ref{transitions}.) The characteristic magnetic field
$B_{\rm c}$ for breaking of time-reversal symmetry is $B_{\rm
c}=a^{-1}\sqrt{N}$.

\begin{figure}[tb]
\centerline{
\psfig{figure=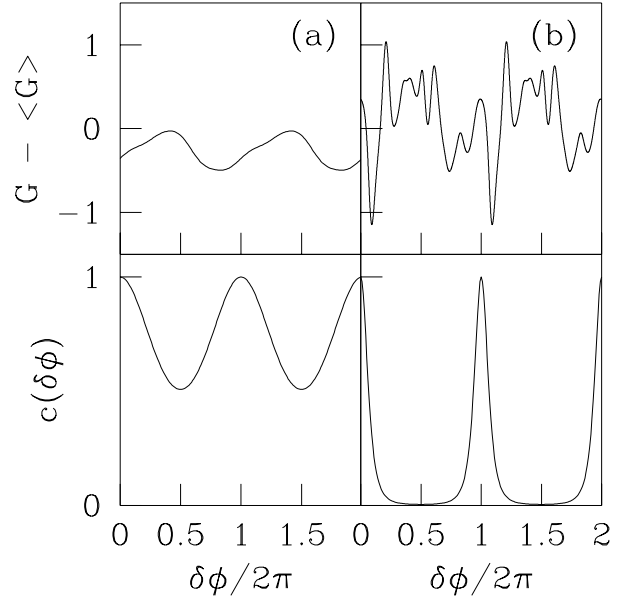,width=
8cm,bbllx=67pt,bblly=160pt,bburx=528pt,bbury=618pt}
}%
\medskip
\caption[]{
Conductance fluctuations in a Josephson junction as a function of the phase
difference between the superconductors, computed for a three-terminal
configuration without time-reversal symmetry. Top panels: Conductance minus the
ensemble average (in units of $2e^{2}/h$);  Bottom panels: Normalized
correlator $c(\delta\phi)=C(0,\delta\phi)/C(0,0)$, with $C(\delta
B,\delta\phi)$ defined in Eq.\ (\protect\ref{Gcorrelator1}). The parameters for
column (a) are $N_{1}=120$, $N_{3}=N_{4}=30$; The parameters for column (b) are
$N_{1}=10$, $N_{3}=N_{4}=80$. ($N_{1}$ denotes the number of modes coupling to
the normal metal, and $N_{3}+N_{4}$ is the total number of modes coupling to
the superconductors.) After Brouwer and Beenakker (1996b).
}\label{fig_correlator}
\end{figure}

The correlator of the conductance can now be calculated by averaging $U$ over
the COE. This can be done perturbatively if $N_{1}$ and $N_{3}$ are both $\gg
1$, for any ratio of $N_{1}$ and $N_{3}$. The result for $N_{1}\gg N_{3}$ is
\begin{mathletters}
\label{eq:NS1}
\begin{eqnarray}
C_{0}(\delta B)&=&\frac{96(N_{3}/N_{1})^{2}}{[1+(\delta B/B_{\rm
c})^{2}]^{2}},\label{eq:NS1a}\\
C_{\phi}(\delta B,\delta\phi)&=&\case{1}{3}C_{0}(\delta
B)\cos\delta\phi,\label{eq:NS1b}
\end{eqnarray}
\end{mathletters}%
whereas for $N_{1}\ll N_{3}$ one has
\begin{mathletters}
\label{eq:NS2}
\begin{eqnarray}
C_{0}&=&\frac{1}{4}\sqrt{\frac{N_{1}}{N_{3}}}\left[1+\frac{2N_{3}}{N_{1}}
\left(\frac{\delta B}{B_{\rm c}}\right)^{2}\right]^{-3/2},\label{eq:NS2a}\\
C_{\phi}&=&\frac{1}{2}\left[1+\frac{2N_{3}}{N_{1}}\left(\frac{\delta B}{B_{\rm
c}}\right)^{2}+\frac{N_{3}}{4N_{1}}(\delta\phi)^{2}\right]^{-2}.\label{eq:NS2b}
\end{eqnarray}
\end{mathletters}%
The difference between the two limiting regimes is illustrated in Fig.\
\ref{fig_correlator}. The ``sample-specific'' curves in the upper panels were
computed by randomly drawing a matrix $S$ from the CUE. The correlators in the
lower panels were computed using large-$N$ perturbation theory. The qualitative
difference between $N_{1}\gtrsim N_{3}$ (Fig.\ \ref{fig_correlator}a) and
$N_{1}\ll N_{3}$ (Fig.\ \ref{fig_correlator}b) is clearly visible.

\subsection{Shot noise}
\label{shotnoiseNS}

If the transmission of an elementary charge $e$ can be regarded as a sequence
of uncorrelated events, then the shot-noise power $P$ equals the value
$2eI\equiv P_{\rm Poisson}$ of a Poisson process (see Sec.\ \ref{shotnoiseN}).
In this subsection we discuss the enhancement of shot noise in an NS junction,
following De Jong and Beenakker (1994). The enhancement originates from the
fact that the current in the superconductor is carried by Cooper pairs in units
of $2e$. However, as we will see, a simple factor-of-two enhancement applies
only in certain limiting cases.

In the normal state, the shot-noise power (at zero temperature and
infinitesimal applied voltage) is given by (B\"{u}ttiker, 1990)
\begin{equation}
P_{\rm N}=P_{0}{\rm Tr}\,tt^{\dagger}(1-tt^{\dagger})
=P_{0}\sum_{n=1}^{N}T_{n}(1-T_{n}),\label{e2}
\end{equation}
with $P_{0}\equiv 2eV(2e^{2}/h)$. Closed ($T_{n}=0$) as well as open
($T_{n}=1$) scattering channels do not fluctuate and therefore give no
contribution to the shot noise. The analogue of Eq.\ (\ref{e2}) for the
shot-noise power of an NS junction is
\begin{equation}
P_{\rm NS}=4P_{0} {\rm Tr}\,
s_{\rm he}^{\vphantom{\dagger}}s_{\rm he}^{\dagger}
(1-s_{\rm he}^{\vphantom{\dagger}}s_{\rm he}^{\dagger})=P_{0}\sum_{n=1}^{N}
\frac{16T_{n}^{2}(1 - T_{n})}{(2-T_{n})^{4}},\label{e17}
\end{equation}
where we have used Eq.\ (\ref{sehheeehh}) (with $\varepsilon=0$) to relate the
scattering matrix $s_{\rm he}$ for Andreev reflection to the transmission
eigenvalues $T_{n}$ of the normal region. This requires zero magnetic field. As
in the normal state, scattering channels which have $T_{n}=0$ or $T_{n}=1$ do
not contribute to the shot noise. However, the way in which partially
transmitting channels contribute is entirely different from the normal state
result (\ref{e2}).

Consider first an NS junction without disorder, but with an arbitrary
transmission probability $\Gamma$ per mode of the interface. In the normal
state, Eq.\ (\ref{e2}) yields $P_{\rm N}=(1-\Gamma)P_{\rm Poisson}$, implying
full Poisson noise for a high tunnel barrier ($\Gamma\ll 1$). For the NS
junction we find from Eq.\ (\ref{e17})
\begin{equation}
P_{\rm NS}=P_{0}N\frac{16\Gamma^{2}(1-\Gamma)}{(2-\Gamma)^{4}}=
\frac{8(1-\Gamma)}{(2-\Gamma)^{2}}P_{\rm Poisson},\label{e18}
\end{equation}
where in the second equality we have used Eq.\ (\ref{keyzero}). This agrees
with results obtained by Khlus (1987), Muzykantski\u{\i} and Khmel'nitski\u{\i}
(1994), Martin (1996), and Anantram and Datta (1996), using different methods.
If $\Gamma<2(\sqrt{2}-1)\approx 0.83$, one observes a shot noise above the
Poisson noise. For $\Gamma\ll 1$ one has
\begin{equation}
P_{\rm NS}=4eI=2 P_{\rm Poisson},\label{e19}
\end{equation}
which is a doubling of the shot-noise power divided by the current with respect
to the normal-state result. This can be interpreted as uncorrelated current
pulses of $2e$-charged particles.

Consider next an NS junction with a disordered normal region, but with an ideal
interface ($\Gamma=1$). We may then apply the formula (\ref{fTaverage}) for the
average of a linear statistic on the transmission eigenvalues to Eqs.\
(\ref{keyzero}) and (\ref{e17}). The result is
\begin{equation}
\frac{\left\langle P_{\rm NS}\right\rangle}
{\left\langle G_{\rm NS}\right\rangle}=
\frac{2}{3}\,\frac{P_{0}}{2e^{2}/h}\;\Rightarrow\;\left\langle P_{\rm
NS}\right\rangle=
{\textstyle\frac{4}{3}}eI={\textstyle\frac{2}{3}}P_{\rm Poisson}.
\label{e20a}
\end{equation}
Equation (\ref{e20a}) is twice the result in the normal state, but still
smaller than the Poisson noise. Corrections to (\ref{e20a}) are of lower order
in $N$ and due to quantum-interference effects.

\begin{figure}[tb]
\centerline{
\psfig{figure=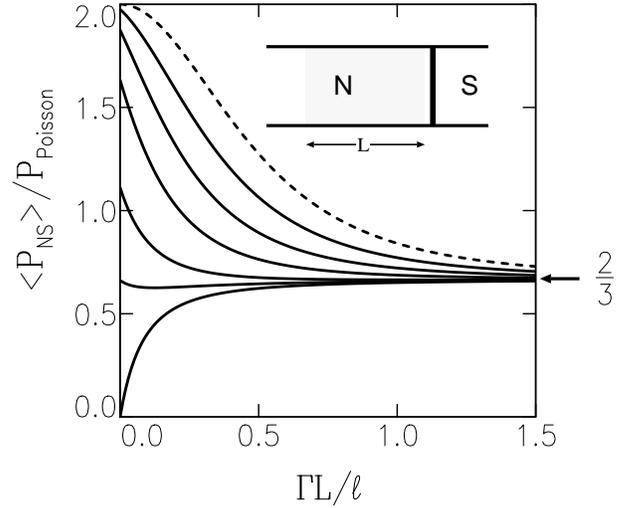,width= 8cm}
}%
\medskip
\caption[]{
The shot-noise power of an NS junction (in units of $P_{\rm Poisson}\equiv
2eI$) as a function of the length $L$ (in units of $l/\Gamma$), for barrier
transparencies  $\Gamma= 1,0.9,0.8,0.6,0.4,0.2$ from bottom to top. The dashed
curve gives the limiting result for $\Gamma\ll 1$. For $L=0$ the noise power
varies as a function of $\Gamma$ according to Eq.\ (\protect\ref{e18}), between
doubled shot noise ($\langle P_{\rm NS}\rangle=4eI$) for a high barrier
($\Gamma\ll 1$) and zero in the absence of a barrier ($\Gamma=1$). For
$L\rightarrow\infty$ the noise power approaches the limiting value $\langle
P_{\rm NS}\rangle=\frac{4}{3}eI$ for each $\Gamma$. After De Jong and Beenakker
(1994).
}\label{fig_shotnoise}
\end{figure}

Finally, consider an NS junction which contains a disordered normal region
(length $L$, mean free path $l$) as well as a non-ideal interface. The scaling
theory of subsection \ref{scaling} can be applied to this case. Results are
shown in Fig.\ \ref{fig_shotnoise}, where $\langle P_{\rm NS}\rangle/P_{\rm
Poisson}$ is plotted against $\Gamma L/l$ for various $\Gamma$. Note the
crossover from the ballistic result (\ref{e18}) to the diffusive result
(\ref{e20a}). For a high barrier ($\Gamma\ll 1$), the shot noise decreases from
twice the Poisson noise to two-thirds the Poisson noise as the amount of
disorder increases.

\section{Conclusion}
\label{conclusion}

We conclude by identifying some open problems and directions for future
research.

\subsection{Higher-dimensional geometries}
\label{higherd}

We have reviewed the random-matrix theory of quantum transport for two
geometries: a quantum dot and a disordered wire. These are, essentially, zero-
and one-dimensional. What about higher dimensionalities? What is the statistics
of the transmission eigenvalues for a two-dimensional thin film or a
three-dimensional cube? Here we summarize what is known.

We recall the result (\ref{rhoTresultP}) for the density of transmission
eigenvalues in the metallic regime,
\begin{equation}
\rho_{0}(T)=\frac{Nl}{2L}\frac{1}{T\sqrt{1-T}}, \label{rhoTresult}
\end{equation}
corresponding to the uniform density (\ref{rho0result}) in the parameterization
$T=1/\cosh^{2}x$. (The density has a cutoff at exponentially small transmission
$T\approx\exp(-2L/l)$, which is irrelevant for transport properties.) The
derivation of Sec.\ \ref{Tdensity} was based on the DMPK equation, and hence
restricted to a wire geometry (length $L$ much greater than width $W$). An
alternative derivation by Nazarov (1994a) shows that the density
(\ref{rhoTresult}) applies also to higher-dimensional geometries in the
metallic regime. Whether or not the conductor is metallic does depend on the
dimensionality. For a wire of length $L$ the condition is $L\ll Nl$, while for
an $L\times L$ square it is $L\ll l\exp(2\pi l/\lambda_{\rm F})$ (Vollhardt and
W\"{o}lfle, 1992). In both cases weak disorder is assumed, meaning that the
mean free path $l$ is much greater than the Fermi wave length $\lambda_{\rm
F}$. An $L\times L\times L$ cube remains in the metallic regime for weak
disorder, regardless of how large $L$ becomes. The dimensionality-independence
of Eq.\ (\ref{rhoTresult}) implies, for example, that the one-third suppression
of the shot-noise discussed in Sec.\ \ref{shotnoiseN} is not restricted to a
wire geometry (Nazarov, 1994a; Altshuler, Levitov, and Yakovets, 1994).

A more general statement is that the non-linear scaling equation
(\ref{rho0eqlambda}) for the eigenvalue density, derived for a wire geometry,
does {\em not\/} in fact require $L\gg W$ for its validity. It is sufficient
that the conductor is in the metallic regime. We do not know of an analytical
proof of this statement, but the numerical evidence for it is quite strong
(Beenakker, Rejaei, and Melsen, 1994). An implication is that the scaling
theory of reflectionless tunneling of Sec.\ \ref{scalingtheory}, which is based
on Eq.\ (\ref{rho0eqlambda}), is not restricted to a wire geometry.

Corrections $\delta\rho(T)$ to the density (\ref{rhoTresult}) due to weak
localization are different for different dimensionalities. We recall the result
(\ref{DELTARHO}) for a wire geometry,
\begin{eqnarray}
\delta\rho(T)&=&\case{1}{4}(1-2/\beta)\bigl[\delta(T-1+0^{+})+
2T^{-1}(1-T)^{-1/2}
\nonumber\\
\mbox{}&&\times(4\ln^{2}[\sqrt{1/T}+\sqrt{1/T-1}]
+\pi^{2})^{-1}\bigr],\label{deltarhoTresult}
\end{eqnarray}
which contains a short-range (delta-function) term at unit transmission plus a
long-range contribution extending down to exponentially small transmission.
Nazarov (1995b) has computed $\delta\rho(T)$ for higher dimensionalities. The
short-range delta-function term is always the same, but the long-range
contribution depends on the geometry. This long-range contribution ensures that
the weak-localization correction to the average conductance acquires the
geometry dependence known from diagrammatic perturbation theory (Lee and
Ramakrishnan, 1985).

Nazarov (1996) has also shown how the geometry dependence of the variance of
the conductance arises from the geometry dependence of the two-point
correlation function $K(T,T')$. For $T,T'$ both close to 1 the correlation
function has the dimensionality-independent form
\begin{eqnarray}
&&\lim_{T,T'\rightarrow
1}K(T,T')=-\frac{1}{\beta\pi^{2}}\frac{\partial}{\partial
T}\frac{\partial}{\partial T'}\nonumber\\
&&\hspace{2cm}\mbox{}\times\ln\left|\frac{\sqrt{1/T-1}-\sqrt{1/T'-1}}
{\sqrt{1/T-1}+\sqrt{1/T'-1}}\right|, \label{KTnear1}
\end{eqnarray}
corresponding to a logarithmic eigenvalue repulsion,
\begin{equation}
\lim_{T,T'\rightarrow 1}u(T,T')=-\ln|T-T'|\label{uTnear1}
\end{equation}
[{\em cf.} Eqs.\ (\ref{Kuinvrelation}) and (\ref{lnuinv2})]. If $T$ and $T'$
are not both close to 1 then the two-point correlation function and the
interaction potential acquire contributions which depend on the dimensionality.

Nazarov's theory is a perturbation theory in the metallic regime: It requires
that the conductance $G$ is much greater than $e^{2}/h$. There exists no theory
for the statistics of transmission eigenvalues in a square or cube geometry
which extends to the insulating regime. This is the outstanding open problem of
the field. So far, limited progress has been made on the extension of the DMPK
equation towards higher dimensionalities (Mello and Tomsovic, 1991, 1992;
Chalker and Bernhardt, 1993; Endesfelder and Kramer, 1993; Tartakovski, 1995;
Endesfelder, 1996).

\subsection{Localization of interacting particles}
\label{interactions}

The interplay of interactions and localization is a formidable problem in
solid-state physics. (For a review, see Belitz and Kirkpatrick, 1994.)
Random-matrix techniques have given insight into the simplest case of two
interacting particles (an ``exciton'') in a one-dimensional random potential.

Dorokhov (1990) considered a pair of harmonically bound particles moving along
a disordered chain of length $L$. In the absence of disorder the total energy
$2E$ of the pair consists of kinetic energy of the center of mass plus the
binding energy\footnote{
Here it is assumed that the two particles are distinguishable, so that the
parity of the wave function under exchange is irrelevant. If the two particles
are identical, then $\varepsilon_{n}=(2n-\frac{3}{2})\hbar\omega$ for two
bosons or for a singlet pair of electrons, while
$\varepsilon_{n}=(2n-\frac{1}{2})\hbar\omega$ for a triplet pair of electrons.}
$\varepsilon_{n}=(n-\frac{1}{2})\hbar\omega$, $n=1,2,\ldots N$, of the harmonic
interaction (frequency $\omega$). The integer $N$ is the largest $n$ such that
$\varepsilon_{n}<2E$. The disorder potential (mean free path $l$) is assumed to
be weak enough that it can be treated perturbatively. This requires $kl\gg N$,
with $\hbar k$ the momentum of a free particle at energy $E$. Another way of
stating this requirement is that the average separation $\bar{d}$ of the two
particles should be much less than the mean free path.

The scattering problem of the pair, involving one propagating mode for the
center-of-mass motion and $N$ bound states for the relative motion, can be
mapped onto the scattering problem of a free particle with $N$ propagating
modes. The probability distribution of the scattering amplitudes of the pair
evolves with increasing $L$ according to a Fokker-Planck equation, analogous to
the DMPK equation for a free particle. For $N=1$ the localization length
$\xi_{\rm pair}$ of the pair is of the order of the mean free path $l$, which
is the single-particle localization length $\xi$ in one dimension. (The length
$\xi_{\rm pair}$ is smaller than $l$ near the ground state of the harmonic
interaction and becomes larger than $l$ on approaching the first excited
state.) For $N\gg 1$ the pair-localization length is greater than the mean free
path by a factor $N$, analogously to Eq.\ (\ref{localizationlength}):
\begin{equation}
\xi_{\rm pair}\simeq Nl\simeq k\bar{d}\xi.\label{xienhanced}
\end{equation}
The maximal enhancement $\xi_{\rm pair}/\xi\simeq kl$ is reached when $\bar{d}$
becomes comparable to $l$.

Shepelyansky (1994) studied the same problem for a weak interaction of
arbitrary sign, treating the two-particle Hamiltonian as a banded random
matrix. Disorder prevents the pair to diffuse apart by more than the
single-particle localization length, regardless of whether the interaction is
attractive or repulsive. The pair-localization length is greater than the
single-particle localization length by a factor $\xi_{\rm pair}/\xi\lesssim
kl$, independent of the sign of the interaction. This surprising result has
generated a great deal of interest in the coherent propagation of correlated
electron pairs. We refer to the proceedings of a recent conference for an
overview (Martin, Montambaux, and Tr\^an Thanh V\^an, 1996). Much of the work
is directed towards an extension of the phenomenon of interaction-assisted
diffusion to spatial dimensions greater than one and to more than two
particles. Whether the phenomenon is relevant for the metal-insulator
transition in a disordered metal remains an open question.

\subsection{Localization of light}
\label{light}

The propagation of electromagnetic waves through a wave guide is the optical
analogue of conduction through a wire. The analogy can be made more precise if
the vector character of the light does not play a role, which is the case in a
two-dimensional geometry with perpendicular polarization. Consider a
monochromatic electric field $\vec{\cal E}(\vec{r},t)=\hat{z}\,{\rm
Re}\,E(x,y)\exp({\rm i}\omega t)$ (frequency $\omega$, wave number
$k=\omega/c$), which varies only in the $x$-$y$ plane and is polarized in the
$z$-direction.The complex scalar field $E(x,y)$ satisfies the Helmholtz
equation
\begin{equation}
\bigl[\nabla^{2}+(\omega/c)^{2}\varepsilon(x,y)\bigr]E(x,y)=0,
\label{Helmholtz}
\end{equation}
with boundary condition $E=0$ at a metal surface. The (relative) dielectric
constant $\varepsilon(x,y)=1+\delta\varepsilon(x,y)$ fluctuates due to disorder
in the wave guide. Equation (\ref{Helmholtz}) is analogous to the
Schr\"{o}dinger equation for the wave function $\psi$ at the Fermi level of a
two-dimensional electron gas. The boundary condition $\psi=0$ applies to an
infinitely high potential barrier. The wave length $\lambda=2\pi c/\omega$
corresponds to the Fermi wave length $\lambda_{\rm F}$, and
$(\omega/c)^{2}\delta\varepsilon(x,y)$ corresponds to the electrostatic
potential $V(x,y)$ times $-2m/\hbar^{2}$. There is also a precise
correspondence between the expressions for the current density, which is
$\propto{\rm Re}\,E^{\ast}\nabla E$ and $\propto{\rm
Re}\,\psi^{\ast}\nabla\psi$ in the optical and electronic case, respectively.

The problem of localization by strong disorder is different in the two cases
(John, 1984), because of the restriction $\delta\varepsilon>-1$ in the optical
problem. Potentials $V$ greater than the Fermi energy have no optical analogue.
As a consequence, the mean free path $l$ for light can not be much smaller than
its wave length $\lambda$. In the case $l\gg\lambda$ of weak disorder, however,
optical and electronic localization are analogous. This is the relevant case
for a wave guide geometry.

One new aspect of the optical problem is the absence of a conservation law if
the dielectric constant has a non-zero imaginary part. The intensity of the
radiation which has propagated without reflection over a distance $L$ is then
multiplied by a factor ${\rm e}^{\sigma L}$, with $\sigma$ negative (positive)
for absorption (amplification). The growth or decay rate $\sigma$ is related to
the dielectric constant by $\sigma=-2k\,{\rm Im}\,\sqrt{\varepsilon}$.
Absorption is present to some degree in any optical system. Amplification
arises as a result of stimulated emission in a laser.

The effect of absorption or amplification on localization of light in a wave
guide can be studied by a generalization of the DMPK equation. The
generalization is simplest for the distribution of the reflection eigenvalues
$R_{n}\equiv\lambda_{n}/(1+\lambda_{n})$ (eigenvalues of the reflection-matrix
product $rr^{\dagger}$). With increasing length $L$ of the wave guide, the
distribution $P(\lambda_{1},\lambda_{2},\ldots\lambda_{N},L)$ evolves according
to
\begin{mathletters}
\label{FPsigma}
\begin{eqnarray}
l\frac{\partial P}{\partial L}&=&
\frac{2}{\beta N+2-\beta}\sum_{n=1}^{N}
\frac{\partial}{\partial\lambda_{n}}\lambda_{n}(1+\lambda_{n}) \nonumber\\
&&\mbox{}\times\left[ J\frac{\partial}{\partial\lambda_{n}}\frac{P}{J}
-\sigma l(\beta N+2-\beta)P\right],\label{FPsigmaa}\\
J&=&\prod_{i<j}^{N} |\lambda_{j}-\lambda_{i}|^{\beta}.\label{FPsigmab}
\end{eqnarray}
\end{mathletters}%
The symmetry index $\beta$ equals $1$, unless time-reversal symmetry is broken
by some magneto-optical effect. For $\sigma=0$, Eq.\ (\ref{FPsigma}) reduces to
the DMPK equation (\ref{DMPK}). The Fokker-Planck equation (\ref{FPsigma}) was
derived for $N=1$ (when $J\equiv 1$) by Gertsenshtein and Vasil'ev (1959),
Kohler and Pa\-pa\-ni\-co\-laou (1976), and Pradhan and Kumar (1994), and for
$N\geq 1$ by Beenakker, Paasschens, and Brouwer (1996), and Bruce and Chalker
(1996). The solution in the limit $L\rightarrow\infty$ is
\begin{equation}
P_{\infty}\propto\prod_{i}\exp\Bigl(\sigma l(\beta N+2-\beta)\lambda_{i}\Bigr)
\prod_{i<j}|\lambda_{j}-\lambda_{i}|^{\beta}.\label{Pinfty}
\end{equation}
Eq.\ (\ref{Pinfty}) holds for both positive and negative $\sigma$, but the
support of $P_{\infty}$ depends on the sign of $\sigma$: All $\lambda$'s have
to be $<-1$ for $\sigma>0$ (amplification) and $>0$ for $\sigma<0$
(absorption). The distribution (\ref{Pinfty}) is known in random-matrix theory
as the Laguerre ensemble .

When $\sigma\neq 0$, the transmission and reflection eigenvalues are no longer
related. The evolution with increasing $L$ of the $R_{n}$'s decouples from that
of the $T_{n}$'s --- but not vice versa. In fact, the evolution of the
transmission eigenvalues depends not just on the reflection eigenvalues, but on
all matrix elements of $rr^{\dagger}$. This is a substantial complication, and
analytical progress has so far been limited to the single-mode case (Rammal and
Doucot, 1987; Freilikher, Pustilnik, and Yurkevich, 1994). One exact result is
the equality of the Lyapunov exponents for absorption and amplification:
$\xi_{n}(\sigma)=\xi_{n}(-\sigma)$ (Paasschens, Misirpashaev, and Beenakker,
1996). (We recall that the Lyapunov exponents determine the decay of the
transmission eigenvalues $T_{n}\propto\exp(-2L/\xi_{n})$ in the limit
$L\rightarrow\infty$.) In the single-mode case, one has
$\xi_{1}=2l(1+|\sigma|l)^{-1}$ (Zhang, 1995). The multi-mode case has not yet
been solved.

Another new aspect of the optical problem is the frequency dependence of the
term $(\omega/c)^{2}\varepsilon$ in the Helmholtz equation, which for electrons
would correspond to an energy-dependent potential. The frequency dependence is
irrelevant for the transmission of a monochromatic wave, but does affect the
propagation of non-monochromatic radiation. In particular, the velocity of
propagation of a light pulse is greatly reduced near a resonance (Van Albada
{\em et al.}, 1991). The theory has been reviewed by Lagendijk and Van Tiggelen
(1996). For a random-matrix approach to resonant multiple scattering, see
Elattari, Kagalovsky, and Weidenm\"{u}ller (1996).

\subsection{Quantum Hall effect}
\label{QHeffect}

The quantum Hall effect occurs in a two-dimensional electron gas in a strong
perpendicular magnetic field (Prange and Girvin, 1990). (The field $B$ should
be sufficiently strong that the cyclotron frequency $\omega_{\rm c}=eB/m$ is
much greater than the elastic scattering rate $1/\tau$, so that the width
$\hbar/\tau$ of the Landau levels is much smaller than their spacing
$\hbar\omega_{\rm c}$.) With increasing $B$, the Hall conductance decreases
stepwise by $e^{2}/h$, each time a Landau level crosses the Fermi level. Each
step is associated with an insulator-metal-insulator transition: The wave
functions are localized on the plateaus of constant Hall conductance and
extended in between the plateaus. The magnetic field dependence of the
localization length in the transition is known from numerical simulations, but
not yet analytically (Huckestein, 1995). An intriguing link with the
random-matrix theory of a chaotic cavity has been suggested by Cobden and Kogan
(1996; see also Wang, Jovanovi\'{c}, and Lee, 1996; Cho and Fisher, 1997).

A three-dimensional system can exhibit the quantum Hall effect if it consists
of a stack of weakly coupled layers perpendicular to the magnetic field. The
Bechgaard salts are a naturally occurring example. One can also grow
semiconductor heterostructures containing multiple layers. This highly
anisotropic system has been called a ``chiral metal'' (Chalker and Dohmen,
1995; Balents and Fisher, 1996; Balents, Fisher, and Zirnbauer, 1997).
Conduction perpendicular to the layers occurs via overlapping states at the
edges of the layers. Unlike the problem of conduction parallel to the layers,
the problem of perpendicular conduction is tractable analytically. A transition
from extended to localized states occurs as the number of layers is increased.
Gruzberg, Read, and Sachdev (1997) have shown that this transition is governed
by the DMPK equation.

\acknowledgments
I am indebted to A. D. Stone for motivating me to write this review. My own
research on random-matrix theory was done in collaboration with P. W. Brouwer,
M. B\"{u}ttiker, M. J. M. de Jong, K. M. Frahm, R. A. Jalabert, I. K.
Marmorkos, J. A. Melsen, T. Sh.\ Misirpashaev, J. C. J. Paasschens, J.-L.
Pichard, B. Rejaei, and S. A. van Langen. While writing this review, I
benefitted greatly from the expert advice of P. W. Brouwer. Support by the
``Ne\-der\-land\-se or\-ga\-ni\-sa\-tie voor We\-ten\-schap\-pe\-lijk
On\-der\-zoek'' (NWO) and by the ``Stich\-ting voor Fun\-da\-men\-teel
On\-der\-zoek der Ma\-te\-rie'' (FOM) is gratefully acknowledged.

\appendix

\section{Integral equation for the eigenvalue density}
\label{Dysonexpansion}

Dyson (1972) derived the integral equation
\begin{eqnarray}
\int\!d\lambda'\,\rho(\lambda')\ln|\lambda-\lambda'|+
\case{1}{2}(1-2/\beta)\ln\rho(\lambda)\nonumber\\
=V(\lambda)+{\rm constant}\label{Dysonint}
\end{eqnarray}
for the eigenvalue density
$\rho(\lambda)=\langle\sum_{i}\delta(\lambda-\lambda_{i})\rangle$ in the
Wigner-Dyson ensemble (\ref{PGibbsab}). The term proportional to $\ln\rho$ is
an order $N^{-1}\ln N$ smaller than the other terms, and terms of still higher
order in $N^{-1}$ are neglected. If the $\ln\rho$ term is neglected as well,
then Eq.\ (\ref{Dysonint}) reduces to Wigner's integral equation
(\ref{wignerint}). Equation (\ref{Dysonint}) holds for a logarithmic eigenvalue
repulsion $u(\lambda,\lambda')=-\ln|\lambda-\lambda'|$. In this Appendix we
will generalize it to a non-logarithmic interaction.

We consider a probability distribution of the form
\begin{mathletters}
\label{Gibbs}
\begin{eqnarray}
&&P\propto{\rm e}^{-\beta W},\;\;
W=\sum_{i<j}u(\lambda_{i},\lambda_{j})+\sum_{i}V(\lambda_{i}),\label{Gibbsa}\\
&&u(\lambda,\lambda')=-\ln|\lambda-\lambda'|+\delta
u(\lambda,\lambda'),\label{Gibbsb}
\end{eqnarray}
\end{mathletters}%
and assume that the limit $\lambda\rightarrow\lambda'$ of $\delta
u(\lambda,\lambda')$ exists. Note that $P$ satisfies (for each $i=1,2,\ldots
N$) the differential equation
\begin{equation}
\frac{\partial}{\partial \lambda_{i}}P+\beta P\frac{\partial}{\partial
\lambda_{i}}W=0. \label{Pdiff1}
\end{equation}
Multiply both sides by $\delta(\lambda-\lambda_{i})$, sum over $i$, and
integrate over $\lambda_{1},\lambda_{2},\ldots \lambda_{N}$. The result is
\begin{eqnarray}
&&\frac{d}{d\lambda}\rho(\lambda)+
\beta\rho(\lambda)\frac{d}{d\lambda}V(\lambda)=\beta
I(\lambda),\label{Pdiff2a}\\
&&I(\lambda)=-\int\!d\lambda'\,\rho_{2}(\lambda,\lambda')\frac{\partial}
{\partial \lambda}u(\lambda,\lambda').\label{Pdiff2b}
\end{eqnarray}
We have defined the pair density
\begin{eqnarray}
\rho_{2}(\lambda,\lambda')&\equiv&\left\langle\sum_{i\neq
j}\delta(\lambda-\lambda_{i})\delta(\lambda'-\lambda_{j})\right\rangle
\nonumber\\
&=&\rho(\lambda)\rho(\lambda')-\rho(\lambda)\delta(\lambda-\lambda')
+K(\lambda,\lambda'). \label{rho2def}
\end{eqnarray}
The two-point correlation function $K(\lambda,\lambda')$ was defined in Eq.\
(\ref{Kdef}).

We proceed with a bit of formal manipulation:
\begin{eqnarray}
&&\int\!d\lambda'\,\rho(\lambda)\delta(\lambda-\lambda')
\frac{\partial}{\partial \lambda}u(\lambda,\lambda')\nonumber\\
&&\hspace*{5mm}=\int\!d\lambda'\rho(\lambda)\delta(\lambda-\lambda')
\frac{\partial}{\partial \lambda}[-\ln|\lambda-\lambda'|+\delta
u(\lambda,\lambda')]\nonumber\\
&&\hspace*{5mm}=\case{1}{2}\rho(\lambda)\frac{d}{d\lambda}\delta
u(\lambda,\lambda)-\int\!d\lambda'\,\case{1}{2}[\rho(\lambda)+\rho(\lambda')]
\frac{\delta(\lambda-\lambda')}{\lambda-\lambda'}\nonumber\\
&&\hspace*{5mm}=\case{1}{2}\rho(\lambda)\frac{d}{d\lambda}\delta
u(\lambda,\lambda)+\case{1}{2}\frac{d}{d\lambda}\rho(\lambda).\label{formal}
\end{eqnarray}
Substitution into Eq.\ (\ref{Pdiff2b}) yields
\begin{eqnarray}
\frac{I(\lambda)}{\rho(\lambda)}&=&\case{1}{2}\frac{d}{d\lambda}
[\ln\rho(\lambda)+ \delta u(\lambda,\lambda)]-\int\!d\lambda'\,\rho(\lambda')
\frac{\partial}{\partial \lambda}u(\lambda,\lambda')\nonumber\\
&&-\int\!d\lambda'\,\frac{K(\lambda,\lambda')}{\rho(\lambda)}
\frac{\partial}{\partial \lambda}u(\lambda,\lambda').\label{Pdiff3}
\end{eqnarray}
Since $K(\lambda,\lambda')$ is of order $N^{0}$, the last term in Eq.\
(\ref{Pdiff3}) is a factor $N$ smaller than the other terms. We neglect this
last term, substitute Eq.\ (\ref{Pdiff3}) into Eq.\ (\ref{Pdiff2a}), divide by
$\beta\rho(\lambda)$, and integrate once over $\lambda$. The result is the
required generalization of Eq.\ (\ref{Dysonint}) to a non-logarithmic
interaction:
\begin{eqnarray}
&&-\int\!d\lambda'\,\rho(\lambda')
u(\lambda,\lambda')+\case{1}{2}(1-2/\beta)\ln\rho(\lambda)+\case{1}{2}\delta
u(\lambda,\lambda)\nonumber\\
&&\hspace{4cm}=V(\lambda)+{\rm constant}.\label{Vurelation2}
\end{eqnarray}

\section{Integration over the unitary group}
\label{UNintegrate}

Averages over the unitary group appear throughout this review. Here we collect
a few results we will need repeatedly. For more extensive treatments we refer
to Creutz (1978), Samuel (1980), Mello (1990), Argaman and Zee (1996), and
Brouwer and Beenakker (1996a).

Let $U$ be an $N\times N$ matrix which is uniformly distributed over the group
${\cal U}(N)$ of $N\times N$ unitary matrices. (This is the circular unitary
ensemble of Sec.\ \ref{circular}.) Averages over ${\cal U}(N)$ are defined as
an integration with the invariant measure $d\mu(U)$,
\begin{equation}
\langle f(U)\rangle=\int d\mu(U)\,f(U),\label{fUaverage}
\end{equation}
normalized such that $\int d\mu(U)=1$. The invariance property means that
\begin{equation}
\langle f(UU_{0})\rangle=\langle f(U_{0}U)\rangle=\langle
f(U)\rangle,\label{fUU0average}
\end{equation}
for any fixed matrix $U_{0}\in{\cal U}(N)$.

The average of a polynomial function
\begin{equation}
f(U)=U^{\vphantom{\ast}}_{\alpha_{1}a_{1}}U^{\vphantom{\ast}}_{\alpha_{2}a_{2}}
\cdots U^{\vphantom{\ast}}_{\alpha_{p}a_{p}}
U^{\ast}_{\beta_{1}b_{1}}U^{\ast}_{\beta_{2}b_{2}}\cdots
U^{\ast}_{\beta_{q}b_{q}} \label{fUdef}
\end{equation}
is zero unless $p=q$ {\em and\/} the sets $\{\alpha_{n}\}=\{\beta_{n}\}$ of
left indices coincide {\em and\/} the sets $\{a_{n}\}=\{b_{n}\}$ of right
indices coincide. The expressions for $p=1$ and 2 are
\begin{eqnarray}
&&\langle U^{\vphantom{\ast}}_{\alpha a}U^{\ast}_{\beta
b}\rangle=\frac{1}{N}\delta_{\alpha\beta}\delta_{ab},\label{UU}\\
&&\langle U^{\vphantom{\ast}}_{\alpha a}U^{\vphantom{\ast}}_{\alpha'
a'}U^{\ast}_{\beta b}U^{\ast}_{\beta' b'}\rangle=\nonumber\\
&&\hspace{1cm}\frac{1}{N^{2}-1}\bigl(
\delta_{\alpha\beta}\delta_{ab}\delta_{\alpha'\beta'}\delta_{a'b'}+
\delta_{\alpha\beta'}\delta_{ab'}\delta_{\alpha'\beta}\delta_{a'b}\bigr)
\nonumber\\
&&\mbox{}-\frac{1}{N(N^{2}-1)}\bigl(
\delta_{\alpha\beta}\delta_{ab'}\delta_{\alpha'\beta'}\delta_{a'b}+
\delta_{\alpha\beta'}\delta_{ab}\delta_{\alpha'\beta}\delta_{a'b'}\bigr).
\label{UUUU}
\end{eqnarray}
The leading order term in powers of $1/N$ in Eq.\ (\ref{UUUU}) is the Gaussian
approximation, which consists in replacing the real and imaginary parts of the
elements of $U$ by independent Gaussian variables with zero mean and variance
$1/2N$. More generally, the Gaussian approximation is the leading order term in
the average
\begin{equation}
\langle f(U)\rangle=N^{-p}\delta_{pq}\sum_{P}\prod_{j=1}^{p}
\delta_{\alpha_{j}\beta_{P(j)}}\delta_{a_{j}b_{P(j)}}+{\cal
O}(N^{-p-1}),\label{fUGaussian}
\end{equation}
where the sum is over all permutations $P$ of the numbers $1,2,\ldots p$.

\section{How to derive Eq.\ (\protect\ref{DELTARHO}) from Eq.\
(\protect\ref{EXACTSOLUTION3X})}
\label{Tdensity2}

The probability distribution (\ref{EXACTSOLUTION3X}) is of the form
(\ref{Gibbs}), with
\begin{mathletters}
\label{deltauV}
\begin{eqnarray}
u(x,x')&=&-\ln|x-x'|+\delta u(x,x'),\label{deltauVa}\\
\delta u(x,x')&=&-\case{1}{2}\ln|(x-x')^{-1}\sinh(x-x')|\nonumber\\
&&\mbox{}-\case{1}{2}\ln|(x+x')\sinh(x+x')|,\label{deltauVb}\\
V(x)&=&\case{1}{2}(N-1+2/\beta)s^{-1}x^{2}+\beta^{-1}\delta
u(x,x).\label{deltauVc}
\end{eqnarray}
\end{mathletters}%
(Instead of $\lambda$, we use here the variable $x\geq 0$.) The density
$\rho(x)$ is determined to order $N^{0}$ by the integral equation
(\ref{Vurelation2}), which in view of Eq.\ (\ref{deltauVc}) takes the form
\begin{eqnarray}
&&-\int_{0}^{\infty}\!dx'\,\rho(x')u(x,x')+\case{1}{2}(1-2/\beta)
[\ln\rho(x)+\delta u(x,x)\nonumber\\
&&\hspace{2cm}\mbox{}+s^{-1}x^{2}]=\case{1}{2}Ns^{-1}x^{2}+{\rm
constant}.\label{Vurelationx}
\end{eqnarray}
We write $\rho(x)=\rho_{0}(x)+\delta\rho(x)$, with
$\rho_{0}(x)=Ns^{-1}\theta(s-x)$ and $\delta\rho(x)$ a correction of order
$N^{0}$. One can verify by substitution that $\rho_{0}$ satisfies Eq.\
(\ref{Vurelationx}) to order $N$,
\begin{equation}
-\int_{0}^{\infty}\!dx'\,\rho_{0}(x')u(x,x')=\case{1}{2}Ns^{-1}x^{2}+{\rm
constant},\label{Vurelationx0}
\end{equation}
for $s\gg 1$, $s\gg x$. Linearization of Eq.\ (\ref{Vurelationx}) around
$\rho_{0}$ yields an equation for $\delta\rho$,
\begin{eqnarray}
&&\int_{0}^{\infty}\!dx'\,\delta\rho(x')u(x,x')=
\case{1}{2}(1-2/\beta)[\ln\rho_{0}(x)+\delta u(x,x)\nonumber\\
&&\hspace{4cm}\mbox{}+s^{-1}x^{2}]+{\rm constant}.\label{Vurelation3}
\end{eqnarray}
We substitute Eq.\ (\ref{deltauV}) and extend $\rho(x)$ symmetrically to
negative $x$: $\rho(-x)\equiv\rho(x)$. Equation (\ref{Vurelation3}) becomes
\begin{eqnarray}
&&-\int_{-\infty}^{\infty}\!dx'\,\delta\rho(x')\ln|(x-x')\sinh(x-x')|
=(1-2/\beta)\nonumber\\
&&\hspace{1cm}\mbox{}\times[\ln\rho_{0}(x)-\case{1}{2}\ln|x\sinh
2x|+s^{-1}x^{2}]+{\rm constant}.\nonumber\\
\label{Vurelation4}
\end{eqnarray}
For $s\gg x$ the term $s^{-1}x^{2}$ may be neglected and the term $\ln\rho_{0}$
may be absorbed into the additive constant. The remaining convolution is
readily inverted by Fourier transformation,
\begin{eqnarray}
&&\delta\rho(k)\frac{2\pi}{|k|}\bigl(1-{\rm
e}^{-\pi|k|}\bigr)^{-1}=(1-2/\beta)\frac{\pi}{|k|}\bigl(1-{\rm
e}^{-\frac{1}{2}\pi|k|}\bigr)^{-1}\nonumber\\
&&\Rightarrow\delta\rho(k)=\case{1}{2}(1-2/\beta)\bigl(1+{\rm
e}^{-\frac{1}{2}\pi|k|}\bigr).\label{deltarhok}
\end{eqnarray}
The inverse Fourier transform of $\delta\rho(k)$ is Eq.\ (\ref{DELTARHO}).

\section{Calculation of the weak-localization corrections in Table
\protect\ref{TABLEDELTAGNS} }
\label{deltaGNSdetails}

In this Appendix we show how the weak-localization corrections $\delta G_{\rm
NS}$ in a normal-metal--superconductor junction, listed in Table
\ref{TABLEDELTAGNS}, are obtained. We first consider a system without
spin-orbit scattering (Brouwer and Beenakker, 1995b), and then discuss the
effect of strong spin-orbit scattering (Slevin, Pichard, and Mello, 1996).
Starting point of the calculation is Eq.\ (\ref{key}). We assume that the
length $L$ of the disordered normal region is much greater than the
superconducting coherence length $\xi\simeq(\hbar v_{\rm
F}l/{\mit\Delta_{0}})^{1/2}$ (with $v_{\rm F}$ the Fermi velocity and $l$ the
mean free path in the normal metal). This implies that the Thouless energy
$E_{\rm c}\simeq\hbar v_{\rm F}l/L^{2}$ is much smaller than the
superconducting energy gap ${\mit\Delta_{0}}$. In the voltage range $V\lesssim
E_{\rm c}/e$ we may therefore assume that $eV\ll{\mit\Delta_{0}}$, hence
$\alpha\equiv\exp[-{\rm i}\arccos(\varepsilon/{\mit\Delta_{0}})]\rightarrow
-{\rm i}$. Using the polar decomposition (\ref{polarS}) of the transmission and
reflection matrices, Eq.\ (\ref{keyb}) can be replaced by
\begin{eqnarray}
m(\varepsilon)&=&\sqrt{{\cal T}(\varepsilon)}\left[1+u(\varepsilon) \sqrt{{\cal
R}(-\varepsilon)}u^*(-\varepsilon)\sqrt{{\cal R}(\varepsilon)}\right]^{-1}
\nonumber\\
&&\mbox{}\times u(\varepsilon)\sqrt{{\cal
T}(-\varepsilon)},\;u(\varepsilon)\equiv
V'^{\vphantom{*}}(\varepsilon)V^*(-\varepsilon),\label{keyc}
\end{eqnarray}
where ${\cal T}$ is the matrix of transmission eigenvalues and ${\cal
R}=1-{\cal T}$. In the presence of time-reversal symmetry, $V'^{\vphantom T} =
V^{\rm T}$. If time-reversal symmetry is broken, $V$ and $V'$ are independent.
In the case of electron-hole degeneracy, the difference between $+\varepsilon$
and $-\varepsilon$ may be neglected. If electron-hole degeneracy is broken, the
scattering matrices at $\pm\varepsilon$ are independent. Of the four entries in
Table \ref{TABLEDELTAGNS}, the case that both time-reversal symmetry and
electron-hole degeneracy are present is the easiest, because then $u=1$ and
Eq.\ (\ref{key}) simplifies to the linear statistic (\ref{keyzero}). The result
for $\delta G_{\rm NS}$ is Eq.\ (\ref{deltaGNSresult}). The three other entries
are more difficult because we need to average over the unitary matrices as well
as over the transmission eigenvalues. We will discuss the three cases
separately.

\subsubsection*{Broken time-reversal symmetry}

We first consider the case that electron-hole degeneracy is present but
time-reversal symmetry is broken. According to the isotropy assumption in a
wire geometry ({\rm cf.} Sec.\ \ref{Brownian}), $V$ and $V'$, and hence $u$,
are uniformly distributed in the unitary group ${\cal U}(N)$. We may perform
the average $\langle\cdots\rangle$ over the ensemble of scattering matrices in
two steps: $\langle\cdots\rangle =\langle\langle\cdots\rangle_{u}\rangle_{\cal
T}$, where $\langle\cdots\rangle_{u}$ and $\langle\cdots\rangle_{\cal T}$ are,
respectively, the average over the unitary matrix ${u}$ and over the
transmission eigenvalues $T_{i}$. We compute $\langle\cdots\rangle_{u}$ by an
expansion in powers of $N^{-1}$. To integrate the rational function
(\ref{keyc}) of $u$ over ${\cal U}(N)$, we first expand it into a geometric
series and then use the general rules for the integration of polynomials of $u$
({\em cf.} Appendix \ref{UNintegrate}). The polynomials we need are
\begin{mathletters}
\label{GNSu}
\begin{eqnarray}
&&\langle G_{\rm NS}\rangle_{u}
=\frac{4e^{2}}{h}\sum_{p,q=0}^{\infty}(-1)^{p+q}M_{pq},\label{GNSua}\\
&& M_{pq}=\langle{\rm Tr}\,{\cal T}(u\sqrt{{\cal R}}u^{*}\sqrt{{\cal
R}})^{p}u{\cal T}u^{\dagger}(\sqrt{{\cal R}}u^{\rm T}\sqrt{{\cal
R}}u^{\dagger})^{q}\rangle_{u}.\nonumber\\
\label{GNSub}
\end{eqnarray}
\end{mathletters}%
Neglecting terms of order $N^{-1}$, we find
\begin{equation}
M_{pq}=\left\{\begin{array}{l}
N\tau_{1}^{2}(1-\tau_{1}^{\vphantom{2}})^{2p}\;{\rm if}\;p=q,\smallskip\\
\tau_{1}(\tau_{1}^{2}+\tau_{1}^{\vphantom{2}}-
2\tau_{2}^{\vphantom{2}})(1-\tau_{1}^{\vphantom{2}})^{p+q-1}\\
-2\,{\rm min}(p,q)\tau_{1}^{2}(\tau_{1}^{2}-\tau_{2}^{\vphantom{2}})
(1-\tau_{1}^{\vphantom{2}})^{p+q-2}\\
\hspace{3cm}{\rm if}\;|p-q|\;{\rm odd},\smallskip\\
0\;{\rm else},
\end{array}\right.\label{Mpqresult}
\end{equation}
where we have defined the moment $\tau_{k}=N^{-1}\sum_{i}T_{i}^{k}$. The
summation over $p$ and $q$ leads to
\begin{equation}
\frac{h}{4e^2}\,\langle G_{\rm NS}\rangle_{u}=\frac{N\tau_{1}}{2-\tau_{1}}-
\frac{4\tau_{1}-2\tau_{1}^{2}+2\tau_{1}^{3}-4\tau_{2}}{\tau_{1}
(2-\tau_{1})^{3}}. \label{GNSavgu}
\end{equation}

It remains to average over the transmission eigenvalues. Since $\tau_{k}$ is a
linear statistic, we know that its sample-to-sample fluctuations are an order
$1/N$ smaller than the average. Hence
\begin{equation}
\langle f(\tau_{k})\rangle_{\cal
T}=f\biglb(\langle\tau_{k}\rangle\bigrb)[1+{\cal O}(N^{-2})],\label{Nofluct}
\end{equation}
which implies that we may replace the average of the rational function
(\ref{GNSavgu}) of the $\tau_{k}$'s by the rational function of the average
$\langle\tau_{k}\rangle$. This average has the $1/N$ expansion
\begin{equation}
\langle\tau_{k}\rangle=\langle\tau_{k}\rangle_{0}+{\cal
O}(N^{-2}),\label{TauAvgExpand}
\end{equation}
where $\langle\tau_{k}\rangle_{0}$ is ${\cal O}(N^{0})$. There is no term of
order $N^{-1}$ in the absence of time-reversal symmetry. From Eqs.\
(\ref{GNSavgu})--(\ref{TauAvgExpand}) we obtain the $1/N$ expansion of the
average conductance,
\begin{eqnarray}
&&\frac{h}{4e^{2}}\,\langle G_{\rm
NS}\rangle=\frac{N\langle\tau_{1}\rangle_{0}}{2-\langle\tau_{1}\rangle_{0}}
\nonumber\\
&&\;\;\;\;\mbox{}-\frac{4\langle\tau_{1}\rangle_{0}-2\langle\tau_{1}
\rangle_{0}^{2}+ 2\langle\tau_{1}\rangle_{0}^{3}-4\langle\tau_{2}\rangle_{0}}
{\langle\tau_{1}\rangle_{0}(2-\langle\tau_{1}\rangle_{0})^{3}}+{\cal
O}(N^{-1}).\label{GNSavgbeta2}
\end{eqnarray}

Equation (\ref{GNSavgbeta2}) is generally valid for any distribution of the
transmission eigenvalues. We apply it to the case of a disordered wire in the
limit $N\rightarrow\infty$, $l/L\rightarrow 0$ at constant $Nl/L$. The moments
$\langle\tau_{k}\rangle_{0}$ are given by
\begin{equation}
\langle\tau_{k}\rangle_{0}=\frac{l}{L}\int_{0}^{\infty}\!
\frac{dx}{\cosh^{2k}x}\Rightarrow\langle\tau_{1}\rangle_{0}=\frac{l}{L},
\;\langle\tau_{2}\rangle_{0}=\frac{2l}{3L}.\label{tau120mean}
\end{equation}
Substitution into Eq.\ (\ref{GNSavgbeta2}) yields the weak-localization
correction $\delta G_{\rm NS}=-\case{2}{3}e^{2}/h$, {\em cf.} Table
\ref{TABLEDELTAGNS}.

\subsubsection*{Broken electron-hole degeneracy}

If time-reversal symmetry is present but electron-hole degeneracy is broken,
then one has $u^{\dagger}(-eV)=u(eV)$, with $u(eV)$ uniformly distributed in
${\cal U}(N)$. A calculation similar to that in the previous subsection yields
for the average over $u$:
\begin{eqnarray}
\frac{h}{4e^{2}}\,\langle G_{\rm
NS}\rangle_{u}&=&N\tau_{1+}^{\vphantom{2}}\tau_{1-}^{\vphantom{2}}
(\tau_{1+}^{\vphantom{2}}+\tau_{1-}^{\vphantom{2}}-\tau_{1+}^{\vphantom{2}}
\tau_{1-}^{\vphantom{2}})^{-1}\nonumber\\
&&\mbox{}+(\tau_{1+}^{\vphantom{2}}+
\tau_{1-}^{\vphantom{2}}-\tau_{1+}^{\vphantom{2}}\tau_{1-}^{\vphantom{2}})^{-3}
\nonumber\\
&&\mbox{}\times\bigl[2\tau_{1+}^{2}\tau_{1-}^{2}-\tau_{1+}^{3}\tau_{1-}^{2}-
\tau_{1+}^{2}\tau_{1-}^{3}-\tau_{2+}^{\vphantom{2}}\tau_{1-}^{2}\nonumber\\
&&\mbox{}-\tau_{1+}^{2}
\tau_{2-}^{\vphantom{2}}+\tau_{2+}^{\vphantom{2}}\tau_{1-}^{3}+\tau_{1+}^{3}
\tau_{2-}^{\vphantom{2}}\bigr],\label{GNSavgunoD}
\end{eqnarray}
where we have abbreviated $\tau_{k\pm}=\tau_k(\pm eV)$. The next step is the
average over the transmission eigenvalues. We may still use Eq.\
(\ref{Nofluct}), and we note that
$\langle\tau_{k}(\varepsilon)\rangle\equiv\langle\tau_{k}\rangle$ is
independent of $\varepsilon$. (The energy scale for variations in
$\langle\tau_{k}(\varepsilon)\rangle$ is $E_{\rm F}$, which is much greater
than the energy scale of interest $E_{\rm c}$.) Instead of Eq.\
(\ref{TauAvgExpand}) we now have the $1/N$ expansion
\begin{equation}
\langle\tau_{k}\rangle=\langle\tau_{k}\rangle_{0}+N^{-1}\delta\tau_{k}+{\cal
O}(N^{-2}),\label{TauAvgExpandnoD}
\end{equation}
which contains also a term of order $N^{-1}$ because of the presence of
time-reversal symmetry. The $1/N$ expansion of $\langle G_{\rm NS}\rangle$
becomes
\begin{eqnarray}
\frac{h}{4e^{2}}\,\langle G_{\rm
NS}\rangle&=&\frac{N\langle\tau_{1}\rangle_{0}}
{2-\langle\tau_{1}\rangle_{0}}+\frac{2\delta\tau_{1}}
{(2-\langle\tau_{1}\rangle_{0})^{2}}\nonumber\\
&&\mbox{}+\frac{2\langle\tau_{1}\rangle_{0}^{2}-
2\langle\tau_{1}\rangle_{0}^{3}-2\langle\tau_{2}\rangle_{0}+
2\langle\tau_{1}\rangle_{0}\langle\tau_{2}\rangle_{0}}
{\langle\tau_{1}\rangle_{0}(2-\langle\tau_{1}\rangle_{0})^{3}}\nonumber\\
&&\mbox{}+{\cal O}(N^{-1}).\label{GNSavgbeta1V} \end{eqnarray}
For the application to a disordered wire we use again Eq.\ (\ref{tau120mean})
for the moments $\langle\tau_{k}\rangle_{0}$, which do not depend on whether
time-reversal symmetry is broken or not. We also need $\delta\tau_{1}$, which
in the presence of time-reversal symmetry is given by
\begin{equation}
\delta\tau_{k}=\int_{0}^{\infty}\frac{\delta\rho(x)dx}{\cosh^{2k}x}
\Rightarrow\delta\tau_{1}=-\case{1}{3}.\label{deltauk}
\end{equation}
Substitution into Eq.\ (\ref{GNSavgbeta1V}) yields $\delta G_{\rm
NS}=-\case{4}{3}e^{2}/h$, {\em cf.} Table \ref{TABLEDELTAGNS}.

\subsubsection*{Both symmetries broken}

In the absence of both time-reversal symmetry and electron-hole degeneracy, the
two matrices $u(eV)$ and $u(-eV)$ are independent, each with a uniform
distribution in ${\cal U}(N)$. Carrying out the two averages over ${\cal U}(N)$
we find
\begin{equation}
\frac{h}{4e^{2}}\,\langle G_{\rm NS}\rangle_{u}=\frac{N\tau_{1+}^{\vphantom{2}}
\tau_{1-}^{\vphantom{2}}}
{\tau_{1+}^{\vphantom{2}}+\tau_{1-}^{\vphantom{2}}-\tau_{1+}^{\vphantom{2}}
\tau_{1-}^{\vphantom{2}}}.\label{GNSavgub2V}
\end{equation}
The average over the transmission eigenvalues becomes
\begin{equation}
\frac{h}{4e^{2}}\,\langle G_{\rm NS}\rangle=\frac{N\langle\tau_{1}\rangle_{0}}
{2-\langle\tau_{1}\rangle_{0}}+{\cal O}(N^{-1}),\label{GNSavgbeta2V}
\end{equation}
where we have used that $\delta\tau_{1}=0$ in the absence of time-reversal
symmetry. We conclude that $\delta G_{\rm NS}=0$ in this case, as indicated in
Table \ref{TABLEDELTAGNS}.

\subsubsection*{Effect of spin-orbit scattering}

In the presence of spin-orbit scattering, the scattering matrix elements are
quaternion numbers. Since a quaternion can be represented by a $2\times 2$
matrix, we can represent the $N\times N$ matrix $V$ with quaternion elements by
a $2N\times 2N$ matrix $v$ with complex elements. We denote this representation
by $V\cong v$. In view of the definitions (\ref{qastdef}) and
(\ref{qdaggerdef}) of complex conjugation and Hermitian conjugation, one has
\begin{equation}
V^{\ast}\cong -{\cal C}v^{\ast}{\cal C},\;\;V^{\dagger}\cong
v^{\dagger},\label{Vvcong}
\end{equation}
where ${\cal C}$ was defined in Eq.\ (\ref{calCdef}). In this notation, the
conductance is given by
\begin{mathletters}
\label{keyspinorbit}
\begin{eqnarray}
G_{\rm NS}&=&2G_{0}{\rm Tr}\,m(eV)m^{\dagger}(eV),\;
G_{0}=e^{2}/h,\label{keyspinorbita}\\
m(\varepsilon)&=&\sqrt{{\cal T}(\varepsilon)}\left[1-u(\varepsilon) \sqrt{{\cal
R}(-\varepsilon)}u^*(-\varepsilon)\sqrt{{\cal R}(\varepsilon)}\right]^{-1}
\nonumber\\
&&\mbox{}\times u(\varepsilon)\sqrt{{\cal
T}(-\varepsilon)},\;u(\varepsilon)\equiv v'^{\vphantom{*}}(\varepsilon){\cal
C}v^*(-\varepsilon).\label{keyspinorbitb}
\end{eqnarray}
\end{mathletters}%
The conductance quantum $G_{0}$ is half as small as in the absence of
spin-orbit scattering, while the dimensionality of the matrices ${\cal T}$ and
${\cal R}=1-{\cal T}$ of transmission and reflection eigenvalues has doubled.
Furthermore, the term $1+u$ in Eq.\ (\ref{keyc}) is replaced by $1-u$ in Eq.\
(\ref{keyspinorbitb}), as a result of the minus sign in the definition
(\ref{Vvcong}) of complex conjugation. The calculations of the previous
subsections can now be repeated starting from Eq.\ (\ref{keyspinorbit}) instead
of from Eq.\ (\ref{keyc}). The result is that each entry in Table
\ref{TABLEDELTAGNS} is to be multiplied by a factor of $-1/2$ (Slevin, Pichard,
and Mello, 1996).


\newpage

\begin{references}
\bibitem{} Abrahams, E., P. W. Anderson, D. C. Licciardello, and T. V.
Ramakrishnan, 1979, Phys.\ Rev.\ Lett.\ {\bf 42}, 673.
\bibitem{} Abrikosov, A. A., 1981, Solid State Comm.\ {\bf 37}, 997.
\bibitem{} Agra\"{\i}t, N., J. G. Rodrigo, and S. Vieira, 1992, Phys.\ Rev.\ B
{\bf 46}, 5814.
\bibitem{} Aleiner, I. L., and L. I. Glazman, 1996, Phys.\ Rev.\ Lett.\ {\bf
77}, 2057.
\bibitem{} Aleiner, I. L., and A. I. Larkin, 1996 (Los Alamos preprint archive,
cond-mat/9610034).
\bibitem{} Alhassid, Y., and H. Attias, 1996, Phys.\ Rev.\ Lett.\ {\bf 76},
1711.
\bibitem{} Alhassid, Y., and C. H. Lewenkopf, 1995, Phys.\ Rev.\ Lett.\ {\bf
75}, 3922.
\bibitem{} Allsopp, N. K., J. Sanchez Canizares, R. Raimondi, and C. J.
Lambert, 1996, J. Phys.\ Condens.\ Matter {\bf 8}, L377.
\bibitem{} Altland, A., 1991, Z. Phys.\ B {\bf 82}, 105.
\bibitem{} Altland, A., and M. R. Zirnbauer, 1996a, Phys.\ Rev.\ Lett.\ {\bf
76}, 3420.
\bibitem{} Altland, A., and M. R. Zirnbauer, 1996b (Los Alamos preprint
archive, cond-mat/9602137).
\bibitem{} Altshuler, B. L., 1985, Pis'ma Zh.\ Eksp.\ Teor.\ Fiz.\ {\bf 41},
530 [JETP Lett.\ {\bf 41}, 648].
\bibitem{} Altshuler, B. L., and A. G. Aronov, 1985, in {\em Electron-Electron
Interactions in Disordered Systems}, edited by A. L. Efros and M. Pollak
(North-Holland, Amsterdam): p.\ 1.
\bibitem{} Altshuler, B. L., Y. Gefen, A. Kamenev, and L. S. Levitov, 1996 (Los
Alamos preprint archive, cond-mat/9609132).
\bibitem{} Altshuler, B. L., D. E. Khmel'nitski\u{\i}, and B. Z. Spivak, 1983,
Solid State Comm.\ {\bf 48}, 841.
\bibitem{} Altshuler, B. L., P. A. Lee, and R. A. Webb, editors, 1991, {\em
Mesoscopic Phenomena in Solids\/} (North-Holland, Amsterdam): p.\ 1.
\bibitem{} Altshuler, B. L., L. S. Levitov, and A. Yu.\ Yakovets, 1994, Pis'ma
Zh.\ Eksp.\ Teor.\ Fiz.\ {\bf 59}, 821 [JETP Lett.\ {\bf 59}, 857].
\bibitem{} Altshuler, B. L., and B. I. Shklovski\u{\i}, 1986, Zh.\ Eksp.\
Teor.\ Fiz.\ {\bf 91}, 220 [Sov.\ Phys.\ JETP {\bf 64}, 127].
\bibitem{} Altshuler, B. L., and B. D. Simons, 1995, in {\em Mesoscopic Quantum
Physics}, edited by E. Akkermans, G. Montambaux, J.-L. Pichard, and J.
Zinn-Justin (North-Holland, Amsterdam).
\bibitem{} Altshuler, B. L., and B. Z. Spivak, 1987, Zh.\ Eksp.\ Teor.\ Fiz.\
{\bf 92}, 609 [Sov.\ Phys.\ JETP {\bf 65}, 343].
\bibitem{} Ambj\o{}rn, J., and G. Akemann, 1996 (Los Alamos preprint archive,
cond-mat/9606129).
\bibitem{} Ambj\o{}rn, J., J. Jurkiewicz, and Yu.\ M. Makeenko, 1990, Phys.\
Lett.\ B {\bf 251}, 517.
\bibitem{} Ambj\o{}rn, J., and Yu.\ M. Makeenko, 1990, Mod.\ Phys.\ Lett.\ A
{\bf 5}, 1753.
\bibitem{} Anantram, M. P., and S. Datta, 1996, Phys.\ Rev.\ B {\bf 53}, 16390.
\bibitem{} Anderson, P. W., E. Abrahams, and T. V. Ramakrishnan, 1979, Phys.\
Rev.\ Lett.\ {\bf 43}, 718.
\bibitem{} Anderson, P. W., D. J. Thouless, E. Abrahams, and D. S. Fisher,
1980, Phys.\ Rev.\ B {\bf 22}, 3519.
\bibitem{} Andreev, A. F., 1964, Zh.\ Eksp.\ Teor.\ Fiz.\ {\bf 46}, 1823 [Sov.\
Phys.\ JETP {\bf 19}, 1228].
\bibitem{} Andreev, A. F., 1966, Zh.\ Eksp.\ Teor.\ Fiz.\ {\bf 51}, 1510 [Sov.\
Phys.\ JETP {\bf 24}, 1019 (1967)].
\bibitem{} Andreev, A. V., O. Agam, B. D. Simons, and B. L. Altshuler, 1996,
Phys.\ Rev.\ Lett.\ {\bf 76}, 3947.
\bibitem{} Argaman, N., 1995, Phys.\ Rev.\ Lett.\ {\bf 75}, 2750.
\bibitem{} Argaman, N., 1996, Phys.\ Rev.\ B {\bf 53}, 7035.
\bibitem{} Argaman, N., 1997 (Los Alamos preprint archive, cond-mat/9608084).
\bibitem{} Argaman, N., Y. Imry, and U. Smilansky, 1993, Phys.\ Rev.\ B {\bf
47}, 4440.
\bibitem{} Argaman, N., and A. Zee, 1996, Phys.\ Rev.\ B {\bf 54}, 7406.
\bibitem{} Artemenko, S. N., A. F. Volkov, and A. V. Za\u{\i}tsev, 1979, Solid
State Comm.\ {\bf 30}, 771.
\bibitem{} Ashcroft, N. W., and N. D. Mermin, 1976, {\em Solid State Physics\/}
(Holt, Rinehart and Winston, New York).
\bibitem{} Averin, D. V., A. N. Korotkov, and K. K. Likharev, 1991, Phys.\
Rev.\ B {\bf 44}, 6199.
\bibitem{} Averin, D. V., and K. K. Likharev, 1991, in {\em Mesoscopic
Phenomena in Solids}, edited by B. L. Altshuler, P. A. Lee, and R. A. Webb
(North-Holland, Amsterdam): p.\ 173.
\bibitem{} Averin, D. V., and Yu.\ V. Nazarov, 1990, Phys.\ Rev.\ Lett.\ {\bf
65}, 2446.
\bibitem{} Bakker, S. J. M., E. van der Drift, T. M. Klapwijk, H. M. Jaeger,
and S. Radelaar, 1994, Phys.\ Rev.\ B {\bf 49}, 13275.
\bibitem{} Balents, L., and M. P. A. Fisher, 1996, Phys.\ Rev.\ Lett.\ {\bf
76}, 2782.
\bibitem{} Balents, L., M. P. A. Fisher, and M. R. Zirnbauer, 1997 (Los Alamos
preprint archive, cond-mat/9608049).
\bibitem{} Balian, R., 1968, Nuovo Cimento {\bf 57}, 183.
\bibitem{} Baranger, H. U., 1996, in {\em Nano-Science and Technology}, edited
by G. Timp (American Institute of Physics, New York, in press).
\bibitem{} Baranger, H. U., R. A. Jalabert, and A. D. Stone, 1993a, Phys.\
Rev.\ Lett.\ {\bf 70}, 3876.
\bibitem{} Baranger, H. U., R. A. Jalabert, and A. D. Stone, 1993b, Chaos, {\bf
3}, 665.
\bibitem{} Baranger, H. U., and P. A. Mello, 1994, Phys.\ Rev.\ Lett.\ {\bf
73}, 142.
\bibitem{} Baranger, H. U., and P. A. Mello, 1995, Phys.\ Rev.\ B {\bf 51},
4703.
\bibitem{} Baranger, H. U., and P. A. Mello, 1996a, Europhys.\ Lett.\ {\bf 33},
465.
\bibitem{} Baranger, H. U., and P. A. Mello, 1996b (Los Alamos preprint
archive, cond-mat/9607155).
\bibitem{} Basor, E. L., and C. A. Tracy, 1993, J. Stat.\ Phys.\ {\bf 73}, 415.
\bibitem{} Beenakker, C. W. J., 1991, Phys.\ Rev.\ B {\bf 44}, 1646.
\bibitem{} Beenakker, C. W. J., 1992a, Phys.\ Rev.\ B {\bf 46}, 12841.
\bibitem{} Beenakker, C. W. J., 1992b, in {\em Transport Phenomena in
Mesoscopic Systems}, edited by H. Fukuyama and T. Ando (Springer, Berlin): p.\
235.
\bibitem{} Beenakker, C. W. J., 1993a, Phys.\ Rev.\ Lett.\ {\bf 70}, 1155.
\bibitem{} Beenakker, C. W. J., 1993b, Phys.\ Rev.\ Lett.\ {\bf 70}, 4126.
\bibitem{} Beenakker, C. W. J., 1993c, Phys.\ Rev.\ B {\bf 47}, 15763.
\bibitem{} Beenakker, C. W. J., 1994a, Nucl.\ Phys.\ B {\bf 422}, 515.
\bibitem{} Beenakker, C. W. J., 1994b, Phys.\ Rev.\ B {\bf 49}, 2205.
\bibitem{} Beenakker, C. W. J., 1995, in {\em Mesoscopic Quantum Physics},
edited by E. Akkermans, G. Montambaux, J.-L. Pichard, and J. Zinn-Justin
(North-Holland, Amsterdam): p.\ 279.
\bibitem{} Beenakker, C. W. J., and M. B\"{u}ttiker, 1992, Phys.\ Rev.\ B {\bf
46}, 1889.
\bibitem{} Beenakker, C. W. J., and J. A. Melsen, 1994, Phys.\ Rev.\ B {\bf
50}, 2450.
\bibitem{} Beenakker, C. W. J., J. A. Melsen, and P. W. Brouwer, 1995, Phys.\
Rev.\ B {\bf 51}, 13883.
\bibitem{} Beenakker, C. W. J., J. C. J. Paasschens, and P. W. Brou\-wer, 1996,
Phys.\ Rev.\ Lett.\ {\bf 76}, 1368.
\bibitem{} Beenakker, C. W. J., and B. Rejaei, 1993, Phys.\ Rev.\ Lett.\ {\bf
71}, 3689.
\bibitem{} Beenakker, C. W. J., and B. Rejaei, 1994a, Phys.\ Rev.\ B {\bf 49},
7499.
\bibitem{} Beenakker, C. W. J., and B. Rejaei, 1994b, Physica A {\bf 203}, 61.
\bibitem{} Beenakker, C. W. J., B. Rejaei, and J. A. Melsen, 1994, Phys.\ Rev.\
Lett.\ {\bf 72}, 2470.
\bibitem{} Beenakker, C. W. J., and H. van Houten, 1991, Sol.\ State Phys.\
{\bf 44}, 1.
\bibitem{} Belitz, D., and T. R. Kirkpatrick, 1994, Rev.\ Mod.\ Phys.\ {\bf
66}, 261.
\bibitem{} Bergmann, G., 1984, Phys.\ Rep.\ {\bf 107}, 1.
\bibitem{} Berkovits, R., and S. Feng, 1992, Phys.\ Rev.\ B {\bf 45}, 97.
\bibitem{} Berkovits, R., and S. Feng, 1994, Phys.\ Rep.\ {\bf 238}, 135.
\bibitem{} Berry, M. V., 1985, Proc.\ R. Soc.\ London A {\bf 400}, 229.
\bibitem{} Birge, N. O., B. Golding, and W. H. Haemmerle, 1989, Phys.\ Rev.\
Lett.\ {\bf 62}, 195.
\bibitem{} Blonder, G. E., M. Tinkham, and T. M. Klapwijk, 1982, Phys.\ Rev.\ B
{\bf 25}, 4515.
\bibitem{} Bl\"{u}mel, R., and U. Smilansky, 1988, Phys.\ Rev.\ Lett.\ {\bf
60}, 477.
\bibitem{} Bl\"{u}mel, R., and U. Smilansky, 1990, Phys.\ Rev.\ Lett.\ {\bf
64}, 241.
\bibitem{} Bohigas, O., 1990, in {\em Chaos and Quantum Physics}, edited by
M.-J. Giannoni, A. Voros, and J. Zinn-Justin (North-Holland, Amsterdam): p.\
87.
\bibitem{} Bohigas, O., M.-J. Giannoni, and C. Schmit, 1984, Phys.\ Rev.\
Lett.\ {\bf 52}, 1.
\bibitem{} Bohigas, O., M.-J. Giannoni, A. M. Ozorio de Almeida, and C. Schmit,
1995, Nonlinearity, {\bf 8}, 203.
\bibitem{} Boutet de Monvel, A., L. Pastur, and M. Shcherbina, 1995, J. Stat.\
Phys.\ {\bf 79}, 585.
\bibitem{} Br\'{e}zin, E., and A. Zee, 1993, Nucl.\ Phys.\ B {\bf 402}, 613.
\bibitem{} Br\'{e}zin, E., and A. Zee, 1994, Phys.\ Rev.\ E {\bf 49}, 2588.
\bibitem{} Brezini, A., and N. Zekri, 1992, Phys.\ Stat.\ Sol.\ B {\bf 169},
253.
\bibitem{} Brody, T. A., J. Flores, J. B. French, P. A. Mello, A. Pandey, and
S. S. M. Wong, 1981, Rev.\ Mod.\ Phys.\ {\bf 53}, 385.
\bibitem{} Brouwer, P. W., 1994 (unpublished).
\bibitem{} Brouwer, P. W., 1995, Phys.\ Rev.\ B {\bf 51}, 16878.
\bibitem{} Brouwer, P. W., 1997 (unpublished).
\bibitem{} Brouwer, P. W., and C. W. J. Beenakker, 1994, Phys.\ Rev.\ B {\bf
50}, 11263.
\bibitem{} Brouwer, P. W., and C. W. J. Beenakker, 1995a, Phys.\ Rev.\ B {\bf
51}, 7739.
\bibitem{} Brouwer, P. W., and C. W. J. Beenakker, 1995b, Phys.\ Rev.\ B {\bf
52}, 3868.
\bibitem{} Brouwer, P. W., and C. W. J. Beenakker, 1995c, Phys.\ Rev.\ B {\bf
52}, 16772.
\bibitem{} Brouwer, P. W., and C. W. J. Beenakker, 1996a, J. Math.\ Phys.\ {\bf
37}, 4904.
\bibitem{} Brouwer, P. W., and C. W. J. Beenakker, 1996b, Phys.\ Rev.\ B {\bf
54}, 12705.
\bibitem{} Brouwer, P. W., and C. W. J. Beenakker, 1997a, Phys.\ Rev.\ B (in
press).
\bibitem{} Brouwer, P. W., and C. W. J. Beenakker, 1997b, Chaos, Solitons and
Fractals (in press).
\bibitem{} Brouwer, P. W., and M. B\"{u}ttiker, 1997 (Los Alamos preprint
archive, cond-mat/9610144).
\bibitem{} Brouwer, P. W., and K. M. Frahm, 1996, Phys.\ Rev.\ B {\bf 53},
1490.
\bibitem{} Bruce, N. A., and J. T. Chalker, 1996, J. Phys.\ A {\bf 29}, 3761.
\bibitem{} Bruun, J., V. C. Hui, and C. J. Lambert, 1994, Phys.\ Rev.\ B {\bf
49}, 4010.
\bibitem{} Bruus, H., C. H. Lewenkopf, and E. R. Mucciolo, 1996, Phys.\ Rev.\ B
{\bf 53}, 9968.
\bibitem{} B\"{u}ttiker, M., 1986a, Phys.\ Rev.\ B {\bf 33}, 3020.
\bibitem{} B\"{u}ttiker, M., 1986b, Phys.\ Rev.\ Lett.\ {\bf 57}, 1761.
\bibitem{} B\"{u}ttiker, M., 1988a, IBM J. Res.\ Dev.\ {\bf 32}, 63.
\bibitem{} B\"{u}ttiker, M., 1988b, IBM J. Res.\ Dev.\ {\bf 32}, 317.
\bibitem{} B\"{u}ttiker, M., 1990, Phys.\ Rev.\ Lett.\ {\bf 65}, 2901.
\bibitem{} B\"{u}ttiker, M., 1992, Phys.\ Rev.\ B {\bf 46}, 12485.
\bibitem{} B\"{u}ttiker, M., 1993, J. Phys.\ Condens.\ Matter {\bf 5}, 9361.
\bibitem{} B\"{u}ttiker, M., and T. Christen, 1996, in {\em Quantum Transport
in Semiconductor Submicron Structures}, edited by B. Kramer, NATO ASI Series
E326 (Kluwer, Dordrecht): p.\ 263.
\bibitem{} B\"{u}ttiker, M., A. Pr\^{e}tre, and H. Thomas, 1993, Phys.\ Rev.\
Lett.\ {\bf 70}, 4114.
\bibitem{} Calogero, F., 1969, J. Math.\ Phys.\ {\bf 10}, 2191.
\bibitem{} Caselle, M., 1995, Phys.\ Rev.\ Lett.\ {\bf 74}, 2776.
\bibitem{} Chalker, J. T., and M. Bernhardt, 1993, Phys.\ Rev.\ Lett.\ {\bf
70}, 982.
\bibitem{} Chalker, J. T., and A. Dohmen, 1995, Phys.\ Rev.\ Lett.\ {\bf 75},
4496.
\bibitem{} Chalker, J. T., and A. M. S. Mac\^{e}do, 1993, Phys.\ Rev.\ Lett.\
{\bf 71}, 3693.
\bibitem{} Chan, I. H., R. M. Clarke, C. M. Marcus, K. Campman, and A. C.
Gossard, 1995, Phys.\ Rev.\ Lett.\ {\bf 74}, 3876.
\bibitem{} Chang, A. M., H. U. Baranger, L. N. Pfeiffer, and K. W. West, 1994,
Phys.\ Rev.\ Lett.\ {\bf 73}, 2111.
\bibitem{} Chang, A. M., H. U. Baranger, L. N. Pfeiffer, K. W. West, and T. Y.
Chang, 1996, Phys.\ Rev.\ Lett.\ {\bf 76}, 1695.
\bibitem{} Charlat, P., H. Courtois, Ph.\ Gandit, D. Mailly, A. F. Volkov, and
B. Pannetier, 1996 (Los Alamos preprint archive, cond-mat/9605021).
\bibitem{} Chen, Y., M. E. H. Ismail, and K. A. Muttalib, 1992, J. Phys.\
Condens.\ Matter {\bf 4}, L417.
\bibitem{} Cho, S., and M. P. A. Fisher, 1997, Phys.\ Rev.\ B (in press).
\bibitem{} Claughton, N. R., R. Raimondi, and C. J. Lambert, 1996, Phys.\ Rev.\
B {\bf 53}, 9310.
\bibitem{} Cobden, D. H., and E. Kogan, 1996, Phys.\ Rev.\ B (in press).
\bibitem{} Courtois, H., Ph.\ Gandit, D. Mailly, and B. Pannetier, 1996, Phys.\
Rev.\ Lett.\ {\bf 76}, 130.
\bibitem{} Creutz, M., 1978, J. Math.\ Phys.\ {\bf 19}, 2043.
\bibitem{} Crisanti, A., G. Paladin, and A. Vulpiani, 1993, {\em Products of
Random Matrices in Statistical Physics\/} (Springer, Berlin).
\bibitem{} Datta, S., 1995, {\em Electronic Transport in Mesoscopic Systems\/}
(Cambridge University Press, Cambridge).
\bibitem{} Debray, P., J.-L. Pichard, J. Vicente, and P. N. Tung, 1989, Phys.\
Rev.\ Lett.\ {\bf 63}, 2264.
\bibitem{} De Gennes, P. G., 1966, {\em Superconductivity of Metals and
Alloys\/} (Benjamin, New York).
\bibitem{} De Jong, M. J. M., 1994, Phys.\ Rev.\ B {\bf 49}, 7778.
\bibitem{} De Jong, M. J. M., and C. W. J. Beenakker, 1992, Phys.\ Rev.\ B {\bf
46}, 13400.
\bibitem{} De Jong, M. J. M., and C. W. J. Beenakker, 1994, Phys.\ Rev.\ B {\bf
49}, 16070.
\bibitem{} De Jong, M. J. M., and C. W. J. Beenakker, 1995, Phys.\ Rev.\ B {\bf
51}, 16867.
\bibitem{} De Jong, M. J. M., and C. W. J. Beenakker, 1996, Physica A {\bf
230}, 219.
\bibitem{} De Jong, M. J. M., and C. W. J. Beenakker, 1997, in {\em Mesoscopic
Electron Transport}, edited by L. P. Kouwenhoven, G. Sch\"{o}n, and L. L. Sohn,
NATO ASI Series E (Kluwer, Dordrecht, in press).
\bibitem{} Den Hartog, S. G., C. M. A. Kapteyn, B. J. van Wees, T. M. Klapwijk,
W. van der Graaf, and G. Borghs, 1996, Phys.\ Rev.\ Lett.\ {\bf 76}, 4592.
\bibitem{} De Vegvar, P. G. N., T. A. Fulton, W. H. Mallison, and R. E. Miller,
1994, Phys.\ Rev.\ Lett.\ {\bf 73}, 1416.
\bibitem{} Dimoulas, A., J. P. Heida, B. J. van Wees, T. M. Klapwijk, W. van de
Graaf, and G. Borghs, 1995, Phys.\ Rev.\ Lett.\ {\bf 74}, 602.
\bibitem{} Dorokhov, O. N., 1982, Pis'ma Zh.\ Eksp.\ Teor.\ Fiz.\ {\bf 36}, 259
[JETP Lett.\ {\bf 36}, 318].
\bibitem{} Dorokhov, O. N., 1983, Zh.\ Eksp.\ Teor.\ Fiz.\ {\bf 85}, 1040
[Sov.\ Phys.\ JETP {\bf 58}, 606].
\bibitem{} Dorokhov, O. N., 1984, Solid State Comm.\ {\bf 51}, 381.
\bibitem{} Dorokhov, O. N., 1988, Phys.\ Rev.\ B {\bf 37}, 10526.
\bibitem{} Dorokhov, O. N., 1990, Zh.\ Eksp.\ Teor.\ Fiz.\ {\bf 98}, 646 [Sov.\
Phys.\ JETP {\bf 71}, 360].
\bibitem{} Doron, E. and U. Smilansky, 1992, Nucl.\ Phys.\ A {\bf 545}, 455c.
\bibitem{} Dupuis, N., and G. Montambaux, 1991, Phys.\ Rev.\ B {\bf 43}, 14390.
\bibitem{} Dyson, F. J. 1962a, J. Math.\ Phys.\ {\bf 3}, 140.
\bibitem{} Dyson, F. J. 1962b, J. Math.\ Phys.\ {\bf 3}, 157.
\bibitem{} Dyson, F. J., 1962c, J. Math.\ Phys.\ {\bf 3}, 1191.
\bibitem{} Dyson, F. J., 1962d, J. Math.\ Phys.\ {\bf 3}, 1199.
\bibitem{} Dyson, F. J., 1972, J. Math.\ Phys.\ {\bf 13}, 90.
\bibitem{} Dyson, F. J., and M. L. Mehta, 1963, J. Math.\ Phys.\ {\bf 4}, 701.
\bibitem{} Edwards, J. T., and D. J. Thouless, 1972, J. Phys.\ C {\bf 5}, 807.
\bibitem{} Efetov, K. B., 1982, Zh.\ Eksp.\ Teor.\ Fiz.\ {\bf 83}, 833 [Sov.\
Phys.\ JETP {\bf 56}, 467].
\bibitem{} Efetov, K. B., 1983, Adv.\ Phys.\ {\bf 32}, 53.
\bibitem{} Efetov, K. B., 1995, Phys.\ Rev.\ Lett.\ {\bf 74}, 2299.
\bibitem{} Efetov, K. B., 1996, {\em Supersymmetry in Disorder and Chaos\/}
(Cambridge University Press, Cambridge).
\bibitem{} Efetov, K. B., and A. I. Larkin, 1983, Zh.\ Eksp.\ Teor.\ Fiz.\ {\bf
85}, 764 [Sov.\ Phys.\ JETP {\bf 58}, 444].
\bibitem{} Elattari, B., V. Kagalovsky, and H. A. Weidenm\"{u}ller, 1996,
Nucl.\ Phys.\ A {\bf 606}, 86.
\bibitem{} Endesfelder, D., and B. Kramer, 1993, Phys.\ Rev.\ E {\bf 48}, 3225.
\bibitem{} Endesfelder, D., 1996, Phys.\ Rev.\ B {\bf 53}, 16555.
\bibitem{} Erd\"{o}s, P., and R. C. Herndon, 1982, Adv.\ Phys.\ {\bf 31}, 65.
\bibitem{} Eynard, B., 1994 (Los Alamos preprint archive, hep-th/9401165).
\bibitem{} Fal'ko, V. I., 1995, Phys.\ Rev.\ B {\bf 51}, 5227.
\bibitem{} Fertig, H. A., and S. Das Sarma, 1989, Phys.\ Rev.\ B {\bf 40},
7410.
\bibitem{} Fisher, D. S., and P. A. Lee, 1981, Phys.\ Rev.\ B {\bf 23}, 6851.
\bibitem{} Folk, J. A., S. R. Patel, S. F. Goddijn, A. G. Huibers, S. M.
Cronenwett, C. M. Marcus, K. Campman, and A. C. Gossard, 1996, Phys.\ Rev.\
Lett.\ {\bf 76}, 1699.
\bibitem{} Forrester, P. J., 1995, Nucl.\ Phys.\ B {\bf 435}, 421.
\bibitem{} Frahm, K. M., 1995a, Phys.\ Rev.\ Lett.\ {\bf 74}, 4706.
\bibitem{} Frahm, K. M., 1995b, Europhys.\ Lett.\ {\bf 30}, 457.
\bibitem{} Frahm, K. M., P. W. Brouwer, J. A. Melsen, and C. W. J. Beenakker,
1996, Phys.\ Rev.\ Lett.\ {\bf 76}, 2981.
\bibitem{} Frahm, K. M., and A. M\"{u}ller-Groeling, 1996, J. Phys.\ A {\bf
29}, 5313.
\bibitem{} Frahm, K. M., and J.-L. Pichard, 1995a, J. Phys.\ I France {\bf 5},
847.
\bibitem{} Frahm, K. M., and J.-L. Pichard, 1995b, J. Phys.\ I France {\bf 5},
877.
\bibitem{} Freidrig, H., and D. Wintgen, 1989, Phys.\ Rep.\ {\bf 183}, 37.
\bibitem{} Freilikher, V., E. Kanzieper, and I. Yurkevich, 1996, Phys.\ Rev.\ E
{\bf 53}, 2200.
\bibitem{} Freilikher, V., M. Pustilnik, and I. Yurkevich, 1994, Phys.\ Rev.\
Lett.\ {\bf 73}, 810.
\bibitem{} Friedman, W. A., and P. A. Mello, 1985a, Ann.\ Phys.\ (N.Y.) {\bf
161}, 276.
\bibitem{} Friedman, W. A., and P. A. Mello, 1985b, J. Phys.\ A {\bf 18}, 425.
\bibitem{} Fyodorov, Y. V., and A. D. Mirlin, 1991, Phys.\ Rev.\ Lett.\ {\bf
67}, 2405.
\bibitem{} Fyodorov, Y. V., and A. D. Mirlin, 1994, Int.\ J. Mod.\ Phys.\ B
{\bf 8}, 3795.
\bibitem{} Gazaryan, Yu.\ L., 1969, Zh.\ Eksp.\ Teor.\ Fiz.\ {\bf 56}, 1856
[Sov.\ Phys.\ JETP {\bf 29}, 996].
\bibitem{} Gertsenshtein, M. E., and V. B. Vasil'ev, 1959, Teor.\ Veroyatn.\
Primen. {\bf 4}, 424; {\bf 5}, 3(E) (1960) [Theor.\ Probab.\ Appl. {\bf 4}, 391
(1959); {\bf 5}, 340(E) (1960)].
\bibitem{} Ghenim, L., D. L. Sivco, A. Y. Cho, and G. Hill, 1996, Phys.\ Rev.\
B {\bf 54}, 11479.
\bibitem{} Goldberg, J., U. Smilansky, M. V. Berry, W. Schweizer, G. Wunner,
and G. Zeller, 1991, Nonlinearity {\bf 4}, 1.
\bibitem{} Golubov, A. A., and M. Yu.\ Kupriyanov, 1995, Pis'ma Zh.\ Eksp.\
Teor.\ Fiz.\ {\bf 61}, 830 [JETP Lett.\ {\bf 61}, 851].
\bibitem{} Golubov, A. A., F. K. Wilhelm, and A. D. Zaikin, 1996 (Los Alamos
preprint archive, cond-mat/9605113).
\bibitem{} Gopar, V. A., M. Mart\'{\i}nez, and P. A. Mello, 1995, Phys.\ Rev.\
B {\bf 51}, 16917.
\bibitem{} Gopar, V. A., M. Mart\'{\i}nez, P. A. Mello, and H. U. Baranger,
1996, J. Phys.\ A {\bf 29}, 881.
\bibitem{} Gopar, V. A., P. A. Mello, and M. B\"{u}ttiker, 1996, Phys.\ Rev.\
Lett.\ {\bf 77}, 3005.
\bibitem{} Gor'kov, L. P., and G. M. Eliashberg, 1965, Zh.\ Eksp.\ Teor.\ Fiz.\
{\bf 48}, 1407 [Sov.\ Phys.\ JETP {\bf 21}, 940].
\bibitem{} Gor'kov, L. P., A. I. Larkin, and D. E. Khmel'nitski\u{\i}, 1979,
Pis'ma Zh.\ Eksp.\ Teor.\ Fiz.\ {\bf 30}, 248 [JETP Lett.\ {\bf 30}, 228].
\bibitem{} Gruzberg, I. A., N. Read, and S. Sachdev, 1997 (Los Alamos preprint
archive, cond-mat/9612038).
\bibitem{} Guhr, T., 1996, Phys.\ Rev.\ Lett.\ {\bf 76}, 2258.
\bibitem{} Gutzwiller, M. C., 1990, {\em Chaos in Classical and Quantum
Mechanics\/} (Springer, Berlin).
\bibitem{} Haake, F., 1992, {\em Quantum Signatures of Chaos\/} (Springer,
Berlin).
\bibitem{} Hackenbroich, G., and H. A. Weidenm\"{u}ller, 1995, Phys.\ Rev.\
Lett.\ {\bf 74}, 4118.
\bibitem{} Hamermesh, M., 1962, {\em Group Theory and its Applications to
Physical Problems\/} (Addison-Wesley, Reading).
\bibitem{} Hastings, M. B., A. D. Stone, and H. U. Baranger, 1994, Phys.\ Rev.\
B {\bf 50}, 8230.
\bibitem{} Hekking, F. W. J., and Yu.\ V. Nazarov, 1993, Phys.\ Rev.\ Lett.\
{\bf 71}, 1625.
\bibitem{} Hekking, F. W. J., and Yu.\ V. Nazarov, 1994, Phys.\ Rev. B {\bf
49}, 6847.
\bibitem{} Holweg, P. A. M., J. A. Kokkedee, J. Caro, A. H. Verbruggen, S.
Radelaar, A. G. M. Jansen, and P. Wyder, 1991, Phys.\ Rev.\ Lett.\ {\bf 67},
2549.
\bibitem{} Hua, L. K., 1963, {\em Harmonic Analysis of Functions of Several
Complex Variables in the Classical Domains\/} (American Mathematical Society,
Providence).
\bibitem{} Huckestein, B., 1995, Rev.\ Mod.\ Phys.\ {\bf 67}, 357.
\bibitem{} H\"{u}ffmann, A., 1990, J. Phys.\ A {\bf 23}, 5733.
\bibitem{} Iida, S., H. A. Weidenm\"{u}ller, and J. A. Zuk, 1990a, Phys.\ Rev.\
Lett.\ {\bf 64}, 583.
\bibitem{} Iida, S., H. A. Weidenm\"{u}ller, and J. A. Zuk, 1990b, Ann.\ Phys.\
(N.Y.) {\bf 200}, 219.
\bibitem{} Imry, Y., 1986a, Europhys.\ Lett.\ {\bf 1}, 249.
\bibitem{} Imry, Y., 1986b, in {\em Directions in Condensed Matter Physics},
edited by G. Grinstein and G. Mazenko (World Scientific, Singapore): p.\ 101.
\bibitem{} Imry, Y., 1995, in {\em Mesoscopic Quantum Physics}, edited by E.
Akkermans, G. Montambaux, J.-L. Pichard, and J. Zinn-Justin (North-Holland,
Amsterdam): p.\ 181.
\bibitem{} Imry, Y., 1996, {\em Introduction to Mesoscopic Physics\/} (Oxford
University Press, Oxford).
\bibitem{} Ishio, H., 1995, J. Phys.\ A {\bf 28}, L469.
\bibitem{} Itzykson, C., and J.-B. Zuber, 1980, {\em Quantum Field Theory\/}
(McGraw-Hill, New York).
\bibitem{} Jalabert, R. A., H. U. Baranger, and A. D. Stone, 1990, Phys.\ Rev.\
Lett.\ {\bf 65}, 2442.
\bibitem{} Jalabert, R. A., and J.-L. Pichard, 1995, J. Phys.\ I France {\bf
5}, 287.
\bibitem{} Jalabert, R. A., J.-L. Pichard, and C. W. J. Beenakker, 1993,
Europhys.\ Lett.\ {\bf 24}, 1.
\bibitem{} Jalabert, R. A., J.-L. Pichard, and C. W. J. Beenakker, 1994,
Europhys.\ Lett.\ {\bf 27}, 255.
\bibitem{} Jalabert, R. A., A. D. Stone, and Y. Alhassid, 1992, Phys.\ Rev.\
Lett.\ {\bf 68}, 3468.
\bibitem{} Jancovici, B., and P. J. Forrester, 1994, Phys.\ Rev.\ B {\bf 50},
14599.
\bibitem{} John, S., 1984, Phys.\ Rev.\ Lett.\ {\bf 53}, 2169.
\bibitem{} Kadigrobov, A., A. Zagoskin, R. I. Shekhter, and M. Jonson, 1995,
Phys.\ Rev.\ B {\bf 52}, 8662.
\bibitem{} Kamenev, A., and Y. Gefen, 1995, Europhys.\ Lett.\ {\bf 29}, 413.
\bibitem{} Kamien, R. D., H. D. Politzer, and M. B. Wise, 1988, Phys.\ Rev.\
Lett.\ {\bf 60}, 1995.
\bibitem{} Kastalsky, A., A. W. Kleinsasser, L. H. Greene, R. Bhat, F. P.
Milliken, and J. P. Harbison, 1991, Phys.\ Rev.\ Lett.\ {\bf 67}, 3026.
\bibitem{} Keldysh, L. V., 1964, Zh.\ Eksp.\ Teor.\ Fiz.\ {\bf 47}, 1515 [Sov.\
Phys.\ JETP {\bf 20}, 1018].
\bibitem{} Keller, M. W., O. Millo, A. Mittal, D. E. Prober, and R. N. Sacks,
1994, Surf.\ Sci.\ {\bf 305}, 501.
\bibitem{} Keller, M. W., A. Mittal, J. W. Sleight, R. G. Wheeler, D. E.
Prober, R. N. Sacks, and H. Shtrikmann, 1996, Phys.\ Rev.\ B {\bf 53}, 1693.
\bibitem{} Khlus, V. A., 1987, Zh.\ Eksp.\ Teor.\ Fiz.\ {\bf 93}, 2179 [Sov.\
Phys.\ JETP {\bf 66}, 1243].
\bibitem{} Khmel'nitski\u{\i}, D. E., 1984, Physica B {\bf 126}, 235.
\bibitem{} Kirkman, P. D., and J. B. Pendry, 1984, J. Phys.\ C {\bf 17}, 5707.
\bibitem{} Klapwijk, T. M., 1994, Physica B {\bf 197}, 481.
\bibitem{} Kobayakawa, T. S., Y. Hatsugai, M. Kohmoto, and A. Zee, 1995, Phys.\
Rev.\ E {\bf 51}, 5365.
\bibitem{} Kohler, W., and G. C. Papanicolaou, 1976, SIAM J.\ Appl.\ Math.\
{\bf 30}, 263.
\bibitem{} Kozub, V. I., J. Caro, and P. A. M. Holweg, 1996, Physica B {\bf
218}, 89.
\bibitem{} Kozub, V. I., and A. M. Rudin, 1995, Phys.\ Rev.\ B {\bf 52}, 7853.
\bibitem{} Kramer, B., and A. MacKinnon, 1993, Rep.\ Prog.\ Phys.\ {\bf 56},
1469.
\bibitem{} Krieger, T. J., 1965, Ann.\ Phys.\ (N.Y.) {\bf 42}, 375.
\bibitem{} Kulik, I. O., and A. N. Omel'yanchuk, 1984, Fiz.\ Nizk.\ Temp.\ {\bf
10}, 305 [Sov.\ J. Low Temp.\ Phys.\ {\bf 10}, 158].
\bibitem{} Kulik, I. O., and R. I. Shekhter, 1975, Zh.\ Eksp.\ Teor.\ Fiz.\
{\bf 68}, 623 [Sov.\ Phys.\ JETP {\bf 41}, 308].
\bibitem{} Kumar, N., 1985, Phys.\ Rev.\ B {\bf 31}, 5513.
\bibitem{} Lagendijk, A., and B. A. van Tiggelen, 1996, Phys.\ Rep.\ {\bf 270},
143.
\bibitem{} Lambert, C. J., 1991, J. Phys.\ Condens.\ Matter {\bf 3}, 6579.
\bibitem{} Lambert, C. J., 1993, J. Phys.\ Condens.\ Matter {\bf 5}, 707.
\bibitem{} Lambert, C. J., V. C. Hui, and S. J. Robinson, 1993, J. Phys.\
Condens.\ Matter {\bf 5}, 4187.
\bibitem{} Landauer, R., 1957, IBM J.\ Res.\ Dev.\ {\bf 1}, 223.
\bibitem{} Landauer, R., 1970, Phil.\ Mag.\ {\bf 21}, 863.
\bibitem{} Landauer, R., 1987, Z. Phys.\ B {\bf 68}, 217.
\bibitem{} Larkin, A. I., and Yu.\ V. Ovchinnikov, 1975, Zh.\ Eksp.\ Teor.\
Fiz.\ {\bf 68}, 1915 [Sov.\ Phys.\ JETP {\bf 41}, 960].
\bibitem{} Larkin, A. I., and Yu.\ V. Ovchinnikov, 1977, Zh.\ Eksp.\ Teor.\
Fiz.\ {\bf 73}, 299 [Sov.\ Phys.\ JETP {\bf 46}, 155].
\bibitem{} Lee, P. A., and T. V. Ramakrishnan, 1985, Rev.\ Mod.\ Phys.\ {\bf
57}, 287.
\bibitem{} Lee, P. A., and A. D. Stone, 1985, Phys.\ Rev.\ Lett.\ {\bf 55},
1622.
\bibitem{} Lenssen, K.-M. H., P. C. A. Jeekel, C. J. P. M. Harmans, J. E.
Mooij, M. R. Leys, J. H. Wolter, and M. C. Holland, 1994, in {\em Coulomb and
Interference Effects in Small Electronic Structures}, edited by D. C. Glattli,
M. Sanquer, and J. Tr\^an Thanh V\^an (Editions Fronti\`eres, Gif-sur-Yvette):
p.\ 63.
\bibitem{} Lenz, G., and F. Haake, 1990, Phys.\ Rev.\ Lett.\ {\bf 65}, 2325.
\bibitem{} Leo, J., and A. H. MacDonald, 1990, Phys.\ Rev.\ Lett.\ {\bf 64},
817.
\bibitem{} Lerner, I. V., and Y. Imry, 1995, Europhys.\ Lett.\ {\bf 29}, 49.
\bibitem{} Lerner, I. V., and M. E. Raikh, 1992, Phys.\ Rev.\ B {\bf 45},
14036.
\bibitem{} Lesovik, G. B., 1989, Pis'ma Zh.\ Eksp.\ Teor.\ Fiz.\ {\bf 49}, 513
[JETP Lett.\ {\bf 49}, 592].
\bibitem{} Lesovik, G. B., A. L. Fauch\`{e}re, and G. Blatter, 1997 (Los Alamos
preprint archive, cond-mat/9610175).
\bibitem{} Lewenkopf, C. H., and H. A. Weidenm\"{u}ller, 1991, Ann.\ Phys.\
(N.Y.) {\bf 212}, 53.
\bibitem{} Liefrink, F., J. I. Dijkhuis, M. J. M. de Jong, L. W. Molenkamp, and
H. van Houten, 1994, Phys.\ Rev.\ B {\bf 49}, 14066.
\bibitem{} Likharev, K. K., 1979, Rev.\ Mod.\ Phys.\ {\bf 51}, 101.
\bibitem{} Mac\^{e}do, A. M. S., 1994a, Phys.\ Rev.\ B {\bf 49}, 1858.
\bibitem{} Mac\^{e}do, A. M. S., 1994b, Phys.\ Rev.\ B {\bf 49}, 16841.
\bibitem{} Mac\^{e}do, A. M. S., 1996, Phys.\ Rev.\ B {\bf 53}, 8411.
\bibitem{} Mac\^{e}do, A. M. S., and J. T. Chalker, 1992, Phys.\ Rev.\ B {\bf
46}, 14985.
\bibitem{} Mac\^{e}do, A. M. S., and J. T. Chalker, 1994, Phys.\ Rev.\ B {\bf
49}, 4695.
\bibitem{} Magn\'{e}e, P. H. C., N. van der Post, P. H. M. Kooistra, B. J. van
Wees, and T. M. Klapwijk, 1994, Phys.\ Rev.\ B {\bf 50}, 4594.
\bibitem{} Mahaux, C., and H. A. Weidenm\"{u}ller, 1969, {\em Shell-Model
Approach to Nuclear Reactions\/} (North-Holland, Amsterdam).
\bibitem{} Mailly, D., and M. Sanquer, 1992, J. Phys.\ I France {\bf 2}, 357.
\bibitem{} Makeenko, Yu.\, 1991, Mod.\ Phys.\ Lett.\ A {\bf 6}, 1901.
\bibitem{} Mani, R. G., L. Ghenim, and T. N. Theis, 1992, Phys.\ Rev.\ B {\bf
45}, 12098.
\bibitem{} Marcus, C. M., 1996 (unpublished).
\bibitem{} Marcus, C. M., A. J. Rimberg, R. M. Westervelt, P. F. Hopkins, and
A. C. Gossard, 1992, Phys.\ Rev.\ Lett.\ {\bf 69}, 506.
\bibitem{} Marcus, C. M., R. M. Westervelt, P. F. Hopkins, and A. C. Gossard,
1993, Phys.\ Rev.\ B {\bf 48}, 2460.
\bibitem{} Marmorkos, I. K., C. W. J.  Beenakker, and R. A.  Jalabert, 1993,
Phys.\ Rev.\ B {\bf 48}, 2811.
\bibitem{} Martin, Th., 1994, in {\em Coulomb and Interference Effects in Small
Electronic Structures}, edited by D. C. Glattli, M. Sanquer, and J. Tr\^an
Thanh V\^an (Editions Fronti\`eres, Gif-sur-Yvette): p.\ 405.
\bibitem{} Martin, Th., 1996, Phys.\ Lett.\ A {\bf 220}, 137.
\bibitem{} Martin, Th., and R. Landauer, 1992, Phys.\ Rev.\ B {\bf 45}, 1742.
\bibitem{} Martin, Th., G. Montambaux, and J. Tr\^an Thanh V\^an, editors,
1996, {\em Correlated Fermions and Transport in Mesoscopic Systems\/} (Editions
Fronti\`eres, Gif-sur-Yvette).
\bibitem{} Maslov, D. L., C. Barnes, and G. Kirczenow, 1993a, Phys.\ Rev.\
Lett.\ {\bf 70}, 1984.
\bibitem{} Maslov, D. L., C. Barnes, and G. Kirczenow, 1993b, Phys.\ Rev.\ B
{\bf 48}, 2543.
\bibitem{} Maxwell, J. C., 1891, {\em A Treatise on Electricity and
Magnetism\/} (Dover, New York).
\bibitem{} McCann, E., and I. V. Lerner, 1996, J. Phys.\ Condens.\ Matter {\bf
8}, 6719.
\bibitem{} Mehta, M. L., 1991, {\em Random Matrices\/} (Academic, New York).
\bibitem{} Mehta, M. L., and A. Pandey, 1983, J. Phys.\ A {\bf 16}, L601.
\bibitem{} Meir, Y., and O. Entin-Wohlman, 1993, Phys.\ Rev.\ Lett.\ {\bf 70},
1988.
\bibitem{} Meir, Y., N. S. Wingreen, O. Entin-Wohlman, and B. L. Altshuler,
1991, Phys.\ Rev.\ Lett.\ {\bf 66}, 1517.
\bibitem{} Meir, Y., N. S. Wingreen, and P. A. Lee, 1991, Phys.\ Rev.\ Lett.\
{\bf 66}, 3048.
\bibitem{} Meirav, U., and E. B. Foxman, 1995, Semicond.\ Sci.\ Technol.\ {\bf
10}, 255.
\bibitem{} Mello, P. A., 1986, J. Math.\ Phys.\ {\bf 27}, 2876.
\bibitem{} Mello, P. A., 1988, Phys.\ Rev.\ Lett.\ {\bf 60}, 1089.
\bibitem{} Mello, P. A., 1990, J. Phys.\ A {\bf 23}, 4061.
\bibitem{} Mello, P. A., 1995, in {\em Mesoscopic Quantum Physics}, edited by
E. Akkermans, G. Montambaux, J.-L. Pichard, and J. Zinn-Justin (North-Holland,
Amsterdam): p.\ 435.
\bibitem{} Mello, P. A., E. Akkermans, and B. Shapiro, 1988, Phys.\ Rev.\
Lett.\ {\bf 61}, 459.
\bibitem{} Mello, P. A., P. Pereyra, and N. Kumar, 1988, Ann.\ Phys.\ (N.Y.)
{\bf 181}, 290.
\bibitem{} Mello, P. A., P. Pereyra, and T. H. Seligman, 1985, Ann.\ Phys.\
(N.Y.) {\bf 161}, 254.
\bibitem{} Mello, P. A., and J.-L. Pichard, 1989, Phys.\ Rev.\ B {\bf 40},
5276.
\bibitem{} Mello, P. A., and J.-L. Pichard, 1991, J. Phys.\ I France {\bf 1},
493.
\bibitem{} Mello, P. A., and B. Shapiro, 1988, Phys.\ Rev.\ B {\bf 37}, 5860.
\bibitem{} Mello, P. A., and A. D. Stone, 1991, Phys.\ Rev.\ B {\bf 44}, 3559.
\bibitem{} Mello, P. A., and S. Tomsovic, 1991, Phys.\ Rev.\ Lett.\ {\bf 67},
342.
\bibitem{} Mello, P. A., and S. Tomsovic, 1992, Phys.\ Rev.\ B {\bf 46}, 15963.
\bibitem{} Mel'nikov, V. I., 1981, Fiz.\ Tverd.\ Tela {\bf 23}, 782 [Sov.\
Phys.\ Solid State {\bf 23}, 444].
\bibitem{} Melsen, J. A., and C. W. J. Beenakker, 1994, Physica B {\bf 203},
219.
\bibitem{} Melsen, J. A., and C. W. J. Beenakker, 1995, Phys.\ Rev.\ B {\bf
51}, 14483.
\bibitem{} Melsen, J. A., P. W. Brouwer, K. M. Frahm, and C. W. J. Beenakker,
1996, Europhys.\ Lett.\ {\bf 35}, 7.
\bibitem{} Mikhlin, S. G., 1964, {\em Integral Equations\/} (Pergamon, New
York).
\bibitem{} Millo, O., S. J. Klepper, M. W. Keller, D. E. Prober, S. Xiong, A.
D. Stone, and R. N. Sacks, 1990, Phys.\ Rev.\ Lett.\ {\bf 65}, 1494.
\bibitem{} Mirlin, A. D., 1994, Phys.\ Rev.\ Lett.\ {\bf 72}, 3437.
\bibitem{} Mirlin, A. D., A. M\"{u}ller-Groeling, and M. R. Zirnbauer, 1994,
Ann.\ Phys.\ (N.Y.) {\bf 236}, 325.
\bibitem{} Morita, Y., Y. Hatsugai, and M. Kohmoto, 1995 (Los Alamos preprint
archive, cond-mat/9409110).
\bibitem{} Mucciolo, E. R., V. N. Prigodin, and B. L. Altshuler, 1995, Phys.\
Rev.\ B {\bf 51}, 1714.
\bibitem{} Muttalib, K. A., 1990, Phys.\ Rev.\ Lett.\ {\bf 65}, 745.
\bibitem{} Muttalib, K. A., 1995, J. Phys.\ A {\bf 28}, L159.
\bibitem{} Muttalib, K. A., Y. Chen, M. E. H. Ismail, and V. N. Nicopoulos,
1993, Phys.\ Rev.\ Lett.\ {\bf 71}, 471.
\bibitem{} Muttalib, K. A., J.-L. Pichard, and A. D. Stone, 1987, Phys.\ Rev.\
Lett.\ {\bf 59}, 2475.
\bibitem{} Muzykantski\u{\i}, B. A., and D. E. Khmel'nitski\u{\i}, 1994,
Physica B {\bf 203}, 233.
\bibitem{} Muzykantski\u{\i}, B. A., and D. E. Khmel'nitski\u{\i}, 1995, Pis'ma
Zh.\ Eksp.\ Teor.\ Fiz.\ {\bf 62}, 68 [JETP Lett.\ {\bf 62}, 76].
\bibitem{} Nagaev, K. E., 1992, Phys.\ Lett.\ A {\bf 169}, 103.
\bibitem{} Nagaev, K. E., 1995, Phys.\ Rev.\ B {\bf 52}, 4740.
\bibitem{} Nazarov, Yu.\ V., 1994a, Phys.\ Rev.\ Lett.\ {\bf 73}, 134.
\bibitem{} Nazarov, Yu.\ V., 1994b, Phys.\ Rev.\ Lett.\ {\bf 73}, 1420.
\bibitem{} Nazarov, Yu.\ V., 1995a, in {\em Quantum Dynamics of Submicron
Structures}, edited by H. A. Cerdeira, B. Kramer, and G. Sch\"{o}n, NATO ASI
Series E291 (Kluwer, Dordrecht): p.\ 687.
\bibitem{} Nazarov, Yu.\ V., 1995b, Phys.\ Rev.\ B {\bf 52}, 4720.
\bibitem{} Nazarov, Yu.\ V., 1996, Phys.\ Rev.\ Lett.\ {\bf 76}, 2129.
\bibitem{} Nazarov, Yu.\ V., and T. H. Stoof, 1996, Phys.\ Rev.\ Lett.\ {\bf
76}, 823.
\bibitem{} Ng, T. K., and P. A. Lee, 1988, Phys.\ Rev.\ Lett.\ {\bf 61}, 1768.
\bibitem{} Nguyen, C., H. Kroemer, and E. L. Hu, 1992, Phys.\ Rev.\ Lett.\ {\bf
69}, 2847.
\bibitem{} Nguyen, V. I., B. Z. Spivak, and B. I. Shklovski\u{\i}, 1985, Pis'ma
Zh.\ Eksp.\ Teor.\ Fiz.\ {\bf 41}, 35 [JETP Lett.\ {\bf 41}, 42].
\bibitem{} Nieuwenhuizen, Th.\ M., and J. M. Luck, 1993, Phys.\ Rev.\ E {\bf
48}, 569.
\bibitem{} Nishioka, H., and H. A. Weidenm\"{u}ller, 1985, Phys.\ Lett.\ B {\bf
157}, 101.
\bibitem{} Oseledec, V. I., 1968, Trans.\ Mosc.\ Math.\ Soc.\ {\bf 19}, 197.
\bibitem{} Paasschens, J. C. J., T. Sh.\ Misirpashaev, and C. W. J. Beenakker,
1996, Phys.\ Rev.\ B {\bf 54}, 11887.
\bibitem{} Pandey, A., and M. L. Mehta, 1983, Commun.\ Math.\ Phys.\ {\bf 87},
449.
\bibitem{} Pandey, A., and P. Shukla, 1991, J. Phys.\ A {\bf 24}, 3907.
\bibitem{} Papanicolaou, G. C., 1971, SIAM J. Appl.\ Math.\ {\bf 21}, 13.
\bibitem{} Pastur, L. A., 1992, Lett.\ Math.\ Phys.\ {\bf 25}, 259.
\bibitem{} Pendry, J. B., 1994, Adv.\ Phys.\ {\bf 43}, 461.
\bibitem{} Pendry, J. B., A. MacKinnon, and P. J. Roberts, 1992, Proc.\ R.
Soc.\ London A {\bf 437}, 67.
\bibitem{} Pereyra, P., and P. A. Mello, 1983, J. Phys.\ A {\bf 16}, 237.
\bibitem{} Petrashov, V. T., V. N. Antonov, P. Delsing, and T. Claeson, 1995,
Phys.\ Rev.\ Lett.\ {\bf 74}, 5268.
\bibitem{} Pichard, J.-L., 1984, Ph.D. Thesis (University of Paris at Orsay,
unpublished).
\bibitem{} Pichard, J.-L., 1991, in {\em Quantum Coherence in Mesoscopic
Systems}, edited by B. Kramer, NATO ASI Series B254 (Plenum, New York): p.\
369.
\bibitem{} Pichard, J.-L., and G. Andr\'{e}, 1986, Europhys.\ Lett.\ {\bf 2},
477.
\bibitem{} Pichard, J.-L., M. Sanquer, K. Slevin, and Ph.\ Debray, 1990, Phys.\
Rev.\ Lett.\ {\bf 65}, 1812.
\bibitem{} Pichard, J.-L., and G. Sarma, 1981, J. Phys.\ C {\bf 14}, L127.
\bibitem{} Pichard, J.-L., N. Zanon, Y. Imry, and A. D. Stone, 1990, J. Phys.\
France {\bf 51}, 587.
\bibitem{} Pluha\u{r}, Z., H. A. Weidenm\"{u}ller, J. A. Zuk, and C. H.
Lewenkopf, 1994, Phys.\ Rev.\ Lett.\ {\bf 73}, 2115.
\bibitem{} Pluha\u{r}, Z., H. A. Weidenm\"{u}ller, J. A. Zuk, C. H. Lewenkopf,
and F. J. Wegner, 1995, Ann.\ Phys.\ (N.Y.) {\bf 243}, 1.
\bibitem{} Poilblanc, D., T. Ziman, J. Bellissard, F. Mila, and G. Montambaux,
1993, Europhys.\ Lett.\ {\bf 22}, 537.
\bibitem{} Poirier, W., D. Mailly, and M. Sanquer, 1996, in {\em Correlated
Fermions and Transport in Mesoscopic Systems}, edited by Th.\ Martin, G.
Montambaux, and J. Tr\^{a}n Thanh V\^{a}n (Editions Fronti\`eres,
Gif-sur-Yvette).
\bibitem{} Politzer, H. D., 1989, Phys.\ Rev.\ B {\bf 40}, 11917.
\bibitem{} Porter, C. E., 1965, {\em Statistical Theories of Spectra:
Fluctuations\/} (Academic, New York).
\bibitem{} Pothier, H., S. Gu\'{e}ron, D. Esteve, and M. H. Devoret, 1994,
Phys.\ Rev.\ Lett.\ {\bf 73}, 2488.
\bibitem{} Pradhan, P., and N. Kumar, 1994, Phys.\ Rev.\ B {\bf 50}, 9644.
\bibitem{} Prange, R. E., and S. M. Girvin, editors, 1990, {\em The Quantum
Hall Effect\/} (Springer, New York).
\bibitem{} Prigodin, V. N., B. L. Altshuler, K. B. Efetov, and S. Iida, 1994,
Phys.\ Rev.\ Lett.\ {\bf 72}, 546.
\bibitem{} Prigodin, V. N., K. B. Efetov, and S. Iida, 1993, Phys.\ Rev.\
Lett.\ {\bf 71}, 1230.
\bibitem{} Prigodin, V. N., K. B. Efetov, and S. Iida, 1995, Phys.\ Rev.\ B
{\bf 51}, 17223.
\bibitem{} Raikh, M. E., and I. M. Ruzin, 1991, in {\em Mesoscopic Phenomena in
Solids}, edited by B. L. Altshuler, P. A. Lee, and R. A. Webb (North-Holland,
Amsterdam): p.\ 315.
\bibitem{} Rammal, R., and B. Doucot, 1987, J.\ Phys.\ France {\bf 48}, 509.
\bibitem{} Rau, J., 1995, Phys.\ Rev.\ B {\bf 51}, 7734.
\bibitem{} Rejaei, B., 1996, Phys.\ Rev.\ B {\bf 53}, 13235.
\bibitem{} Samuel, S., 1980, J. Math.\ Phys.\ {\bf 21}, 2695.
\bibitem{} Sharvin, Yu.\ V., 1965, Zh.\ Eksp.\ Teor.\ Fiz.\ {\bf 48}, 984
[Sov.\ Phys.\ JETP {\bf 21}, 655].
\bibitem{} Shekhter, R. I., 1972, Zh.\ Eksp.\ Teor.\ Fiz.\ {\bf 63}, 1410
[Sov.\ Phys.\ JETP {\bf 36}, 747 (1973)].
\bibitem{} Shelankov, A. L., 1980, Pis'ma Zh.\ Eksp.\ Teor.\ Fiz.\ {\bf 32},
122 [JETP Lett.\ {\bf 32}, 111].
\bibitem{} Shelankov, A. L., 1984, Fiz.\ Tverd.\ Tela {\bf 26}, 1615 [Sov.\
Phys.\ Solid State {\bf 26}, 981].
\bibitem{} Shepelyansky, D. L., 1994, Phys.\ Rev.\ Lett.\ {\bf 73}, 2607.
\bibitem{} Shklovski\u{\i}, B. I., and B. Z. Spivak, 1990, in {\em Hopping
Conduction in Semiconductors}, edited by M. Pollak and B. I. Shklovski\u{\i}
(North-Holland, Amsterdam).
\bibitem{} Simons, B. D., and B. L. Altshuler, 1993, Phys.\ Rev.\ Lett.\ {\bf
70}, 4063.
\bibitem{} Sivan, U., Y. Imry, and A. G. Aronov, 1994, Europhys.\ Lett.\ {\bf
28}, 115.
\bibitem{} Sivan, U., F. P. Milliken, K. Milkove, S. Rishton, Y. Lee, J. M.
Hong, V. Boegli, D. Kern, and M. DeFranza, 1994, Europhys.\ Lett.\ {\bf 25},
605.
\bibitem{} Slevin, K., and T. Nagao, 1993, Phys.\ Rev.\ Lett.\ {\bf 70}, 635.
\bibitem{} Slevin, K., and T. Nagao, 1994, Phys.\ Rev.\ B {\bf 50}, 2380.
\bibitem{} Slevin, K., J.-L. Pichard, and P. A. Mello, 1991, Europhys.\ Lett.\
{\bf 16}, 649.
\bibitem{} Slevin, K., J.-L. Pichard, and P. A. Mello, 1996, J. Phys.\ I France
{\bf 6}, 529.
\bibitem{} Slevin, K., J.-L. Pichard, and K. A. Muttalib, 1993, J.\ Phys.\ I
France {\bf 3}, 1387.
\bibitem{} Smilansky, U., 1990, in {\em Chaos and Quantum Physics}, edited by
M.-J. Giannoni, A. Voros, and J. Zinn-Justin (North-Holland, Amsterdam): p.\
371.
\bibitem{} Spivak, B. Z., and D. E. Khmel'nitski\u{\i}, 1982, Pis'ma Zh.\
Eksp.\ Teor.\ Fiz.\ {\bf 35}, 334 [JETP Lett.\ {\bf 35}, 412].
\bibitem{} Steinbach, A. H., J. M. Martinis, and M. H. Devoret, 1995, Bull.\
Am.\ Phys.\ Soc.\ {\bf 40}, 400.
\bibitem{} Steinbach, A. H., J. M. Martinis, and M. H. Devoret, 1996, Phys.\
Rev.\ Lett.\ {\bf 76}, 3806.
\bibitem{} Stone, A. D., P. A. Mello, K. A. Muttalib, and J.-L. Pichard, 1991,
in {\em Mesoscopic Phenomena in Solids}, edited by B. L. Altshuler, P. A. Lee,
and R. A. Webb (North-Holland, Amsterdam): p.\ 369.
\bibitem{} Stone, A. D., and A. Szafer, 1988, IBM J. Res.\ Dev.\ {\bf 32}, 384.
\bibitem{} Stoof, T. H., and Yu.\ V. Nazarov, 1996a, Phys.\ Rev.\ B {\bf 53},
1050.
\bibitem{} Stoof, T. H., and Yu.\ V. Nazarov, 1996b, Phys.\ Rev.\ B {\bf 54},
772.
\bibitem{} Sutherland, B., 1971, J. Math.\ Phys.\ {\bf 12}, 246.
\bibitem{} Sutherland, B., 1972, Phys.\ Rev.\ A {\bf 5}, 1372.
\bibitem{} Szafer, A., and B. L. Altshuler, 1993, Phys.\ Rev.\ Lett.\ {\bf 70},
587.
\bibitem{} Szafer, A., and A. D. Stone, 1989, Phys.\ Rev.\ Lett.\ {\bf 62},
300.
\bibitem{} Takane, Y., and H. Ebisawa, 1991, J.\ Phys.\ Soc.\ Japan {\bf 60},
3130.
\bibitem{} Takane, Y., and H. Ebisawa, 1992a, J. Phys.\ Soc.\ Jpn.\ {\bf 61},
1685.
\bibitem{} Takane, Y., and H. Ebisawa, 1992b, J. Phys.\ Soc.\ Jpn.\ {\bf 61},
2858.
\bibitem{} Takane, Y., and H. Ebisawa, 1993, J. Phys.\ Soc.\ Jpn.\ {\bf 62},
1844.
\bibitem{} Takane, Y., and H. Otani, 1994, J.\ Phys.\ Soc.\ Japan {\bf 63},
3361.
\bibitem{} Takayanagi, H., E. Toyoda, and T. Akazaki, 1996 (unpublished).
\bibitem{} Tartakovski, A. V., 1995, Phys.\ Rev.\ B {\bf 52}, 2704.
\bibitem{} Thouless, D. J., 1977, Phys.\ Rev.\ Lett.\ {\bf 39}, 1167.
\bibitem{} Van Albada, M. P., B. A. van Tiggelen, A. Lagendijk, and A. Tip,
1991, Phys.\ Rev.\ Lett.\ {\bf 66}, 3132.
\bibitem{} Van Houten, H, C. W. J. Beenakker, and A. A. M. Staring, 1992,
in: {\em Single Charge Tunneling}, edited by H. Grabert and M. H. Devoret
(Plenum, New York): p.\ 167.
\bibitem{} Van Kampen, N. G., 1981, {\em Stochastic Processes in Physics and
Chemistry\/} (North-Holland, Amsterdam).
\bibitem{} Van Langen, S. A., P. W. Brouwer, and C. W. J. Beenakker, 1996,
Phys.\ Rev.\ E {\bf 53}, 1344.
\bibitem{} Van Son, P. C., H. van Kempen, and P. Wyder, 1987, Phys.\ Rev.\
Lett.\ {\bf 59}, 2226.
\bibitem{} Van Son, P. C., H. van Kempen, and P. Wyder, 1988, J. Phys.\ F {\bf
18}, 2211.
\bibitem{} Van Wees, B. J., P. de Vries, P. Magn\'{e}e, and T. M. Klapwijk,
1992, Phys.\ Rev.\ Lett.\ {\bf 69}, 510.
\bibitem{} Verbaarschot, J. J. M., H. A. Weidenm\"{u}ller, and M. R. Zirnbauer,
1985, Phys.\ Rep.\ {\bf 129}, 367.
\bibitem{} Volkov, A. F., 1994, Phys.\ Lett.\ A {\bf 187}, 404.
\bibitem{} Volkov, A. F., N. Allsopp, and C. J. Lambert, 1996, J. Phys.\
Condens.\ Matter {\bf 8}, L45.
\bibitem{} Volkov, A. F., and A. V. Za\u{\i}tsev, 1996, Phys.\ Rev.\ B {\bf
53}, 9267.
\bibitem{} Volkov, A. F., A. V. Za\u{\i}tsev, and T. M. Klapwijk, 1993, Physica
C {\bf 210}, 21.
\bibitem{} Vollhardt, D., and P. W\"{o}lfle, 1992, in {\em Electronic Phase
Transitions}, edited by W. Hanke and Yu.\ V. Kopaev (North-Holland, Amsterdam):
p.\ 1.
\bibitem{} Wang, Z., B. Jovanovi\'{c}, and D.-H. Lee, 1996, Phys.\ Rev.\ Lett.\
{\bf 77}, 4426.
\bibitem{} Washburn, S., and R. A. Webb, 1986, Adv.\ Phys.\ {\bf 35}, 375.
\bibitem{} Wegner, F., 1979, Z. Phys.\ B {\bf 35}, 207.
\bibitem{} Weidenm\"{u}ller, H. A., 1990, Physica A {\bf 167}, 28.
\bibitem{} Westervelt, R. M., 1996, in {\em Nano-Science and Technology},
edited by G. Timp (American Institute of Physics, New York, in press).
\bibitem{} Wigner, E. P., 1957, in {\em Proc.\ Canadian Mathematical
Congress\/} (Univ.\ of Toronto Press, Toronto): p.\ 174; reprinted in Porter
(1965): p.\ 188.
\bibitem{} Wigner, E. P., 1967, SIAM Rev.\ {\bf 9}, 1.
\bibitem{} Xiong,  P., G. Xiao, and R. B. Laibowitz, 1993, Phys.\ Rev.\ Lett.\
{\bf 71}, 1907.
\bibitem{} Yang, X., H. Ishio, and J. Burgd\"{o}rfer, 1995, Phys.\ Rev.\ B {\bf
52}, 8219.
\bibitem{} Yip, S., 1995, Phys.\ Rev.\ B {\bf 52}, 15504.
\bibitem{} Yurke, Y., and G. P. Kochanski, 1990, Phys.\ Rev.\ B {\bf 41}, 8184.
\bibitem{} Za\u{\i}tsev, A. V., 1980, Zh.\ Eksp.\ Teor.\ Fiz.\ {\bf 78}, 221;
{\bf 79}, 2016(E) [Sov.\ Phys.\ JETP {\bf 51}, 111; {\bf 52}, 1018(E)].
\bibitem{} Za\u{\i}tsev, A. V., 1984, Zh.\ Eksp.\ Teor.\ Fiz.\ {\bf 86}, 1742
[Sov.\ Phys.\ JETP {\bf 59}, 1015].
\bibitem{Zai94} Za\u{\i}tsev, A. V., 1994, Phys.\ Lett.\ A {\bf 194}, 315.
\bibitem{} Zanon, N., and J.-L. Pichard, 1988, J. Phys. France {\bf 49}, 907.
\bibitem{} Zhang, Z. Q., 1995, Phys.\ Rev.\ B, {\bf 52}, 7960.
\bibitem{} Zhou, F., and B. Spivak, 1996 (Los Alamos preprint archive,
cond-mat/9604185).
\bibitem{} Zhou, F., B. Spivak, and A. Zyuzin, 1995, Phys.\ Rev.\ B {\bf 52},
4467.
\bibitem{} Zirnbauer, M. R., 1992, Phys.\ Rev.\ Lett.\ {\bf 69}, 1584.
\bibitem{} Zirnbauer, M. R., 1993, Nucl.\ Phys.\ A {\bf 560}, 95.

\end{references}
\end{document}